\documentclass[twoside,12pt]{article}
\usepackage{amsmath,amsfonts,amssymb,bm}
\usepackage{color}
\usepackage{dsfont}
\usepackage{graphicx}
\usepackage{pzccal}
\usepackage[comma,square,sort,compress,numbers]{natbib}

\usepackage{amsmath}
\usepackage{amsfonts,amssymb,bm}
\usepackage{color}
\usepackage{graphicx}
\usepackage{pzccal}

\usepackage{mathtools}
\usepackage{amssymb,amsfonts}
\usepackage{bm}
\usepackage{slashed}

\usepackage[svgnames]{xcolor}
\usepackage[english]{babel}
\usepackage{blindtext}
\usepackage{microtype}
\usepackage{tikz}
\usepackage{dcolumn}
\usepackage{multirow}
\usepackage[]{units}

\usepackage{graphicx}
\usepackage[caption=false]{subfig}

\usepackage{ifpdf}

\graphicspath{{.}{figures/}}

\def\a{\alpha}

\def\g{\gamma}

\def\ve{\varepsilon}

\def\s{\sigma}

\def\D{\Delta}

\def\hs{\hspace}

\def\ol{\overline}

\def\lf{\left}
\def\rg{\right}
\def\la{\langle}
\def\ra{\rangle}

\def\lra{\longrightarrow}

\definecolor{darkgreen}{rgb}{0,0.5,0}
\definecolor{purple}{rgb}{0.5,0,0.5}
\definecolor{nblue}{rgb}{0.0,0.0,0.50}
\definecolor{scarlet}{rgb}{1.0,0.2,0}

\newcommand{\lsim}{\mathrel{\rlap{\lower4pt\hbox{\hskip0pt$\sim$}}
\raise1pt\hbox{$<$}}}           
\newcommand{\gsim}{\mathrel{\rlap{\lower4pt\hbox{\hskip0pt$\sim$}}
\raise1pt\hbox{$>$}}}           

\numberwithin{equation}{section}
\numberwithin{figure}{section}
\numberwithin{table}{section}

\begin{document}

\title{Explanation and Prediction of Observables \\using Continuum Strong QCD}

\author{Ian C. Clo\"et and Craig D. Roberts\\[2ex]
Physics Division, Argonne National Laboratory, \\
Argonne, Illinois 60439, USA
}

\maketitle

\begin{abstract}
%
The last five years have brought considerable progress in the study of the bound-state problem in continuum quantum field theory.  We highlight a subset of that progress; viz., that made within the context of Dyson Schwinger equation analyses of cold, sparse hadrons.  Our focus is primarily on advances in the reliable computation, explanation and prediction of quantities that are truly measurable; but we also review aspects of a new paradigm that has condensates contained within hadrons, and explain that the asymptotic form of parton distribution amplitudes and functions are practically unreachable with terrestrial facilities.  Given the pace of expansion in experiment and improvement in theory, it appears possible that the next five years will bring profound growth in our store of knowledge about hadrons and nuclei.

\begin{center}
\parbox{0.95\linewidth}{\emph{Keywords}:
confinement,
dynamical chiral symmetry breaking,
Dyson-Schwinger equations,
EMC effect,
hadron spectrum,
hadron elastic and transition form factors,
in-hadron condensates,
Nambu--Jona-Lasinio model,
NuTeV anomaly,
parton distribution amplitudes (PDAs) and functions (PDFs),
transverse momentum dependent PDFs}
\end{center}
\end{abstract}

\tableofcontents


\section{Introduction}
Hadron physics is an international research endeavour of remarkable scope.  Indeed, before the end of this decade the field will be operating a host of upgraded or new facilities and detectors.
An illustrative list may readily be compiled: Beijing's electron-positron collider; FAIR, the Facility for Antiproton and Ion Research, under construction near Darmstadt, Germany;
J-PARC, the Japan Proton Accelerator Research Complex, 150km NE of Tokyo;
the ALICE and COMPASS detectors at CERN;
the Nuclotron based Ion Collider fAcility (NICA), under development in Dubna, Russia;
and in the USA, both RHIC (Relativistic Heavy Ion Collider) at Brookhaven National Laboratory, which focuses primarily on the strong-interaction phase transition, physics just 10$\mu$s after the Big Bang,
and the Thomas Jefferson National Accelerator Facility (JLab), exploring the nature of cold hadronic matter.  Regarding the upgrade of JLab \cite{Dudek:2012vr}, commissioning of Hall-A and Hall-D will take place in 2014, a process that is itself expected to yield new physics results, and project completion is expected in 2017.
In addition to these existing facilities and those under construction, excitement is also attached to the discovery potential of proton-nucleus collisions at the Large Hadron Collider (LHC) \cite{Salgado:2011wc} and new machines, such as an electron ion collider (EIC) \cite{Accardi:2012hwp}, in China or the USA.

Experiments at these facilities will explore a diverse range of scientific challenges \cite{Dudek:2012vr,Salgado:2011wc,Accardi:2012hwp}:
\begin{itemize}
\item They will hunt for exotic hadrons -- states whose quantum numbers cannot be supported by quantum mechanical quark-antiquark systems; and hybrid hadrons -- states with quark-model quantum numbers but a non-quark-model decay pattern.  Both systems are suspected to possess valence-gluon content, which translates into a statement that they are expected to have a large overlap with interpolating fields that explicitly contain gluon fields.  The discovery of such states would force a dramatic reassessment of the idea that it is possible to draw a distinction between matter fields and force fields, a notion which has existed since the time of Maxwell and before, because such exotic and hybrid matter can conceivably be composed solely of force fields.

\item The facilities will also provide new data on nucleon elastic and transition form factors, which will provide many opportunities for theory.  For example, viewed appropriately, such data can assist in charting the pointwise behaviour at infrared momenta of QCD's running coupling and dressed-masses; it will enable theory to reveal those correlations that are key to nucleon structure; and, indeed, assist the community to expose the facts and fallacies in modern descriptions of nucleon structure.

\item Precision experimental study of the valence region within hadrons, and theoretical computation of distribution functions and distribution amplitudes are also anticipated.  In this connection, computation is critical.  Without it, no amount of data will reveal anything about the theory underlying the phenomena of strong interaction physics.

\item We should also see significant experimental and theoretical progress on questions at the very heart of nuclear physics; namely, how do nuclei emerge from QCD and are there ascertainable remnants of this emergence in nuclear structure?  The exploration of such themes brings one into contact with a host of problems.  Preeminent amongst them is the EMC effect.  This dramatic modification of the structure functions of bound nucleons was discovered thirty years ago \cite{Aubert:1983xm}; and yet there is still no widely accepted explanation.  It is certain, however, that revelation of the EMC effect destroyed a particle physics paradigm regarding QCD and nuclear structure: it made plain that valence quarks in a nucleus carry less momentum than valence quarks in a nucleon.   No explanation of the emergence of nuclei from QCD will be complete if it does not simultaneously solve the puzzle posed by the EMC effect.

\item Finally, the modernised and new facilities will continue to search for and exploit opportunities to use precise measurements of strong-interaction phenomena as a means by which to search for physics beyond the Standard Model.  For example, with precise measurements of hadron electroweak properties one may place hard lower-bounds on the scale at which new physics might begin to have an impact.  In fact, experiment and theory constraints on the strangeness content of the nucleon already place tight bounds on dark-matter--hadron cross-sections \cite{Giedt:2009mr}; and high-luminosity lepton-beam hadron-physics machines can scan a plausible mass region for dark photons, which are a possible explanation for the muon ``$g-2$ anomaly'' \cite{Bennett:2006fi}.  Such experiments are also necessary in order to resolve perceived discrepancies between contemporary data and the Standard Model, as is the case with the NuTeV result \cite{Zeller:2001hh} for the electroweak mixing angle.
\end{itemize}

The anticipated wealth of new experimental data will pose many challenges to which theory will need to respond.  The response must be fluid, it must rapidly provide an intuitive understanding, and it must define a path toward answers and new discoveries.  In this milieu, notwithstanding its steady progress toward results with input parameters that approximate the real world, the numerical simulation of lattice-regularised QCD will not suffice.  Approaches formulated in the continuum, inspired by, based upon, or connected directly with QCD are necessary.  In our view, indeed, attention must focus upon the continuum bound-state problem in quantum field theory because this approach offers the possibility of posing and answering the questions at the heart of hadron physics: 
\begin{itemize}
\item what is confinement; 
\item what is dynamical chiral symmetry breaking; 
\item and how are they related?  
\end{itemize}
It appears inconceivable to us now that two phenomena, so critical in the Standard Model and tied to the dynamical generation of a single mass-scale, can have different origins and fates.

The continuum bound-state problem has long been studied in quantum mechanics, with constituent-quark and potential models having a distinguished history, see, e.g., Refs.\,\cite{Capstick:2000qj,Aznauryan:2012ba,Crede:2013kia} and references therein and thereto.  Typically, however, such approaches are unable to unify the physics of light-quark mesons and baryons \cite{Swanson:2012zz}.  Approaches connected with QCD sum rules \cite{Aznauryan:2012ba,Colangelo:2000dp} do not have this difficulty but they are plagued with what can rapidly become an overwhelming number of parameters: the so-called vacuum condensates, whose constraint benefits from comparison with lattice-QCD or fitting to experiment.  More recently, a combination of ideas from the light-front formulation of quantum field theory and models derived from the notion of gauge-gravity duality has vigorously been pursued \cite{Aznauryan:2012ba,Broadsky:2012rw,Brodsky:2013ar}.  However, connecting the parameters deployed in this approach with those of QCD poses a problem.  

A widely used alternative to these methods is provided by the Dyson-Schwinger equations (DSEs).  This approach may quite properly be considered to encompass a diverse range of Hamiltonian and Lagrangian based methods, with varying degrees of connection to QCD.  At the highest level, the DSEs enable veracious relationships to be drawn between empirical observables and real features of QCD.  At a different level, they provide the means by which to anticipate and elucidate the possible quantitative impacts on observables of properties that may qualitatively be associated with QCD.  In both guises the DSEs provide a powerful tool, which has been employed with marked success to connect QCD with predictions of hadron observables.  This will be emphasised and exemplified in this review, which describes selected recent progress in the study of hadrons and nuclei.  It complements other such efforts \cite{Roberts:1994dr,Roberts:2000aa,Maris:2003vk,Pennington:2005be,Holl:2006ni,%
Fischer:2006ub,Roberts:2007jh,Roberts:2007ji,Holt:2010vj,Swanson:2010pw,Chang:2011vu,%
Boucaud:2011ug,Roberts:2012sv,Bashir:2012fs}.

\section{Emergence of Scale}
\subsection{Conformal anomalies}
The action that defines the theory of massless (chiral) quantum chromodynamics (QCD) is conformally invariant.  Associated with this feature are a dilatation current, which is conserved in a classical (unquantised) treatment of the theory, and an array of related Ward-Green-Takahashi (WGT) identities \cite{Ward:1950xp,Green:1953te,Takahashi:1957xn,Takahashi:1985yz} between the theory's Schwinger functions.  Were these identities to remain valid in a complete treatment of the Standard Model, then the natural hadronic mass scale would be zero and all Schwinger functions would be homogeneous, with naive scaling degree.  This is plainly not the case empirically.

The conundrum is resolved by noting that classical WGT identities are derived without accounting for the effect of regularisation and renormalisation in four-dimensional quantum field theory.  This procedure leads to scale anomalies in the WGT identities originating with the dilatation current \cite{Collins:1976yq,Nielsen:1977sy,tarrach}.  Therefore, a dynamically generated mass-scale, typically denoted $\Lambda_{\rm QCD}$, is connected with \emph{quantum} chromodynamics.  The value of $\Lambda_{\rm QCD}$ must be determined empirically.

It has long been recognised that the quantum breaking of classical QCD's conformal invariance has far-reaching consequences in the analysis of high-energy processes  \cite{Brodsky:1980ny,Braun:2003rp}.  On the other hand, whilst these and related observations are instructive in principle, and motivate a class of contemporary models (see, e.g., Refs.\,\cite{Choi:2008yj,Grigoryan:2008cc,Chabysheva:2012fe,Brodsky:2013npa,Brodsky:2013ar}), they provide little in the way of explanation for the vast array of nonperturbative strong interaction phenomena.  Knowing that scale invariance is broken by QCD dynamics is not the same as explaining how a proton, constituted from nearly massless current-quarks, itself acquires a mass $m_p \sim 1\,$GeV which is contained within a confinement domain whose radius is $r_{\rm c} \sim 1/\sigma_{\rm c}$, with $\sigma_{\rm c} \sim 0.25\,$GeV$\,\sim \Lambda_{\rm QCD}$.  Such questions can only be answered within a framework that enables the computation of bound-state properties from quantised chromodynamics.  This is highlighted further by observing that quantum electrodynamics also possesses a scale anomaly \cite{Adler:1976zt} but lies within a class of theories whose dynamical content and predictions are completely different.

Two \emph{a priori} independent, emergent mass scales are identified in the preceding passage; namely, the scale associated with QCD's confinement length and that associated with dynamical chiral symmetry breaking (DCSB), which is responsible for constituent-like behaviour of low-momentum dressed-quarks \cite{Roberts:2007ji}.  Following the introductory discussion, it appears probable that these scales are both intimately connected and originate in the same dynamics that explain the difference between the scale anomalies in QCD and QED.
However, this is not proven and the questions of whether confinement can exist without DCSB in QCD, or \emph{vice-versa}, remain open.  This fact is emphasised by the ongoing debate about coincidence of the deconfinement and chiral symmetry restoring transitions of chiral QCD in-medium 
(see, e.g., Refs.\,\cite{Braun:2011fw,Bashir:2012fs,Petreczky:2012rq,Fischer:2012vc}). Of these two mass scales, that associated with confinement is the most problematic.

\subsection{Confinement}
\label{sec:confinement}
There are two important aspects to the question of confinement.  One is folkloric and summarised well in the conceptual design report for Hall-D at JLab \cite{CDRHallD}:\\[1ex]
%
%
\centerline{\parbox{0.9\linewidth}{``[\,\ldots] the color field lines between a quark and an anti-quark [\,\ldots] form flux tubes.  A unit area placed midway between the quarks and perpendicular to the line connecting them intercepts a constant number of field lines, independent of the distance between the quarks.  This leads to a constant force between the quarks -- and a large force at that, equal to about 16 metric tons.  The potential associated with this constant force is linear and grows with increasing distance.  It takes infinite energy to separate the quarks to infinity and thus, qualitatively at least, this accounts for confinement.''}}

\medskip

\hspace*{-\parindent}The crippling flaw in this argument is that 16 metric tons of force produces a lot of pions; and including that effect completely destroys the picture.

This brings us to the other aspect of confinement, which is fact, based firmly in quantum field theory and a real world that contains quarks with light current-quark masses.  This is distinct from the artificial universe of pure-gauge QCD without dynamical quarks, studies of which often tend merely to focus on achieving an area law for a Wilson loop and hence are irrelevant to the question of light-quark confinement.  The point is that the potential between infinitely-heavy quarks measured in numerical simulations of quenched lattice-regularised QCD -- the so-called static potential -- is simply \emph{irrelevant} to the question of confinement in a universe in which light quarks are ubiquitous.  In fact, it is a basic feature of QCD that light-particle creation and annihilation effects are essentially nonperturbative and therefore it is impossible in principle to compute a (non light-front) quantum mechanical potential between two light quarks \cite{Bali:2005fu,Chang:2009ae}.  This means that there is no flux tube in a universe with light quarks and consequently that the flux tube is not the correct paradigm for confinement.

Some will object to this statement, citing, in defence of flux tubes and linear potentials, the ``empirical fact of linear Regge trajectories''; i.e., linear, parallel trajectories they perceive in the $(J,M^2)$ and $(n,M^2)$ planes, where $M^2$ represents hadron masses-squared, $J$ is a quantum-mechanical constituent-quark orbital angular momentum and $n$ is a constituent-quark radial quantum number.  To such objections, one may first respond with an observation \cite{Veltmann:2003}:\\[1ex]
\centerline{\parbox{0.9\linewidth}{``Nowadays the Regge trajectories have largely disappeared, not in the least because these higher spin bound states are hard to find experimentally.  At the peak of the Regge fashion (around 1970) theoretical physics produced many papers containing families of Regge trajectories, with the various (hypothetically straight) lines based on one or two points only!''}}

\medskip

\hspace*{-\parindent}That has not changed.  In fact, many modern, theoretically predicted Regge trajectories have no empirical masses on them.  One can also cite a systematic analysis of the hadron spectrum, which shows \cite{Tang:2000tb}: \\[1ex]
\centerline{\parbox{0.9\linewidth}{``\ldots that meson trajectories are non-linear and intersecting.  [and] \ldots that all current meson Regge trajectories models are ruled out by data.''}}

\medskip

\hspace*{-\parindent}This clear and concrete conclusion from a careful analysis of the spectrum has recently acquired additional strength via an independent analysis of the spectrum, using different methods, that arrives at a similar judgement \cite{Masjuan:2012gc}.  Furthermore, analyses of lattice-QCD results contradict predictions of flux tube models on that domain of current-quark mass where proponents expected them to be valid \cite{Dudek:2012ag}.  Hence nowadays, as noted above, there is no evidence that can properly be said to support the flux tube paradigm for confinement.

To explain an alternative, consider the following.  We know with certainty that QCD is an asymptotically free gauge theory \cite{Politzer:2005kc,Wilczek:2005az,Gross:2005kv}.  As such, it is potentially the only known instance of a theory that can rigorously be defined nonperturbatively.\footnote{Asymptotic freedom is a necessary condition.  It is certain that the Abelian gauge theory of quantum electrodynamics (QED) cannot be defined nonperturbatively.  The Landau pole in Abelian gauge theories lies at asymptotically large momenta; and this guarantees incompleteness of the theory at large energy scales.}  QCD is perturbatively renormalisable, with seven distinct renormalisation constants.  It will exist nonperturbatively if these renormalisation constants are computable and produce finite ratios at any two renormalisation scales, including those in the infrared.  In that case, the question of confinement can properly be posed and, perhaps, answered.

\begin{figure}[t]
\centerline{\includegraphics[clip,width=0.66\textwidth]{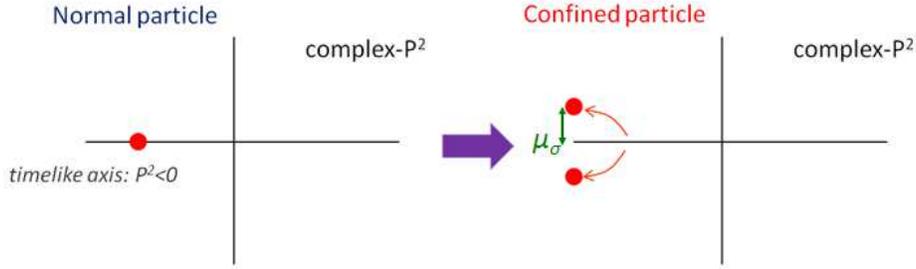}}
\caption{\label{fig:confinement} \emph{Left panel} -- An observable particle is associated with a pole at timelike-$P^2$, which becomes a branch point if, e.g., the particle is dressed by photons.
\emph{Right panel} -- When the dressing interaction is confining, the real-axis mass-pole splits, moving into a pair (or pairs) of complex conjugate singularities,
or is otherwise altered similarly, with the critical outcome being that the associated spectral density is no longer positive definite.
A particle whose propagator is characterised by such a spectral density cannot possess a mass-shell.
A dynamically generated mass-scale, $\mu_\sigma$, is connected with this change in the spectral structure.  In our practical illustration, it is the imaginary part of the smallest magnitude singularity.  The inverse of this scale, $d_\sigma=1/\mu_\sigma$, is a measure of the four-vector distance over which the dressed parton may propagate before losing its identity; i.e., the dressed-parton's fragmentation length.  Typically, $d_\sigma \lesssim 0.5\,$fm \cite{Bhagwat:2002tx}. }
\end{figure}

That answer is plausibly connected with the image depicted in Fig.\,\ref{fig:confinement}; viz., it is possible that the emergent phenomenon of confinement is associated with dramatic, dynamically-driven changes in the analytic structure of QCD's propagators and vertices (QCD's Schwinger functions) \cite{Gribov:1999ui,Munczek:1983dx,Stingl:1983pt,Cahill:1988zi,Krein:1990sf,%
Efimov:1993zg,Dokshitzer:2004ie}.  For example, it can be read from the reconstruction theorem \cite{SW80,GJ81} that the only Schwinger functions which can be associated with expectation values in the Hilbert space of observables; namely, the set of measurable expectation values, are those that satisfy the axiom of reflection positivity.  This is an extremely tight constraint.  It can be shown to require as a necessary condition that the Fourier transform of the momentum-space Schwinger function is a positive-definite function of its Poincar\'e-invariant arguments.  This condition suggests a practical confinement test, which can be used with numerical solutions of the DSEs (see, e.g., Sec.\,III.C of Ref.\,\cite{Hawes:1993ef} and Sec.\,IV of Ref.\,\cite{Maris:1995ns}).  Some implications and uses of reflection positivity are discussed and illustrated in Sec.~2 of Ref.\,\cite{Roberts:2007ji}.  This translation of the confinement problem brings it into a domain that can be addressed via a concerted effort in experiment and theory: theory can identify signatures for such effects in observables and experiment can test the predictions.

\begin{figure}[t]

\centerline{
\includegraphics[clip,width=0.5\textwidth]{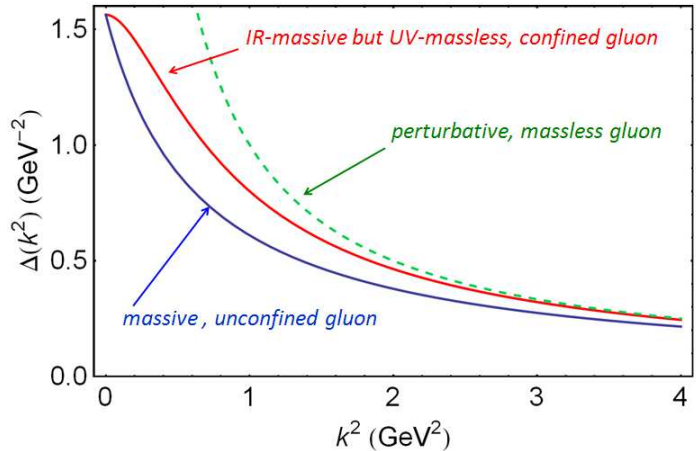}}

\caption{\label{fig:gluonrp}
$\Delta(k^2)$, the function that describes dressing of a Landau-gauge gluon propagator, plotted for three distinct cases.
A bare gluon is described by $\Delta(k^2) = 1/k^2$ (the dashed line), which is plainly convex on $k^2\in (0,\infty)$.  Such a propagator has a representation in terms of a non-negative spectral density.
In some theories, interactions generate a mass in the transverse part of the gauge-boson propagator, so that $\Delta(k^2) = 1/(k^2+M_g^2)$, with $M_g$ constant, which can also be represented in terms of a non-negative spectral density.
In QCD, however, self-interactions generate a momentum-dependent mass for the gluon, which is large at infrared momenta but vanishes in the ultraviolet \protect\cite{Boucaud:2011ug}.  This is illustrated by the curve labelled ``IR-massive but UV-massless.''  With the generation of a mass-\emph{function}, $\Delta(k^2)$ exhibits an inflexion point and hence cannot be expressed in terms of a non-negative spectral density.
}
\end{figure}

Whilst on this subject, it is notable that any 2-point Schwinger function with an inflexion point at $p^2 > 0$ must breach the axiom of reflection positivity, so that a violation of positivity can be determined by inspection of the pointwise behaviour of the Schwinger function in momentum space (Ref.\,\cite{Bashir:2008fk}, Sec.\,IV.B).
Consider then $\Delta(k^2)$, which is the single scalar function that describes the dressing of a Landau-gauge gluon propagator.  A large body of work has focused on exposing the behaviour of $\Delta(k^2)$ in the pure Yang-Mills sector of QCD.  These studies are reviewed in Ref.\,\cite{Boucaud:2011ug}.
A connection with the expression and nature of confinement in the Yang-Mills sector is indicated in Fig.\,\ref{fig:gluonrp}.  The appearance of an inflexion point in the two-point function generated by the gluon's momentum-dependent mass-function is impossible to overlook.  Hence this gluon cannot appear in the Hilbert space of observable states.  (The inflexion point possessed by $M(p^2)$, visible in Fig.\,\ref{gluoncloud}, conveys the same properties on the dressed-quark propagator.)

Numerical simulations of lattice-QCD confirm the appearance of an inflexion point in both the dressed-gluon and -quark propagators (e.g., see Ref.\,\cite{Boucaud:2011ug} and Fig.\,\ref{gluoncloud}).  The signal is clearest for the gluon owing to the greater simplicity of simulations in the pure Yang-Mills sector \cite{Skullerud:2000un,Bonnet:2000kw,Bowman:2004jm,Bowman:2007du,Kamleh:2007ud,Sternbeck:2007ug}.

The possibility that QCD might be rigorously well defined is one of its deepest fascinations.  In that case, irrespective of its connection with the Standard Model and a description of Nature, QCD might stand alone as an archetype -- the only internally consistent quantum field theory which is defined at all energy scales.  (Notably, there is no confirmed breakdown of QCD over an enormous energy range: $0<E<8\,$TeV.)  This is a remarkable possibility, and one with wide-ranging consequences and opportunities.  For example, it means that QCD-like theories provide a viable paradigm for extending the Standard Model to greater scales than those already probed.  Contemporary research in this direction is typified by the notion of extended technicolour \cite{Andersen:2011yj,Sannino:2013wla}, in which electroweak symmetry breaks via a fermion bilinear operator in a strongly-interacting non-Abelian theory; and the Higgs sector of the Standard Model becomes an effective description of a more fundamental fermionic theory, similar to the Ginzburg-Landau theory of superconductivity.

\subsection{Dynamical chiral symmetry breaking}
\label{sec:dcsb}
This is the other fundamental emergent phenomenon in QCD.  Dynamical chiral symmetry breaking (DCSB); namely, the generation of mass \emph{from nothing} is a theoretically established nonperturbative feature of QCD \cite{national2012Nuclear}.  We insist on the term ``dynamical'' as distinct from spontaneous because nothing is added to QCD in order to effect this remarkable outcome.  Instead, simply through quantising the classical chromodynamics of massless gluons and quarks, a large mass-scale is generated.  This phenomenon is the most important mass generating mechanism for visible matter in the Universe, being responsible for approximately 98\% of the proton's mass.

The reality of DCSB means the Standard Model's Higgs mechanism is largely irrelevant to the bulk of normal matter in the Universe.  There is a caveat; namely, as so often, the pion is exceptional.  Its mass is given by the simple product of two terms, one
of which is the ratio of two order parameters for DCSB whilst the other is determined by the current-quark mass (see Sec.\,\ref{sec:BSE}).  Hence the pion would be massless in the absence of a mechanism that can generate a current-mass for at least one light-quark.  The impact of a massless, strongly-interacting particle on the physics of the Universe would be dramatic, and best avoided, so that something like the Higgs mechanism must be part of the Standard Model.

\begin{figure}[t]
\vspace*{1ex}

\centerline{
\includegraphics[clip,width=0.5\textwidth]{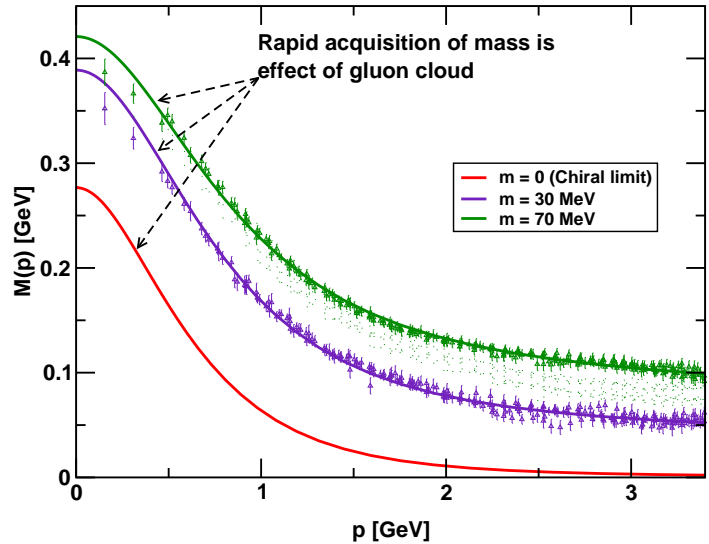}}

\caption{\label{gluoncloud}
Dressed-quark mass function, $M(p)$ in Eq.\,(\protect\ref{SgeneralN}): \emph{solid curves} -- DSE results, explained in Refs.\,\protect\cite{Bhagwat:2003vw,Bhagwat:2006tu}, ``data'' -- numerical simulations of lattice-regularised QCD \protect\cite{Bowman:2005vx}.  (NB.\ $m=70\,$MeV is the uppermost curve and current-quark mass decreases from top to bottom.)  One observes the current-quark of perturbative QCD evolving into a constituent-quark as its momentum becomes smaller.  The constituent-quark mass arises from a cloud of low-momentum gluons attaching themselves to the current-quark.  This is dynamical chiral symmetry breaking (DCSB): an essentially nonperturbative effect that generates a quark mass \emph{from nothing}; namely, it occurs even in the chiral limit.  
}
\end{figure}

The most fundamental expression of DCSB is the behaviour of the dressed-quark mass-function, $M(p)$, which is a basic element in the dressed-quark propagator\footnote{Our Euclidean metric conventions are outlined in App.\,\protect\ref{sec:Euclidean}.  An explanation of the need for Euclidean space in nonperturbative studies of QCD may be found in Sec.\,1.3 of Ref.\,\protect\cite{Roberts:2012sv}.}
\begin{equation}
\label{SgeneralN}
S(p) = -i\gamma\cdot p \, \sigma_V(p^2) + \sigma_S(p^2) = \frac{1}{i\gamma\cdot p A(p^2) + B(p^2)} = \frac{Z(p^2)}{i\gamma\cdot p + M(p^2)}
\end{equation}
that may be obtained as a solution to QCD's fermion gap equation.  The mass function is illustrated in Fig.\,\ref{gluoncloud}.  It arises primarily because a dense cloud of gluons comes to clothe a low-momentum quark and explains how an almost-massless, parton-like quark at high energies transforms, at low energies, into a constituent-like quark that possesses an effective ``spectrum mass'' $M_D \sim m_p/3$.  Consequently, the proton's mass would remain almost unchanged even if the current-quarks were truly massless.

This is a particular feature of the mass function.  It is a single curve that connects the infrared and ultraviolet regimes of the theory, and establishes that the constituent-quark and current-quark masses are simply two connected points separated by a large momentum interval.  The curve shows that QCD's dressed-quark behaves as a constituent-quark, a current-quark, or something in between, depending on the momentum of the probe which explores the bound-state containing the dressed-quark.  It follows that calculations of hadron properties that treat momentum transfers $Q^2 \gsim M_H^2$, where $M_H$ is the mass of the hadron involved, require a Poincar\'e-covariant approach that can veraciously realise quantum field theoretical effects \cite{Cloet:2008re}.  Owing to the vector-exchange character of QCD, covariance also guarantees the existence of nonzero quark orbital angular momentum in a hadron's rest-frame \cite{Brodsky:1980zm,Bhagwat:2006xi,Bhagwat:2006pu,Cloet:2007pi}.

DCSB is expressed in numerous aspects of the spectrum and interactions of hadrons; e.g., the large splitting between parity partners \cite{Chang:2011ei,Chen:2012qr} and the existence and location of a zero in some hadron form factors \cite{Wilson:2011aa,Cloet:2013gva}.  Indeed, as we shall see herein, the dressed-quark mass function has a remarkable capacity to correlate and to contribute significantly in explaining a wide range of diverse phenomena.

Whilst canvassing the topic of dynamical mass generation, it should be noted that gluons, too, acquire a mass \cite{Cornwall:1981zr,Mandula:1987rh}.  The gluon self-energy may also be obtained from a gap equation.  That equation expresses the effects of both colour-charge antiscreening (by self interactions amongst the gluons) and colour-charge screening (by quark loops).  Analyses of the gluon gap equation show that the Landau-gauge dressed-gluon propagator\footnote{As explained elsewhere \protect\cite{Bashir:2008fk,Bashir:2009fv,Raya:2013ina}, Landau gauge is used for many reasons.  It is, \emph{inter alia}: a fixed point of the renormalisation group; that gauge for which sensitivity to model-dependent differences between \emph{Ans\"atze} for the fermion--gauge-boson vertex are least noticeable; and a covariant gauge, which is readily implemented in numerical simulations of lattice regularised QCD (see, e.g., Refs.\,\protect\cite{Bonnet:2000kw,Bowman:2004jm,Bowman:2007du,Kamleh:2007ud,
Sternbeck:2007ug,Boucaud:2011ug,Cucchieri:2007md,Cucchieri:2011um,Cucchieri:2011aa}, and citations therein and thereto).  Importantly, an intelligent capitalisation on the gauge covariance of Schwinger functions obviates any question about the gauge dependence of gauge invariant quantities.}
is characterised by a single scalar function, $\Delta(k^2)$, that is bounded, regular and monotonic at spacelike momenta, $k^2>0$, and achieves its maximum value on this domain at $k^2=0$ \cite{Boucaud:2008ji,Boucaud:2008ky,Aguilar:2009nf,Aguilar:2010gm,LlanesEstrada:2012my,%
Ayala:2012pb,Ibanez:2012zk}.  (See, also, Fig.\,\ref{fig:gluonrp}.)  Such behaviour is confirmed by numerical simulations of lattice-regularised QCD \cite{Bonnet:2000kw,Bowman:2004jm,Bowman:2007du,Kamleh:2007ud,Sternbeck:2007ug,Boucaud:2011ug}.

\begin{figure}[t]
\leftline{\includegraphics[clip,width=0.45\textwidth]{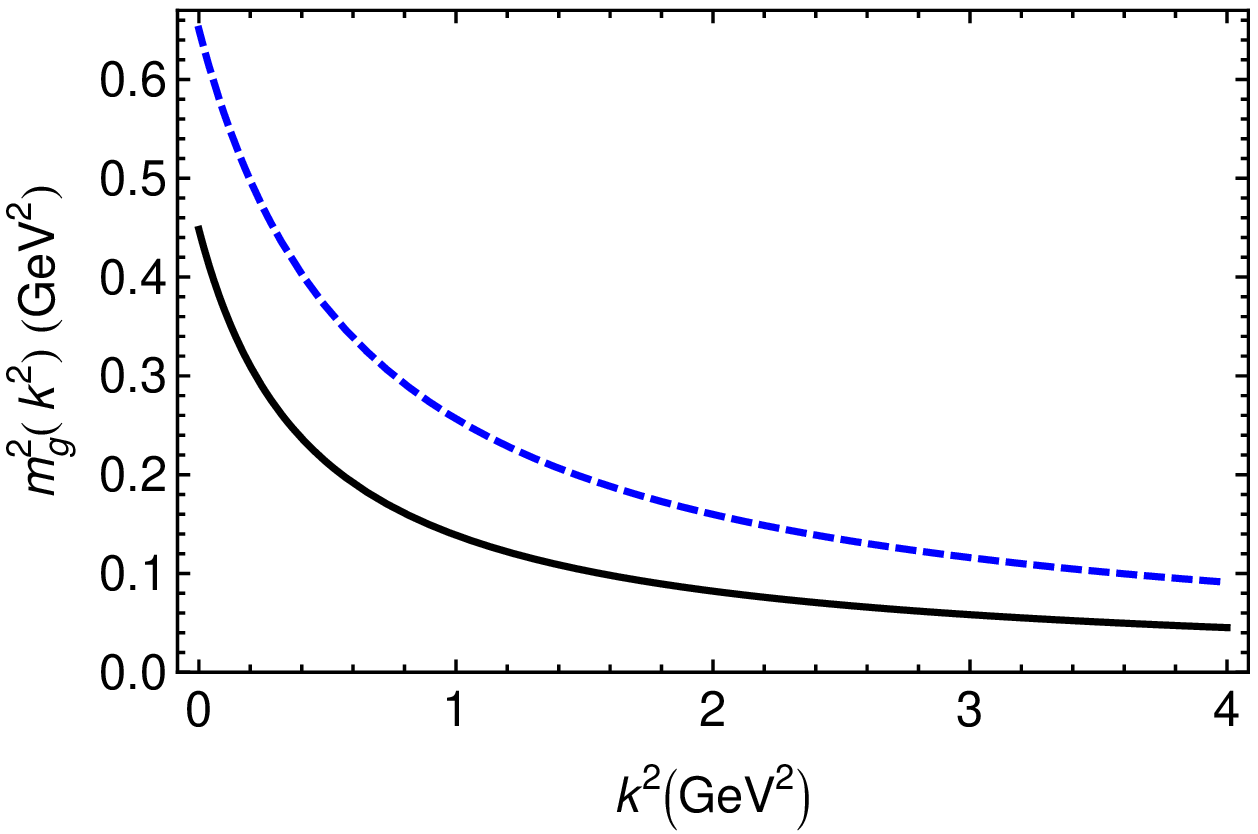}}
\vspace*{-13.4em}

\rightline{\includegraphics[clip,width=0.462\textwidth]{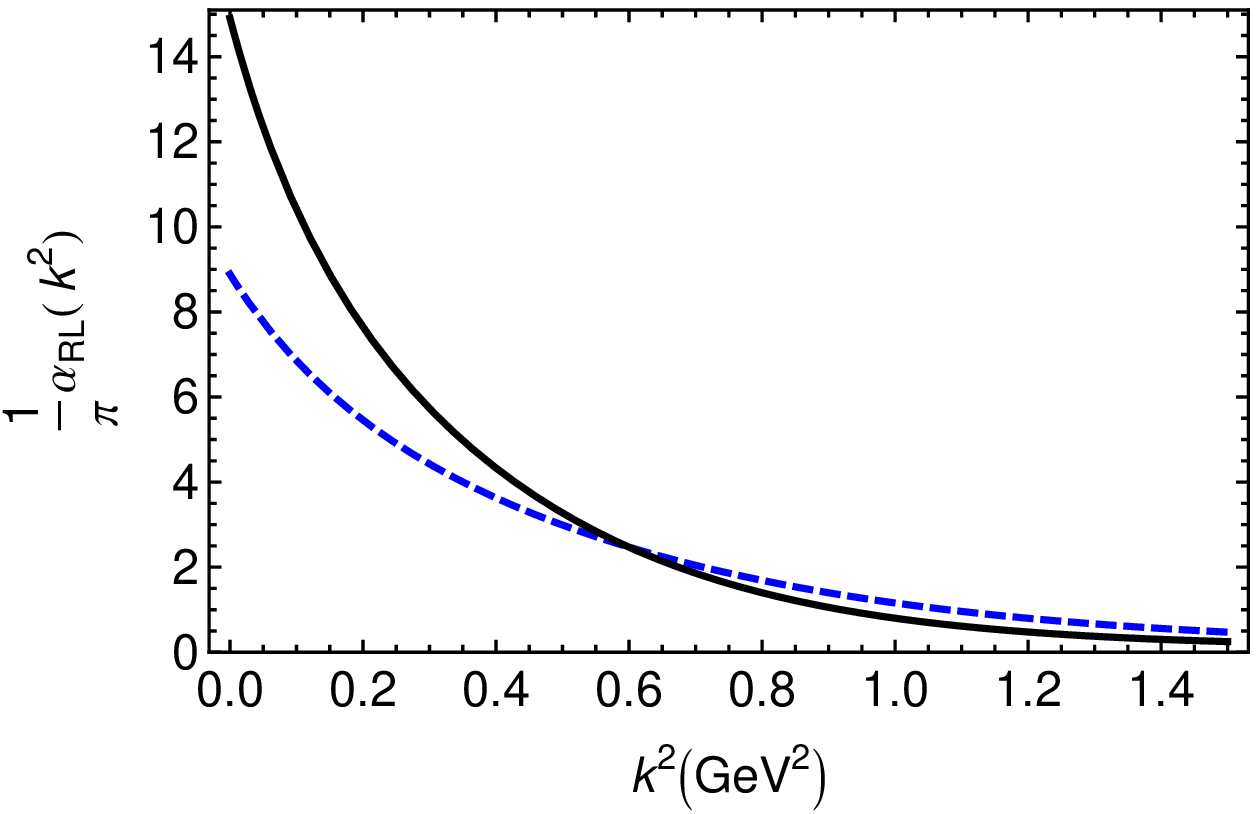}}

\caption{\label{fig:gluonrunning}
\emph{Left panel} -- gluon running-mass; and \emph{Right panel} -- effective running-coupling.
Matching related curves in each panel (solid to solid or dashed to dashed) gives an equivalent description of observables within the rainbow-ladder truncation \protect\cite{Qin:2011dd}.  (``Rainbow-ladder'' is the leading order contribution in the systematic, symmetry-preserving DSE truncation scheme introduced in Refs.\,\protect\cite{Munczek:1994zz,Bender:1996bb}.)
}
\end{figure}

It is possible to extract an approximation to the pointwise behaviour of the gluon's running mass from a numerical solution for $\Delta(k^2)$.  This is not an unambiguous process because  $\Delta(k^2)$ contains two pieces of information; namely, the running coupling and the running mass.  The contemporary procedure is to write \cite{Aguilar:2009nf,Aguilar:2010gm,Oliveira:2010xc}
\begin{equation}
\label{fitalpha}
\Delta(k^2) \approx \frac{4\pi \alpha(k^2)}{k^2 + m_g^2(k^2)}\,,\quad
m_g^2(k^2) = \frac{M_g^4}{M_g^2+k^2},
\end{equation}
and to fit a functional form for the running coupling and the dressed-gluon mass-scale, $M_g$.  In this way one infers $M_g \approx 0.4\,$--$\,0.6\,$GeV; i.e., a gluon mass-scale that is 1.5-2.0\,times larger than the scale associated with the dressed-quark mass in the infrared, see Fig.\,\ref{gluoncloud}.  The pointwise behaviour of $\alpha(k^2)$ and $m_g(k^2)$ obtained in this way is illustrated in Fig.\,\ref{fig:gluonrunning}.

It is worth remarking that the running gluon mass in Eq.\,\eqref{fitalpha} is essentially nonperturbative.  This is readily made apparent.  Consider that the one-loop running coupling in QCD is \cite{Beringer:1900zz}
\begin{equation}
\label{alphaOneLoop}
\alpha_s(Q^2) \stackrel{Q^2 > 10\,\Lambda_{\rm QCD}^2}{\approx} \frac{4 \pi}{\beta_0\,\ln[Q^2/\Lambda_{\rm QCD}^2]},
\end{equation}
with $\beta_0 = 11 - (2/3) n_f$, where $n_f$ is the number of active flavours; i.e., the number of quark flavours whose current-quark mass is less than the momentum scale relevant to the process under consideration.  It follows that
\begin{equation}
m_g^2(k^2) \stackrel{k^2\gg M_g^2}{=} \frac{M_g^4}{k^2} =
\frac{M_g^4}{\Lambda_{\rm QCD}^2} \, {\rm e}^{ -\frac{4\pi}{\beta_0 \alpha(k^2)}}\,;
\end{equation}
viz., on a domain of intermediate momenta, the dressed-gluon mass is accurately approximated by a function that exhibits an essential singularity in the running coupling.  No perturbative analysis can produce such structure.  It is a general feature of QCD that the appearance of power-law behaviour in the evolution of a Schwinger function is an unambiguous signal for the dominance of nonperturbative dynamics on the associated domain.  (This power-law momentum-dependence is typically augmented by an additional logarithm associated with the Schwinger function's anomalous dimension.)  It is evident from Figs.\,\ref{fig:gluonrp}, \ref{gluoncloud}, \ref{fig:gluonrunning} that such behaviour typically begins as one decreases momentum arguments below $k^2 = \Lambda_\chi^2 \sim 2\,$GeV$^2$.

\section{Basic equations for hadron physics}
\subsection{Gap equation}
\label{sec:GapEquation}
\begin{figure}[tbp]
\leftline{%
\includegraphics[clip,width=0.25\textwidth,angle=-90]{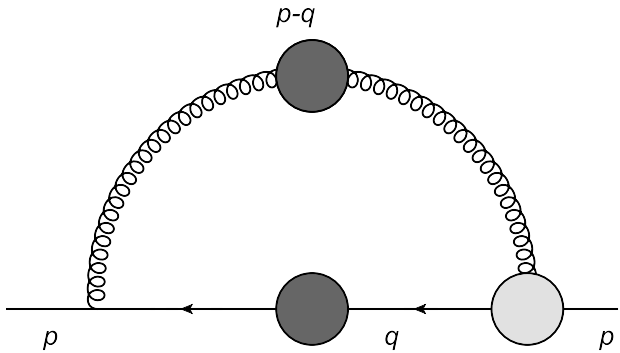}}
\vspace*{-26ex}

\rightline{\parbox{22em}{\caption{\label{FigSigma} Dressed-quark self energy in Eq.\,\protect\eqref{gendseN}.  The kernel is composed from the dressed-gluon propagator (spring with dark circle) and the dressed-quark-gluon vertex (light-circle), and the equation is nonlinear owing to the appearance of the dressed-quark propagator (line with dark circle).  This image encodes every imaginable, valid Feynman diagram.  (Momentum flows from right-to-left.)}}}

\vspace*{2ex}

\end{figure}

The insights described above may reasonably be described as a culmination of fifty years research into the nature and solutions of the gap equation in gauge theories, which began with a study of strong-coupling QED \cite{Johnson:1964da}.  The gap equation in QCD is\footnote{As explained elsewhere \protect\cite{Williams:2003du}, notwithstanding the appearance of Gribov copies in a covariant-gauge formulation of QCD, the standard forms of the DSEs remain valid.  Notably, too, where a comparison is possible, solutions obtained in nonlocal, Gribov-copy-free gauges do not differ in any material way from those determined in a related covariant gauge \protect\cite{Zhang:2004gv}.
}
\begin{equation}
S_f(p)^{-1} = Z_2 \,(i\gamma\cdot p + m_f^{\rm bm}) + \Sigma(p)\,,\quad
\Sigma(p) = Z_1 \int^\Lambda_{dq}\!\! g^2 D_{\mu\nu}(p-q)\frac{\lambda^a}{2}\gamma_\mu S_f(q) \frac{\lambda^a}{2}\Gamma^f_\nu(q,p) , \rule{1em}{0ex}
\label{gendseN}
\end{equation}
where the self energy is illustrated in Fig.\,\ref{FigSigma} and: $f$ is a quark flavour label, $D_{\mu\nu}$ is the gluon propagator; $\Gamma^f_\nu$, the quark-gluon vertex; $\int^\Lambda_{dq}$, a symbol that represents a Poincar\'e invariant regularisation of the four-dimensional Euclidean integral, with $\Lambda$ the regularisation mass-scale;\footnote{A Pauli-Villars-like scheme is usually adequate.  See, e.g., Refs.\,\protect\cite{Holl:2005vu,Chang:2008ec}.} $m_f^{\rm bm}(\Lambda)$, the current-quark bare mass; and $Z_{1,2}(\zeta^2,\Lambda^2)$, respectively, the vertex and quark wave-function renormalisation constants, with $\zeta$ the renormalisation point -- dependence upon which is not usually made explicit.  Notwithstanding this, the gap equation is only completely defined once a renormalisation condition is specified.

The gap equation's solution takes the form in Eq.\,\eqref{SgeneralN}.  However, the $\zeta$-dependence is suppressed therein, so here we note that, amongst the equivalent functions introduced in Eq.\,\eqref{SgeneralN}, only the mass function, $M(p^2)=B(p^2,\zeta^2)/A(p^2,\zeta^2)$, is truly independent of the renormalisation point.  It is worth noting that the renormalised current-quark mass,
\begin{equation}
\label{mzeta}
m_f^\zeta = Z_m^{-1}(\zeta,\Lambda) \, m^{\rm bm}(\Lambda) = Z_4^{-1} Z_2\, m_f^{\rm bm},
\end{equation}
wherein $Z_4$ is the renormalisation constant associated with the Lagrangian's mass-term, is simply the dressed-quark mass function evaluated at one particular deep spacelike point; viz,
\begin{equation}
m_f^\zeta = M_f(\zeta^2)\,.
\end{equation}
The renormalisation-group invariant current-quark mass may be inferred via
\begin{equation}
\label{mfhat}
\hat m_f = \lim_{p^2\to\infty} \left[\frac{1}{2}\ln \frac{p^2}{\Lambda^2_{\rm QCD}}\right]^{\gamma_m} M_f(p^2)\,,
\end{equation}
where $\gamma_m = 4/\beta_0$.  The chiral limit is expressed by
\begin{equation}
\hat m_f = 0\,.
\end{equation}
Moreover,
\begin{equation}
\forall \zeta \gg \Lambda_{\rm QCD}, \;
\frac{m_{f_1}^\zeta}{m^\zeta_{f_2}}=\frac{\hat m_{f_1}}{\hat m_{f_2}}\,.
\end{equation}
This relationship is broken by nonperturbative dynamics, however, so that
\begin{equation}
\frac{m_{f_1}^{\zeta=p^2}}{m^{\zeta=p^2}_{f_2}}=\frac{M_{f_1}(p^2)}{M_{f_2}(p^2)}
\end{equation}
is not independent of $p^2$: in the infrared; i.e., $\forall p^2 \lesssim \Lambda_\chi^2$, it then expresses a ratio of constituent-like quark masses, which, for light quarks, are two orders-of-magnitude larger than their current-masses and nonlinearly related to them \cite{Flambaum:2005kc,Holl:2005st}.

The features and flaws of each DSE are evident in the gap equation.  It is a nonlinear integral equation for the dressed-quark propagator and hence can yield much-needed nonperturbative information.  However, the kernel involves the two-point function $D_{\mu\nu}$ and the three-point function $\Gamma^f_\nu$.  The gap equation is therefore coupled to the DSEs satisfied by these functions, which in turn involve higher $n$-point functions.  Hence the DSEs are a tower of coupled integral equations, with a tractable problem obtained only once a truncation scheme is specified.  The best known truncation scheme is the weak coupling expansion, which reproduces every diagram in perturbation theory.  This scheme is systematic and valuable in the analysis of large momentum transfer phenomena because QCD is asymptotically free but it precludes any possibility of obtaining nonperturbative information.

The first systematic, symmetry-preserving DSE truncation scheme that is applicable nonperturbatively was introduced in Refs.\,\cite{Munczek:1994zz,Bender:1996bb}.  Its mere existence enabled the proof of exact nonperturbative results in QCD.\footnote{See, e.g., Sec.\,V in Ref.\,\cite{Bashir:2012fs}.  It is also worth remarking that so long as the truncation used to define the Gap, Bethe-Salpeter and Faddeev equations is symmetry preserving, then all low-energy theorems are readily reproduced.}
The scheme remains the most widely used today.  Its leading-order term provides the rainbow-ladder (RL) truncation, which is accurate for ground-state vector- and isospin-nonzero-pseudoscalar-mesons \cite{Maris:2003vk,Chang:2011vu,Bashir:2012fs}, and nucleon and $\Delta$ properties \cite{Eichmann:2011ej,Chen:2012qr,Segovia:2013rca,Segovia:2013uga} because corrections in these channels largely cancel, owing to parameter-free preservation of the Ward-Takahashi identities.  However, they do not cancel in other channels \cite{Roberts:1996jx,Roberts:1997vs,Bender:2002as,Bhagwat:2004hn}.  Hence studies based on the rainbow-ladder truncation, or low-order improvements thereof, have usually provided poor results for scalar- and axial-vector-mesons \cite{Cloet:2007pi,Burden:1996nh,Watson:2004kd,Maris:2006ea,Fischer:2009jm,%
Krassnigg:2009zh}, produced masses for exotic states that are too low in comparison with other estimates \cite{Cloet:2007pi,Qin:2011dd,Burden:1996nh,Qin:2011xq,Krassnigg:2009zh}, and exhibit gross sensitivity to model parameters for tensor-mesons \cite{Krassnigg:2010mh} and excited states \cite{Qin:2011dd,Qin:2011xq,Holl:2004fr,Holl:2004un}.  In these circumstances one must conclude that physics important to these states is omitted.

Fortunately, a recently developed truncation scheme overcomes these difficulties \cite{Chang:2009zb} and is beginning to have a material impact.  (An overview is presented in Sec.\,VI of Ref.\,\cite{Bashir:2012fs}.)  This scheme is also symmetry preserving.  Its additional strengths, however, are the capacities to work with an arbitrary dressed-quark gluon vertex and express DCSB nonperturbatively in the Bethe-Salpeter kernels.  The new scheme has enabled a range of novel nonperturbative features of QCD to be demonstrated.  For example, the existence of dressed-quark anomalous chromo- and electro-magnetic moments \cite{Chang:2010hb} and the key role they play in determining observable quantities \cite{Chang:2011tx}; and, in addition, elucidation of the causal connection between DCSB and the splitting between vector and axial-vector mesons \cite{Chang:2011ei} and the impact of this splitting on the baryon spectrum \cite{Chen:2012qr}.\footnote{In connection with the RL truncation, it is sometimes noted that a similar procedure in atomic physics fails in connection with the hydrogen atom \protect\cite{IZ80}.  An analogue in QCD is provided by heavy-light mesons, in the study of which the cancellations required to the preserve accuracy of the RL truncation are blocked because of the vastly different current-masses of the valence-quarks involved.  These issues are canvassed in Sec.\,9.D of Ref.\,\protect\cite{Bashir:2012fs}, wherein it is also noted that the scheme introduced in Ref.\,\protect\cite{Chang:2009zb} can remedy the problem.
}

The kernel of the gap equation, Eq.\,\eqref{gendseN}, involves the dressed-gluon propagator and the dressed-quark-gluon vertex.  As noted above, the gluon propagator may be obtained from its own gap equation.  The qualitative nature of the solution is known (see the discussion associated with Fig.\,\ref{fig:gluonrp}) and significant effort is currently being expended on acquiring quantitatively reliable results (see, e.g., Refs.\,\cite{Ayala:2012pb,Cucchieri:2011ig,Aguilar:2012rz,Dudal:2012zx,Strauss:2012dg,Zwanziger:2012xg,Blossier:2013te}).  Fortunately, the remaining uncertainty is restricted to the far infrared, which is a domain of support that has little impact on most hadron observables.  Hence, the qualitatively and semiquantitively reliable information available today is sufficient to make useful predictions for hadron observables.

The quark-gluon vertex, $\Gamma_\mu^f$, satisfies a Bethe-Salpeter equation.  However, that equation is quite complex (see, e.g., Fig.\,2.6 in Ref.\,\cite{Roberts:1994dr}) 
and hence little has thus far been obtained directly from its analysis.
The common approach to describing $\Gamma_\mu^f$ in gauge theories is to develop \emph{Ans\"atze}  constrained by all available, reliable information.  This is illustrated, e.g., in Refs.\,\cite{Chang:2011ei,Bashir:2011dp,Aguilar:2013ac,Qin:2013mta,Rojas:2013tza,He:2013jaa}.  Employed judiciously, such \emph{Ans\"atze} can be very effective, especially since the alternative is to use the bare vertex, which is seldom satisfactory.

\subsection{Bethe-Salpeter equations}
\label{sec:BSE}
The pion occupies a special place in nuclear and particle physics.  It is the archetype for meson-exchange forces \cite{Yukawa:1935xg} and hence, even today, plays a critical role as an elementary field in the Hamiltonian that describes nuclear structure \cite{Pieper:2001mp,Machleidt:2011zz}.  On the other hand, following introduction of the constituent-quark model \cite{GellMann:1964nj,Zweig:1981pd}, the pion came to be considered as an ordinary quantum mechanical bound-state of a constituent-quark and constituent-antiquark.  In that approach, however, explaining its properties requires a finely tuned potential \cite{Godfrey:1985xj}.

As we shall see in this subsection, the modern paradigm views the pion in a very different manner \cite{Maris:1997hd}: it is both a conventional bound-state in quantum field theory and the Goldstone mode associated with DCSB in QCD.  Given this apparent dichotomy, fine tuning should not play any role in a veracious explanation of pion properties.  The pion's peculiarly low (lepton-like) mass, its strong couplings to baryons, and numerous other characteristics are all unavoidable consequences of chiral symmetry and the pattern by which it is broken in the Standard Model.  Therefore, descriptions of the pion within frameworks that cannot faithfully express symmetries and their breaking patterns (such as constituent-quark models) are unreliable.  A natural method for computing properties of the pion and other mesons in quantum field theory is provided by the Bethe-Salpeter equation (BSE).

The analysis of BSEs in colour singlet channels has been providing valuable information for hadron physics for the last twenty years.  This is illustrated, e.g., by the early studies in Refs.\,\cite{Jain:1993qh,Maris:1997tm,Maris:1999nt}.  (It is notable that these examples all employ the rainbow-ladder truncation.)  The inhomogeneous BSE for a quark and antiquark of total momentum $P$, interacting in a channel characterised by the Dirac matrix ${\cal m}_{\bar q q}$, may be written:
\begin{equation}
[\Gamma_{{\cal m}}(k;P)]_{tu} = Z_{\cal m} \, [{\cal m}_{\bar q q}]_{tu}
+ \int_{dq}^\Lambda [ S(q_\eta) \Gamma_{{\cal m}}(q;P) S(q_{\bar\eta}) ]_{sr} K_{tu}^{rs}(q,k;P)\,,
\label{bsetextbook}
\end{equation}
in which the solution, $\Gamma_{{\cal m}}(k;P)$, is the Bethe-Salpeter amplitude, $Z_{\cal m}$ is the relevant renormalisation constant, $q_\eta = q+ \eta P$ and $q_{\bar\eta} = q- (1-\eta) P$, $K$ is the fully-amputated quark-antiquark scattering kernel, and the colour-, Dirac- and flavour-matrix structure of the elements in the equation is  denoted by the indices $r,s,t,u$.\footnote{N.B.\ By definition, $K$ does not contain quark-antiquark to single gauge-boson annihilation diagrams, nor diagrams that become disconnected by cutting one quark and one antiquark line.  Furthermore, in a symmetry preserving treatment of Eq.\,\eqref{bsetextbook} no observable can depend on $\eta\in [0,1]$; i.e., on the definition of the relative momentum between the quark and antiquark.}  Equations of this type describe, e.g., how the pion appears as a Goldstone mode \cite{Maris:1997hd} and the photon couples to a dressed-quark \cite{Maris:1999bh}, and produce the mass and Bethe-Salpeter amplitude for meson bound states.  Without such information, a continuum study of the properties of mesonic bound-states within the Standard Model is impossible.

\begin{figure}[tbp]
\centerline{%
\includegraphics[clip,width=0.65\textwidth]{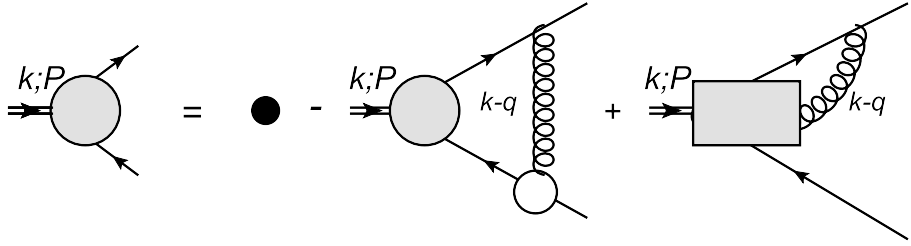}}
\vspace*{-10ex}

\caption{\label{FigBSE} Illustration of the Bethe-Salpeter equation expressed in Eq.\,\protect\eqref{genbse}: Bethe-Salpeter amplitude, $\Gamma(k;P)$, which is the quantity sought -- shaded circle; and vertex inhomogeneity, which defines the channel -- filled circle.  Elements of the kernel: dressed-quark propagator -- oriented line; $\Lambda(k,q;P)$, which depends implicitly on the Bethe-Salpeter amplitude -- shaded box; dressed-quark-gluon vertex -- open circle; and dressed-gluon propagator -- spring.}
\end{figure}

The problem with Eq.\,\eqref{bsetextbook} is that given a dressed-quark-gluon vertex in the gap equation, Eq.\,\eqref{gendseN}, whose diagrammatic content is unknown, then it is generally impossible to construct the kernel $K$ such that the combination of gap and Bethe-Salpeter equations satisfies the relevant WGT identities.  This is bad because one cannot claim a connection with quantum field theory if those identities are broken.  The key step in overcoming this difficulty was a realisation that Eq.\,\eqref{bsetextbook} can be expressed differently \cite{Chang:2009zb}; i.e., as depicted in Fig.\,\ref{FigBSE}.  Owing to its connection with the appearance of Goldstone bosons in the Standard Model, we provide a concrete mathematical realisation of Fig.\,\ref{FigBSE} via the pseudovector channel:
\begin{eqnarray}
\nonumber
\Gamma_{5\mu}^{fg}(k;P) &=& Z_2 \gamma_5\gamma_\mu - Z_1 \! \int_{dq}^\Lambda g^2 D_{\alpha\beta}(k-q)\,
\frac{\lambda^a}{2}\,\gamma_\alpha S_f(q_\eta) \Gamma_{5\mu}^{fg}(q;P) S_g(q_{\bar\eta}) \frac{\lambda^a}{2}\,\Gamma_\beta^g(q_{\bar\eta},k_{\bar\eta}) \\
& &+ Z_1\!\int^\Lambda_{dq} g^2 D_{\alpha\beta}(k-q)\, \frac{\lambda^a}{2}\,\gamma_\alpha S_f(q_\eta) \frac{\lambda^a}{2} \Lambda_{5\mu\beta}^{fg}(k,q;P),\label{genbse}
\end{eqnarray}
where $\Lambda_{5\mu\beta}^{fg}$ is a 4-point Schwinger function.  The content of the right-hand-side is completely equivalent to that of Eq.\,(\ref{bsetextbook}).  However, in striking qualitative opposition to that textbook equation, Eq.\,(\ref{genbse}) partly embeds the solution vertex in the four-point function, $\Lambda$, whilst simultaneously explicating a part of the effect of the dressed-quark-gluon vertex.  This has the invaluable consequence of enabling the derivation of both an integral equation for the new Bethe-Salpeter kernel, $\Lambda$, in which the driving term is the dressed-quark-gluon vertex \cite{Bender:2002as}, and a WGT identity relating $\Lambda$ to that vertex \cite{Chang:2009zb}.  In the pseudovector channel that identity is
\begin{equation}
P_\mu \Lambda_{5\mu\beta}^{fg}(k,q;P) + i [m_f(\zeta)+m_g(\zeta)] \Lambda_{5\beta}^{fg}(k,q;P)
= \Gamma_\beta^f(q_\eta,k_\eta) \, i\gamma_5+ i\gamma_5 \, \Gamma_\beta^g(q_{\bar\eta},k_{\bar\eta}) \,, \label{LavwtiGamma}
\end{equation}
where $\Lambda_{5\beta}^{fg}$ is the analogue of $\Lambda_{5\mu\beta}^{fg}$ in the inhomogeneous pseudoscalar BSE.

Using Eq.\,\eqref{gendseN}, Eq.\,\eqref{genbse} and its pseudoscalar analogue, and Eq.\,\eqref{LavwtiGamma}, one may readily verify that
\begin{equation}
P_\mu \Gamma_{5\mu}^{fg}(k;P) + \, i\,[m_f(\zeta)+m_g(\zeta)] \,\Gamma_5^{fg}(k;P)
= S_f^{-1}(k_\eta) i \gamma_5 +  i \gamma_5 S_g^{-1}(k_{\bar\eta}) \,.
\label{avwtimN}
\end{equation}
This is the flavour-nonsinglet axial-vector WGT identity, which expresses chiral symmetry and the pattern by which it is broken in the Standard Model.  (There are important differences in treating the flavourless channel, which are discussed in Ref.\,\cite{Bhagwat:2007ha} and references therein.)

If one now makes use of the fact that the longitudinal part of the inhomogeneous axial-vector vertex and the pseudoscalar vertex both exhibit a pole at $P^2+m_{H^-}^2=0$, where $m_{H^-}$ is the mass of any given pseudoscalar bound-state, and that the residue of those poles involve the bound-state's Bethe-Salpeter amplitude, $\Gamma_{H_{0^-}}$, then one obtains the following mass formula for pseudoscalar mesons:
\begin{equation}
\label{mrtrelation}
f_{H_{0^-}} m_{H_{0^-}}^2 = (m_{f}^\zeta + m_{g}^\zeta)\, \rho_{H_{0^-}}^\zeta,
\end{equation}
where, with the trace over colour and spinor indices,
\begin{eqnarray}
i f_{H_{0^-}} P_\mu  &=& \langle 0 | \bar q_{g} \gamma_5 \gamma_\mu q_{f} |H_{0^-}\rangle
= Z_2\; {\rm tr}_{\rm CD}
\int_{dq}^\Lambda i\gamma_5\gamma_\mu S_{f}(q_\eta) \Gamma_{H_{0^-}}(q;P) S_{g}(q_{\bar\eta})\,, \label{fpigen} \\
i\rho_{H_{0^-}} &=& -\langle 0 | \bar q_{g} i\gamma_5 q_{f} |H_{0^-} \rangle
= Z_4\; {\rm tr}_{\rm CD}
\int_{dq}^\Lambda \gamma_5 S_{f}(q_\eta) \Gamma_{H_{0^-}}(q;P) S_{g}(q_{\bar\eta}) \,.
\label{rhogen}
\end{eqnarray}
The Bethe-Salpeter amplitude for a pseudoscalar meson bound-state has the form \cite{LlewellynSmith:1969az}:
\begin{equation}
\Gamma_{H_{0^-}}(k;P) = \gamma_5 \left[
i E_{H_{0^-}}(k;P) + \gamma\cdot P F_{H_{0^-}}(k;P)  + \gamma\cdot k \, G_{H_{0^-}}(k;P) - \sigma_{\mu\nu} k_\mu P_\nu H_{H_{0^-}}(k;P)
\right],
\label{genGpi}
\end{equation}
which is determined from the associated homogeneous BSE.  (The parallel of Eq.\,\eqref{mrtrelation} for scalar mesons is presented in Ref.\,\cite{Chang:2011mu}.)  The quantity
\begin{equation}
\chi_{H_{0^-}}(k;P) = S_{f}(k_\eta) \Gamma_{H_{0^-}}(k;P) S_{g}(k_{\bar\eta})
\label{chipi}
\end{equation}
is the meson's Bethe-Salpeter wave-function.  It is the analogue in quantum field theory of the Schr\"odinger wave function in quantum mechanics and, whenever a nonrelativistic limit makes sense, they become the same in that limit \cite{Salpeter:1951sz,Gross:1982nz}.

In QCD, the quark wave-function and Lagrangian mass renormalisation constants, $Z_{2,4}(\zeta,\Lambda)$, respectively, depend on the gauge parameter in precisely the manner needed to ensure that the right-hand sides of Eqs.\,(\ref{fpigen}), (\ref{rhogen}) are gauge-invariant.  Moreover, $Z_2(\zeta,\Lambda)$ ensures that the right-hand side of Eq.\,(\ref{fpigen}) is independent of both $\zeta$ and $\Lambda$, so that $f_{H_{0^-}}$ is truly an observable; and $Z_4(\zeta,\Lambda)$ ensures that $\rho_{H_{0^-}}^\zeta$ is independent of $\Lambda$ and evolves with $\zeta$ in just the way necessary to guarantee that the product $m^\zeta \rho_{H_{0^-}}^\zeta$ is renormalisation-point-independent.  In addition, Eq.\,(\ref{mrtrelation}) is valid for every pseudoscalar meson and for any value of the current-quark masses; viz., $\hat m_{f,g} \in [ 0,\infty)$.  This includes arbitrarily large values and also the chiral limit, in whose neighbourhood Eq.\,(\ref{mrtrelation}) can be shown \cite{Maris:1997hd} to reproduce the familiar Gell-Mann--Oakes--Renner relation \cite{GellMann:1968rz}.  Notably, the associated derivation shows that what has conventionally been identified with a ``vacuum quark condensate'' is actually a quality contained wholly within the pion itself (see Sec.\,\ref{sec:CCC}).

The axial-vector Ward-Takahashi identity, Eq.\,(\ref{avwtimN}), is a crucial bridge to Eqs.\,(\ref{mrtrelation}) -- (\ref{rhogen}); and on the way one can also prove the following quark-level Goldberger-Treiman relations \cite{Maris:1997hd}:
\begin{eqnarray}
\label{gtlrelE}
f_{H_{0^-}}^0 E^0_{H_{0^-}}(k;0) &=& B^0(k^2)\,,\\
\label{gtlrelF}
F^0_R(k;0) + 2 f_{H_{0^-}}^0 F^0_{H_{0^-}}(k;0) &=& A^0(k^2)\,,\\
\label{gtlrelG}
G^0_R(k;0) + 2 f_{H_{0^-}}^0 G^0_{H_{0^-}}(k;0) &=& \frac{d}{dk^2}A^0(k^2)\,,\\
\label{gtlrelH}
H^0_R(k;0) + 2 f_{H_{0^-}}^0 H^0_{H_{0^-}}(k;0) &=& 0\,,
\end{eqnarray}
wherein the superscript indicates that the associated quantity is evaluated in the chiral limit, and $F_R$, $G_R$, $H_R$ are analogues in the inhomogeneous axial-vector vertex of the scalar functions in the $H_{0^-}$-meson's Bethe-Salpeter amplitude.

These identities are of critical importance in QCD.  Combined with Eq.\,\eqref{mrtrelation}, they have numerous empirical consequences.  Those uncovered already are demonstrated and explained in Refs.\,%
\cite{Holl:2005vu,Holl:2004fr,Maris:1998hc,GutierrezGuerrero:2010md,Roberts:2010rn,%
Roberts:2011wy,Nguyen:2011jy,Chang:2012cc,Chen:2012txa}.

We will highlight just one in particular; viz., Eq.\,\eqref{gtlrelE} can be used to prove that a massless pseudoscalar meson appears in the chiral-limit hadron spectrum if, and only if, chiral symmetry is dynamically broken.  This is Goldstone's theorem \cite{Goldstone:1961eq,Goldstone:1962es}.  Equation~\eqref{gtlrelE} shows, in addition, that DCSB has a much deeper and farther reaching impact on physics within the strong interaction sector of the Standard Model.  Namely, the not so widely known fact that Goldstone's theorem is fundamentally an expression of equivalence between the one-body problem and the two-body problem in the pseudoscalar channel: the solution of the two-body pseudoscalar bound-state problem is almost completely known once the one-body problem is solved for the dressed-quark propagator, with the relative momentum within the bound-state identified unambiguously with the momentum of the dressed-quark.  This latter emphasises that Goldstone's theorem has a pointwise expression in QCD; and, furthermore, that pion properties are an almost direct measure of the mass function depicted in Fig.\,\ref{gluoncloud}.  Thus, enigmatically, properties of the massless pion are the cleanest expression of the mechanism that is responsible for almost all the visible mass in the universe.

\subsection{Faddeev equations}
\label{subsec:FE}
The Bethe-Salpeter equations provide information about quark-antiquark vertices and bound-states.  The analogues for three-quark systems are the Poincar\'e covariant Faddeev equations.  One may derive these equations by considering that a baryon appears as a pole in a six-point quark Green function, with the residue proportional to the baryon's Faddeev amplitude.  The Faddeev equation then sums all possible exchanges and interactions that can take place between three dressed-quarks.

\begin{figure}[t]
\centerline{%
\includegraphics[clip,width=0.60\textwidth]{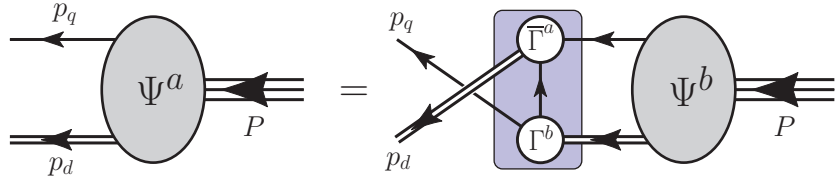}}
\caption{\label{figFaddeev} Poincar\'e covariant Faddeev equation.  $\Psi$ is the Faddeev amplitude for a baryon of total momentum $P= p_q + p_d$.  The shaded rectangle demarcates the kernel of the Faddeev equation: \emph{single line}, dressed-quark propagator; $\Gamma$,  diquark correlation amplitude; and \emph{double line}, diquark propagator.  The Faddeev amplitude expresses the relative momentum correlation between the dressed-quark and -diquarks within the baryon.}
\end{figure}

This equation was first considered in Ref.\,\cite{Cahill:1988dx}, which presented a tractable simplification, illustrated in Fig.\,\ref{figFaddeev}, that was founded on the observation that an interaction which describes colour-singlet mesons also generates nonpointlike quark-quark (diquark) correlations in the colour-$\bar 3$ (antitriplet) channel \cite{Cahill:1987qr}.  As experience with this equation has grown, it has become apparent that the dominant correlations for ground state octet and decuplet baryons are scalar ($0^+$) and axial-vector ($1^+$) diquarks because, for example, the associated mass-scales are smaller than the baryons' masses \cite{Burden:1996nh,Maris:2002yu} and their parity matches that of these baryons.  It follows that only they need be retained in approximating the quark-quark scattering matrix which appears as part of the Faddeev equation \cite{Cloet:2008re,Eichmann:2008ef,Roberts:2011cf}.  On the other hand, pseudoscalar ($0^-$) and vector ($1^-$) diquarks dominate in the parity-partners of ground state octet and decuplet baryons \cite{Roberts:2011cf}.  Use of the Faddeev equation allows one to treat mesons and baryons on the same footing and, in particular, enables the impact of DCSB to be expressed in the prediction of baryon properties.

It is valuable to highlight that diquark correlations are not inserted ``by hand'' into the Faddeev equation.  Both the appearance of such correlations and their importance are a dynamical consequence of the strong coupling in QCD and a further manifestation of the crucial role of DCSB.  Whether one exploits this feature in developing an approximation to the quark-quark scattering matrix within the Faddeev equation \cite{Cloet:2008re,Chang:2011tx,Cloet:2011qu}, as illustrated in Fig.\,\ref{figFaddeev}, or chooses instead to eschew the simplification it offers, the outcome, when known, is the same \cite{Eichmann:2009qa}.  Notably, empirical evidence in support of the presence of diquarks in the proton is accumulating \cite{Wilson:2011aa,Close:1988br,Cloet:2005pp,Cates:2011pz,Cloet:2012cy,Qattan:2012zf}.

It should also be stressed that these dynamically generated correlations are not the pointlike diquarks of yore, which were introduced \cite{Lichtenberg:1967zz,Lichtenberg:1968zz} in order to simplify the study of systems constituted from three constituent-quarks.  The modern dynamical diquark correlation is nonpointlike, with the charge radius of a given diquark being typically 10\% larger than its mesonic analogue \cite{Roberts:2011wy}.  Hence, diquarks are soft components within baryons.

\section{Confinement contains condensates}
\label{sec:CCC}
\subsection{Grave puzzle}
As we have emphasised, DCSB is a crucial emergent phenomenon in the Standard Model, which is very clearly expressed in the dressed-quark mass function of Fig.\,\ref{gluoncloud}.  This understanding is relatively recent, however, and some contemporary textbooks and hadro-particle physics practitioners continue to conflate DCSB with the existence of a spacetime-independent $\bar q q$ condensate that permeates the Universe.  This notion was born with the introduction of QCD sum rules as a theoretical artifice to estimate nonperturbative strong-interaction matrix elements \cite{Shifman:1978bx} and is typically tied to a belief that the QCD vacuum is characterised by numerous distinct, spacetime-independent condensates.

This belief is harmless unless one imagines that the theory of gravity is understood well enough so that it may be coupled to quantum field theory.  Subscribers to this view argue \cite{Turner:2001yu,Bass:2011zz} that the energy-density of the Universe must receive a contribution from such vacuum condensates and that the only possible covariant form for the energy of the (quantum) vacuum; viz.,
\begin{equation}
T_{\mu\nu}^{\rm VAC} = \rho_{\rm VAC}\, \delta_{\mu\nu}\,,
\end{equation}
is mathematically equivalent to the cosmological constant.  From this perspective, the quantum vacuum is \cite{Turner:2001yu} ``\ldots a perfect fluid and precisely spatially uniform \ldots'' so that ``Vacuum energy is almost the perfect candidate for dark energy.''  Now, if the ground state of QCD is really expressed in a nonzero spacetime-independent expectation value $\langle\bar q q\rangle$, then the energy difference between the symmetric and broken phases is of order $M_{\rm QCD} \sim 0.3\,$GeV, as indicated by Fig.\,\ref{gluoncloud}.  One obtains therefrom:
\begin{equation}
\rho_\Lambda^{\rm QCD} = 10^{46} \rho_\Lambda^{\rm obs};
\end{equation}
i.e., the contribution from the QCD vacuum to the energy density associated with the cosmological constant exceeds the observed value by forty-six orders-of-magnitude.  In fact, the discrepancy is far greater if the Higgs vacuum expectation value is treated in a similar manner.

This mismatch has been called \cite{Zee:2008zza} ``\ldots one of the gravest puzzles of theoretical physics.''  However, the problem vanishes if one discards the notion that condensates have a physical existence, which is independent of the hadrons that express QCD's asymptotically realisable degrees of freedom \cite{Brodsky:2009zd}; namely, if one accepts that such condensates are merely mass-dimensioned parameters in one or another theoretical computation and truncation scheme.  This appears mandatory in a confining theory \cite{Chang:2011mu,Brodsky:2010xf,Brodsky:2012ku}, a perspective one may embed in a broader context by considering just what is observable in quantum field theory \cite{Weinberg:1978kz}: ``\ldots although individual quantum field theories have of course a good deal of content, quantum field theory itself has no content beyond analyticity, unitarity, cluster decomposition and symmetry.''  If QCD is a confining theory, then the principle of cluster decomposition is only realised for colour singlet states \cite{Krein:1990sf} and all observable consequences of the theory, including its ground state, can be expressed via an hadronic basis.  This is quark-hadron duality.

The new hypothesis laid out in Refs.\,\cite{Chang:2011mu,Brodsky:2009zd,Brodsky:2010xf,Brodsky:2012ku} can therefore be succinctly expressed as follows: ``If quark-hadron duality is a reality in QCD, then condensates, those quantities that have commonly been viewed as constant empirical mass-scales that fill all spacetime, are instead wholly contained within hadrons; i.e., they are a property of hadrons themselves and expressed, e.g., in their Bethe-Salpeter or light-front wave functions.''  
Given the importance of this shift in perspective, it is worth recapitulating herein upon its foundations and consequences.

One may begin with a realisation that in Dirac's light-front form of relativistic dynamics \cite{Dirac:1949cp}, the ground state of the theory is a structureless Fock-space vacuum without a $\langle \bar q q \rangle$ condensate, or anything else of this nature.  Furthermore, as was first argued using the light-front framework in Ref.\,\cite{Casher:1974xd}, DCSB and the associated quark condensate must be a property of hadron wave functions, not of the vacuum.  This thesis has also been explored in Refs.\,\cite{Burkardt:1998dd,Glazek:2011vg}.  One subtlety in characterising the formal quantity $\langle 0 | {\cal O} | 0 \rangle$, where ${\cal O}$ is a product of quantum field operators, is evident when one recalls that this can automatically be rendered zero by normal-ordering the operator ${\cal O}$.  This subtlety is especially delicate in a confining theory because the vacuum state in such a theory is not defined relative to the fields in the Lagrangian -- gluons and quarks -- but to the actual physical, colour-singlet states.

In a rigorous statistical mechanical treatment of a phase transition such as that involving magnetism or superconductivity, the transition occurs only in the infinite-volume limit, and the order parameter, e.g., magnetisation or Cooper pair condensate, is a constant that extends throughout spacetime.  However, as emphasised in Ref.\,\cite{Brodsky:2009zd,Brodsky:2008be}, experimentally one always observes magnetism and superconductivity in finite samples, and the magnetisation or Cooper pair condensates are constants only within the material that supports them, not throughout an infinite volume.  In a similar manner, particularly because of confinement, one may argue that QCD condensates are completely contained within that domain which permits the propagation of the gluons and quarks that produce them; namely, inside hadrons.

\subsection{Role of confinement}
The so-called vacuum condensates, which are written as vacuum expectation values of local operators, are phenomenological parameters that were introduced at a time of limited computational resources in order to assist with the theoretical estimation of essentially nonperturbative strong-interaction matrix elements \cite{Shifman:1978bx}.  A universality of the condensates was assumed; namely, that the properties of all hadrons could be expanded in terms of the same condensates.  Whilst this helps to retard proliferation, there are nevertheless infinitely many distinct condensates.  As qualities associated with an unmeasurable state (the vacuum) such condensates do not admit direct measurement.  Practitioners have attempted to assign values to them via an internally consistent treatment of many separate empirical observables.  However, only one, the quark condensate, is attributed a value with any confidence.  The difficulties and capacities of the sum rules approach are detailed elsewhere \cite{Leinweber:1995fn}.

In tackling a problem as difficult as determining the truly observable predictions of nonperturbative QCD, theory has naturally employed artifices.  Problems arise only when notional elements in the computational structure are erroneously imbued with an empirical nature.  This is the case with the QCD vacuum condensates: from being merely mass-dimensioned parameters in a theoretical truncation scheme, with no existence independent of hadrons, in the minds of some they have been transformed into measurable spacetime-independent vacuum configurations of QCD's elementary degrees-of-freedom.  In the presence of confinement, the
latter is impossible, and the measurable impact of the so-called condensates is expressed entirely in the properties of QCD's asymptotically realisable states; namely hadrons.  Faith in empirical vacuum condensates may be compared with an earlier misguided conviction that the universe was filled with a luminiferous aether, which was not overturned before completion of a renowned experiment \protect\cite{MichelsonMorley}.

As explained in Sec.\,\ref{sec:confinement}, confinement is a statement about real-world QCD, in which light-quarks are ubiquitous and pions are light.  It is equivalent to exact quark-hadron duality; namely, that all observable consequences of QCD can be computed using an hadronic basis.  Equivalently, the Hilbert space associated with the measurable Hamiltonian of QCD is spanned by colour-singlet state-vectors; viz.,
\begin{equation}
{\cal H}_{\rm QCD} = \sum_n \, E_n |H^{1_c}_n \rangle \langle H^{1_c}_n|\,,
\end{equation}
where $|H^{1_c}_n \rangle$ are colour singlets.  Causality entails that QCD possesses a state of lowest observable energy, which one can choose to be $E_0=0$.  The state associated with this energy is the vacuum.  It is the state with zero hadrons.

A precise definition of the vacuum is only possible if one has a nonperturbative definition of the field variable associated with the asymptotic one-particle state, for then the vacuum is that state obtained when the field annihilation operator acts on the asymptotic one-particle state, which is unambiguous.  This is closely connected with the point about normal-ordering.  One may visualise the creation and annihilation operators for such states as rigorously defined via smeared sources on a spacetime lattice.  The ground-state is defined with reference to such operators, employing, e.g., the Gell-Mann--Low theorem \cite{GellMann:1951rw}, which is applicable in this case because there are well-defined asymptotic states and associated annihilation and creation operators.

The notion of a structured vacuum in QCD involves an analogy drawn between DCSB in the strong interaction and the BCS-theory of superconductivity \cite{Nambu:2011zz}.  The BCS approach is a mean-field theory based on a Hamiltonian expressed in terms of well-defined quasiparticle operators.  There is a known relation between the bare-particle and quasiparticle operators and, under certain conditions, the latter can possess a nonzero expectation value in the vacuum defined via the bare-particle annihilation operator.  Owing to confinement, these steps are impossible in QCD.
Furthermore, the BCS-based analysis is subject to the comment reiterated above \cite{Brodsky:2009zd,Brodsky:2008be}; namely, that although formally a phase transition in statistical mechanics requires an infinite-volume limit and the resulting order parameter (here the Cooper pair condensate) is a constant throughout infinite space, one experimentally observes the Cooper pair condensate to exist only inside finite pieces of superconducting materials, not to be a constant extending throughout infinite space.  In statistical mechanics and condensed matter discussions of real phase transitions and critical phenomena, one is careful to distinguish between the idealised infinite-volume limit and actual experimental observations on finite samples.  One must at least be equally careful in gauge theories.

Amongst the consequences of confinement is the absence of asymptotic gluon and quark states.  It is therefore impossible to write a valid nonperturbative definition of a single gluon or quark annihilation operator.  To do so would be to answer the question: What is the operator that annihilates a state which is unmeasurable?  So although one can define a perturbative (bare) vacuum for QCD, it is impossible to rigorously define a ground state for QCD upon a foundation of gluon and quark (quasiparticle) operators.  Likewise, it is impossible to construct an interacting vacuum -- a BCS-like trial state -- and hence DCSB in QCD cannot rigorously be expressed via a spacetime-independent coherent state built upon the ground state of perturbative QCD (pQCD).  Whilst this does not prevent one from following this path to build models for use in hadron physics phenomenology (Ref.\,\cite{Finger:1981gm} is a pertinent example), it does invalidate any claim that theoretical artifices in such models are accurate descriptions of QCD.

\subsection{Gell-Mann--Oakes--Renner (GMOR) relation}
The arguments detailed in Refs.\,\cite{Chang:2011mu,Brodsky:2009zd,Brodsky:2010xf} proceed via the proof of exact and hence model-independent results in QCD, amongst them: the chiral-limit vacuum quark condensate is equivalent to the pseudoscalar meson leptonic decay constant, in the sense that they are both obtained as the chiral-limit value of well-defined gauge-invariant hadron-to-vacuum transition amplitudes that possess a spectral representation in terms of the current-quark mass \cite{Brodsky:2010xf}; the same is true in the scalar channel \cite{Chang:2011mu}; and in-hadron quark condensates can be represented through a given hadron's scalar form factor at zero momentum transfer \cite{Chang:2011mu}.

It is appropriate here to exemplify these notions via Eqs.\,\eqref{mrtrelation}--\eqref{rhogen}; viz., defining
\begin{equation}
\label{kappazeta}
\kappa^\zeta_{P_{f g}} := \rho_{P_{f g}}^\zeta f_{P_{f g}}=: [\chi_{P_{f g}}^\zeta]^3\,,
\end{equation}
then it is an exact result in QCD, valid for arbitrarily small or large current-quark masses and for both ground- and excited-states \cite{Holl:2004fr}, that
\begin{equation}
\label{GMORP}
f_{P_{f g}}^2 m_{P_{f g}}^2 = [m_{f}^\zeta +m_{g}^\zeta]\, \kappa^\zeta_{P_{fg}}
= [\hat m_{f} + \hat m_{g}]\, \hat \kappa_{P_{f g}},
%
%
\end{equation}
where, as usual, the circumflex indicates a renormalisation-group-invariant quantity.  Moreover \cite{Maris:1997hd}
\begin{equation}
\lim_{\hat m\to 0} \kappa^\zeta_{P_{f g}}
= - \lim_{\hat m\to 0} f_\pi \langle 0 | \bar q i\gamma_5 q |\pi \rangle
=
Z_4 \, {\rm tr}_{\rm CD}\int_{dq}^\Lambda S^0(q;\zeta) =  -\langle \bar q q \rangle_\zeta^0\,;
\label{qbqpiqbq0}
\end{equation}
namely, the so-called vacuum quark condensate is, in fact, the chiral-limit value of the in-meson condensate; i.e., it describes a property of the chiral-limit pseudoscalar meson.  This condensate is therefore no more a property of the ``vacuum'' than the pseudoscalar meson's chiral-limit leptonic decay constant.  Moreover, given that Eq.\,\eqref{qbqpiqbq0} is an identity in QCD, any veracious calculation of $\langle \bar q q \rangle_\zeta^0$ is the computation of a gauge-invariant property of the pion's wave-function.
(We return to this point in Sec.\,\ref{SecInsightLQCD}.)

It is perhaps even more valuable to highlight the precise form of the Gell-Mann--Oakes--Renner (GMOR) relation; viz., Eq.\,(3.4) in Ref.\,\cite{GellMann:1968rz}:
\begin{equation}
\label{gmor}
m_\pi^2 = \lim_{P^\prime \to P \to 0} \langle \pi(P^\prime) | {\cal H}_{\chi{\rm sb}}|\pi(P)\rangle\,,
\end{equation}
where $m_\pi$ is the pion's mass and ${\cal H}_{\chi{\rm sb}}$ is that part of the hadronic Hamiltonian density which explicitly breaks chiral symmetry.  It is crucial to observe that the operator expectation value in Eq.\,(\ref{gmor}) is evaluated between pion states.
Moreover, the virtual low-energy limit expressed in Eq.\,\eqref{gmor} is purely formal.  It does not describe an achievable empirical situation, as is explained in connection with Eq.\,\eqref{eq:constituents} below.

In terms of QCD quantities, Eq.\,(\ref{gmor}) entails
\begin{equation}
\label{gmor1}
\forall m_{ud} \sim 0\,,\;  m_{\pi^\pm}^2 =  m_{ud}^\zeta \, {\cal S}_\pi^\zeta(0)\,,
\quad
{\cal S}_\pi^\zeta(0) = - \langle \pi(P) | \mbox{\small $\frac{1}{2}$}(\bar u u + \bar d d) |\pi(P)\rangle\,,
\end{equation}
where $m_{ud}^\zeta = m_u^\zeta+m_d^\zeta$ and ${\cal S}^\zeta(0)$ is the pion's scalar form factor at zero momentum transfer, $Q^2=0$.  The right-hand-side (rhs) of Eq.\,(\ref{gmor1}) is proportional to the pion $\sigma$-term (see, e.g., Ref.\,\cite{Flambaum:2005kc}).  Consequently, using the connection between the $\sigma$-term and the Feynman-Hellmann theorem, Eq.\,(\ref{gmor}) is actually the statement
\begin{equation}
\label{pionmass2}
\forall m_{ud} \simeq 0\,,\; m_\pi^2 = m_{ud}^\zeta \frac{\partial }{\partial m^\zeta_{ud}} m_\pi^2.
\end{equation}

Now, using Eq.\,\eqref{GMORP}, one obtains
\begin{equation}
\label{gmor2}
{\cal S}_\pi^\zeta(0)
= \frac{\partial }{\partial m^\zeta_{ud}} m_\pi^2
=\frac{\partial }{\partial m^\zeta_{ud}} \left[ m_{ud}^\zeta\frac{\rho_\pi^\zeta}{f_\pi}\right].
\end{equation}
Equation~(\ref{gmor2}) is valid for any values of $m_{u,d}$, including the neighborhood of the chiral limit, wherein
\begin{equation}
\label{gmor3}
{\cal S}_\pi^\zeta(0)= \frac{\partial }{\partial m^\zeta_{ud}} \left[ m_{ud}^\zeta\frac{\rho_\pi^\zeta}{f_\pi} \right]_{m_{ud} = 0}
= \frac{\rho_\pi^{\zeta 0}}{f_\pi^0}\,.
\end{equation}
The superscript ``0'' indicates that the quantity is computed in the chiral limit.  With Eqs.\,\eqref{kappazeta}, \eqref{qbqpiqbq0}, (\ref{gmor1}), (\ref{gmor2}), (\ref{gmor3}), one has shown that in the neighborhood of the chiral limit
\begin{equation}
m_{\pi^\pm}^2 =  -m_{ud}^\zeta  \frac{\langle \bar q q \rangle^{\zeta 0}}{(f_\pi^0)^2} + {\rm O}(m_{ud}^2).
\label{truegmor}
\end{equation}
This is a QCD derivation of the commonly recognised form of the GMOR relation.  Neither PCAC nor soft-pion theorems were employed in analysing the rhs of Eqs.\,\,\eqref{gmor}, (\ref{gmor1}).  In addition, the derivation shows that in the chiral limit, the matrix element describing the pion-to-vacuum transition through the pseudscalar vertex is equal to the expectation value of the explicit chiral symmetry breaking term in the QCD Lagrangian \emph{computed between pion states} after normalisation by the pion's decay constant.

This recapitulation of the analysis in Ref.\,\cite{Chang:2011mu} emphasises anew that any connection between the pion mass and a ``vacuum'' quark condensate is purely a theoretical artifice.  The true connection is that which one would expect; viz., the pion's mass is a property of the pion, determined by the interactions between its constituents.  One may further highlight the illogical nature of the vacuum connection by considering the expectation value
\begin{equation}
\label{HOH}
\langle H(P) | \bar q {\cal O} q | H(P)\rangle \,,
\end{equation}
where $H(P)$ represents some hadron with total momentum $P$.  If one chooses ${\cal O}=\gamma_\mu$, then all readers will accept that Eq.\,\eqref{HOH} yields the electric charge of the hadron $H$.  The choice ${\cal O}=\gamma_5\gamma_\mu$ in Eq.\,\eqref{HOH} will yield the axial charge of the hadron $H$; ${\cal O}=\sigma_{\mu\nu}$ will yield the tensor charge of the hadron $H$; and the choice ${\cal O}=\mathbf I$ will yield the scalar charge of the hadron $H$.  All these statements are true irrespective of the hadron involved, so that the scalar charge of the pion; i.e., ${\cal S}_\pi^\zeta(0)=\kappa_\pi^0/(f_\pi^0)^2$, the ``B-parameter'' in chiral perturbation theory, is no more a property of the vacuum than is the pion's electric charge.

\subsection{Empirical consistency}
It will be plain to circumspect practitioners that the notion of condensates being contained within hadrons does not contradict any empirical observation.  An attachment to convention, however, may obscure this simple fact.  
For example, we disagree with a suggestion made elsewhere \cite{Reinhardt:2012xs} that containing condensates within hadrons entails that the lowest excitations in the pseudoscalar and scalar channels are degenerate.  In QCD, as illustrated in Fig.\,\ref{gluoncloud}, DCSB is expressed in the dressed-quark mass-function through the material enhancement of $M(p)$ on the domain of infrared momenta, $p\lesssim 5\Lambda_{\rm QCD}$.  In modern approaches to the bound-state problem, such as DSE- and lattice-QCD, bound-states are constituted using such propagators.  Bound-states therefore express the features realised in these propagators.  It is possible to illustrate this using the simplest of confining models.  With a symmetry-preserving and confining regularisation of a vector-vector contact-interaction, the dressed-quark is described by a dynamically generated mass, $M$, which is large in the chiral limit.
At lowest-order in a symmetry-preserving truncation of the model's DSEs \cite{Bender:1996bb}, one finds algebraically that $m_\pi= 0$, $m_\sigma = 2 M$ \cite{Roberts:2011wy}.  Corrections to the leading order truncation do not change $m_\pi$ but markedly increase $m_\sigma$ \cite{Chang:2009zb}.  Degeneracy of the lowest excitations in the pseudoscalar and scalar channels is only achieved when $M=0$; i.e., if chiral symmetry is not dynamically broken.  The splitting between vector and axial-vector mesons and parity partners in the baryon spectrum are explained in the same manner \cite{Chang:2011ei,Chen:2012qr}.

The spectrum of QCD exhibits a large splitting between parity partners because of DCSB, which is manifest in the Schwinger functions of the excitations confined within hadrons.  It should be recognised that the computation of colour-nonsinglet Schwinger functions in isolation is an artifice.  In the fully self-consistent treatment of bound-states, the dressing phenomenon illustrated in Fig.\,\ref{gluoncloud} takes place in the background field generated by the other constituents: the field's influence is concentrated in the far infrared, $p\lesssim \Lambda_{\rm QCD}$, and its presence ensures the manifestations of gluon- and quark-dressing are gauge invariant.


It is also a misapprehension to suggest \cite{Lee:2012ga} that the hypothesis of in-hadron condensates entails a violation of Bose-Einstein statistics owing to a possibility that the condensates in any two pions may have different chiral orientations.  There is no such possibility.  This suggestion is analogous to the already falsified claim that $\sigma$ and $\pi$ mesons would be indistinguishable when condensates are confined within hadrons.

It is nevertheless worth expounding upon this point.  The so-called orientation of the chiral condensate, whether in-hadron or otherwise, is fixed, once and for all, as soon as the gap equation is solved and the choice made to define the solution in the form of Eq.\,\eqref{SgeneralN}.  Owing, for example, to the hierarchical nature of the DSEs, the structure of Eq.\,\eqref{SgeneralN} is communicated to \emph{all} Schwinger functions.  This is not different in principal from choosing the form of the mass-term in the QCD action.  This fact is explicit in perturbation theory, the QCD sum-rules method and, indeed, in any approach that uses Schwinger functions when computing interactions involving one or more hadrons.  For example, one does not compute $\pi \pi$ scattering by using dressed-quark Schwinger functions with arbitrary, mismatched chiral orientations in the asymptotic states.  Instead, one arrives at predictions based in principle upon analyses of the $q \bar q q \bar q$ four-point function in the appropriate channels, expressed in terms of Schwinger functions whose chiral orientation is uniform and fixed according to Eq.\,\eqref{SgeneralN}.

The paradigm of in-hadron condensates has also been declared to be incompatible with chiral Lagrangian models \cite{Lee:2012ga}: the centre of this objection is Eq.\,(6) therein.  However, that equation is nothing more than Eq.\,\eqref{qbqpiqbq0} above and the associated musings in Ref.\,\cite{Lee:2012ga} are rendered moot by the arguments and information described in connection with Eqs.\,\eqref{qbqpiqbq0}-\eqref{truegmor} above.

Equally, we disagree with a claim \cite{Reinhardt:2012xs} that the containment
of condensates within hadrons precludes chiral symmetry restoration at nonzero
baryon density.  As reviewed in Ref.\,\cite{Roberts:2000aa}, a nonzero chemical potential, $\mu$, has a dramatic impact on the dressed-quark propagator once $\mu$ arrives at the vicinity of a critical value.  At zero temperature that value is $\mu_{\rm cr} \approx 0.3\,$GeV \cite{Qin:2010nq}.  This is also true of nonzero temperature; and as the realisation of DCSB in the dressed-quark propagator is suppressed and finally eliminated by growth of these intensive thermodynamic parameters, so do parity partners become degenerate.  Concrete examples in  confining models are detailed in Refs.\,\cite{Maris:2000ig,Wang:2013wk}.


%

Another point worth elucidating relates to the pion's charge radius.  The electromagnetic radius of any hadron which couples to pseudoscalar mesons must diverge in the chiral limit.  This long-known effect arises because the propagation of \emph{massless} on-shell colour-singlet pseudoscalar mesons is undamped \cite{Beg:1973sc,Pervushin:1974nm,Gasser:1983yg,Alkofer:1993gu}.  Since the pion couples to itself, one might suggest that, owing to the divergence of its charge radius, each pion will grow to fill the universe, so that, in this limit, the in-pion condensate reproduces the conventional paradigm.

Confinement, again enables one to refute this objection.  As noted above, general arguments, as well as DSE- and lattice-QCD studies, indicate that confinement entails dynamical mass generation for gluons and quarks.  The zero-momentum value of the momentum-dependent dynamical quark masses $M(0)$ and effective gluon mass $m_g(0)$ remain large in the limit of vanishing current-quark mass.  In fact, these values are almost independent of the current-quark mass in the neighborhood of the chiral limit.  (This is apparent in Fig.\,\ref{gluoncloud}.)
As a consequence, one can argue that the quark-gluon containment-radius of all hadrons is finite in the chiral limit.  Indeed it is almost insensitive to the magnitude of the current-quark mass because the dynamical masses of the hadron's constituents are frozen at large values; viz.,
\begin{equation}
\label{eq:constituents}
M(0) \lesssim m_g(0) =: m_{\rm c} \sim 2\Lambda_{\rm QCD}-3 \Lambda_{\rm QCD}
\,.
\end{equation}
These considerations indicate that the divergence of the pion's electromagnetic radius does not correspond to expansion of a condensate from within the pion but rather to the copious production and subsequent propagation of composite pions, each of which contains a condensate whose value is essentially unchanged from its nonzero current-quark mass value within a containment-domain whose size is similarly unaffected.  That domain is specified by a radius
$r_{\rm c} \sim 1/m_{\rm c}$.

There is more to be said in connection with the definition and consequences of a chiral limit.  Nambu-Goldstone bosons are weakly interacting in the infrared limit.  However, at nonzero energies, their interactions are, in general, strong, and they always couple strongly, e.g., to the nucleon.  Plainly, the existence of strongly-interacting massless composites would have an enormous impact on the evolution of the universe; and it is na\"{\i}ve to imagine that one can simply set $\hat m_{u,d}=0$ and consider a circumscribed range of manageable consequences whilst ignoring the wider implications for hadrons, the Standard Model and beyond.  For example, with all else held constant, Big Bang Nucleosynthesis is very sensitive to the value of the pion-mass \cite{Flambaum:2007mj,Bedaque:2010hr}.  We are fortunate that the absence of quarks with zero current-quark mass has produced a universe in which we exist so that we may carefully ponder the alternative.

As mentioned above, a universality of condensates was assumed in order to slow growth in the number of undetermined parameters that appear in the sum rules scheme.  However, with the appreciation that condensates are contained within hadrons, the assumption of universality is seen to be quantitatively false \cite{Chang:2011mu}.  This is similar, in fact, to the assumption of vacuum saturation for the four-quark condensate, which underestimates the correct result by $\sim 65$\% \cite{Nguyen:2010yj}.  It is nonetheless interesting that the magnitude of the in-hadron quark condensate is only weakly sensitive to the host state \cite{Chang:2011mu,Roberts:2011ea}.  This, too, is tied to the preeminent role played by the dressed-gluon and -quark propagators in producing bound-states and their masses.

%
\subsection{Insights into simulations of lattice-QCD}
\label{SecInsightLQCD}
There are many articles in the hadron physics literature that describe numerical estimates of the so-called vacuum quark condensate using simulations of lattice-QCD,  Refs.\,\cite{Burger:2012ti,McNeile:2012xh} are amongst the most recent.  In order to understand the connection between such results and the paradigm of in-hadron condensates it is important to consider Ref.\,\cite{Langfeld:2003ye}, which demonstrated that the quantity extracted in lattice simulations and identified with the vacuum quark condensate is, instead, precisely the in-pion condensate.  The analysis in Ref.\,\cite{Langfeld:2003ye} establishes that, in the chiral limit, the Banks-Casher formula \cite{Banks:1979yr}, the OPE condensate \cite{Lane:1974he,Politzer:1976tv} and the trace of the dressed-quark propagator are all the same; and they are all equal to the in-pion condensate and only the in-pion condensate.  We recapitulate here upon important elements of Ref.\,\cite{Langfeld:2003ye}.

In making the connection with simulations of lattice-QCD, one is concerned in the continuum with the quantity in Eq.\,(31) of Ref.\,\cite{Langfeld:2003ye}:
\begin{equation}
\tilde\sigma(m) := N_c {\rm tr}_{\rm D}\!\int_{dq}^\Lambda \tilde S_m(q)\,,
\end{equation}
where $m$ is the current-quark bare mass.  As usual herein, the integral sign means a translationally invariant regularisation of the integral with a regularization scale $\Lambda$.  In the context of lattice-QCD investigations, one must assume that they recover translational invariance if appropriate care is taken with limits following a carefully planned and systematic collection of lattice studies.  This being the case, one may equate $\Lambda$ with $1/a$, where ``$a$'' is the lattice spacing.

Now, in order to give meaning to $\tilde\sigma(m)$ in a lattice simulation, a great deal of care must be taken.  For example, no lattice simulation can directly consider the chiral limit.  Hence, there will always be a bare mass with which to contend: $m(\Lambda)$.  Suppose now that one were able to keep all other things fixed and take the limit $a\to 0$, then one would find
\begin{equation}
\label{UVcutoffDominance}
\tilde\sigma(m) \stackrel{a\simeq 0}{=} \frac{3}{4\pi^2} \frac{m(\Lambda)}{a^2}\,;
\end{equation}
i.e., the simulation would yield no signal for a dynamical OPE condensate.  It would instead produce just the usual perturbative divergence from the trace of the propagator.  That divergence is eliminated in the continuum through normal ordering.  In lattice simulations, a carefully controlled subtraction must be employed.

\begin{figure}[t]
\leftline{\includegraphics[clip,width=0.47\textwidth]{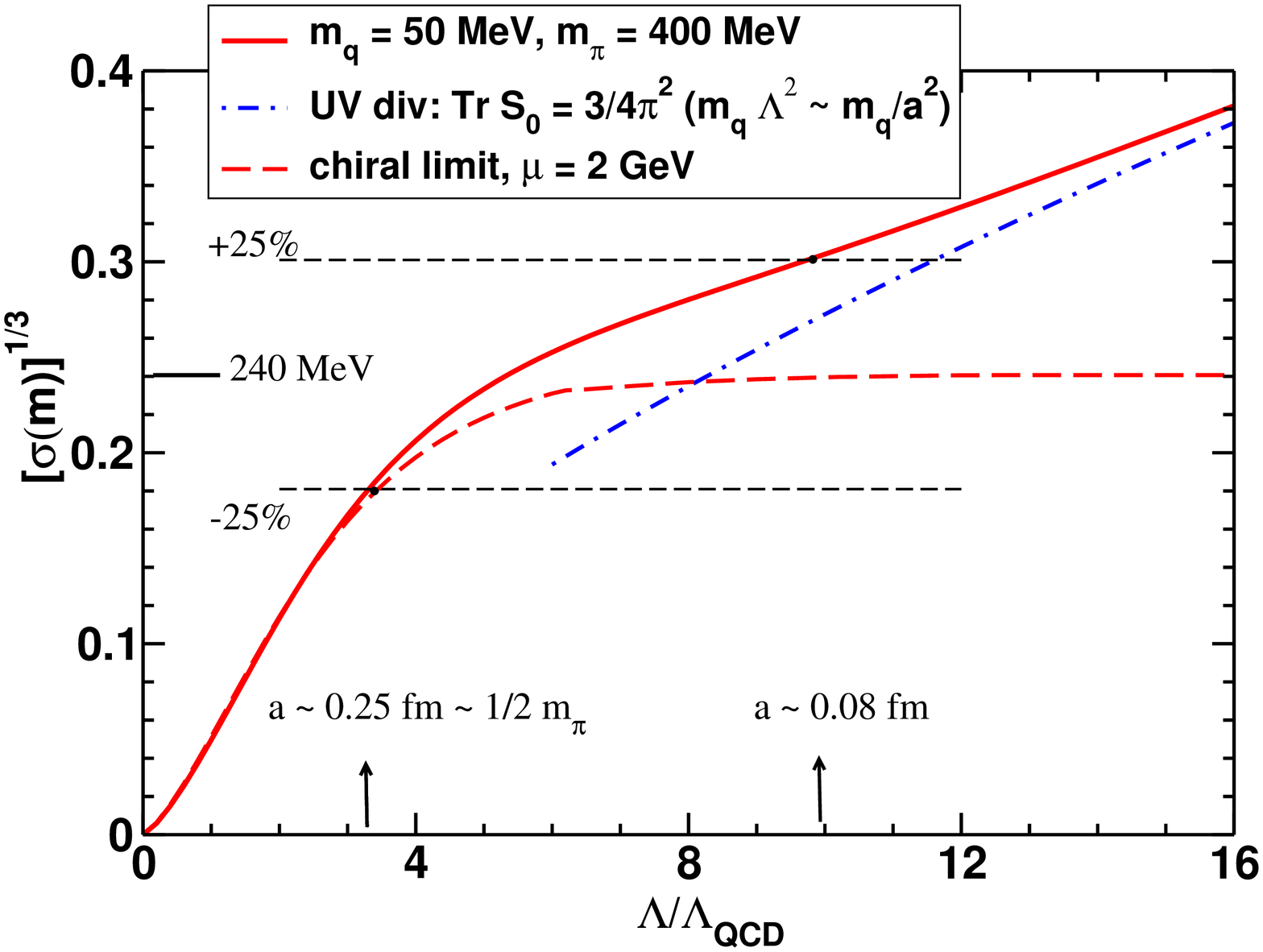}}
\vspace*{-34ex}

\rightline{\parbox{22em}{\caption{\label{FigPCTandy}
The OPE condensate has a value of roughly $240\,$MeV.  The thin dashed lines mark a window of $\pm 25$\% around this value.  The solid curve depicts a realistic result for $\tilde\sigma(m)$, taken from Ref.\,\protect\cite{Langfeld:2003ye}, and the dashed-curve is the chiral limit result.  The dot-dashed curve is the ultraviolet-cutoff-dominated result in Eq.\,\protect\eqref{UVcutoffDominance}.  A lattice-QCD simulation is sensitive to the signal of the OPE condensate for $\Lambda/\Lambda_{\rm QCD} \sim 4-9$, which corresponds to a lattice spacing in the range $a \in (0.1,0.25)\,$\,fm.}}}
\vspace*{2ex}

\end{figure}

The nature of such subtractions is critical.  Given that the OPE condensate has a value of just $(0.24\,{\rm GeV})^3$, then, with a bare mass $m(a=1/\Lambda)=50\,$MeV, the quantity $m(\Lambda)/a^2$ is greater than the OPE condensate for $a<0.1\,$fm or $(1/a)>2\,$GeV.  At the other extreme, suppose that the lattice spacing, $a$, is large; e.g., $a>1/[2m_\pi(a)]$, where $m_\pi(a)$ is the mass of the pion in the simulation under consideration, then the simulation cannot produce any signal for the collective effect of dynamical chiral symmetry breaking.  In a simulation with $m(a)=50\,$MeV, the pion mass would be $0.4\,$GeV so that $a>1/[2m_\pi(a)]=0.25\,$fm is ``large''.  In a lattice simulation, therefore, one is constrained to a narrow domain, within which one may find an identifiable signal for the OPE condensate:
\begin{equation}
0.1\,{\rm fm} \lesssim a \lesssim 0.25\,{\rm fm}.
\end{equation}
This is made plain in Fig.\,1 of Ref.\,\cite{Langfeld:2003ye} and illustrated in another manner by Fig.\,\ref{FigPCTandy}: any subtraction must reliably extract the dashed curve from the solid curve, whose behaviour comes rapidly to be dominated by the divergence expressed in Eq.\,\eqref{UVcutoffDominance} and illustrated by the dot-dashed curve.

Suppose now that one has achieved simulation parameters for which a signal of the OPE condensate is obtained.  What does that signal describe?  Well, it is widely assumed to be the ``vacuum quark condensate''.  That, however, is a misapprehension.  The critical statement is Eq.\,\eqref{gtlrelE} herein; namely, in the chiral limit the dressed-quark mass function is \emph{identical} to the pseudoscalar component of the pion's canonically normalised Bethe-Salpeter amplitude, where the relative momentum within the bound-state is rigorously identified with the momentum of the dressed-quark.  Thus a properly tuned lattice simulation that obtains what some consider to be a signal for the OPE condensate has actually arrived at a lattice determination of the in-pion condensate.  This is one part of the content of Eq.\,\eqref{qbqpiqbq0}.

It is now natural to ask after a so-called gluon condensate.  In this connection we note that lattice practitioners search for what is, if anything, a small signal by computing the expectation value of an elementary plaquette and subtracting the very large perturbative contribution that dominates the expectation value \cite{Banks:1981zf,DiGiacomo:1981wt}.  Whilst this pattern follows that used to obtain a signal for the in-pion condensate, the difficulties are greater in this instance because the plaquette operator has higher mass-dimension and the analysis requires calculation of the perturbative contribution to an extraordinary level of accuracy; i.e., to very high order: twelve-loops at least \cite{Rakow:2005yn,Horsley:2012ra}.  Since the large-order behavior of the perturbative expansion is expected to be strongly influenced by infrared renormalons \cite{Beneke:1998ui}, which are responsible for factorial growth of coefficients in a continuum perturbative series, there are potentially serious additional complications in computing the subtraction terms.  Indeed, it has been suggested that any ``gluon condensate'' thus obtained is a procedure-dependent concept \cite{Ji:1995fe}.  Renormalons might not be a problem in lattice-QCD, owing to the infrared and ultraviolet cutoffs \cite{Horsley:2012ra}.  Notwithstanding this, few lattice-QCD results are available.  They are just in quenched QCD: a theory with only loose connections to the Standard Model's QCD, which possesses confined gluons and light-quarks; and the positive value is a factor of five larger than inferred via QCD sum rules phenomenology.  In the latter connection, it should be noted that contemporary estimates are compatible with zero, within errors \cite{Ioffe:2002be}.  This possibility is emphasised by the negative value obtained for this sum-rules parameter in an analysis of hadronic $\tau$ decays \cite{Davier:2005xq} and the absence of a signal for a dimension-four condensate in a lattice-QCD determination of $\Lambda_{\rm QCD}$ from the ghost-gluon coupling \cite{Blossier:2011tf}.
Finally, since one may view a ``gluon condensate'', of any dimension \cite{Blossier:2013te}, as part of the dressing that clothes the nonperturbative gluon propagator and because such gluons are confined within hadrons, then all associated ``condensates'', too, are contained within hadrons. 

\subsection{Light-front view}
There is merit now in returning to the point of normal ordering, whose importance in discussing the connection between physical and vacuum matrix elements, especially for condensates, is emphasised in Ref.\,\cite{Brodsky:2009zd}.  In general, normal-ordering in the equal-time second-quantised formulation of  a quantum field theory which exhibits essentially nonperturbative phenomena is an ill-defined operation because, e.g., the exponentiation involved in writing the Heisenberg field operator induces all orders of bare parton creation and annihilation processes, and no finite sum can recover a nonperturbative effect.  As indicated above, a mean-field approximation to a non-confining theory can be used to define a nonperturbative but truly approximate set of states that provides a diagonal basis and associated single quasiparticle creation and annihilation operators.  However, this is impossible for the confined gluons and quarks of QCD.

Of course, the question of normal-ordering is eliminated if one employs the light-front formulation of quantum field theory.  In that case the vacuum is defined as the lowest-mass eigenstate of the associated light-front Hamiltonian by quantising at fixed $\tau= t-z$; and this vacuum is remarkably simple because the kinematic restriction to $k^+=k^0+k^3>0$ ensures that the ground-state of the interacting Hamiltonian is the same as that of the free Hamiltonian.  There are other advantages, too.  The front-form vacuum and its eigenstates are Lorentz invariant, whereas the instant-form vacuum depends on the observer's Lorentz frame.  Moreover, the instant-form vacuum is a state defined at the same time, $t$, at all spatial points in the universe, whereas the front-form vacuum senses only those phenomena which are causally connected; i.e., within an observer's light-cone.

This last point ensures that the front-form is well-suited to computation of the cosmological constant because the constant is a property of the Universe measured within the causal horizon; i.e., it is expressed in the matrix element of the energy-momentum tensor in the background universe, which is completely determined by events that occur within a causally connected domain.  It is practically impossible, on the other hand, to obtain a reliable result using instant-form dynamics since the truncations necessary in order to obtain a result will generally violate Lorentz invariance.  Hence one should not be surprised when expectations based on assumed properties of the vacuum associated with a truncated instant-form Hamiltonian are misleading.

\begin{figure}[t]
\includegraphics[clip,width=0.32\textwidth]{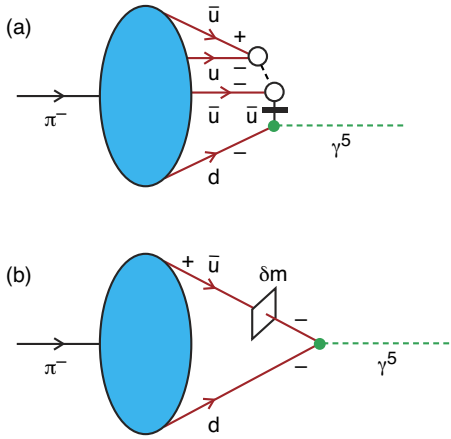}
\vspace*{-33ex}

\rightline{\parbox{27em}{\caption{\label{figinstantaneous}
Light-front contributions to $\rho_\pi=-\langle 0| \bar q \gamma_5 q |\pi\rangle$, $f_\pi \rho_\pi$ is the in-pion condensate.
\emph{Upper panel} -- A non-valence piece of the meson's light-front wave-function, whose contribution to $\rho_\pi$ is mediated by the light-front instantaneous quark propagator (vertical crossed-line).  The ``$\pm$'' denote parton helicity.
\emph{Lower panel} -- There are infinitely many such diagrams, which can introduce chiral symmetry breaking in the light-front wave-function in the absence of a current-quark mass.
(The case of $f_\pi$, which is also an order parameter for DCSB, can be viewed analogously when the $P^-$ current is employed to express this observable.)}}}
\end{figure}

With the shift to a paradigm in which DCSB is expressed as an in-hadron property, one can readily visualize a mechanism that might produce DCSB within the light-front formulation of QCD.  As illustrated in Fig.\,\ref{figinstantaneous}, the light-front-instantaneous quark propagator can mediate a contribution from higher Fock-state components to the matrix elements
\begin{eqnarray}
f_\pi P^- & = & 2 \sqrt N_c \, Z_2 \int_0^1\! dx \! \int\mbox{\footnotesize $\displaystyle\frac{d^2 k_\perp}{16\pi^3}$} \, \psi(x,k_\perp) \frac{k_\perp^2\! + m^{\zeta\,2}}{P^+\,x(1-x)} + \mbox{instantaneous}\,,
\label{LFfpiO} \\
\rho_\pi & = & \sqrt N_c \, Z_2 \int_0^1\! dx\!
\int\mbox{\footnotesize $\displaystyle\frac{d^2 k_\perp}{16\pi^3}$} \,
\psi(x,k_\perp) \, \frac{ m^\zeta }{x (1-x)} + \mbox{instantaneous}\,,
\label{LFrhopiO}
\end{eqnarray}
where $P=(P^+,P^-=m_\pi^2/P^+,\vec{0}_\perp)$ and both currents receive contributions from the ``instantaneous'' part of the quark propagator ($\sim \gamma^+/k^+$) and the associated gluon emission, which are not written explicitly.  In Eqs.\,(\ref{LFfpiO}), (\ref{LFrhopiO}), $\psi(x,k_\perp)$ is the valence-only Fock state of the pion's light-front wave-function.  Diagrams such as those in Fig.\,\ref{figinstantaneous} connect dynamically-generated chiral-symmetry breaking components of the meson's light-front wave-function to the matrix elements in Eqs.\,(\ref{LFfpiO}), (\ref{LFrhopiO}).  There are infinitely many contributions of this type and they do not depend sensitively on the current-quark mass in the neighborhood of the chiral limit.  This mechanism is kindred to that discussed in Ref.\,\cite{Casher:1974xd}.

In closing this section we reiterate that absolute confinement of gluons and quarks is a prerequisite for the containment of condensates within hadrons.  Hence, no model without confinement, even if treated correctly, can undermine the foundations of the in-hadron condensate paradigm in QCD.  The implications of this hypothesis are significant and wide-ranging.  For example, in connection with the cosmological constant, putting QCD condensates back into hadrons reduces the mismatch between experiment and theory by a factor of $10^{46}$.  Furthermore, if technicolour-like theories \cite{Andersen:2011yj,Sannino:2013wla} are the correct scheme for explaining electroweak symmetry breaking, then the impact of the notion of in-hadron condensates is far greater still \cite{Brodsky:2009zd}.

\section{Pion's light-front wave functions}
\label{SecPLFWF}
\subsection{Probability interpretation}
As we noted in connection with Eq.\,\eqref{chipi}, e.g., a meson's Bethe-Salpeter wave function is the quantum field theory analogue of the Schr\"odinger wave function that would describe the system if it were simply quantum mechanical, and whenever a nonrelativistic limit makes sense, the Bethe-Salpeter and Schr\"odinger wave functions become the same in that limit \cite{Salpeter:1951sz,Gross:1982nz}.  For those desiring a probability interpretation of wave functions, this would be reassuring except, of course, for the fact that a nonrelativistic limit is never appropriate when solving continuum bound-state equations for the mass of composite systems containing the light $u$-, $d$- and $s$-quarks.

To explain this comment, consider that the momentum-space wave function for a nonrelativistic quantum mechanical system, $\psi(p,t)$, is a probability amplitude, such that $|\psi(p,t)|^2$ is a non-negative density which expresses the probability that the system is described by momenta $p$ at a given equal-time instant $t$.  Although the replacement of certainty in classical mechanics by probability in quantum mechanics was disturbing for some, the step to relativistic quantum field theory is still more confounding.  Much of the additional difficulty owes to the loss of particle number conservation when this step is made.  Two systems with equal energies need not have the same particle content because that is not conserved by Lorentz boosts and thus interpretation via probability densities is typically lost.  To exemplify: a charge radius cannot generally be defined via the overlap of two wave functions because the initial and final states do not possess the same four-momentum and hence are not described by the same wave function.

Such difficulties may be circumvented by formulating a theory on the light-front because the eigenfunctions of the light-front Hamiltonian are independent of the system's four-momentum \cite{Keister:1991sb,Coester:1992cg,Brodsky:1997de}.  The light-front wave function of an interacting quantum system therefore provides a connection between dynamical properties of the underlying relativistic quantum field theory and notions familiar from nonrelativistic quantum mechanics.  It can translate features that arise purely through the infinitely-many-body nature of relativistic quantum field theory into images whose interpretation is seemingly more straightforward.  Naturally, that is only achieved if the light-front wave function can be calculated.

\subsection{Valence quark distribution amplitude}
\label{sec:vqPDA}
DCSB is a phenomenon for which a quantum mechanical image would be desirable.  Strictly impossible in quantum mechanics with a finite number of degrees-of-freedom, the expression of this striking emergent feature of QCD in the light-front formulation of quantum field theory is only now beginning to be exposed.

As we described in Sec.\,\ref{sec:BSE}, the impact of DCSB is expressed with particular force in properties of the pion.  It is the pseudo-Goldstone boson that emerges when chiral symmetry is dynamically broken, so that its very existence as the lightest hadron is grounded in DCSB.  Consequently, numerous model-independent statements can be made about the pion's Bethe-Salpeter amplitude and its relationship to the dressed-quark propagator, amongst them Eqs.\,\eqref{gtlrelE}--\eqref{gtlrelH}.  Given that the pion's light-front valence-quark distribution amplitude (PDA) can be computed from these two quantities, their calculation provides a means by which to expose DCSB in a wave function with quantum mechanical characteristics.

Consider, therefore, the following projection of the pion's Bethe-Salpeter wave function onto the light-front (we assume isospin symmetry, so that $\hat m_u=\hat m_d=:\hat m$)
\begin{equation}
f_\pi\, \varphi_\pi(x) = {\rm tr}_{\rm CD}
Z_2 \! \int_{dq}^\Lambda \!\!
\delta(n\cdot q_\eta - x \,n\cdot P) \,\gamma_5\gamma\cdot n\, \chi_\pi(q;P)\,,
\label{pionPDA}
\end{equation}
where: $n$ is a light-like four-vector, $n^2=0$; and $P$ is the pion's four-momentum, $P^2=-m_\pi^2$ and $n\cdot P = -m_\pi$, with $m_\pi$ being the pion's mass.  Using Eq.\,\eqref{pionPDA}, one may show that the moments of the distribution; viz., $\langle x^m\rangle := \int_0^1 dx \, x^m \varphi_\pi(x)$, are given by
\begin{equation}
f_\pi (n\cdot P)^{m+1} \langle x^m\rangle =
{\rm tr}_{\rm CD}
Z_2 \! \int_{dq}^\Lambda \!\!
(n\cdot q_\eta)^m \,\gamma_5\gamma\cdot n\, \chi_\pi(q;P)\,.
\label{phimom}
\end{equation}

Significant features of $\varphi_\pi(x)$ in Eq.\,\eqref{pionPDA} can be elucidated algebraically with a simple model before employing numerical solutions for $S(p)$, $\Gamma_\pi$.  To this end, with $\Delta_M(s) = 1/[s+M^2]$ and $\eta = 0$ in Eqs.\,\eqref{chipi}, \eqref{pionPDA}, consider
\begin{eqnarray}
\label{pointS}
S(p) &=& [-i\gamma \cdot p + M] \Delta_M(p^2)\,, \\
%
%
\label{rhoEpi}
\Gamma_\pi(q;P) & = &
i\gamma_5 \frac{M^{1+2\nu}}{f_\pi} \!\! \int_{-1}^{1}\!\! \!dz \,\rho_\nu(z)
\Delta_M^\nu(q_{+z}^2)\,,
\quad \rho_\nu(z) = \frac{1}{\surd \pi}\frac{\Gamma(\nu + 3/2)}{\Gamma(\nu+1)}\,(1-z^2)^\nu\,,
\end{eqnarray}
where $q_{\pm z} = q-(1 \mp z )P/2$.
Inserting Eqs.\,\eqref{pointS}, \eqref{rhoEpi} into Eq.\,\eqref{phimom},
using a Feynman parametrisation to combine denominators,
shifting the integration variable to isolate the integrations over Feynman parameters from that over the four-momentum $q$,
and recognising that $d^4q$-integral as the expression for $f_\pi$, one obtains
\begin{equation}
\label{momnum}
\langle x^m \rangle_\nu =
\frac{\Gamma (2 \nu +2) \Gamma (m+\nu +1)}{\Gamma (\nu +1) \Gamma (m+2 \nu +2)}\,.
\end{equation}

Suppose that $\nu =0$; i.e., as plain from Eq.\,\eqref{rhoEpi}, the pion's Bethe-Salpeter amplitude is independent of momentum and hence describes a point-particle, then Eq.\,\eqref{momnum} yields
\begin{equation}
\label{mom0m}
\langle x^m \rangle_0 =
\frac{\Gamma (2) \Gamma (m+1)}{\Gamma (1) \Gamma (m+2)} = \frac{1}{m+1}\,.
\end{equation}
These are the moments of the distribution amplitude
\begin{equation}
\varphi_\pi(x) = 1\,,
\end{equation}
which is indeed that of a pointlike pion \cite{Roberts:2010rnS}.

Alternatively, consider $\nu=1$.  Then $\Gamma_\pi(k^2) \sim 1/k^2$ for large relative momentum.  This is the behaviour in QCD at $k^2\gg M_g^2$, where $M_g \simeq 0.5\,$GeV is the  dynamically generated gluon mass discussed in connection with Fig.\,\ref{fig:gluonrunning}.  With $\nu=1$, Eq.\,\eqref{momnum} yields
\begin{equation}
\label{mom1m}
\langle x^m \rangle_1 = 
\frac{\Gamma (4) \Gamma (m+2)}{\Gamma (2) \Gamma (m+4)} = \frac{6}{(m+3)(m+2)}\,.
\end{equation}
These are the moments of
\begin{equation}
\varphi_\pi^{\rm asy}(x) = 6 \,x \,(1-x)\,,
\label{PDAasymp}
\end{equation}
which is precisely the asymptotic distribution amplitude for the pion in QCD \cite{Efremov:1979qk,Lepage:1979zb,Lepage:1980fj}.  (An analysis of the domain upon which this form might provide a good approximation to the PDA is presented in Sec.\,\ref{sec:phiasy}.)

It is readily established that with Eqs.\,\eqref{pointS}--\eqref{rhoEpi} in  Eq.\,\eqref{phimom} one obtains the ``asymptotic'' distribution associated with a $(1/k^2)^\nu$ vector-exchange interaction; viz.,
\begin{equation}
\label{asymptdistn}
\varphi_\pi(x) = \frac{\Gamma(2\nu + 2)}{\Gamma(\nu+1)^2}\, x^\nu (1-x)^\nu\,.
\end{equation}
Notably, the $z$-modulated dependence on $q\cdot P$ in Eq.\,\eqref{rhoEpi}
is the critical factor in obtaining the results described here.  To illustrate, if one uses $\nu=1$ but $2\rho_\nu(z) = \delta(1-z)+\delta(1+z)$, then point-particle moments, Eq.\,\eqref{mom0m}, are obtained even though $\Gamma_\pi(k^2) \sim 1/k^2$ for $k^2 \gg M_g^2$.  There is a natural explanation.  Namely, with such a form for $\rho_\nu(z)$ one assigns equal probability to two distinct configurations: valence-quark with all the pion's momentum and valence-antiquark with none or antiquark with all the momentum and quark with none.  In assigning equal weight to these two extreme configurations one has defined a bound-state with point-particle-like characteristics.  It follows that deviations from the asymptotic distribution may be expressed through $\rho_\nu(z)$.

In Ref.\,\cite{Chang:2013pq} the gap and pion Bethe-Salpeter equations were solved numerically using the interaction in Ref.\,\cite{Qin:2011dd}, which preserves the one-loop renormalisation group behaviour of QCD and guarantees that the quark mass-function, $M(p^2)= B(p^2,\zeta^2)/A(p^2,\zeta^2)$, is independent of the renormalisation point, which was chosen to be $\zeta=2\,$GeV.  In completing the gap and Bethe-Salpeter kernels,  Ref.\,\cite{Chang:2013pq} employed two different procedures and compared their results: RL (rainbow-ladder) truncation, the most widely used DSE computational scheme in hadron physics, detailed in App.\,A.1 of Ref.\,\cite{Chang:2012cc}; and the DCSB-improved (DB) kernels detailed in App.\,A.2 of Ref.\,\cite{Chang:2012cc}, which are the most refined kernels currently available.  Both schemes are symmetry-preserving and hence ensure Eq.\,\eqref{gtlrelE}; but the latter incorporates essentially nonperturbative effects associated with DCSB into the kernels, which are omitted in rainbow-ladder truncation and any stepwise improvement thereof \cite{Chang:2009zb}.

\begin{figure}[t]
\leftline{\includegraphics[width=0.45\linewidth]{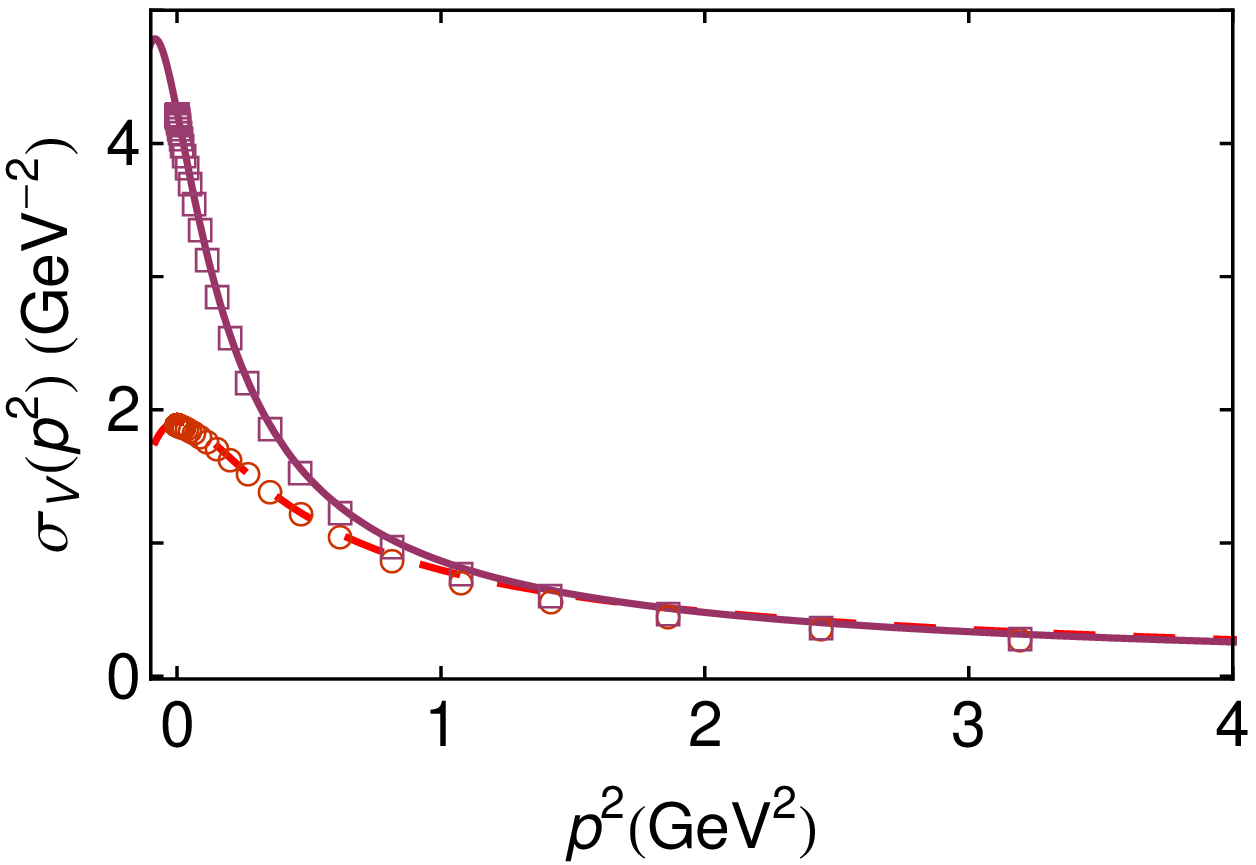}}

\vspace*{-34ex}

\rightline{\includegraphics[width=0.45\linewidth]{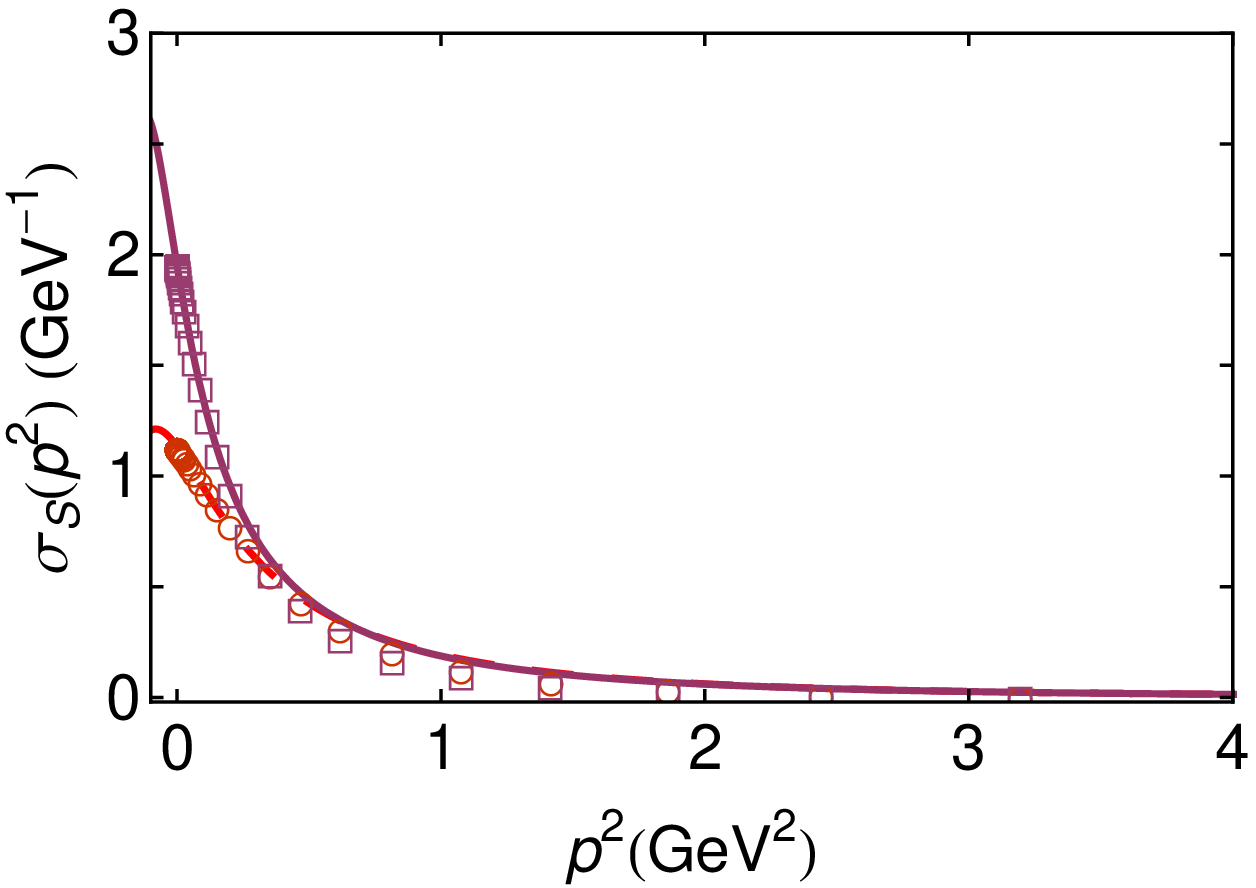}}

\caption{Functions characterising the dressed quark propagator.
\emph{Left panel}. $\sigma_V(p^2)$ -- RL kernel: solution (open circles) and interpolation function (long-dashed curve); and DB kernel: solution (open squares) and interpolation function (solid curve).
\emph{Right panel}. $\sigma_S(p^2)$, with same legend.
In the chiral limit at large $p^2$, $\sigma_V(p^2)\sim 1/p^2$ and $\sigma_S(p^2) \sim 1/p^4$.
\label{fig:Splot}}
\end{figure}

A particular feature of Ref.\,\cite{Chang:2013pq} is a novel technique for what amounts to solving the practical problem of continuing from Euclidean metric to Minkowski space, which is where the light-front is defined.  It was noted therein that solutions of the gap and Bethe-Salpeter equations are typically obtained as matrices and computation of the moments in Eq.\,\eqref{phimom} is cumbersome with such input.  In fact, it is practically impossible.  Thus, Ref.\,\cite{Chang:2013pq} employed algebraic parametrisations of each array to serve as interpolations in evaluating the moments.  The scalar functions $\sigma_{V,S}$ in the dressed-quark propagator, Eq.\,\eqref{SgeneralN}, were represented as meromorphic functions with no poles on the real $p^2$-axis \cite{Bhagwat:2002tx}, a feature consistent with confinement, as explained in Sec.\,\ref{sec:confinement}.  Regarding the pion's Bethe-Salpeter amplitude, each scalar function was expressed via a Nakanishi-like representation \cite{Nakanishi:1963zz,Nakanishi:1969ph,Nakanishi:1971}; i.e., through integrals like Eq.\,\eqref{rhoEpi}, with parameters fitted to that function's first two $q\cdot P$ Chebyshev moments.  (Details are presented in App.\,\ref{app:nakanishi}.)  The quality of the description is illustrated via the dressed-quark propagator in Fig.\,\ref{fig:Splot}.

Using Eq.\,\eqref{phimom} and the representations described above, it is straightforward to compute arbitrarily many moments of the pion's PDA, $\{\langle x^m\rangle|m=1,\ldots,m_{\rm max}\}$: typically, Ref.\,\cite{Chang:2013pq} employed $m_{\rm max}=50$.
Since Gegenbauer polynomials of order $\alpha$, $\{C_n^{\alpha}(2 x -1)| n=0,\ldots,\infty\}$, are a complete orthonormal set on $x\in[0,1]$ with respect to the measure $[x (1-x)]^{\alpha_-}$, $\alpha_-=\alpha-1/2$, they enable reconstruction of any function defined on $x\in[0,1]$. (N.B.\, $\varphi_\pi(x)$ is even under $x\leftrightarrow (1-x)=:\bar x$.  It vanishes at the endpoints unless the interaction is momentum-independent.)
One may therefore write
\begin{equation}
\label{PDAGalpha}
\varphi_\pi^{G_s}(x) = N_\alpha [x \bar x]^{\alpha_-}
\bigg[ 1 + \sum_{2,4,\ldots}^{j_s} a_j^\alpha C_j^\alpha(x -\bar x) \bigg],
\end{equation}
where $\alpha_-=\alpha-1/2$ and $N_\alpha = \Gamma(2\alpha+1)/[\Gamma(\alpha+1/2)]^2$, and minimise $\varepsilon_s = \sum_{m=1,\ldots,m_{\rm max}} |\langle x^m\rangle^{G_s}/\langle x^m\rangle-1|$.  It was found that a value of $j_s=2$ ensured mean-\{$|\langle x^m\rangle^{G_{s+2}}/\langle x^m\rangle^{G_{s}}-1||m=1,\ldots,m_{\rm max}\}< 1$\%.
In using Gegenbauer-$\alpha$ polynomials, Ref.\,\cite{Chang:2013pq} allowed the PDA to differ from $\varphi_\pi^{\rm asy}$ for any finite $\zeta$ and thereby accelerated the procedure's convergence by optimising $\alpha$.  One may project the result thus obtained onto a $\{C_n^{3/2}\}$-basis, which is that used by other authors, but this incurs significant costs, as will be explained below.

\begin{figure}[t]
\leftline{\includegraphics[width=0.45\linewidth]{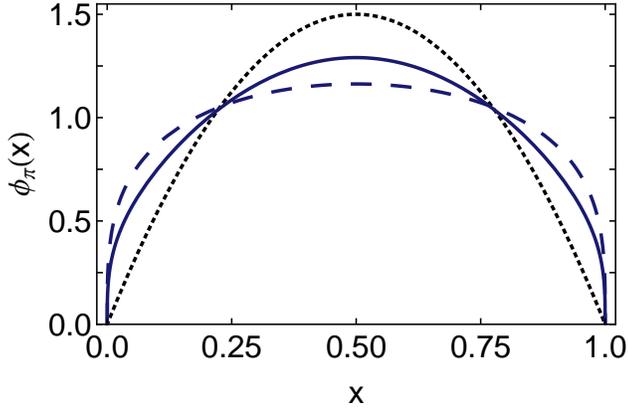}}
\vspace*{-26ex}

\rightline{\parbox{23em}{\caption{
%
Distribution amplitude at $\zeta=2\,$GeV, computed in Ref.\,\protect\cite{Chang:2013pq}.
Curves: solid, DCSB-improved kernel (DB); dashed, rainbow-ladder (RL); and dotted, asymptotic distribution.
\label{fig:phiplot}}}}\vspace*{11ex}
\end{figure}

The dashed curve in Fig.~\ref{fig:phiplot} is the RL result, obtained with $D\omega = (0.87\,{\rm GeV})^3$, $\omega=0.5\,$GeV.  It is described by
\begin{equation}
\label{resphipi2}
\varphi_\pi^{\rm RL}(x) = 1.74 [x \bar x]^{\alpha_-^{\rm RL}} \, [1 + a_2^{\rm RL} C_2^{\alpha_{\rm RL}}(x - \bar x)]\,,
\end{equation}
with $\alpha_{\rm RL} = 0.79$, $a_2^{\rm RL}=0.0029$.  Projected onto a Gegenbauer-$(\alpha=3/2)$ basis, Eq.\,\eqref{resphipi2} corresponds to $a_2^{(3/2)}=0.23$, \ldots, $a_{14}^{(3/2)}=0.022$, etc.  That $j\geq 14$ is required before $a_{j}^{(3/2)}<0.1\,a_2^{(3/2)}$ underscores the merit of reconstruction via Gegenbauer-$\alpha$ polynomials at any reasonable scale, $\zeta$.  The merit is greater still if, as in lattice-QCD, one only has access to a single nontrivial moment.  In seeking an estimate of $\varphi_\pi(x)$, it is better to fit $\alpha$ than to force $\alpha=3/2$ and infer a value for $a_2^{(3/2)}$.

The solid curve in Fig.\,\ref{fig:phiplot}, described by
\begin{equation}
\label{resphipi2DB}
\varphi_\pi^{\rm DB}(x) = 1.81 [x \bar x]^{\alpha_-^{\rm DB}} \, [1 + a_2^{\rm DB} C_2^{\alpha_{\rm DB}}(x - \bar x)]\,,
\end{equation}
$\alpha_{\rm DB} = 0.81$, $a_2^{\rm DB}=-0.12$, was obtained using the most sophisticated symmetry-preserving DSE kernels that are currently available \cite{Chang:2011ei}, with $D\omega = (0.55\,{\rm GeV})^3$ and an anomalous chromomagnetic moment $\tilde\eta=0.6$.  Projected onto a $\{C_n^{3/2}\}$-basis,
Eq.\,\eqref{resphipi2DB} 
corresponds to $a_2^{(3/2)}=0.15$.  Only for $j\geq 14$ is $a_{j}^{(3/2)}<0.1 \, a_2^{(3/2)}$.

By way of context, it is worth noting that a computation using QCD sum rules \cite{Braun:1988qv} produced $\varphi_\pi(x=1/2,\zeta_1) = 1.2 \pm 0.3$, where $\zeta_1=1\,$GeV.  This value may reasonably be compared with:
\begin{equation}
\varphi_\pi^{\rm RL}(1/2,\zeta_2) = 1.16\,,\quad
\varphi_\pi^{\rm DB}(1/2,\zeta_2) = 1.29\,;
\end{equation}
and the limiting value, obtained from the asymptotic form, viz. $\varphi_\pi^{\rm asy}(1/2) = 1.5$.  There are also model predictions; e.g., $\varphi_\pi(1/2)=1.11$ in a relativistic constituent-quark model \cite{Frederico:1994dx}, and  $\varphi_\pi(1/2)=1.27$ in a nonlocal condensate model \cite{Mikhailov:1986be} and an AdS/QCD model \cite{Brodsky:2006uqa}, which both argue in favour of the same distribution; viz.,
\begin{equation}
\label{phiModel}
\varphi_\pi^{M}(x) = \frac{8}{\pi} \sqrt{x \bar x}\,.
\end{equation}
The problem with these and kindred model results is that the scales at which they should hold are unknown in principle.

Another point of comparison is found in a second moment of the distribution.  The DSE distribution amplitudes produce
\begin{equation}
\langle (2x-1)^2 \rangle^{\rm RL} = 0.28\,,\quad
\langle (2x-1)^2 \rangle^{\rm DB} = 0.25\,,
\end{equation}
in agreement with the lattice-QCD result reported in Ref.\,\cite{Braun:2006dg}:
\begin{equation}
\label{latticemoment}
\int_0^1 dx\, (2 x - 1)^2 \, \varphi_\pi^{\rm LQCD}(x,\tau_2) = 0.27\pm 0.04\,,
\end{equation}
obtained at the same renormalisation scale using two flavours of dynamical, $O(a)$-improved Wilson fermions and linearly extrapolating to the empirical pion mass, $\hat m_\pi$, from results at $m_\pi^2/\hat m_\pi^2 = 20,35,50$.\footnote{
At next-to-leading order in chiral perturbation theory, all non-analytic corrections to the relevant matrix element are contained in $f_\pi$ \protect\cite{Chen:2003fp,Chen:2005js}, so a linear-in-$m_\pi^2$ extrapolation of the moment itself can be reliable.}  The uncertainty in Eq.\,\eqref{latticemoment} is large, however, as may be seen by comparing the indicated range with moment values computed using the two limiting extremes $\varphi_\pi = \varphi_\pi^{\rm asy}$ and $\varphi_\pi =\,$constant; viz., $1/5$ and $1/3$, respectively.  A more accurate result would be valuable and is anticipated \cite{braunprivate:2013}.
The models noted above yield $0.28$ \cite{Frederico:1994dx} and $1/4$ \cite{Mikhailov:1986be,Brodsky:2006uqa} but, again, the scales at which these results should individually be relevant are unknown.

It should be plain from these observations that neither a single piece of local nor a single piece of global information, nor even a combination of the two can serve as anything like a tight constraint on the pion's PDA.  One must obtain a complete map of the pointwise behaviour.  This returns us to Fig.\,\ref{fig:phiplot}, from which numerous qualitatively significant results can be read.  Here we stress two.
The most important being that DCSB is expressed in the PDA through a marked broadening with respect to $\varphi_\pi^{\rm asy}$.  This may be claimed because the PDAs in Fig.\,\ref{fig:phiplot} were computed at a low renormalisation scale in the chiral limit, whereat the quark mass function owes entirely to DCSB; and, on the domain $0<p^2<\zeta^2$, the nonperturbative interactions responsible for DCSB produce significant structure in the dressed-quark's self-energy.  The PDA is an integral of the pion's Bethe-Salpeter wave function, whose pointwise behaviour is rigorously connected with that of the quark self-energy [see Eqs.\,\eqref{gtlrelE}--\eqref{gtlrelH}].  Hence, the structure of the pion's distribution amplitude at the hadronic scale is a pure expression of DCSB.  As the scale is removed to extremely large values, phase space growth diminishes the impact of nonperturbative DCSB interactions, so that the PDA relaxes to its asymptotic form.

Significant, too, is the pointwise difference between the DB and RL results.  It is readily understood, bearing in mind that low-$m$ moments are most sensitive to $\varphi_\pi(x)$ in the neighbourhood of $x=1/2$, whereas high-$m$ moments are sensitive to its endpoint behaviour.
RL-kernels ignore DCSB in the quark-gluon vertex.  Therefore, to describe a given body of phenomena, they must shift all DCSB-strength into the infrared behaviour of the quark propagator, whilst nevertheless maintaining perturbative behaviour for $p^2>\zeta^2$.  This requires $B(p^2)$ to be large at $p^2=0$ but drop quickly, behaviour which influences $\varphi_\pi(x)$ via Eq.\,\eqref{gtlrelE}.  The concentration of strength at $p^2\simeq 0$ forces large values for the small-$m$ moments, which translates into a broad distribution.
In contrast, the DB-kernel builds DCSB into the quark-gluon vertex and its impact is therefore shared between more elements of a calculation.  Hence a smaller value of $B(p^2=0)$ is capable of describing the same body of phenomena; and this self-energy need fall less rapidly in order to reach the common asymptotic limit.  (Using Eqs.\,\eqref{SgeneralN}, these remarks become evident in Fig.\,\ref{fig:Splot}.)  It follows that the low-$m$ moments are smaller and the distribution is narrower.
Both PDAs have the same large-$x$ behaviour because the RL and DB kernels agree at ultraviolet momenta.

The DSE PDA computations unify a diverse range of phenomena.  The rainbow-ladder result, e.g., connects directly with \emph{ab initio} predictions for: $\pi \pi$ scattering, and pion electromagnetic elastic and transition form factors \cite{Maris:2003vk}; and nucleon and $\Delta$ properties \cite{Eichmann:2011ej}.  And, as mentioned in Sec.\,\ref{sec:GapEquation}, although use of DCSB-improved kernels is just beginning, the related DSE prediction for the PDA links immediately with analyses showing that DCSB is, e.g., responsible for both a large dressed-quark anomalous magnetic moment \cite{Chang:2010hb,Kochelev:1996pv,Bicudo:1998qb,Diakonov:2002fq} and the splitting between parity partners in the spectrum \cite{Chang:2011ei,Chen:2012qr}.

In closing this subsection it is worth reiterating that the pion's PDA is the closest thing in QCD to a quantum mechanical wave function for the pion.  Its hardness at an hadronic scale is a direct expression of DCSB.

\subsection{PDA evolution and \mbox{$\mathbf \varphi_\pi^{\rm asy}(x)$}}
\label{sec:phiasy}
In the preceding subsection we did not present any details about the dependence of the PDA on the momentum-scale $\zeta$ or, equivalently, the length-scale $\tau=1/\zeta$, which characterises the process in which the pion is involved.  On the domain within which QCD perturbation theory is valid, the equation describing the $\tau$-evolution of $\varphi_\pi(x;\tau)$ is known and has the solution  \cite{Efremov:1979qk,Lepage:1980fj}
\begin{equation}
\label{PDAG3on2}
\varphi_\pi(x;\tau) = \varphi^{\rm asy}_\pi(x)
\bigg[ 1 + \!\! \sum_{j=2,4,\ldots}^{\infty} \!\! \!\! a_j^{3/2}(\tau) \,C_j^{(3/2)}(x -\bar x) \bigg],\;\;\\
\end{equation}
where $\varphi^{\rm asy}_\pi(x)$ is given in Eq.\,\eqref{PDAasymp} and the expansion coefficients $\{a_j^{3/2},j=1,\ldots,\infty\}$ evolve logarithmically with $\tau$: they vanish as $\tau\to 0$.  These features owe to the fact that in the neighbourhood $\tau \Lambda_{\rm QCD} \simeq 0$, QCD is invariant under the collinear conformal group
SL$(2;\mathbb{R})$
\cite{Brodsky:1980ny,Braun:2003rp}.  Indeed, the Gegenbauer-$\alpha=3/2$ polynomials are merely irreducible representations of this group.  A correspondence with the spherical harmonics expansion of the wave functions for $O(3)$-invariant systems in quantum mechanics is plain.

In the absence of additional information, it has commonly been assumed that at any length-scale $\tau$, a useful approximation to $\varphi_\pi(x;\tau)$ is obtained by using just the first few terms of the expansion in Eq.\,\eqref{PDAG3on2}.  (This assumption has led to models for $\varphi_\pi(x)$ whose pointwise behaviour is not concave on $x\in[0,1]$; e.g., to ``humped'' distributions \cite{Chernyak:1983ej}.)  Whilst the assumption is satisfied on $\tau \Lambda_{\rm QCD} \simeq 0$, it is hard to justify at the length-scales available in typical contemporary experiments, which correspond to $\zeta \simeq 2\,$GeV.  This is highlighted by the fact that $\varphi^{\rm asy}_\pi(x)$ can only be a good approximation to the pion's PDA when it is accurate to write $u_{\rm v}^\pi(x) \approx \delta(x)$, where $u_{\rm v}^\pi(x)$ is the pion's valence-quark distribution function \cite{Georgi:1951sr,Gross:1974cs,Politzer:1974fr}.  This is far from valid at currently accessible momentum scales \cite{Holt:2010vj,Nguyen:2011jy,Hecht:2000xa,Wijesooriya:2005ir,Aicher:2010cb}, as we will subsequently make plain.

To illustrate these remarks, consider that a value
\begin{equation}
\label{a2Old}
a_2^{3/2}(\tau_2)=0.201(114)\,,
\end{equation}
$\tau_2=1/[2\,{\rm GeV}]$, was obtained using Eq.\,\eqref{PDAG3on2} as a tool for expressing the result of a numerical simulation of lattice-regularised QCD \cite{Braun:2006dg}.  This indicates a large correction to the asymptotic form, $\varphi^{\rm asy}_\pi(x)$, and gives no reason to expect that the ratio $a_4^{3/2}(\tau_2)/a_2^{3/2}(\tau_2)$ is small.  Now, at leading-logarithmic accuracy the moments in Eq.\,\eqref{PDAG3on2} evolve from $\tau_2\to \tau$ as follows \cite{Efremov:1979qk,Lepage:1980fj}:
\begin{equation}
\label{llevolution}
a_j^{3/2}(\tau) = a_j^{3/2}(\tau_2) \left[\frac{\alpha_s(\tau_2)}{\alpha_s(\tau)}\right]^{\gamma_j^{(0)}/\beta_0},
\end{equation}
where the one-loop strong running-coupling is given in Eq.\,\eqref{alphaOneLoop} and, with $C_F = 4/3$,
\begin{equation}
\gamma_j^{(0)} = C_F\left[3 + \frac{2}{(j+1)\,(j+2)} - 4\,\sum_{k=1}^{j+1}\,\frac{1}{k}\right].
%
\end{equation}
Using $n_f=4$ and $\Lambda_{\rm QCD}=0.234\,$GeV for illustration \cite{Qin:2011dd}, it is necessary to evolve to $\tau_{100}=1/[100\,{\rm GeV}]$, before $a_2^{3/2}(\tau)$ even falls to 50\% of its value in Eq.\,\eqref{a2Old}.  The $a_4^{3/2}$ coefficient still holds 37\% of its value at $\tau_{100}$.  This pattern is qualitatively preserved with higher order evolution \cite{Mikhailov:1984ii,Mueller:1998fv}.  These observations suggest that the asymptotic domain lies at very large momenta indeed.  We will return to this point.

As explained in Sec.\,\ref{sec:vqPDA}, the pion's valence-quark PDA has recently been computed using QCD's DSEs; and at the scale $\zeta=2\,$GeV, $\varphi_\pi(x;\tau_2)$ is much broader than the asymptotic form, $\varphi^{\rm asy}_\pi(x)$ in Eq.\,\eqref{PDAasymp}.  Indeed, the power-law dependence is better characterised by $[x \bar x]^{\alpha_-}$ with $\alpha_- \approx 0.3$, a value very different from that associated with the asymptotic form; viz., $\alpha_-^{\rm asy}=1$.  This dilation is a long-sought and unambiguous expression of dynamical chiral symmetry breaking (DCSB) on the light-front \cite{Chang:2011mu,Brodsky:2010xf,Brodsky:2012ku}.

If one insists on using Eq.\,\eqref{PDAG3on2} to represent such a broad distribution, then, as we saw in connection with Eqs.\,\eqref{resphipi2}, \eqref{resphipi2DB}, $a_{14}^{3/2}$ is the first expansion coefficient whose magnitude is less-than 10\% of $a_{2}^{3/2}$.  This is not surprising for the following reasons.  The polynomials $\{C_j^{(3/2)}(x-\bar x),j=1,\ldots,\infty\}$ are a complete orthonormal set on $x\in[0,1]$ with respect to the measure $x\bar x$.  Just as any attempt to represent a box-like curve via a Fourier series will inevitably lead to slow convergence and spurious oscillations, so does the use of Gegenbauer polynomials of order $\alpha=3/2$ to represent a function better matched to the measure $[x\bar x]^{0.3}$.  This latter measure is actually associated with Gegenbauer polynomials of order $\alpha = 4/5$.  Observations such as these led to the PDA reconstruction method adopted in Ref.\,\cite{Chang:2013pq}.

As a framework within continuum quantum field theory, the DSE study of Ref.\,\cite{Chang:2013pq} was able to reliably compute arbitrarily many moments of the PDA using Eq.\,\eqref{pionPDA} and from those moments reconstruct the distributions in Fig.\,\ref{fig:phiplot}.  The concave nature of the PDA is important and one may readily see that it is natural.  To this end, consider that the pion multiplet contains a charge-conjugation eigenstate.  Therefore, the peak in the leading Chebyshev moment of each of the three significant scalar functions that appear in the expression for $\Gamma_\pi(q;P)$ occurs at $2 k_{\rm rel}:=q_\eta + q_{\bar\eta} = 0$; i.e., at zero relative momentum \cite{Qin:2011xq,Maris:1997tm}.  Moreover, these Chebyshev moments are monotonically decreasing with $k_{\rm rel}^2$.  Such observations suggest that $\varphi_\pi(x)$ should exhibit a single maximum, which appears at $x=1/2$; i.e., $\varphi_\pi(x)$ is a symmetric, concave function on $x\in [0,1]$.

We saw in Sec.\,\ref{sec:vqPDA} that one may efficiently reconstruct $\varphi_\pi(x)$ from its moments by using Gegenbauer polynomials of order $\alpha$, with this order -- the value of $\alpha$ -- determined by the moments themselves, not fixed beforehand.  Namely, using Eq.\,\eqref{PDAGalpha}, very rapid progress from the moments to a converged representation of the PDA was obtained.  Indeed, $j_s=2$ was sufficient, with $j_s=4$ producing no change in a sensibly plotted curve that was greater than the line-width.  Naturally, once obtained in this way, one may project $\varphi_\pi(x;\tau)$ onto the form in Eq.\,\eqref{PDAG3on2}; viz., for $j=2,4,\ldots\,$,
\begin{equation}
\label{projection}
a_j^{3/2} = \frac{2}{3}\ \frac{2\,j+3}{(j+2)\,(j+1)}\int_0^1 dx\, C_j^{(3/2)}(2\,x-1)\,\varphi_\pi(x),
\end{equation}
therewith obtaining all coefficients necessary to represent any computed distribution in the conformal form without ambiguity or difficulty.

It was shown in Ref.\,\cite{Cloet:2013tta} that one can take this approach a step further; viz., adopting it, too, when presented even with only limited information about $\varphi_\pi(x;\tau)$.  In this connection, consider that since discretised spacetime does not possess the full rotational symmetries of the Euclidean continuum, then, with current algorithms, only one nontrivial moment of $\varphi_\pi(x)$ can be computed using numerical simulations of lattice-regularised QCD.  Thus, Eq.\,\eqref{latticemoment} is the lone existing piece of lattice-QCD information on $\varphi_\pi(x)$.  As described in connection with Eq.\,\eqref{a2Old}, it was used in Ref.\,\cite{Braun:2006dg} to constrain $a_2^{3/2}(\tau_2)$ in Eq.\,\eqref{PDAG3on2} and therewith produce a ``double-humped'' PDA.  It is straightforward to establish that a double-humped form lies within the class of distributions produced by a pion Bethe-Salpeter amplitude that may be characterised as vanishing at zero relative momentum, instead of peaking thereat.

\begin{figure}[t]
\leftline{\includegraphics[clip,width=0.47\linewidth]{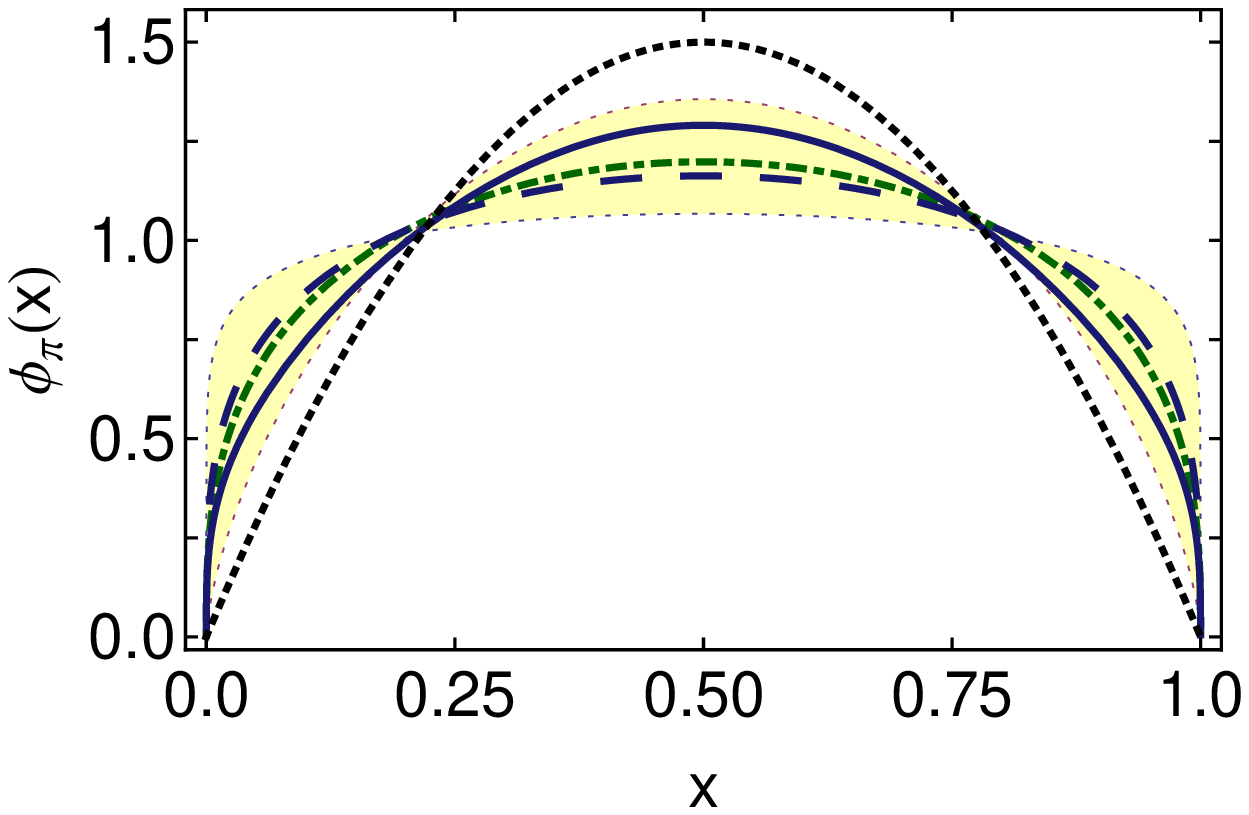}}
\vspace*{-34.3ex}

\rightline{\includegraphics[clip,width=0.47\linewidth]{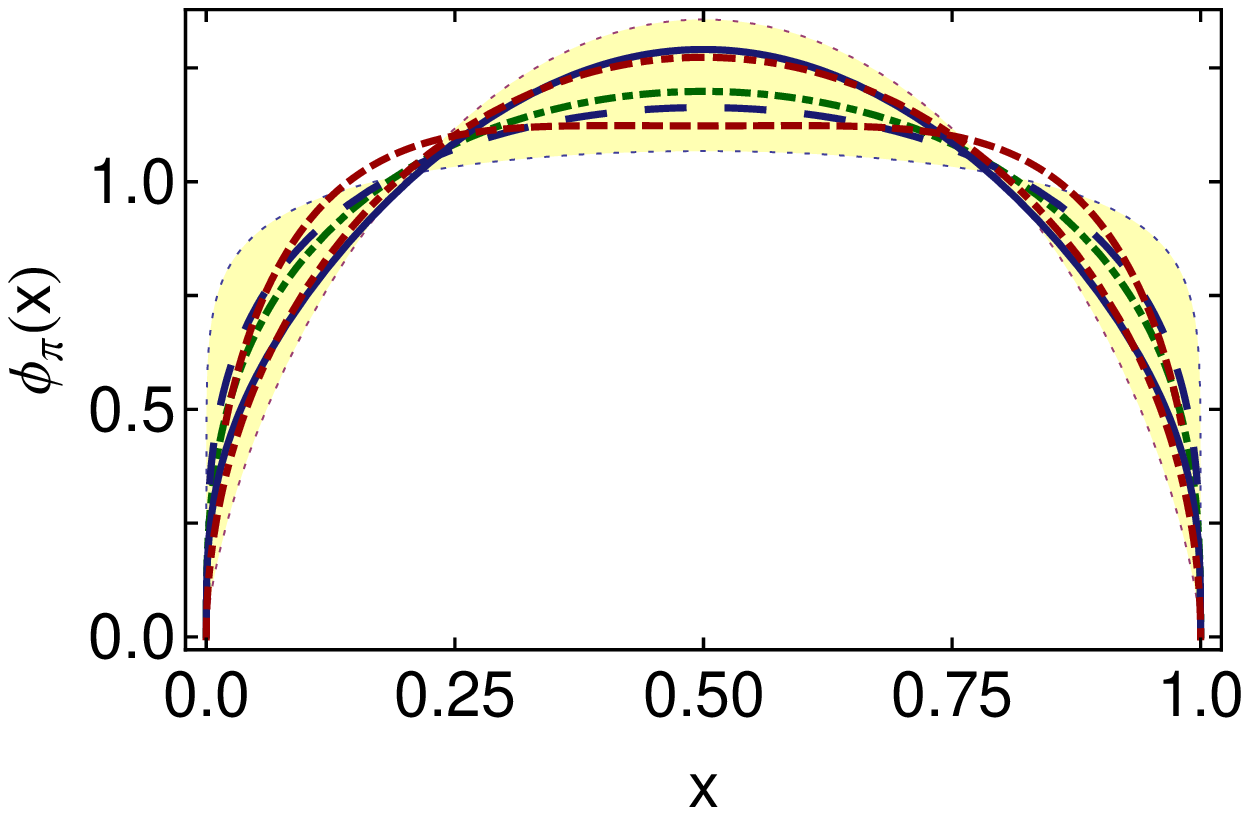}}
\caption{\label{Fig1Gegenbauer}
\underline{Left panel}.  \emph{Dot-dashed curve}, embedded in the shaded region, $\varphi_\pi(x;\tau_2)$ in Eq.\,\protect\eqref{pionPDAlattice}.  The shaded region indicates the extremes allowed by the errors on $\alpha_-$.
For comparison, the DSE results described in Sec.\,\protect\ref{sec:vqPDA} are also depicted: \emph{solid curve}, $\varphi_\pi(x;\tau_2)$ obtained with the best DSE truncation currently available, which includes important features of DCSB in building the kernels; and \emph{dashed curve}, result obtained in rainbow-ladder truncation.
The \emph{dotted curve} is $\varphi_\pi^{\rm asy}(x)$.
\underline{Right panel}.  Same as left panel except that $\varphi_\pi^{\rm asy}(x)$ is omitted and two curves are added: \emph{short dashed}, the constituent-quark model result in Ref.\,\protect\cite{Frederico:1994dx}; and \emph{dot-dot-dash-dash} curve, the result in Eq.\,\protect\eqref{phiModel} \cite{Mikhailov:1986be,Brodsky:2006uqa}.  The latter is practically indistinguishable from the DSE result obtained in the DB truncation whereas the former is inconsistent with all other curves and bounds displayed.
}
\end{figure}

Now, suppose instead that one analyses the single unit of information in Eq.\,\eqref{latticemoment} using Eq.\,\eqref{PDAGalpha} but discarding the sum, a procedure which acknowledges implicitly that the pion's PDA should exhibit a single maximum at $x=1/2$.  Then Eq.\,\eqref{latticemoment} constrains $\alpha$, with the result
\begin{equation}
\label{pionPDAlattice}
\varphi_\pi(x;\tau_2) = N_\alpha \, x^{\alpha_-} (1-x)^{\alpha_-},\;
\alpha_- = 0.35^{+ 0.32 = 0.67}_{-0.24=0.11},
\end{equation}
which is depicted in Fig.\,\ref{Fig1Gegenbauer}.  Employed thus, the lattice-QCD result, Eq.\,\eqref{pionPDAlattice}, produces a concave amplitude in agreement with contemporary DSE studies and confirms that the asymptotic distribution, $\varphi_\pi^{\rm asy}(x)$, is not a good approximation to the pion's PDA at $\zeta=2\,$GeV.

Equation~\eqref{pionPDAlattice} actually favours the DSE result obtained with the interaction of Ref.\,\cite{Qin:2011dd} and the RL truncation  (dashed curve).  The other DSE prediction (solid curve) was obtained with the same interaction but using the DB kernels for the gap and Bethe-Salpeter equations.  The solid curve should therefore provide the more realistic result.  That the PDA inferred from Eq.\,\eqref{latticemoment} is closer to the RL result is nonetheless readily understood.  As just described, RL computations omit important features of DCSB and, in being obtained by linearly extrapolating from large pion masses, so, effectively, does the lattice result.  It should be anticipated that improved lattice simulations will produce a PDA in better agreement with the solid curve in Fig.\,\ref{Fig1Gegenbauer}.

The right panel of Fig.\,\ref{Fig1Gegenbauer} also displays the PDAs computed using the models described in Refs.\,\cite{Frederico:1994dx,Mikhailov:1986be,Brodsky:2006uqa}.  The qualitatively important observation is that each framework produces a broad concave distribution.  In detail, one will observe that the pointwise behaviour of the constituent-quark model result \cite{Frederico:1994dx} is not supported by either the lattice-QCD constraint or the DSE predictions.  On the other hand, there is a remarkable agreement between the DSE-DB result and that in Eq.\,\eqref{phiModel}, obtained in the nonlocal condensate and AdS/QCD models \cite{Mikhailov:1986be,Brodsky:2006uqa}.  Measured via the $L^1$-norm, the difference between these functions is just 2\%, whereas the $L^1$-difference between the DSE-DB result and the asymptotic PDA is 15\%.  (The $L^1$-difference between the constituent-quark model result \cite{Frederico:1994dx} and the DSE-DB curve is 10\%.)

The authors of Refs.\,\cite{Mikhailov:1986be,Brodsky:2006uqa} imagine that their models determine the PDA at a scale $\zeta_0 \lesssim 1\,$GeV.  If one accepts this at face value and assumes that Eq.\,\eqref{phiModel} is valid at $\zeta_0 = 1\,$GeV, then using leading order evolution one obtains
\begin{equation}
\label{phiModel4}
\varphi_\pi^{M}(x;\zeta_2) \approx N_{1.1} [x \bar x]^{0.6}\,.
\end{equation}
The $L^1$-difference between this evolved result and the DB curve in Fig.\,\ref{Fig1Gegenbauer} is 4\%.  Thus a lack of knowledge about the precise scale at which Eq.\,\eqref{phiModel} is valid has no material impact on the picture we have drawn; namely, at a scale relevant to contemporary experiment the leading-order leading-twist PDA is a broad, concave function.

It is perhaps worth emphasising and illustrating that information is gained using the procedure advocated in Ref.\,\cite{Cloet:2013tta} but not lost, so we list the first three Gegenbauer-$\alpha=3/2$ moments computed by reprojecting Eq.\,\eqref{pionPDAlattice} onto the expansion in Eq.\,\eqref{PDAG3on2}, using Eq.\,\eqref{projection}:
\begin{equation}
\label{a246New}
a_2^{3/2}(\tau_2) = 0.20 \pm 0.12 \,,\quad
a_4^{3/2}(\tau_2) = 0.093 \pm 0.064 \,,\quad
a_6^{3/2}(\tau_2) = 0.055 \pm 0.041 \,.
\end{equation}
Naturally, the result for $a_2^{3/2}(\tau_2)$ is equivalent to that in Eq.\,\eqref{a2Old}, but the values of $a_{4,6}^{3/2}(\tau_2)$ provide new information, which might be used to inform other approaches to the problem of computing $\varphi_\pi(x)$; e.g., Refs.\,\cite{Radyushkin:2009zg,Agaev:2010aq,Bakulev:2012nh,ElBennich:2012ij}.  Moreover, with Eq.\,\eqref{pionPDAlattice} one obtains
\begin{equation}
\varphi_\pi(x=1/2;\tau_2) = 1.20^{+0.16=1.36}_{-0.13=1.07}\,,
\end{equation}
which agrees with the result $\varphi_\pi(1/2) = 1.2 \pm 0.3$ obtained using QCD sum rules \cite{Braun:1988qv}.

\begin{figure}[t]
\leftline{\includegraphics[clip,width=0.45\linewidth]{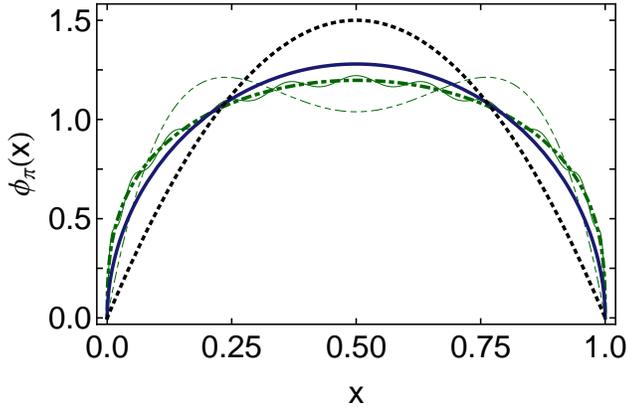}}
\vspace*{-32ex}

\rightline{\parbox{23em}{\caption{\label{Fig2Gegenbauer}
\emph{Dot-dashed curve}, $\varphi_\pi(x;\tau_2)$ in Eq.\,\protect\eqref{pionPDAlattice}; oscillatory \emph{thin solid curve}, Gegenbauer-$\alpha=3/2$ representation obtained with $10$ nontrivial moments ($a_{20}^{3/2}/a_2^{3/2}=0.044$); and \emph{thin dot-dot-dashed curve}, Gegenbauer-$\alpha=3/2$ representation obtained with just $1$ nontrivial moment ($a_{2}^{3/2}=0.20$, Eq.\,\protect\eqref{a246New}.).
\emph{Solid curve}, $\varphi_\pi(x;\tau_{10})$ in Eq.\,\protect\eqref{pionPDAlattice10}; i.e., leading-order evolution of $\varphi_\pi(x;\tau_2)$ to $\tau_{10}=1/[10\,{\rm GeV}]$, which corresponds to a hard scale of $100\,$GeV$^2$.
The dotted curve is $\varphi_\pi^{\rm asy}(x)$ in Eq.\,\protect\eqref{PDAasymp}.}}}
\end{figure}

As noted above, one may accurately compute arbitrarily many Gegenbauer-$\alpha=3/2$ moments by reprojecting the result in Eq.\,\eqref{pionPDAlattice} onto the Gegenbauer-$\alpha=3/2$ basis, Eqs.\,\eqref{PDAG3on2}, \eqref{projection}.  It is therefore straightforward to evolve Eq.\,\eqref{pionPDAlattice} to any scale $\zeta$ that might be necessary in order to consider a given process.  This is illustrated in Fig.\,\ref{Fig2Gegenbauer}.
The figure was prepared \cite{Cloet:2013tta} by expressing Eq.\,\eqref{pionPDAlattice} in the form of Eq.\,\eqref{PDAG3on2} with ten nontrivial moments, $\{a_j^{3/2}(\tau_2),j=2,\ldots,20\}$.
(N.B.\ The double-humped dot-dot-dashed curve, which depicts the result obtained if just the first moment is kept, highlights the limitation inherent in using Eq.\,\eqref{PDAG3on2} with limited information.)
Using the ten-moment expression and the leading-logarithmic formula, Eq.\,\eqref{llevolution}, those moments were evolved from $\zeta=2\,$GeV to $\zeta=10\,$GeV, producing a ten-moment representation of $\varphi_\pi(x;\tau_{10})$.  It, too, oscillates about a concave curve.  Working with the errors indicated in Eq.\,\eqref{pionPDAlattice}, one finds
\begin{equation}
\label{pionPDAlattice10}
\varphi_\pi(x;\tau_{10}) = N_\alpha \, x^{\alpha_-} (1-x)^{\alpha_-},\;
\alpha_- = 0.51^{+ 0.25 = 0.76}_{-0.20=0.31} .
\end{equation}
The ``central'' value of $\alpha_-=0.51$ is used to plot the thick, solid curve in Fig.\,\ref{Fig2Gegenbauer}.  Using Eqs.\,\eqref{a246New} and the comment after Eq.\,\eqref{a2Old}, one finds that it is only for $\zeta\gtrsim 100\,$GeV that $a_2^{3/2}\lesssim 10$\% and $a_4^{3/2}/a_2^{3/2}\lesssim 30$\%.  Evidently, the influence of DCSB, which is the origin of the amplitude's breadth, persists to remarkably small length-scales.

\begin{figure}[t]
\leftline{\includegraphics[clip,width=0.45\linewidth]{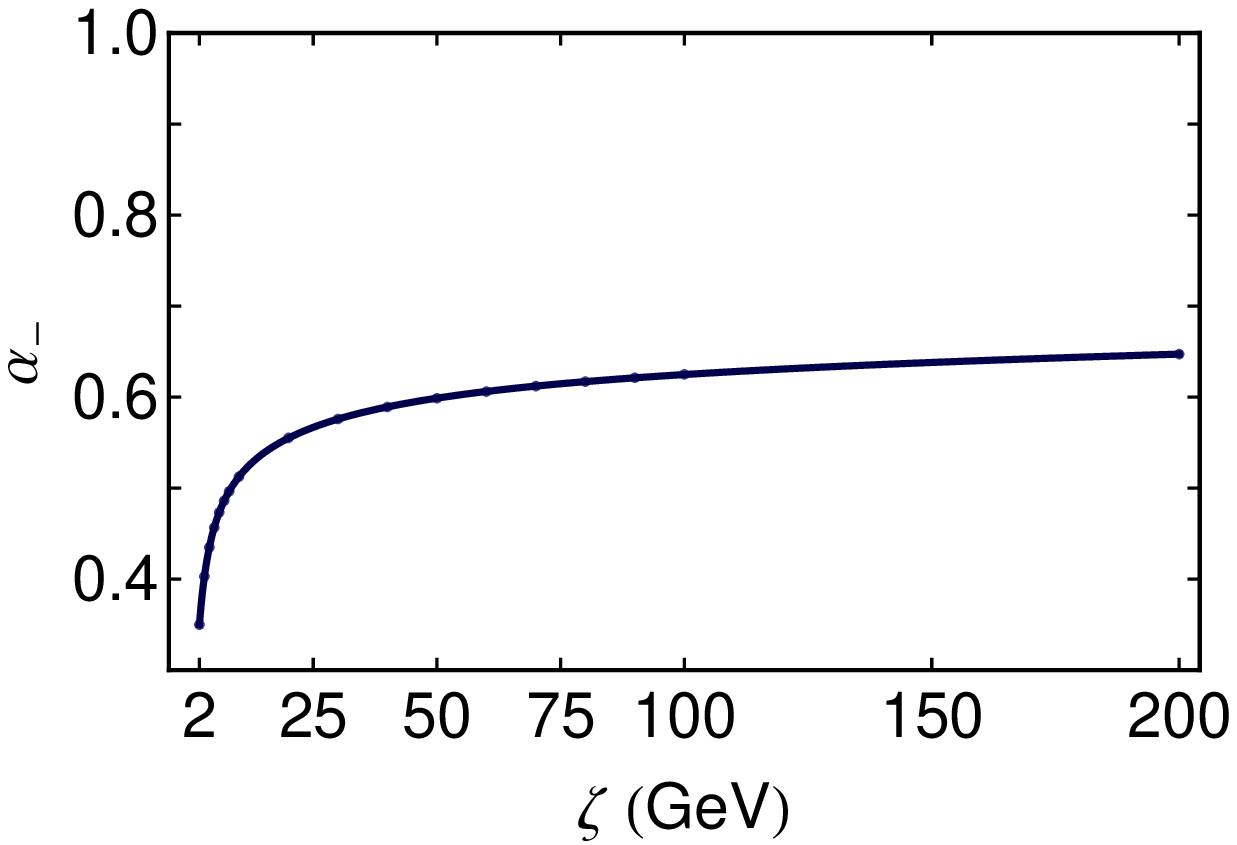}}
\vspace*{-30.5ex}

\rightline{\includegraphics[clip,width=0.45\linewidth]{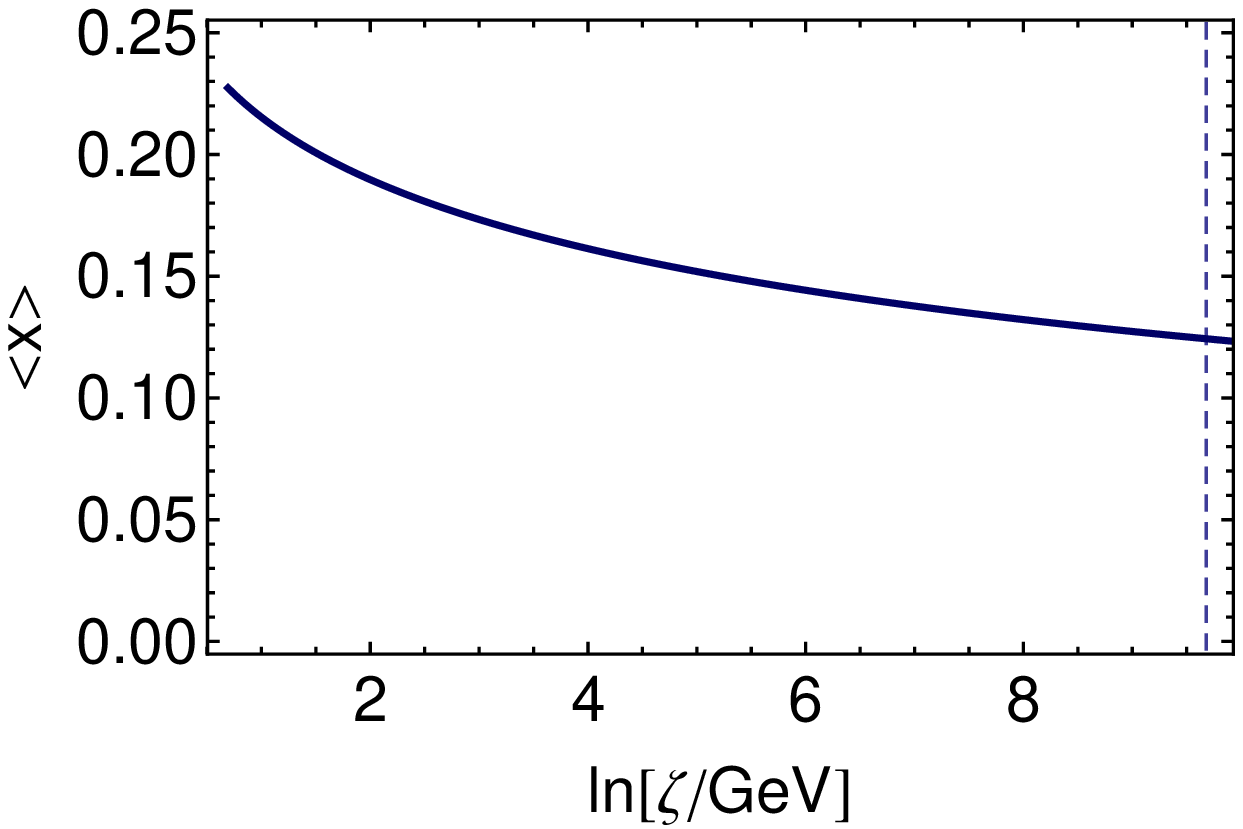}}

\caption{\label{FigalphaEvol}
\underline{Left panel}.  Evolution with momentum scale of the effective exponent that characterises the lattice-QCD PDA: $\alpha_-=0.35$ at $\zeta_2=2\,$GeV, Eq.\,\protect\eqref{pionPDAlattice}.  The lattice-QCD PDA has evolved to the asymptotic form when $\alpha_-(\zeta)=1$.  Evidently, this value is practically unreachable.
\underline{Right panel}.  Evolution with momentum scale of the light-front fraction of the pion's momentum carried by a dressed-valence-quark, defined in Eq.\,\protect\eqref{averagex}.  The vertical dashed line marks $\zeta=16\,$TeV; i.e., an energy scale characteristic of the LHC, at which point the two dressed valence-quarks are still carrying 25\% of the pion's momentum.}
\end{figure}

Another two illustrations are very valuable.  Consider first that using the approach described in the preceding paragraph, the lattice-QCD PDA, Eq.\,\eqref{pionPDAlattice}, can be evolved to any value of $\zeta>\zeta_2$.  So, using a 30-moment Gegenbauer-$\alpha=3/2$ representation of Eq.\,\eqref{pionPDAlattice} and the leading-logarithmic formula, we evolved the lattice-QCD PDA stepwise to $\zeta_{200}=200\,$GeV.  We found that at each and every intermediate value of $\zeta$, the evolved PDA is accurately approximated by the $j_s=0$ version of Eq.\,\eqref{PDAGalpha}.  This procedure therefore provides a curve $\alpha_-(\zeta)$.  One has reached the asymptotic PDA when $\alpha_-(\zeta)=1$.  We depict $\alpha_-(\zeta)$ in Fig.\,\ref{FigalphaEvol}.  Plainly, even at $\zeta=200\,$GeV, $\alpha_-$ is just 65\% of the asymptotic value.  Moreover, the figure shows that the evolution to $\alpha_- = 1$ is painfully slow, as logarithmic changes are, and practically unreachable.

Given that the pion's PDA is difficult to measure directly, it is perhaps better to work with the pion's valence-quark PDF, $u_{\rm v}^\pi(x;\zeta)$, which has been measured and found to be in line with expectations based the DSE prediction \cite{Hecht:2000xa,Aicher:2010cb}.  As we stated at the beginning of this subsection, $\varphi^{\rm asy}_\pi(x)$ can only be a good approximation to the pion's PDA when it is accurate to write $u_{\rm v}^\pi(x;\zeta) \approx \delta(x)$ \cite{Georgi:1951sr,Gross:1974cs,Politzer:1974fr}.  Plainly, in this case:
\begin{equation}
\lim_{\Lambda_{\rm QCD}/\zeta \to 0} \langle x \rangle^\pi_\zeta  = 0\,,\quad
\langle x \rangle^\pi_\zeta :=
\int_0^1 dx\, x \, u_{\rm v}^\pi(x;\zeta) \,;
\label{averagex}
\end{equation}
i.e., the light-front fraction of the pion's momentum carried by dressed valence-quarks is zero.

We will now address the issue of how this compares with empirical reality.  At $\zeta=2\,$GeV one finds that dressed valence-quarks carry 45\% of the pion's light-front momentum; i.e., $2\,\langle x \rangle^\pi_{\zeta_2} = 0.45$ \cite{Hecht:2000xa,Aicher:2010cb}.  The remainder is carried by glue (44\%) and sea-quarks (11\%).\footnote{There are no surprises in these values because it has long been known empirically that $\lesssim 50$\% of the momentum of a hadron is carried by gluons at $\zeta=2\,$GeV; e.g., see Ref.\,\protect\cite{ESW96}, Eq.(4.24).  Indeed, almost coincident with the discovery of asymptotic freedom it was shown that the light-front fraction of a hadron's momentum carried by gluons saturates at a value of approximately one-half in the limit $\Lambda_{\rm QCD}/\zeta\to 0$ \cite{Georgi:1951sr,Gross:1974cs,Politzer:1974fr}.}
The equation describing the $\zeta$-evolution of $u_{\rm v}^\pi(x;\zeta)$ is known and hence one can readily compute $\langle x \rangle^\pi_\zeta$.  We depict the next-to-leading-order evolution \cite{Floratos:1981hs} of the valence-quark's light-front momentum fraction in the right panel of Fig.\,\ref{FigalphaEvol}.  This figure adds enormously to the information provided in the left panel.  It shows that the momentum fraction evolves so slowly that even at an energy scale characteristic of the LHC, 25\% of the pion's momentum is carried by dressed-valence quarks, 21\% is carried by sea-quarks and 54\% is carried by glue.  (Owing to saturation of the gluon fraction, evolution very slowly shifts momentum from the valence-quarks to the sea-quarks.)  Plainly, even at LHC energy scales, nonperturbative effects such as DCSB are playing a crucial role in setting the scales in PDAs and PDFs.

The information presented thus far exposes a critical internal inconsistency in Ref.\,\cite{Aitala:2000hb}, which claims to represent a direct measurement of $\varphi_\pi^2(x)$.  Using the reasoning therein, the two panels in Fig.\,3 of Ref.\,\cite{Aitala:2000hb} correspond to $\zeta \approx 2\,$GeV (left) and $\zeta \approx 3\,$GeV (right).  The left panel depicts a broad distribution, for which Eq.\,\eqref{projection} yields $a_2^{3/2} \approx 0.27$, whereas the right panel is the asymptotic distribution, for which $a_2^{3/2}=0$; and, as we have now seen, it is impossible for QCD evolution from $\zeta=2 \to 3\,$GeV to connect these two curves.  Therefore, they cannot represent the same pion property and it is not credible to assert that $\varphi_\pi(x)$ is well represented by the asymptotic distribution for $\zeta^2 \gtrsim 10\,$GeV$^2$.  The assumptions which underly the claims in Ref.\,\cite{Aitala:2000hb} should be carefully re-examined.

The material presented in this subsection establishes that contemporary DSE- and lattice-QCD computations, at the same scale, agree on the pointwise form of the pion's PDA, $\varphi_\pi(x;\tau)$.  This unification of DSE- and lattice-QCD results expresses a deeper equivalence between them, expressed, in particular, via the common behaviour they predict for the dressed-quark mass-function \cite{Roberts:2007ji,Bhagwat:2003vw,Bowman:2005vx,Bhagwat:2006tu}, which is a definitive signature of DCSB and the origin of the distribution amplitude's dilation.

Furthermore, the associated discussion supports a view that $\varphi_\pi^{\rm asy}(x)$ is a poor approximation to $\varphi_\pi(x;\tau)$ at all momentum-transfer scales that are either now accessible to experiments involving pion elastic or transition processes, or will become so in the foreseeable future \cite{Dudek:2012vr,Huber:2008id,Uehara:2012ag,Holt:2012gg}.  Available information indicates that the pion's PDA is significantly broader at these scales; and hence that predictions of leading-order, leading-twist formulae involving $\varphi^{\rm asy}_\pi(x)$ are a misleading guide to interpreting and understanding contemporary experiments.  At accessible energy scales, a better guide is obtained by using the broad PDA described herein in such formulae.  As we shall see, this is adequate for the charged pion's elastic form factor.  However, it will probably be necessary to explicitly consider higher-twist and higher-order $\alpha$-strong corrections in controversial cases such as the $\gamma^\ast \gamma \to \pi^0$ transition form factor \cite{Roberts:2010rn,Agaev:2010aq,Bakulev:2012nh,Brodsky:2011yv}.

\subsection{Light front distribution of the chiral condensate}
\label{sec:lfchi}
We have already remarked that arguably the most fundamental expression of DCSB is the behaviour of the dressed-quark mass-function, $M(p)$ in Fig.\,\ref{gluoncloud}.  A derived measure of DCSB is the chiral condensate; and, as we explained in Sec.\,\ref{sec:CCC}, this fascinating quantity is confined to the interior of the pion \cite{Chang:2011mu,Brodsky:2010xf,Brodsky:2012ku,Brodsky:2008be,Brodsky:2009zd,Glazek:2011vg}, and hence describes intimate properties of QCD's Goldstone mode that are associated with DCSB.  The chiral condensate is properly defined via Eqs.\,\eqref{kappazeta}, \eqref{qbqpiqbq0}; i.e.,
\begin{equation}
\label{Realgmor}
\kappa_0^\zeta = \lim_{\hat m_{u,d}\to 0}\kappa_\pi^\zeta,\quad
\kappa_\pi^\zeta := f_\pi \rho_\pi^\zeta,
\end{equation}
where $f_\pi$, $\rho_\pi^\zeta$ are defined respectively in Eqs.\,\eqref{fpigen}, \eqref{rhogen}.  The content of Eq.\,(\ref{fpigen}) is well known: $f_\pi$ is the pion's leptonic decay constant, and the rhs of that equation expresses the axial-vector projection of the pion's Bethe-Salpeter wave-function onto the origin in configuration space.  Likewise, Eq.\,(\ref{rhogen}) is this wave-function's pseudoscalar projection onto the origin.  It therefore describes another type of pion decay constant.

The quantities $f_\pi$ and $\rho_\pi$ are both equivalent order parameters for DCSB; and, owing to DCSB, they are related through the mass formula in Eq.\,\eqref{mrtrelation}, which in the present case reads:
\begin{equation}
f_\pi m_\pi^2 = 2\, m^\zeta \rho_\pi^\zeta.
\end{equation}
We reiterate here that $m^\zeta \rho_\pi^\zeta$ is renormalisation point independent and hence the ground-state pseudoscalar meson is massless in the chiral limit \cite{Holl:2004fr}.  It is also important to observe that the pseudovector and pseudoscalar projections of the pion's Bethe-Salpeter wave function onto the origin in configuration space provide the only nonzero results: projections through Dirac scalar, vector or tensor matrices yield zero; viz., with $Z_O$ the appropriate renormalisation constant,
\begin{equation}
Z_O\; {\rm tr}_{\rm CD}
\int_{dq}^\Lambda O_{\rm D} S(q_\eta) \Gamma_{\pi}(q;P) S(q_{\bar\eta}) =0 \,,\quad O_{\rm D} = \mathbf{I}_D,\,\gamma_\mu,\,\sigma_{\mu\nu}\,.
\end{equation}

Building upon the material reviewed hitherto, Ref.\,\cite{Chang:2013epa} proceeded to expose a novel nonperturbative feature of QCD.  To explain, consider the light-front expression for $\rho_\pi^\zeta$ in Eq.\,\eqref{LFrhopiO}.  Given the explicit appearance of the current-quark mass, contributions from the ``instantaneous'' part of the quark propagator ($\sim \gamma^+/k^+$) and the associated gluon emission are critical to producing a nonzero chiral-limit result: one must sum infinitely many nontrivial terms in order to compensate for $\hat m \to 0$.  As illustrated by Fig.\,\ref{figinstantaneous}, these nontrivial terms actually express couplings to higher Fock state components in the pion's light-front wave-function.  Such couplings are absent when one computes the $\gamma_5 \gamma\cdot n$-projection of the pion's wave function, as in Eq.\,\eqref{pionPDA} \cite{Lepage:1980fj}. 
Consequently, $\rho_\pi^\zeta$ and the pseudoscalar projection of the pion's Bethe-Salpeter wave-function onto the light-front both contain essentially new information, exposing process-independent features of the pion that owe to nonvalence Fock states in its light-front wave function.  (The role of such collective behaviour in forming a chiral condensate was anticipated in Ref.\,\cite{Casher:1974xd}.)

Consider therefore the pseudoscalar projection of the pion's Poincar\'e-covariant
Bethe-Salpeter wave-function onto the light-front \cite{Chang:2013epa}:
\begin{equation}
\rho_\pi^\zeta\, \omega_\pi(x) = {\rm tr}_{\rm CD}
Z_4 \! \int_{dk}^\Lambda \!\!
\delta(n\cdot k_\eta - x \,n\cdot P) \,\gamma_5\, \chi_\pi(k;P)\,.
\label{pionPDA5}
\end{equation}
As with $\varphi_\pi(x)$, $\omega_\pi(x)=\omega_\pi(\bar x)$ because the neutral pion is an eigenstate of the charge conjugation operator.  The notational switch $\varphi_\pi \to \omega_\pi$ was made in order to emphasise that the distribution amplitude defined in Eq.\,\eqref{pionPDA5} includes information about all Fock state components of the pion's light-front wave function.  In fact, given that the zeroth moment of the rhs measures the in-pion condensate, then $\omega_\pi(x)$ may be interpreted as describing the light-front distribution of the chiral condensate.   The moments of this distribution are obtained via
\begin{equation}
i \rho_\pi (n\cdot P)^{m} \langle x^m_\omega\rangle =
{\rm tr}_{\rm CD}
Z_4 \! \int_{dk}^\Lambda \!\!
(n\cdot k_\eta)^m \,\gamma_5 \chi_\pi(k;P)\,.
\label{omegamom}
\end{equation}

It is worth noting that $\omega_\pi(x)$ was first considered in Ref.\,\cite{Braun:1989iv}, wherein it was identified as a twist-three two-particle distribution amplitude.  As such, it is important in the analysis of hard exclusive processes and, in particular, the study of $B$-meson pionic decays using light-cone sum rules \cite{Beneke:2003zv}.  QCD sum rules estimates of $\omega_\pi(x)$ are described in Refs.\,\cite{Ball:1998je,Ball:2006wn}.

\begin{figure}[t]
\leftline{\includegraphics[clip,width=0.45\linewidth]{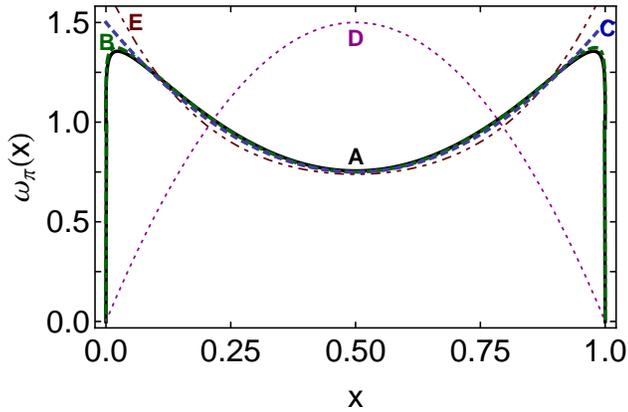}}
\vspace*{-29ex}

\rightline{\parbox{23em}{\caption{\label{Fig1LFchi}
\emph{Solid curve}\,(A) -- $\omega_\pi(x)$ computed at $\zeta_2=2\,$GeV; and \emph{dot-dashed curve}\,(B) -- $\omega_\pi(x)$ computed at $\zeta_{19}=19\,$GeV.
\emph{Dashed curve}\,(C) -- $\omega_\pi^{\rm asy}(x)$ in Eq.\,\protect\eqref{asyomega}, the asymptotic distribution of the chiral condensate within the pion.  \emph{Dotted curve}\,(D) --  for comparison, $\varphi_\pi^{\rm asy}(x)$ in Eq.\,\protect\eqref{PDAasymp}.
\emph{Dot-dot-dashed curve}\,(E) -- sum rules result in Ref.\,\protect\cite{Ball:2006wn}.
}}}
\vspace*{6ex}

\end{figure}

Before describing a numerical computation of this distribution, it is useful to develop intuition about its pointwise behaviour.  To achieve that one may apply the analysis associated with Eqs.\,\eqref{pointS}, \eqref{rhoEpi} to Eq.\,\eqref{omegamom} and thereby obtain
\begin{equation}
\langle x^m_\omega \rangle_\nu = \left[m (1+m) + 2 \nu  (1+m+\nu)\right] \frac{\Gamma (2+2 \nu ) \Gamma (m+\nu )}{2 \Gamma (2+\nu) \Gamma
   (2+ m+2 \nu)}\,,
\end{equation}
from which one can reconstruct the distribution
\begin{equation}
_\nu\omega_\pi(x) = \frac{(1+\nu) \Gamma(2+2\nu)}{2 (1+2\nu) \Gamma(\nu) \Gamma(2+\nu)} \, [x\bar x]^{\nu-1} \bigg[1+
\frac{C_2^{(\nu-1/2)}(x-\bar x )}{(2\nu - 1 )(\nu+1)}  \bigg]\,. \label{omeganu}
\end{equation}
The result for $\nu=1$; i.e.,
\begin{equation}
\label{asyomega}
\omega_\pi^{\rm asy}(x) = \, _1\omega_\pi(x) = 1 + \frac{1}{2} C_2^{(1/2)}(x-\bar  x)\,,
\end{equation}
as curve-C in Fig.\,\ref{Fig1LFchi}.  This is the asymptotic distribution of the chiral condensate within the pion, in exactly the same sense that Eq.\,\eqref{PDAasymp} is the asymptotic form of the pion's valence-quark PDA.  The behaviour of $\omega_\pi^{\rm asy}(x)$ is striking.  It shows that the chiral condensate is primarily located in components of the pion's wave-function that express correlations with large relative momenta.

To understand this feature, it is useful to draw a contrast with the observations made in the paragraph preceding that containing Eq.\,\eqref{projection}; viz., that the peak in $\varphi_\pi^{\rm asy}(x)$ at $x=1/2$ is a consequence of the fact that the leading Chebyshev moment of each of the three significant scalar functions which appear in $\Gamma_\pi(k;P)$, Eq.\,\eqref{genGpi}, occurs at zero relative momentum $(k_{\rm rel} = 0)$ and, moreover, that these Chebyshev moments are monotonically decreasing with $k_{\rm rel}^2$.  In the case of $\omega_\pi(x)$ it is the quark-level Goldberger-Treiman relation, Eq.\,\eqref{gtlrelE}, that is relevant.  From this it follows that the chiral condensate may be read from the large-$k_{\rm rel}^2$ (large relative momentum) behaviour of the dominant term in the pion's Bethe-Salpeter amplitude,  which is just the scalar piece of the self-energy connected with the dressed quark confined within the pion \cite{Lane:1974he,Politzer:1976tv,Langfeld:2003ye}.

Having recapitulated upon the insights developed in Refs.\,\cite{Cloet:2013tta,Chang:2013epa}, it is now appropriate to describe realistic results for $\omega_\pi(x)$.  Such computations may readily be accomplished by applying to Eq.\,\eqref{omegamom} the methods described in Secs.\,\ref{sec:vqPDA} and \ref{sec:phiasy}.  (In this case, Eq.\,\eqref{logfactor} and the associated comments are important.)  The $\hat m=0$ result, curve-A in Fig.\,\ref{Fig1LFchi}, is
\begin{equation}
\omega_\pi(x;\zeta_2) =
N_\alpha [ x \bar x ]^{\alpha_-} [1+ a_2 C_2^{(\alpha)}(x - \bar x)]\,,
\end{equation}
with $\alpha = \nu-1/2$, $\nu=1.05$, $a_2=0.48$.
We note that $50$ moments were used to reconstruct $\omega_\pi(x)$ in Ref.\,\cite{Chang:2013epa}.  It was unnecessary to use more because the same distribution was obtained from $100$ moments.  There is naturally no ambiguity in the result, since the polynomials $\{C_j^{(\alpha)}(x-\bar x),j=1,\ldots,\infty\}$  are a complete orthonormal set on $x\in[0,1]$ with respect to the measure $[x \bar x]^{\alpha_-}$.  Owing to the fact that $\omega_\pi(x)=\omega_\pi(\bar x)$, only polynomials of even degree contribute.  The reconstruction procedure converged at the first step; i.e., a converged result was obtained by terminating the series at $j=2$.  Including the second term, $j=4$, altered nothing by more than $0.1$\%.

The result in Fig.\,\ref{Fig1LFchi} is striking, as we now explain.
The first thing of which to be aware is that only the $E_\pi(k;P)$ term in Eq.\,\eqref{genGpi} provides a nonzero contribution when one removes the regularisation scale $\Lambda\to\infty$.  This is because $\lim_{\Lambda\to\infty} Z_4(\zeta,\Lambda) = 0$; and whilst the integral of the $E_\pi(k;P)$ term diverges with $\Lambda$ at precisely the rate required to produce a finite, nonzero, $\Lambda$-independent result, the terms $F_\pi(k;P)$, $G_\pi(k;P)$, $H_\pi(k;P)$ provide contributions to the integral that are finite as $\Lambda\to\infty$ and hence disappear when multiplied by a renormalisation constant which vanishes in this limit.

This re-emphasises the explanation provided for Eq.\,\eqref{asyomega}.  It also entails that the result is a model-independent feature of QCD.  Since the integral is dominated by the ultraviolet behaviour of the integrand, no difference in Bethe-Salpeter kernels at infrared momenta can have an impact.  Owing to Eq.\,\eqref{gtlrelE}, the chiral-limit result is completely determined by the momentum-dependence of the scalar piece of the self-energy associated with the dressed-quark that is confined within the pion.  This momentum-dependence is the same in all DSE truncation schemes that preserve the one-loop renormalisation group properties of QCD.  This was confirmed in Ref.\,\cite{Chang:2013epa} by computing $\omega_\pi(x)$ in both RL and DB truncation and verifying that the results are identical.

The distribution was also computed $\omega_\pi(x)$ at two different scales; viz., $\zeta_2$ and $\zeta_{19}$.  (The latter value is used commonly in DSE studies that follow Ref.\,\cite{Maris:1997tm}.)  As one should expect and is evident in Fig.\,\ref{Fig1LFchi}, $\omega_\pi(x) \to \omega_\pi^{\rm asy}(x)$ as $\Lambda_{\rm QCD}/\zeta \to 0$.  However, as stressed in Sec.\,\ref{sec:phiasy} via the evolution of $\varphi_\pi(x)$, the rate of approach to the asymptotic form is extremely slow.

With the results in Fig.\,\ref{Fig1LFchi}, Ref.\,\cite{Chang:2013epa} provided a model-independent demonstration that the chiral condensate is primarily located in components of the pion's wave-function that express correlations with large relative momenta, a feature which entails that light-front longitudinal zero modes do not play a material role in forming the chiral condensate.  This consequence may be elucidated by noting that $\omega_\pi(x=0)=0=\omega_\pi(x=1)$ at any finite renormalisation scale and $\omega_\pi(x) = \omega_\pi(\bar x)$.  Hence, the maximal contribution to the chiral condensate is obtained when half the partons carry a near-zero fraction of the pion's light-front momentum but the other half carry a near-unit fraction.  This discourse complements arguments to the same effect in Ref.\,\cite{Brodsky:2010xf}.

The prediction for the light-front distribution of the chiral condensate is a model-independent feature of QCD.  It should be verified.  This is a theoretical challenge because few contemporary techniques with a veracious connection to QCD can provide access to anything other than the pion's valence-quark parton distribution amplitude, whereas the chiral condensate receives contributions from all Fock-states in the pion's light-front wave function.
Lattice-QCD is one applicable tool.  However, with existing algorithms it can only be used to compute one nontrivial moment of $\omega_\pi(x)$.  QCD sum rules might also be employed usefully: indeed, estimates exist \cite{Ball:1998je,Ball:2006wn}.  They, too, typically work with moments of the distribution.  In order to assist practitioners in meeting the theoretical challenge, Ref.\,\cite{Chang:2013epa} presented predictions for the lowest three moments:
\begin{equation}
\label{latticemomentChi}
\int_0^1 \!dx\, (2 x - 1)^{2 j} \, \omega_\pi(x,\zeta_2)
= \left\{
\begin{array}{l}
 0.39\,, j=1\\
 0.25\,, j=2\\
 0.18\,, j=3
 \end{array}\right. ,
\end{equation}
and observed in addition that $\omega_\pi(1/2,\zeta_2)=0.76$ cf.\ $\omega_\pi^{\rm asy}(1/2)=3/4$.

A comparison with the contemporary sum rules estimate \cite{Ball:2006wn} is worthwhile.  That result corresponds to a renormalisation scale of roughly $\zeta_1=1\,$GeV.  It is plotted as curve-E in Fig.\,\,\ref{Fig1LFchi}, produces $j=1,2,3$ moments $0.41$, $0.27$, $0.20$, respectively, and $\omega_\pi^{\rm SR}(1/2,\zeta_1)=0.74$.  The agreement with the prediction of Ref.\,\cite{Chang:2013epa} is plainly very good.  Differences are only marked in the neighbourhood of the endpoints, something one might have anticipated given that just low-order moments can practically be constrained in a sum rules analysis and such moments possess little sensitivity to the behaviour of $\omega_\pi$ in the neighbourhood of the endpoints.  The generally good agreement with the DSE prediction from such limited input provides strong support for the model-independent nature of that result.  This is further emphasised by the fact that the estimate in Ref.\,\cite{Ball:2006wn} improves over an earlier calculation \cite{Ball:1998je} and, as gauged by the $L^1$-norm, the modern refinement shifts the earlier result toward the DSE prediction.

In closing this subsection it is worth reiterating that it is a misapprehension to suppose that the pion's valence-quark parton distribution amplitude expresses all bound-state dynamics associated with the pion.  Definitive features of Goldstone boson structure are expressed in the pseudoscalar projection of the pion's Poincar\'e-covariant Bethe-Salpeter amplitude onto the light-front, which images the light-front distribution of the chiral condensate.

\section{Charged-pion elastic form factor}
\label{sec:pionFF}
\subsection{Background}
The fascination of the pion is compounded by the existence of exact results for both soft and hard processes.  For example, there are predictions for low-energy $\pi \pi$ scattering \cite{Weinberg:1966kf,Colangelo:2001df} and the neutral-pion's two-photon decay \cite{Adler:1969gk,Bell:1969ts}; and, on the other hand, pQCD yields predictions for pion elastic and transition form factors at asymptotically high energies \cite{Efremov:1979qk,Lepage:1979zb,Lepage:1980fj,Farrar:1979aw}.  The empirical verification of the low-energy results \cite{Batley:2010zza,Larin:2010kq} is complemented by a determined experimental effort to test the high-energy form-factor predictions \cite{Huber:2008id,Uehara:2012ag,Aubert:2009mc,Volmer:2000ek,Horn:2006tm,E1206101}.  In contrast to the low-energy experiments, however, which check global symmetries and breaking patterns that might be characteristic of a broad class of theories, the high-energy experiments are a direct probe of QCD itself; and some would argue that QCD has not passed these tests.

We do not share this view, given that QCD's failure was also suggested in connection with measurements of the pion's valence-quark distribution function \cite{Conway:1989fs} and that those claims are now known to be erroneous \cite{Holt:2010vj,Nguyen:2011jy,Hecht:2000xa,Wijesooriya:2005ir,Aicher:2010cb}.  Nevertheless, an explanation is required for the mismatch between extant experiments on the pion's electromagnetic form factor and what is commonly presumed to be the prediction of pQCD.

The QCD prediction can be stated succinctly \cite{Efremov:1979qk,Lepage:1979zb,Farrar:1979aw,Lepage:1980fj}:
\begin{equation}
\label{pionUV}
\exists Q_0>\Lambda_{\rm QCD} \; |\;  Q^2 F_\pi(Q^2) \stackrel{Q^2 > Q_0^2}{\approx} 16 \pi \alpha_s(Q^2)  f_\pi^2 \mathpzc{w}_\varphi^2,
\end{equation}
with
\begin{equation}
\label{wphi}
\mathpzc{w}_\varphi = \frac{1}{3} \int_0^1 dx\, \frac{1}{x} \varphi_\pi(x)\,,
\end{equation}
where $\varphi_\pi(x)$ is the PDA discussed in Sec.\,\ref{sec:vqPDA}.  The value of $Q_0$ is not predicted by pQCD.

Notably, $\mathpzc{w}_\varphi=1$ if one uses the ``asymptotic'' PDA, Eq.\,\eqref{PDAasymp}.  This form for the PDA is certainly valid on the domain $\Lambda_{\rm QCD}^2/Q^2\simeq 0$.  As explained in Sec.\,\ref{sec:phiasy}, however, the domain $\Lambda_{\rm QCD}^2/Q^2\simeq 0$ corresponds to \emph{very} large values of $Q^2$.

The perceived disagreement between experiment and QCD theory is based on an observation that at $Q^2=4\,$GeV$^2$, approximately the midpoint of the domain accessible at the upgraded JLab facility \cite{Dudek:2012vr}, Eqs.\,\eqref{PDAasymp}, \eqref{pionUV} yield
\begin{equation}
\label{pionUV4}
Q^2 F_\pi(Q^2) \stackrel{Q^2=4\,{\rm GeV}^2}{=} 0.15\,,
\end{equation}
where we have used $n_f=4$ and $\Lambda_{\rm QCD}=0.234\,$GeV for illustration \cite{Qin:2011dd}.  The result in Eq.\,\eqref{pionUV4} is a factor of $2.7$ smaller than the empirical value quoted at $Q^2 =2.45\,$GeV$^2$ \cite{Huber:2008id,Horn:2006tm}: $0.41^{+0.04}_{-0.03}$; and a factor of three smaller than that computed at $Q^2 =4\,$GeV$^2$ in Ref.\,\cite{Maris:2000sk}.  Notably, Ref.\,\cite{Maris:2000sk} provided the only prediction for the pointwise behaviour of $F_\pi(Q^2)$ that is both applicable on the entire spacelike domain currently mapped reliably by experiment and confirmed thereby.

In this case the perception of a mismatch and a real discrepancy are not equivalent because, as indicated above, one can convincingly argue that $Q^2=4\,$GeV$^2$ is not within the domain $\Lambda_{\rm QCD}^2/Q^2\simeq 0$ upon which Eq.\,\eqref{PDAasymp} is valid \cite{Cloet:2013tta}.  This being so and given the successful prediction in Ref.\,\cite{Maris:2000sk}, one is naturally led to ask whether the methods used therein can address the issue of the ultimate validity of Eq.\,\eqref{pionUV}.

Until recently, the answer was ``no'', owing to an over-reliance hitherto on brute numerical methods in such computations.  That has now changed, however, with a refinement of known methods \cite{Nakanishi:1963zz,Nakanishi:1969ph,Nakanishi:1971} described in association with a computation of the pion's light-front wave-function in Sec.\,\ref{sec:vqPDA}.  These methods enable reliable computation of the pion's electromagnetic form factor to arbitrarily large-$Q^2$ and the correlation of that result with Eq.\,\eqref{pionUV} using the consistently computed distribution amplitude, $\varphi_\pi(x)$ \cite{Chang:2013pq}.

\subsection{Computing the form factor}
The formulae expressing projection of the pion's Poincar\'e-covariant Bethe-Salpeter amplitudes onto the light-front are exact.  They receive no corrections.  In the case of the pion's elastic and transition electromagnetic form factors, however, a truncation must be decided upon before an expression can be written for the associated matrix elements.  At leading order in the systematic and symmetry-preserving DSE truncation scheme introduced in Refs.\,\cite{Munczek:1994zz,Bender:1996bb}; i.e., in RL truncation, the pion form factor is \cite{Roberts:1994hh}
\begin{equation}
K_\mu F_\pi(Q^2) = N_c {\rm tr}_{\rm D} 
\int\! \frac{d^4 k}{(2\pi)^4}\,
\chi_\mu(k+p_f,k+p_i)  \Gamma_\pi(k_i;p_i)\,S(k)\,\Gamma_\pi(k_f;-p_f)\,, \quad\label{RLFpi}
\end{equation}
where $Q$ is the incoming photon momentum, $p_{f,i} = K\pm Q/2$, $k_{f,i}=k+p_{f,i}/2$.\footnote{ Corrections to the rainbow-ladder truncation are illustrated in Refs.\,\protect\cite{Maris:2000sk,Eichmann:2011ec}.  Their impact is understood \protect\cite{Lepage:1980fj}.  The dominant effect is a modification of the power associated with the logarithmic running in Eq.\,\eqref{pionUV}; i.e., the anomalous dimension.  That running is slow and hence the diagrams omitted have no material impact on the results reported in Ref.\,\protect\cite{Chang:2013nia}.}  The other elements in Eq.\,\eqref{RLFpi} are the dressed-quark propagator, which, to be consistent with Eq.\,\eqref{RLFpi}, is computed from the rainbow-truncation gap equation; the pion Bethe-Salpeter amplitude, computed in rainbow-ladder truncation; and the unamputated dressed-quark-photon vertex, $\chi_\mu(k_f,k_i)$, which should also be computed in rainbow-ladder truncation.

The leading-order DSE result for the pion form factor is now determined once an interaction kernel is specified for the rainbow gap equation.  In common with Ref.\,\cite{Chang:2013pq}, Ref.\,\cite{Chang:2013nia} used the kernel explained in Ref.\,\cite{Qin:2011dd}, which is also described briefly in App.\,\ref{app:nakanishi}.  By using precisely the rainbow-ladder kernel described in Ref.\,\cite{Chang:2013pq}, it was unnecessary to solve numerically for the dressed-quark propagator and pion Bethe-Salpeter amplitude.  Instead, Ref.\,\cite{Chang:2013nia} could employ the generalised Nakanishi representations for $S(p)$ and $\Gamma_\pi(k;P)$ described in App.\,\ref{app:nakanishi}.

That was not the case for $\chi_\mu(k_f,k_i)$, however, because, even now, such a representation is lacking.  The study therefore used the following \emph{Ansatz}, expressed solely in terms of the functions which characterise the dressed-quark propagator ($q=k_f-k_i$)
\begin{eqnarray}
\nonumber
\chi_\mu(k_f,k_i) & = & \gamma_\mu X_1(k_f,k_i) + \gamma\cdot k_{f}\gamma_{\mu}\gamma\cdot k_{i}  \, X_2(k_f,k_i)
+  \, i \, [\gamma\cdot k_f \gamma_\mu + \gamma_\mu \gamma\cdot k_i ] \, X_3(k_f,k_i) \\
&& - \, \tilde\eta\, \sigma_{\mu\nu} q_\nu \, \sigma_S(q^2)\, X_1(k_f,k_i)\,, \label{ChiAnsatz}
\end{eqnarray}
where
$\tilde\eta$ is a parameter and, with $\Delta_{F}(k_f^2,k_i^2)= [F(k_f^2)-F(k_i^2)]/[k_f^2-k_i^2]$:
\begin{equation}
X_1(k_f,k_i) = \Delta_{k^2 \sigma_V}(k_f^2,k_i^2)\,,\quad
X_2(k_f,k_i) = \Delta_{\sigma_V}(k_f^2,k_i^2)\,, \quad
X_3(k_f,k_i) = \Delta_{\sigma_S}(k_f^2,k_i^2)\,.
\end{equation}

Plainly, progress toward computing the spacelike behaviour of $F_\pi(Q^2)$ is expedited by using an \emph{Ansatz} instead of solving the rainbow-ladder Bethe-Salpeter equation for $\chi_\mu(k_f,k_i)$.  It is a valid procedure so long as nothing essential to understanding the form factor is lost thereby.  This is established by listing the following features of the \emph{Ansatz}.
The first line in Eq.\,\eqref{ChiAnsatz} is obtained using the gauge technique \cite{Delbourgo:1977jc}.  Hence, the vertex satisfies the longitudinal Ward-Green-Takahashi (WGT) identity \cite{Ward:1950xp,Green:1953te,Takahashi:1957xn}, is free of kinematic singularities, reduces to the bare vertex in the free-field limit, and has the same Poincar\'e transformation properties as the bare vertex.
With the term in the second line of Eq.\,\eqref{ChiAnsatz}, the \emph{Ansatz} also includes a dressed-quark anomalous magnetic moment, made mandatory by DCSB \cite{Chang:2010hb,Bashir:2011dp,Kochelev:1996pv,Bicudo:1998qb} and the transverse WGT identities \cite{Qin:2013mta}.
Finally, numerical solutions of the rainbow-ladder Bethe-Salpeter equation for the vertex \cite{Maris:1999bh} and algebraic analyses of vertex structure \cite{Chang:2010hb,Bashir:2011dp,Qin:2013mta} show that nonperturbative corrections to the bare vertex are negligible for spacelike momenta $Q^2\gtrsim 1\,$GeV$^2$.
%
A deficiency of Eq.\,\eqref{ChiAnsatz} is omission of nonanalytic structures associated with the $\rho$-meson pole but such features have only a modest impact on $Q^2 r_\pi^2 \lesssim 1$, where $r_\pi$ is the pion's charge radius, and are otherwise immaterial at spacelike momenta \cite{Alkofer:1993gu,Roberts:1994hh,Roberts:2000aa}.

With each of the elements in Eq.\,\eqref{RLFpi} expressed via a generalised spectral representation, as detailed in App.\,\ref{app:nakanishi}, the computation of $F_\pi(Q^2)$ reduces to the act of summing a series of terms, all of which involve a single four-momentum integral.  The integrand denominator in every term is a product of $k$-quadratic forms, each raised to some power.
Within each such term, one employs a Feynman parametrisation in order to combine the denominators into a single quadratic form, raised to the appropriate power.  A suitably chosen change of variables then enables one to readily evaluate the four-momentum integration using standard algebraic techniques.

As remarked in Sec.\,\ref{sec:vqPDA}, this is the paramount advantage of the technique employed in Ref.\,\cite{Chang:2013pq}.  In providing a practical approach to the problem of continuing from Euclidean metric to Minkowski space, it circumvents a longstanding problem.  Namely, as practitioners continue to find, with gap and Bethe-Salpeter equation solutions represented only by arrays of numbers it is nigh impossible to characterise and track complex-valued singularities that move with increasing $Q^2$ into the domain sampled by a numerical Euclidean-momentum integration, so that choosing and following an acceptable integration contour is practically hopeless.

After calculation of the four-momentum integration, evaluation of the individual term is complete after one computes a finite number of simple integrals; namely, the integrations over Feynman parameters and the spectral integral.
The complete result for $F_\pi(Q^2)$ follows after summing the series.

\begin{figure}[t]
\begin{centering}
\leftline{\includegraphics[clip,width=0.45\linewidth]{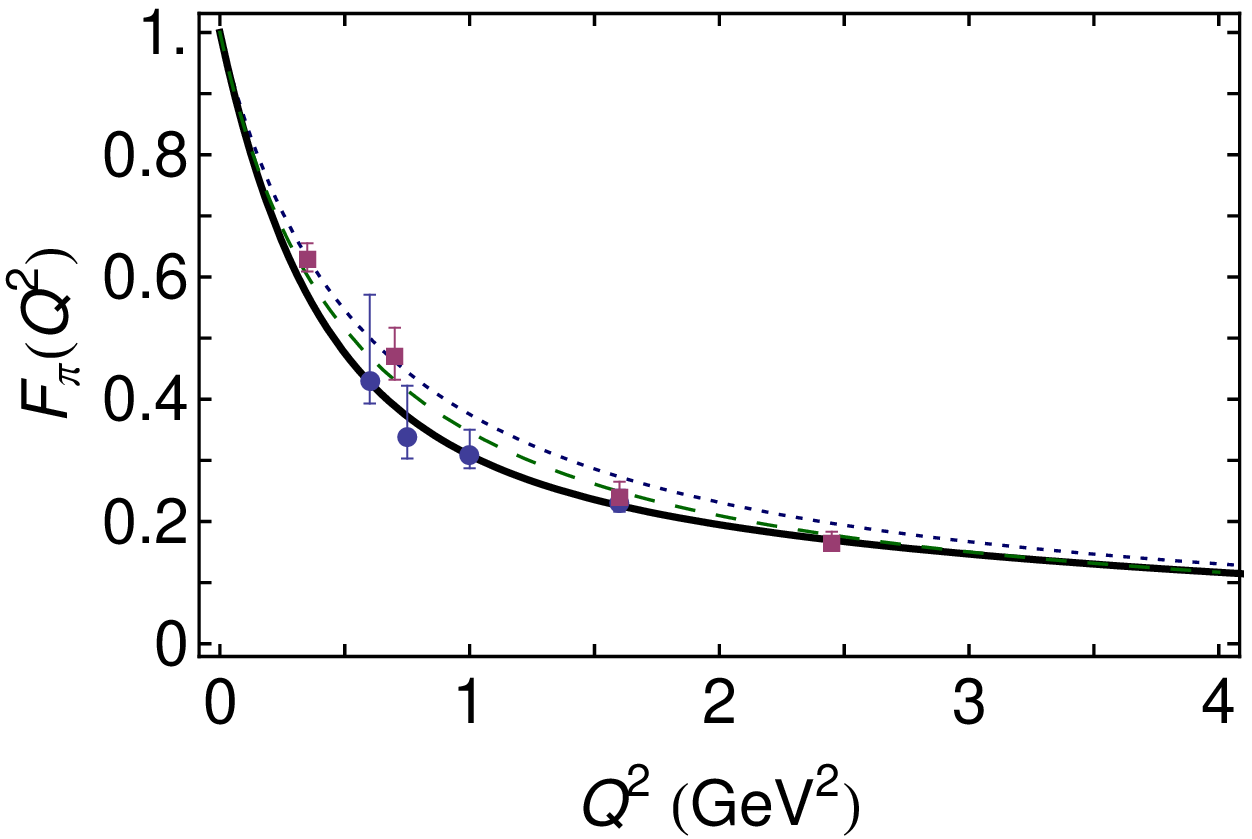}}
\vspace*{-33.4ex}

\rightline{\includegraphics[clip,width=0.45\linewidth]{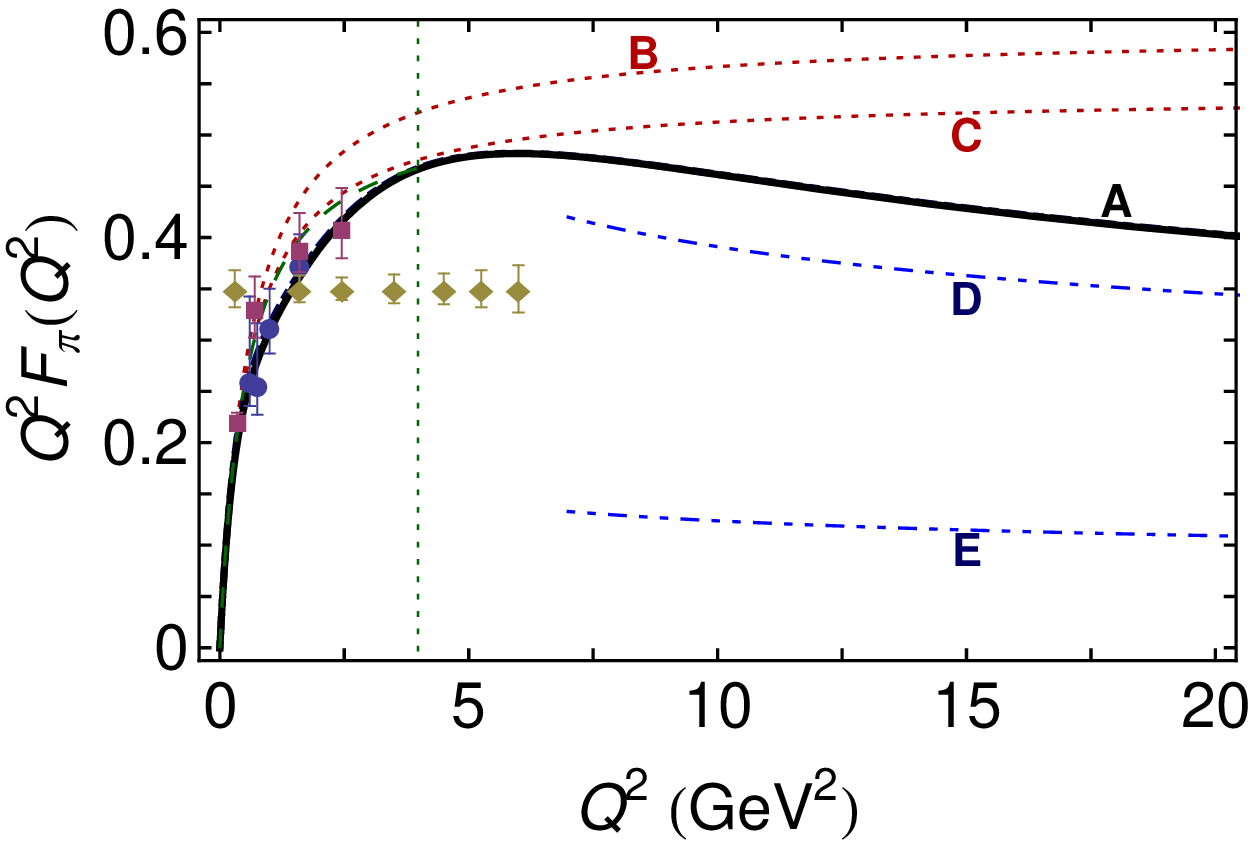}}
\end{centering}
\caption{\label{Fig1CLpion}
\underline{Left panel}.
\emph{Solid curve} -- Charged pion form factor, computed with $\tilde\eta = 0.5$ in Eq.\,\protect\eqref{ChiAnsatz}; \emph{long-dashed curve} -- calculation in Ref.\,\protect\cite{Maris:2000sk}; and \emph{dotted curve} -- monopole form ``$1/(1+Q^2/m_\rho^2)$,'' where $m_\rho=0.775\,$GeV is the $\rho$-meson mass.  In both panels, filled-circles and -squares are the data are described in Ref.\,\protect\cite{Huber:2008id}.
\underline{Right panel}.
$Q^2 F_\pi(Q^2)$.  \emph{Solid curve}\,(A) -- prediction obtained with $\tilde\eta = 0.5$ in Eq.\,\protect\eqref{ChiAnsatz}; \emph{dot-dash curve} -- prediction obtained with $\tilde\eta = 0$; and \emph{long-dashed curve} -- calculation in Ref.\,\protect\cite{Maris:2000sk}, which is limited to the domain $Q^2<4\,$GeV$^2$, whose boundary is indicated by the vertical dotted line.
Remaining curves, from top to bottom: \emph{dotted curve}\,(B) -- monopole form ``$1/(1+Q^2/m_\rho^2)$;'' \emph{dotted curve}\,(C) -- monopole form fitted to data in Ref.\,\protect\cite{Amendolia:1986wj}, with mass-scale $0.74\,$GeV;
\emph{Dot-dot--dashed curve}\,(D) -- Eq.\,\protect\eqref{pionUV} computed with $\varphi_\pi(x)$ in Eq.\,\protect\eqref{phiRLreal}; and \emph{Dot-dot--dashed curve}\,(E) -- Eq.\,\protect\eqref{pionUV} computed with $\varphi_\pi^{\rm asy}(x)$ in Eq.\,\protect\eqref{PDAasymp}.
The filled diamonds indicate the projected reach and accuracy of a forthcoming experiment \protect\cite{E1206101}.
}
\end{figure}

\subsection{Numerical Results}
The electromagnetic pion form factor, computed from Eq.\,\eqref{RLFpi} using the elements and procedures described above, is depicted as curve-A in Fig.\,\ref{Fig1CLpion}.
It is evident from the left panel that this prediction is practically indistinguishable from that described in Ref.\,\cite{Maris:2000sk} on the spacelike domain $Q^2 < 4\,$GeV$^2$, which was the largest value computable reliably in that study.  Critically, however, the new prediction extends to arbitrarily large momentum transfers: owing to the improved algorithms, it describes an unambiguous continuation of the earlier DSE prediction to the entire spacelike domain.  It thereby achieves a longstanding goal.\footnote{The current status of lattice-QCD calculations of $F_\pi(Q^2)$ is described in Refs.\,\protect\cite{Alexandrou:2011iu,Renner:2012yh}.  Results with quantitatively controlled uncertainties are beginning to become available.  Within errors, the estimated charge radius agrees with experiment; and simulations are also exploring a non-zero but still low $Q^2$ domain ($0<Q^2<1\,$GeV$^2$).  Various systematic uncertainties become more challenging with increasing $Q^2$, making a larger domain inaccessible at present.}

The momentum reach of the improved techniques is emphasised by the right panel in Fig.\,\ref{Fig1CLpion}.   The prediction for $F_\pi(Q^2)$ is depicted on the domain $Q^2\in [0,20]\,$GeV$^2$ but was computed in Ref.\,\cite{Chang:2013nia} out to $Q^2=100\,$GeV$^2$.  If necessary, reliable results could readily have been obtained at even higher values.  That is not required, however, because the longstanding questions revolving around $F_\pi(Q^2)$, which we described in opening this subsection, may be answered via Fig.\,\ref{Fig1CLpion}.

Before tackling those issues it is important to note that using $\tilde\eta=0.5$, a value commensurate with contemporary estimates \cite{Chang:2010hb,Bashir:2011dp,Qin:2013mta,Chang:2011tx}, the dressed-quark anomalous magnetic moment term in Eq.\,\eqref{ChiAnsatz} has almost no impact on $F_\pi(Q^2)$: the solid and dot-dashed curves in the right panel of Fig.\,\ref{Fig1CLpion} are essentially indistinguishable.  Indeed, for $Q^2 > 4\,$GeV$^2$ there is no difference and hence, as promised in connection with Eq.\,\eqref{ChiAnsatz}, the dressed-quark anomalous magnetic moment has no bearing on the ultraviolet behaviour of the form factor.  On the other hand, it does modestly influence the pion's charge radius: $r_\pi = 0.64\,$fm with $\tilde\eta=0$; whereas $r_\pi = 0.66\,$fm with $\tilde\eta=0.5$.  (Empirically \cite{Beringer:1900zz}, $r_\pi = 0.672 \pm 0.008\,$fm.)  Notably, the radius continues to grow with increasing $\tilde\eta$.  Thus, even though the pion is a pseudoscalar, the dressed-quark anomalous magnetic moment alters the pion's charge distribution.  This effect may be understood as the result of spin-orbit repulsion between the dressed-quarks within the pion, whose rest-frame wave-function necessarily has $P$-wave components in a Poincar\'e-covariant framework \cite{Bhagwat:2006xi}.

We have stressed that the ultraviolet behaviour of $F_\pi(Q^2)$ is of great contemporary interest.  A key feature of the rainbow-ladder prediction for $Q^2 F_\pi(Q^2)$ is therefore the maximum at $Q^2\approx 6\,$GeV$^2$ that is evident in the right panel of  Fig.\,\ref{Fig1CLpion}.  The domain upon which the flattening of the curve associated with this extremum is predicted to occur will be accessible to next-generation experiments \cite{E1206101}.  Unfortunately, on this domain it will still be difficult to distinguish between the theoretical prediction and the monopole fitted to data in Ref.\,\cite{Amendolia:1986wj}.

\subsection{Drawing connections with perturbative QCD}
A maximum appears necessary if $Q^2 F_\pi(Q^2)$ is ever to approach the value predicted by pQCD, Eq.\,\eqref{pionUV}.  In this connection, too, Ref.\,\cite{Chang:2013nia} had something to add.  The result in Eq.\,\eqref{pionUV4} is associated with curve-E in the right panel of Fig.\,\ref{Fig1CLpion}, which is typically plotted in such figures and described as the prediction of pQCD.  That would be true if, and only if, the pion's valence-quark distribution amplitude were well described by $\varphi_\pi^{\rm asy}(x)$ at the scale $Q^2\sim 4\,$GeV$^2$.  However, that is not the case, as we saw in Sec.\,\ref{sec:vqPDA}.

The correct comparison with pQCD should be drawn as follows.  Using precisely the interaction that was employed to compute $F_\pi(Q^2)$, one obtains the rainbow-ladder truncation result described in Secs.\,\ref{sec:vqPDA}, \ref{sec:phiasy}; viz.,
\begin{equation}
\label{phiRLreal}
\varphi_\pi(x;\zeta=2\,{\rm GeV}) \approx N_\alpha\, x^{\alpha_-} (1-x)^{\alpha_-}\,,
\end{equation}
with $\alpha_- = 0.3$.  This is the amplitude which should be used to calculate the pQCD prediction appropriate for comparison with contemporary experiments.  That computed result is drawn as curve-D in the right panel of Fig.\,\ref{Fig1CLpion};\footnote{One might also include the $Q^2$-evolution of $\varphi_\pi(x;Q^2)$ in curve-D.  However, nonperturbative evolution is slow, being overestimated using the leading-order formula, so that ``freezing'' $\varphi_\pi(x;Q^2)=\varphi_\pi(x;Q^2=4\,{\rm GeV}^2)$ provides a valid approximation on the domain depicted.  However, the $Q^2$-evolution of $\varphi_\pi(x;Q^2)$ must be included in a figure that extends to significantly larger $Q^2$ so that the computed approach of curve-D to curve-E is manifest.
}
i.e., this curve is the pQCD prediction obtained when Eq.\,\eqref{phiRLreal} is used in Eqs.\,\eqref{pionUV}--\eqref{wphi}.

Stated simply, curve-D in the right panel of Fig.\,\ref{Fig1CLpion} is the pQCD prediction obtained when the pion valence-quark PDA has the form appropriate to the scale accessible in modern experiments.  Its magnitude is markedly different from that obtained using the asymptotic PDA in Eq.\,\eqref{PDAasymp}; viz., curve-E, which is only valid at truly asymptotic momenta.
The meaning of ``truly asymptotic'' is readily illustrated.  The PDA in Eq.\,\eqref{phiRLreal} produces $\mathpzc{w}_\varphi^2 = 3.2$, which is to be compared with the value computed using the asymptotic PDA: $\mathpzc{w}^{\rm asy}_\varphi = 1.0$.  Applying leading-order QCD evolution to the PDA in Eq.\,\eqref{phiRLreal}, one must reach momentum transfer scales $Q^2 > 1000\,$GeV$^2$ before $\mathpzc{w}_\varphi^2 < 1.6$; i.e., before $\mathpzc{w}_\varphi^2$ falls below half its original value.  This is yet another expression of the results depicted in Figs.\,\ref{FigalphaEvol}.

\subsection{Remarks}
Given these observations, the near agreement between the pertinent perturbative QCD prediction in Fig.\,\ref{Fig1CLpion} (right panel, curve-D) and the DSE prediction for $Q^2 F_\pi(Q^2)$ (right panel, curve-A) is striking.  It highlights that a single DSE interaction kernel, determined fully by just one parameter and preserving the one-loop renormalisation group behaviour of QCD, has completed the task of unifying the pion's electromagnetic form factor and its valence-quark distribution amplitude; and, indeed, numerous other quantities \cite{Chang:2011vu,Bashir:2012fs,Maris:2003vk,Eichmann:2011ej}.

Moreover, this leading-order, leading-twist QCD prediction, obtained with a pion valence-quark PDA evaluated at a scale appropriate to the experiment, Eq.\,\eqref{phiRLreal}, underestimates the full DSE-RL computation by merely an approximately uniform 15\% on the domain depicted.
The small mismatch is not eliminated by variation of $\Lambda_{\rm QCD}$ within its empirical bounds, which shifts curve-D by only $\pm 3$\% at $Q^2=20\,$GeV$^2$.  It is instead explained by a combination of higher-order, higher-twist corrections to Eq.\,\eqref{pionUV} in pQCD on the one hand, and shortcomings in the rainbow-ladder truncation, which predicts the correct power-law behaviour for the form factor but not precisely the right anomalous dimension in the strong coupling calculation on the other hand.
Hence, as anticipated earlier \cite{Maris:1998hc} and expressing a result that can be understood via the behaviour of the dressed-quark mass-function (see Eq.\,VII.18 in Ref.\,\cite{Bashir:2012fs}), one should expect dominance of hard contributions to the pion form factor for $Q^2\gtrsim 8\,$GeV$^2$.  Expressed differently, on $Q^2\gtrsim 8\,$GeV$^2$, $F_\pi(Q^2)$ will exhibit precisely the momentum-dependence anticipated from QCD, the power-law behaviour plus logarithmic violations of scaling, but with the normalisation fixed by a pion wave function whose dilation with respect to $\varphi_\pi^{\rm asy}(x)$ is a definitive signature of DCSB, which is such a crucial feature of the Standard Model.

\section{Nucleon elastic form factors}
\subsection{Charge distribution of the proton}
\label{secNucleonFF}
As we saw in Sec.\,\ref{sec:pionFF}, elastic form factors provide vital information about the structure and composition of the most basic elements of nuclear physics.  They are a measurable and physical manifestation of the nature of the hadrons' constituents and the dynamics that binds them together.  New, accurate form factor data are driving paradigmatic shifts in our pictures of hadrons and their structure; e.g., the role of orbital angular momentum and nonpointlike diquark correlations, the scale at which pQCD effects become evident, the strangeness content of nonstrange hadrons, the role of a meson-cloud, etc.

This is nowhere more evident than in analyses of experimental data acquired during the last decade, which have imposed a new ideal.  Namely, despite its simple valence-quark content, the internal structure of the nucleon is very complex, with marked differences between the distributions of total charge and magnetisation \cite{Jones:1999rz} and also between the distributions carried by the different quark flavours \cite{Cates:2011pz}.  The challenge now is to explain the observations in terms of elemental nonperturbative features of the strong interaction.

In this connection, we will here review work which shows that the behaviour of the proton's electric form factor in the $6\,$-$10\,$GeV$^2$ range is particularly sensitive to the rate at which the dressed-quark mass runs from the nonperturbative into the perturbative domain of QCD \cite{Cloet:2013gva}.

The proton's momentum-space charge and magnetisation distributions are measured through combinations of the two Poincar\'e-invariant elastic form factors that are required to express the proton's electromagnetic current:
\begin{equation}
i e \, \bar u(p^\prime) \big[ \gamma_\mu F_1(Q^2) +
\frac{Q_\nu}{2m_N}\, \sigma_{\mu\nu}\,F_2(Q^2)\big] u(p)\,,
\end{equation}
where $Q=p^\prime - p$, $u(p)$ and $\bar u(p^\prime)$ are, respectively, spinors describing the incident, scattered proton, and $F_{1,2}(Q^2)$ are the proton's Dirac and Pauli form factors.  The charge and magnetisation distributions \cite{Sachs:1962zzc}
\begin{equation}
\label{GEMpeq}
G_E(Q^2)  =  F_1(Q^2) - \tau F_2(Q^2)\,,\quad
G_M(Q^2)  =  F_1(Q^2) + F_2(Q^2)\,,
\end{equation}
feature in the electron-proton elastic scattering cross-section
\begin{equation}
\left(\frac{d\sigma}{d\Omega}\right) =
\left(\frac{d\sigma}{d\Omega}\right)_{\rm Mott}
\left[ G_E^2(Q^2) + \frac{\tau}{\varepsilon} G_M^2(Q^2)\right]
\frac{1}{1+\tau} \,,
\label{eq:rosenbluthM}
\end{equation}
where $\tau = Q^2/[4 m_N^2]$, $m_N$ is the proton's mass, and $\varepsilon$ is the polarisation of the virtual photon that mediates the interaction in Born approximation.  Equation~\eqref{eq:rosenbluthM} is simply a modification of the Rutherford cross-section \cite{Rutherford:1911zz} so as to include the effects of spin \cite{Mott04061929}.
(A modern view of the relationship between $G_{E,M}$ and configuration-space charge and magnetisation densities is provided elsewhere \cite{Miller:2010nz}.)

The first data on the proton's form factors were made available by the experiments described in Ref.\,\cite{Hofstadter:1955ae}.  In Born approximation one may infer the individual contribution from each form factor to the cross section by using the technique of Rosenbluth separation \cite{Rosenbluth:1950yq}.  Namely, one considers the reduced cross-section, $\sigma_{\rm R}$, defined via:
\begin{equation}
\sigma_{\rm R} \, \left(\frac{d\sigma}{d\Omega}\right)_{\rm Mott}
:= \varepsilon (1+\tau) \, \frac{d\sigma}{d\Omega}\,.
\label{eq:sigmareduced}
\end{equation}
It is plain from Eq.\,\eqref{eq:rosenbluthM} that $\sigma_{\rm R}$ is linearly dependent on $\varepsilon$; and so a linear fit to the reduced cross-section, at fixed $Q^2$ but a range of $\varepsilon$ values, provides $G_E^2(Q^2)$ as the slope and $\tau G_M^2(Q^2)$ as the $\varepsilon=0$ intercept.  Owing to the relative factor of $\tau$, however, the signal for $G_M^2(Q^2)$ is enhanced with increasing momentum transfer, a fact which complicates an empirical determination of the proton's charge distribution for $Q^2\gtrsim 1\,$GeV$^2$.  Notwithstanding this, of necessity the method was employed exclusively until almost the turn of the recent millennium and, on a domain that extends to 6\,GeV$^2$, it produced
\begin{equation}
\label{GEeqGM}
\left. \mu_p\, \frac{G_E(Q^2)}{G_M(Q^2)} \right|_{\rm Rosenbluth} \approx 1\,,
\end{equation}
and hence a conclusion that the distributions of charge and magnetisation within the proton are approximately identical on this domain \cite{Arrington:2006zm,Perdrisat:2006hj}.  Significantly, this outcome is consistent with the, then popular, simple pictures of the proton's internal structure in which, e.g., quark orbital angular momentum and correlations play little role.

The situation changed dramatically when the combination of high energy, current and polarisation at JLab enabled polarisation-transfer reactions to be measured \cite{Jones:1999rz}.  In Born approximation, the scattering of longitudinally polarised electrons results in a transfer of polarisation to the recoil proton with only two nonzero components: P$_\perp$, perpendicular to the proton momentum in the scattering plane; and P$_\parallel$, parallel to that momentum.  The ratio ${\rm P}_\perp/{\rm P}_\parallel$ is proportional to $G_E(Q^2)/G_M(Q^2)$ \cite{Akhiezer:1974em,Arnold:1980zj}.  A series of such experiments
\cite{Jones:1999rz,Gayou:2001qd,Gayou:2001qt,Punjabi:2005wq,Puckett:2010ac,Puckett:2011xg}
has determined that $G_E(Q^2)/G_M(Q^2)$ decreases almost linearly with $Q^2$ and might become negative for $Q^2 \gtrsim 8\,$GeV$^2$.  Such behaviour contrasts starkly with Eq.\,\eqref{GEeqGM}; and since the proton's magnetic form factor is reliably known on a spacelike domain that extends to $Q^2 \approx 30\,$GeV$^2$ \cite{Kelly:2004hm,Bradford:2006yz}, the evolution of this ratio exposes novel features of the proton's charge distribution, as expressed in $G_E(Q^2)$.

An explanation of the discrepancy between the Rosenbluth and polarisation transfer results for the ratio is currently judged to lie in two-photon-exchange corrections to the Born approximation, which affect the polarisation transfer extraction of $G_E(Q^2)/G_M(Q^2)$ far less than they do the ratio inferred via Rosenbluth separation \cite{Arrington:2011dn}.  The last decade has thus forced acceptance of a new paradigm; viz., the nucleon's internal structure must actually be very complex, with marked differences between the distributions of charge and magnetisation.

Given that sixty years of experimental effort has thus far discovered only one hadronic form factor that displays a zero; namely, the Pauli form factor associated with the transition between the proton and its first radial excitation (the Roper resonance), and that this feature was discovered just recently \cite{Dugger:2009pn,Aznauryan:2009mx,Aznauryan:2011td}, the chance that the proton's electric form factor might become negative is fascinating.  It is therefore worth elucidating the conditions under which that outcome is realisable \emph{before} the zero is empirically either located or eliminated as a reasonable possibility.  This is even more valuable if the appearance or absence of a zero is causally connected with a fundamental nonperturbative feature of the Standard Model.

A continuum computation of the proton's elastic form factors was therefore considered in Ref.\,\cite{Cloet:2013gva}.\footnote{The current status of lattice-QCD calculations of nucleon form factors is described in Refs.\,\protect\cite{Alexandrou:2011iu,Renner:2012yh}.  Contemporary computations have produced mixed results, with both some encouraging agreements and some noticeable disagreements.  Results, which may fairly be described as exploratory, are available on $0 \lesssim Q^2\lesssim 1.5\,$GeV$^2$.}
For a proton described by the amplitude in Fig.\,\ref{figFaddeev}, the electromagnetic current is known \cite{Oettel:1999gc}.  The key element in constructing that current is the dressed-quark-photon vertex.  It is plain from a consideration of the Ward-Green-Takahashi identities \cite{Ward:1950xp,Green:1953te,Takahashi:1957xn,Takahashi:1985yz} and the structure of the functions in Eq.\,\eqref{SgeneralN} that the bare vertex ($\gamma_\mu$) is not a good approximation to the dressed vertex for $Q^2 \lesssim 2\,$GeV$^2$, where (as above) $Q$ is the incoming photon momentum.  This has long been clear \cite{Ball:1980ay} and recent years have produced a sophisticated understanding of the coupling between the photon and a dressed-fermion.  Two model-independent results, which have emerged from the vast body of literature, are crucial herein \cite{Chang:2010hb,Qin:2013mta}: the \emph{Ansatz} described in Ref.\,\cite{Ball:1980ay} is the unique form for the solution of the longitudinal Ward-Green-Takahashi identity; and the transverse part of the dressed vertex expresses a dressed-quark anomalous magnetic moment distribution, which is large at infrared momenta.  Stated simply, the photon to dressed-quark coupling is markedly different from that of a pointlike Dirac fermion.

\begin{figure}[t]
\leftline{%
\includegraphics[clip,width=0.45\linewidth]{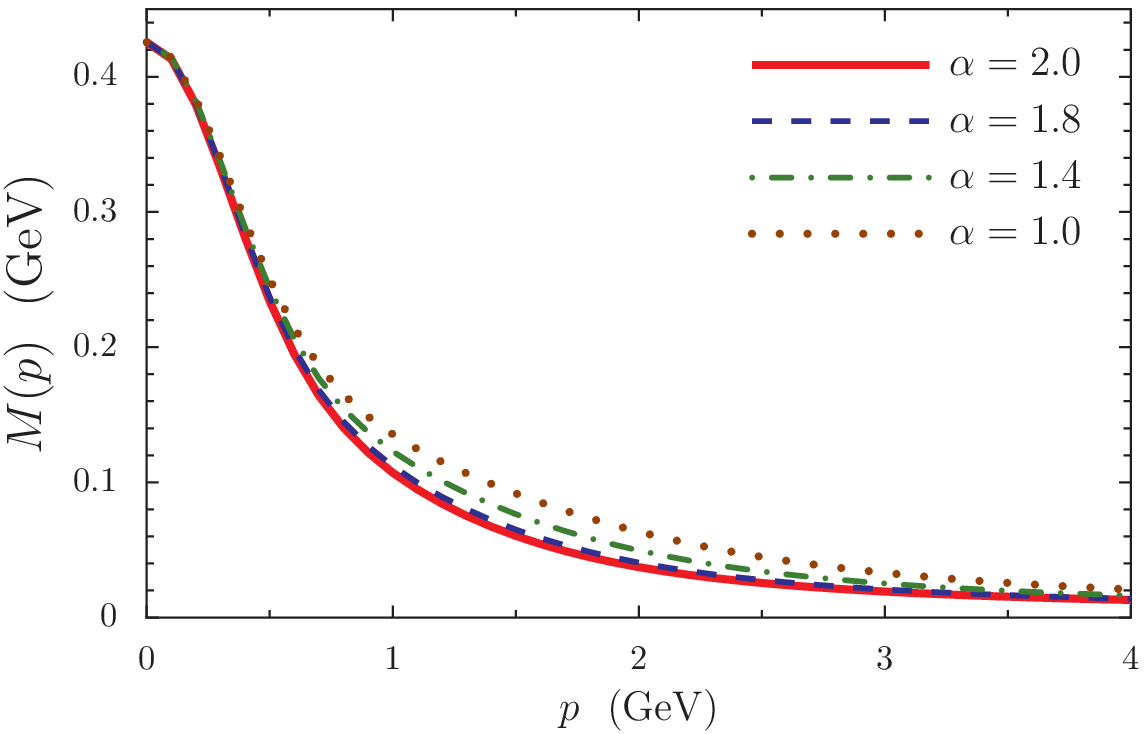}}
\vspace*{-29.5ex}

\rightline{%
\includegraphics[clip,width=0.465\linewidth]{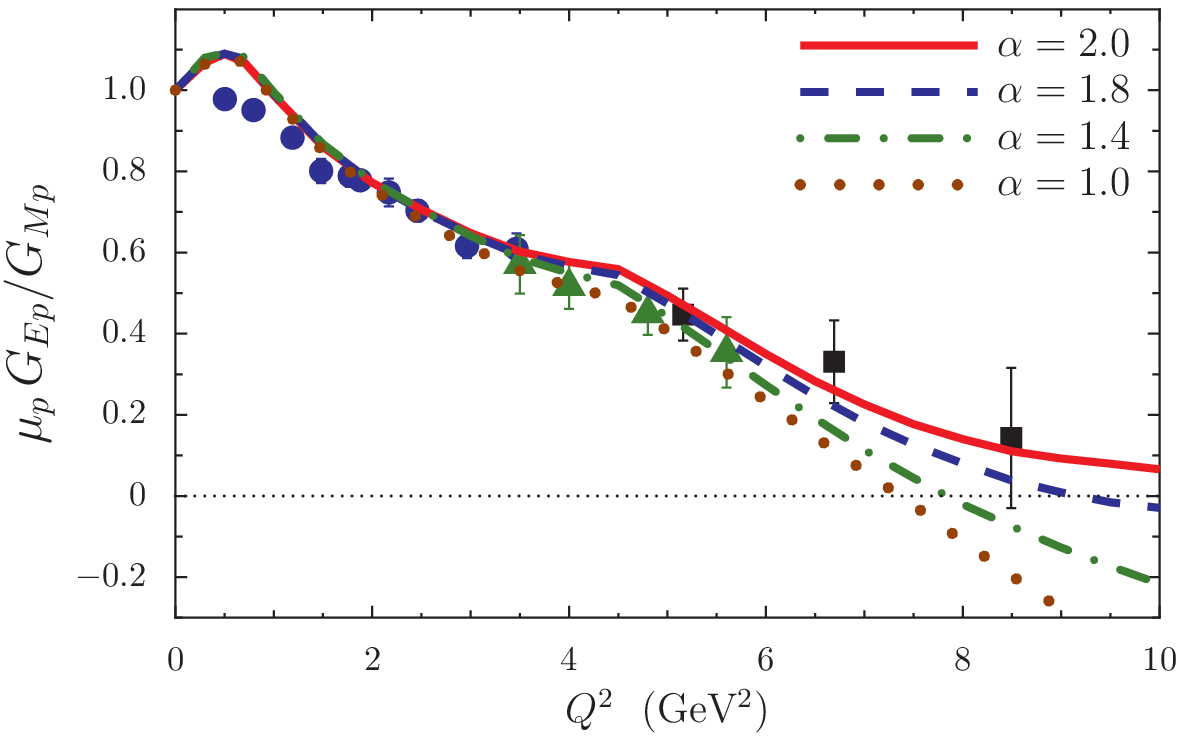}}
\caption{\label{figMp}
\emph{Left panel}. Dressed-quark mass function employed in Ref.\,\protect\cite{Cloet:2013gva}.  $\alpha=1$ specifies the reference form  
and increasing $\alpha$ diminishes the domain upon which DCSB is active.
\emph{Right panel}. Response of $\mu_p G_E/G_M$ to increasing $\alpha$; i.e., to an increasingly rapid transition between constituent- and parton-like behaviour of the dressed-quarks.  Data are from Refs.\,\protect\cite{Jones:1999rz,Gayou:2001qd,Gayou:2001qt,Punjabi:2005wq,Puckett:2010ac,Puckett:2011xg}.}
\end{figure}

The computation of the proton's elastic form factors, using the elements detailed above, is exemplified in Refs.\,\cite{Cloet:2008re,Chang:2011tx,Cloet:2011qu}.  That framework was used in Ref.\,\cite{Cloet:2013gva}, with the dressed-quark mass-function illustrated in the left panel of Fig.\,\ref{figMp}, the associated dressed-quark propagator, and the following dressed-quark--photon vertex:
\begin{eqnarray}
\label{eqVertex}
\Gamma_\mu(k,p) & = & \Gamma_\mu^{\rm BC}(k,p) - \tilde\eta_e \sigma_{\mu\nu} q_\nu \Delta_B(k^2,p^2)\,,
\end{eqnarray}
with $q=k-p$, $t=k+p$, and \cite{Ball:1980ay}
\begin{eqnarray}
\Gamma_\mu^{\rm BC}(k,p) &=& \sum_{j=1}^3 \lambda_j(k,p) \, L^j_\mu(k,p)\,,
\end{eqnarray}
where:
\begin{equation}
L^1_\mu = \gamma_\mu\,, \quad
L^2_\mu= (1/2)\, t_\mu\, \gamma\cdot t\,, \quad
L^3_\mu = - i t_\mu\, \mathbf{I}_{\rm D}\,;
\end{equation}
$\lambda_1 = \Sigma_A(k^2,p^2)$, $\lambda_2 = \Delta_A(k^2,p^2)$, $\lambda_3 = \Delta_B(k^2,p^2)$; and $\Sigma_{\phi}(k^2,p^2) = [\phi(k^2)+\phi(p^2)]/2$,
with $A$, $B$ in Eq.\,\eqref{SgeneralN}.  The second term in Eq.\,\eqref{eqVertex} expresses the momentum-dependent dressed-quark anomalous magnetic moment distribution, with $\tilde\eta_e=0.4$ being the modulating magnitude \cite{Qin:2013mta}.
Notably, $G_{E}(Q^2)$ and $G_M(Q^2)$ are described equally well \cite{Cloet:2008re,Chang:2011tx,Cloet:2011qu}.

In order to highlight a connection between DCSB and the $Q^2$-dependence of proton form factors, Ref.\,\cite{Cloet:2013gva} introduced a damping factor, $\alpha$, into the dressed-quark propagator used for all calculations in Refs.\,\cite{Cloet:2008re}.\footnote{Explicitly, $b_3 \to \alpha b_3$ in Eq.\,(A.19) of Ref.\,\protect\cite{Cloet:2008re}, the effect of which is a modification in Eq.\,\eqref{SgeneralN}
that may be approximated as $B(p)\to B(p)(1+ \alpha f(p))/(1+\alpha^2 f(p))$, $f(p)=2(p/2)^4/(1+(p/2)^6)$.}
The value $\alpha=1$ specifies the reference form of the dressed-quark propagator, which was obtained in a fit to a diverse array of pion properties \protect\cite{Burden:1995ve}.  It produces a chiral-limit condensate \cite{Chang:2011mu,Brodsky:2010xf,Brodsky:2012ku,Chang:2013epa} $\langle \bar q q \rangle^0_\pi = -(0.250{\rm GeV}=:\chi^0_\pi)^3$; and is associated with a prediction of the pion's valence-quark distribution function \protect\cite{Hecht:2000xa} that was recently verified empirically \protect\cite{Aicher:2010cb}.

As $\alpha$ is increased, the rate at which the dressed-quark mass function drops towards its perturbative behaviour is accelerated so that, as evident in the left panel of Fig.\,\ref{figMp}, the strength of DCSB is diminished and the influence of explicit chiral symmetry breaking is exposed at smaller dressed-quark momenta.  This is the qualitative impact of $\alpha$ that was exploited in Ref.\,\cite{Cloet:2013gva}.

At each value of $\alpha$, Ref.\,\cite{Cloet:2013gva} repeated all steps in the computation detailed in Ref.\,\cite{Cloet:2008re}.  Namely, the Faddeev equation was solved to obtain the proton's mass and amplitude, and, using that material the authors constructed the current and computed the proton's elastic form factors.  The scalar and axial-vector diquark masses were held fixed as $\alpha$ was varied, in which case the nucleon mass, $m_N$, drops by $<1$\% as $\alpha$ is increased from $1.0$ to $2.0$.  Since damping was deliberately implemented so that the pointwise evolution of $M(p^2)$ to its ultraviolet asymptote is accelerated without changing $M(p^2=0)$ and because the computed values of masses are primarily determined by the infrared value of mass-functions \cite{Chen:2012qr}, this is a reasonable assumption on the input and an understandable result for $m_N$.

The effect on $G_E(Q^2)/G_M(Q^2)$, produced by suppressing DCSB, is displayed in the right panel of Fig.\,\ref{figMp}.  The impact is striking.  For $\alpha=1$, the result in Ref.\,\cite{Chang:2011tx} is recovered.  It exhibits a zero in $G_E(Q^2)$, and hence in the ratio, at $Q^2 \approx 8\,$GeV$^2$.  However, as $\alpha$ is increased, so that the strength of DCSB is damped, the zero is pushed to larger values of $Q^2$, until it disappears completely at $\alpha=2.0$.  Associating the curves in the left and right panels of the figure, one observes that apparently modest changes in the rate at which the mass function drops toward its ultraviolet asymptote have a dramatic effect on the location and existence of a zero in  $G_E(Q^2)/G_M(Q^2)$.

\begin{figure}[t]
\leftline{%
\includegraphics[clip,width=0.45\linewidth]{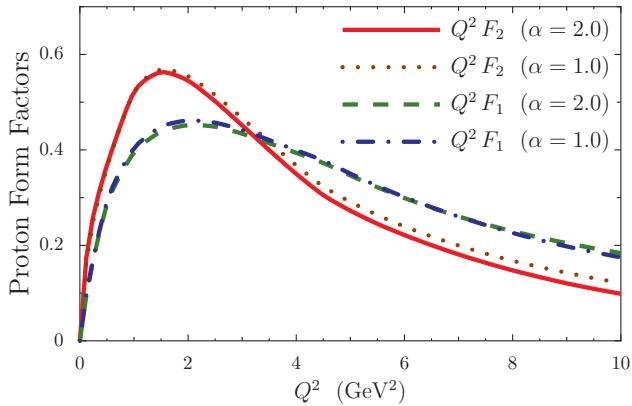}}
\vspace*{-30ex}

\rightline{\parbox{23em}{\caption{\label{figF2}
$Q^2$-weighted proton Dirac ($Q^2 F_1$) and Pauli ($Q^2 F_2$) form factors calculated with $\alpha=1.0$, $2.0$.  $Q^2 F_1$ shows little sensitivity to the rate at which dressed-quarks make the transition between constituent- and parton-like behaviour and hence $F_1$, none at all.  In contrast, $Q^2 F_2$, which also appears in the definition of $G_E$, exhibits a measurable dependence.}}}
\vspace*{4ex}

\end{figure}

In order to explain this remarkable behaviour, it is useful to recall Eqs.\,\eqref{GEMpeq}.  The magnetic form factor is a simple additive linear combination of the proton's Dirac and Pauli form factors.  Therefore, small changes in $F_{1,2}(Q^2)$, arising from the differences displayed in the left panel of Fig.\,\ref{figMp} and illustrated in Fig.\,\ref{figF2}, cannot have a large impact.  On the other hand, the electric form factor is a \emph{difference}, in which changes in the Pauli form factor are amplified with increasing $Q^2$.

Physically, the Pauli form factor is a gauge of the distribution of magnetisation within the proton.  Absent $F_2$, the proton's electromagnetic current would be like that of a Dirac fermion.  The $Q^2=0$ value of the Pauli form factor is the proton's anomalous magnetic moment; and the evolution with $Q^2$ measures the distribution of anomalous magnetisation within the proton bound-state.  In the DSE approach, the proton's magnetisation is carried by dressed-quarks and influenced by correlations amongst them.  The latter are expressed via the Faddeev wave-function, obtained by reattaching the quark lines to the Faddeev amplitude.  This wave function exhibits $S$-, $P$- and $D$-wave quark orbital angular momentum correlations in the proton's rest frame \cite{Cloet:2007pi}.
The resulting nucleon mass is $1.18\,$GeV, a value which accommodates the material negative pion-loop corrections \cite{Cloet:2008re,Suzuki:2009nj}.

Suppose for a moment that quarks are described by a momentum-independent dressed-mass, as in Ref.\,\cite{Wilson:2011aa}.  In that counterpoint to QCD, the dressed-quarks produce hard Dirac and Pauli form factors, which yield a ratio $\mu_p G_E/G_M$ that possesses a zero at $Q^2\lesssim 4\,$GeV$^2$.

Alternatively, consider a proton comprised of dressed-quarks associated with the mass function in the left panel of Fig.\,\ref{figMp}.  This mass function is large at infrared momenta but approaches the current-quark mass as the momentum of the dressed-quark increases.  As we have explained, such is the behaviour in QCD: dressed-quarks are massive in the infrared but become parton-like in the ultraviolet, characterised thereupon by a mass function that is modulated by the current-quark mass.  In this case, the proton's dressed-quarks possess constituent-quark-like masses at small momenta.  Thus, for all considered values of $\alpha$: these quarks possess a large anomalous magnetic moment at infrared momenta (in keeping with their large mass) \cite{Chang:2010hb}; $F_{1,2}(Q^2)$ are insensitive to $\alpha$ on this domain; and hence so is the ratio $\mu_p G_E/G_M$.

On the other hand, as the momentum transfer grows, the structure of the integrands in the computation of the elastic form factors ensures that the dressed-quark mass functions are increasingly sampled within the domain upon which the chiral condensate \cite{Chang:2011mu,Brodsky:2010xf,Brodsky:2012ku,Chang:2013epa} modulates the magnitude of $M(p^2)$.  This corresponds empirically to momentum transfers $Q^2 \gtrsim 5\,$GeV$^2$.  Plainly, as this chiral order parameter becomes smaller, a part of DCSB is suppressed, and the dressed-quarks become increasingly parton-like; viz., they are partially unclothed and come to behave as light fermion degrees of freedom on a larger momentum domain.  
Following in large part, then, from the fact that light-quarks must have a small anomalous magnetic moment \cite{Chang:2010hb}, the proton Pauli form factor generated dynamically therewith drops more rapidly to zero: the quark angular momentum correlations remain but the individual dressed-quark magnetic moments diminish markedly.  This is apparent in Fig.\,\ref{figF2}.

Thus, as a consequence of suppressing the domain upon which DCSB is active, an effect expressed via a suppression of $\chi^0_\pi$ in the model used for this illustration, the zero in the ratio $\mu_p G_E/G_M$ is pushed to larger values of $Q^2$, until it disappears from the currently accessible experimental domain when $\chi^0_\pi$ falls to roughly 80\% of its unperturbed value.  Indeed, in this case there is no zero in the computed result on $Q^2>0$.

\begin{figure}[t]
\leftline{%
\includegraphics[clip,width=0.45\linewidth]{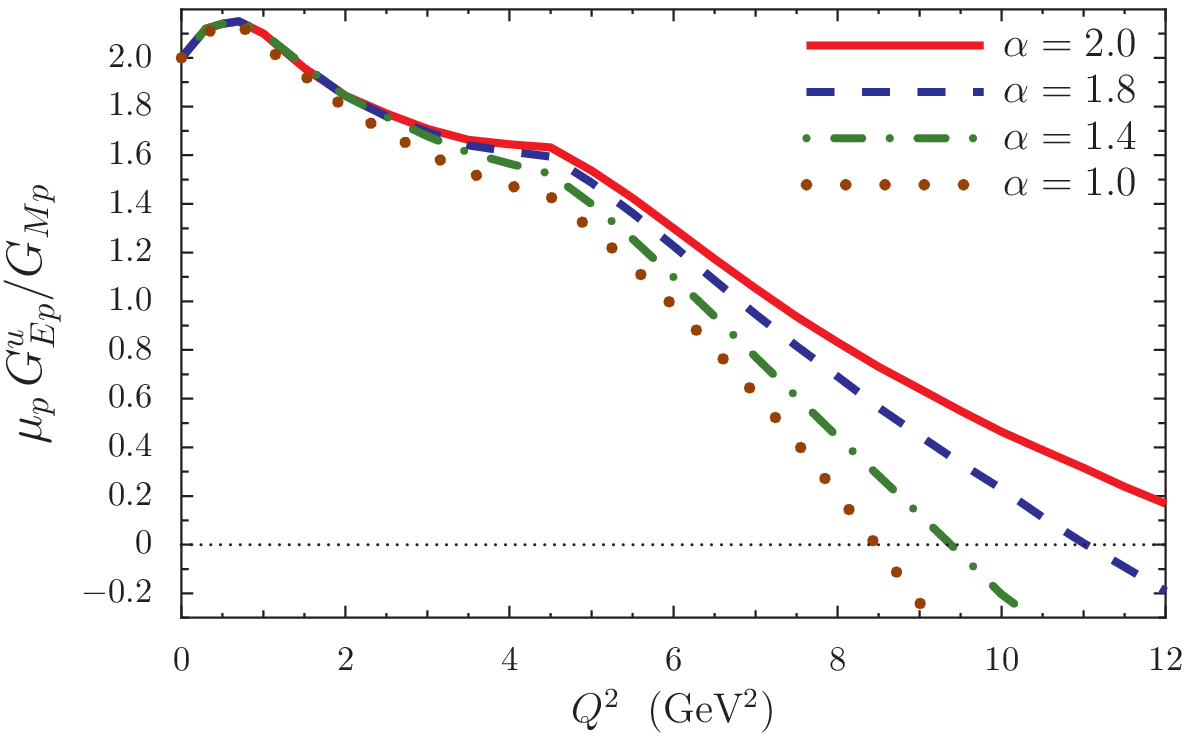}}
\vspace*{-29.5ex}

\rightline{%
\includegraphics[clip,width=0.465\linewidth]{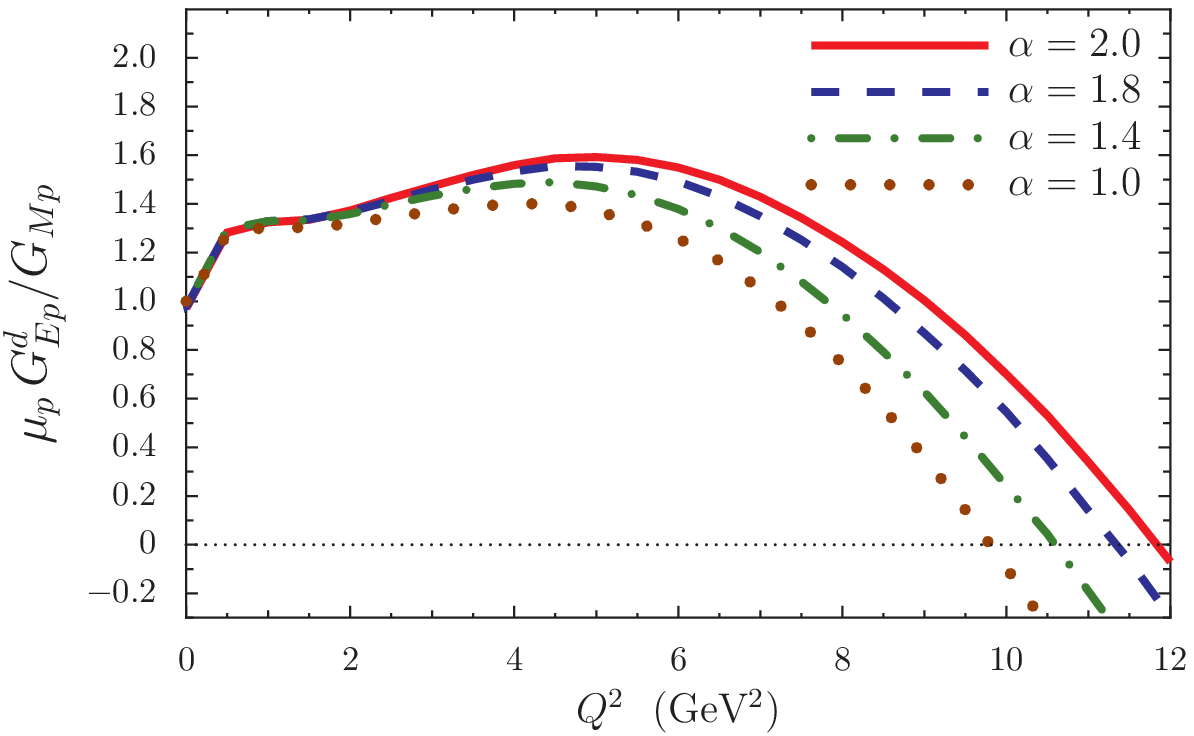}}
\caption{\label{figFlavourSepGEGM}
\emph{Left panel}.  $\alpha$-dependence of the $u$-quark contribution to $\mu_p G_{Ep}(Q^2)/G_{Mp}(Q^2)$.
\emph{Right panel}. $\alpha$-dependence of the $d$-quark contribution to $\mu_p G_{Ep}(Q^2)/G_{Mp}(Q^2)$.
The normalisation is such that $G_{Ep}= e_u G_{Ep}^u+e_d G_{Ep}^d$, where $e_{u,d}$ are the valence-quark electric charges in units of the positron's charge.}
\end{figure}

In Fig.\,\ref{figFlavourSepGEGM} we display the manner by which the distinct valence-quark flavours contribute to this effect.  Owing to the presence of strong diquark correlations, the singly-represented $d$-quark is usually sequestered inside a soft diquark correlation.  So, although it does become parton-like more quickly as $\alpha$ is increased, that change is hidden from view.  The effect in the right panel of Fig.\,\ref{figMp} is therefore driven primarily by the electric form factor of the doubly-represented $u$-quark, which is four times more likely than the $d$-quark to be involved in a hard interaction.  Indeed, it is apparent that $G_{Ep}/G_{Mp} \approx G_{Ep}^u/G_{Mp}$.

An improvement of Ref.\,\cite{Cloet:2013gva} is possible; e.g., via the \emph{ab initio} treatment of the DSEs detailed in Refs.\,\cite{Eichmann:2008ef,Eichmann:2011vu}.  However, close inspection of results already obtained in that more sophisticated approach lends support to the conclusions reached already; namely, as $\chi^0_\pi/M(0)$ decreases, the proton's electric form factor approaches zero less rapidly.

We explained above that the fully-consistent treatment of a quark-quark interaction which yields dressed-quarks with a constant mass-function, produces a zero in $\mu_p G_{Ep}(Q^2)/G_{Mp}(Q^2)$ at a small value of $Q^2$.  At the other extreme, a theory in which the mass-function rapidly becomes partonic -- namely, is very soft -- produces no zero at all.  From a theoretical perspective, there are numerous possibilities in between.
It follows that the possible existence and location of the zero in the ratio of proton elastic form factors [$\mu_p G_{Ep}(Q^2)/G_{Mp}(Q^2)$] are a fairly direct measure of the nature of the quark-quark interaction in the Standard Model.  Like the dilation of the pion's valence-quark PDA, they are a cumulative gauge of the momentum dependence of the interaction, the transition between the associated theory's nonperturbative and perturbative domains, and the width of that domain.  Hence, in extending experimental measurements of this ratio, and thereby the proton's charge form factor, to larger momentum transfers; i.e., in reliably determining the proton's charge distribution, there is an extraordinary opportunity for a constructive dialogue between experiment and theory.  That feedback will assist greatly with contemporary efforts to reveal the character of the strongly interacting part of the Standard Model and its emergent phenomena.

\subsection{Nucleon structure at very high $x$}
A great deal more information is contained in nucleon elastic form factors.  This may be illustrated by exhibiting a connection with valence-quark PDFs at very high $x$.  Since the advent of the parton model and the first deep inelastic scattering (DIS) experiments there has been a determined effort to deduce the PDFs of the most stable hadrons: neutron, proton and pion \cite{Holt:2010vj}.  The behavior of such distributions on the far valence domain (Bjorken-$x> 0.5$) is of particular interest because this domain is definitive of hadrons; e.g., quark content on the far valence domain is how one distinguishes between a neutron and a proton.  Indeed, all Poincar\'e-invariant properties of a hadron: baryon number, charge, flavour content, total spin, etc., are determined by the PDFs which dominate on the far valence domain.

The endpoint of the far valence domain, $x=1$, is especially significant because, whilst all familiar PDFs vanish at $x=1$, ratios of any two need not; and, under DGLAP evolution, the value of such a ratio is invariant \cite{Holt:2010vj}.  Thus, e.g., with $d_v(x)$, $u_v(x)$ the proton's $d$, $u$ valence-quark PDFs, the value of $\lim_{x\to 1} d_v(x)/u_v(x)$ is an unambiguous, scale invariant, nonperturbative feature of QCD.  It is therefore a keen discriminator between frameworks that claim to explain nucleon structure.  Furthermore, Bjorken-$x=1$ corresponds strictly to the situation in which the invariant mass of the hadronic final state is precisely that of the target; viz., elastic scattering.  The structure functions inferred experimentally on the neighborhood $x\simeq 1$ are therefore determined theoretically by the target's elastic form factors.

\begin{table}[t]
\begin{center}
\caption{\label{tab:a}
Selected predictions for the $x=1$ value of the indicated quantities, where the elements are:
$F_2^{p,n}$ -- unpolarised nucleon structure functions;
$u$, $d$ -- unpolarised valence-quark distribution functions;
$\Delta u$, $\Delta d$ -- distribution functions for longitudinally polarised valence-quarks;
$A_1^{p,n}$ -- nucleon longitudinal spin asymmetries.
The DSE results were computed in Ref.\,\protect\cite{Roberts:2013mja}: DSE-1 (also denoted ``DSE realistic'' herein) indicates use of the momentum-dependent dressed-quark mass-function in Ref.\,\protect\cite{Cloet:2008re}; and DSE-2 (also denoted ``DSE contact'') corresponds to predictions obtained with a contact interaction \protect\cite{Roberts:2011wy}.
The next four rows are, respectively, results drawn from Refs.\,\protect\cite{Close:1988br,Cloet:2005pp,Hughes:1999wr,Isgur:1998yb}.
The last row, labeled ``pQCD,'' expresses predictions made in Refs.\,\protect\cite{Farrar:1975yb,Brodsky:1994kg}, which are actually model-dependent: they assume an SU$(6)$ spin-flavour wave function for the proton's valence-quarks and the corollary that a hard photon may interact only with a quark that possesses the same helicity as the target.
}
\begin{tabular}{lrrrrrrr}\hline
    & $\frac{F_2^n}{F_2^p}$ & $\frac{d}{u}$ & $\frac{\Delta d}{\Delta u}$
    & $\frac{\Delta u}{u}$ & $\frac{\Delta d}{d}$ & $A_1^n$ & $A_1^p$\\\hline
%
DSE-1
& $0.49$ & $0.28$ & $-0.11$ & 0.65 & $-0.26$ & 0.17 & 0.59 \\
DSE-2& $0.41$ & $0.18$ & $-0.07$ & 0.88 & $-0.33$ & 0.34 & 0.88\\\hline
$0_{[ud]}^+$ & $\frac{1}{4}$ & 0 & 0 & 1 & 0 & 1 & 1 \\[0.5ex]
NJL & $0.43$ & $0.20$ & $-0.06$ & 0.80 & $-0.25$ & 0.35& 0.77 \\[0.5ex]
SU$(6)$ & $\frac{2}{3}$ & $\frac{1}{2}$ & $-\frac{1}{4}$ & $\frac{2}{3}$ & $-\frac{1}{3}$ & 0 & $\frac{5}{9}$ \\[0.5ex]
CQM & $\frac{1}{4}$ & 0 & 0 & 1 & $-\frac{1}{3}$ & 1 & 1 \\[0.5ex]
pQCD & $\frac{3}{7}$ & $\frac{1}{5}$ & $\frac{1}{5}$ & 1 & 1 & 1 & 1 \\\hline
\end{tabular}
\end{center}
\end{table}

\begin{figure}[t]
\leftline{\includegraphics[clip,angle=-90,width=0.55\textwidth]{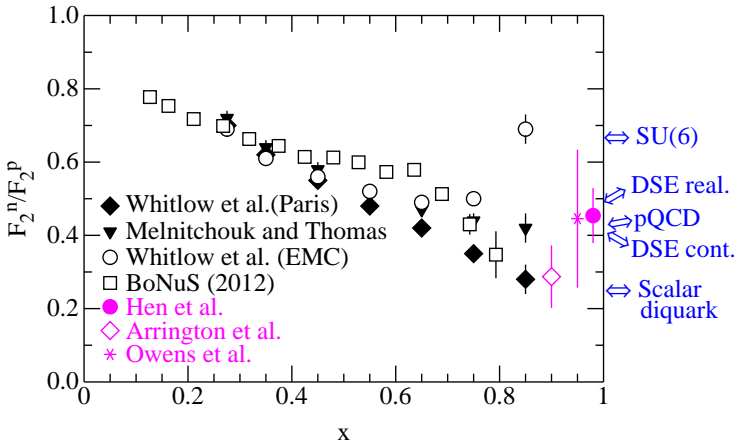}}
\vspace*{-28ex}

\rightline{\parbox{19em}{\caption{\label{f2nratio} $F_2^n/F_2^p$ as a function of $x$.  Results from five extraction methods are shown \protect\cite{Whitlow:1991uw,Melnitchouk:1995fc,Arrington:2008zh,Hen:2011rt,Arrington:2011qt,Owens:2012bv} along with selected predictions from Table~\protect\ref{tab:a}. N.B.\ DSE~realistic corresponds to DSE-1 and DSE~contact to DSE-2.
}}}
\vspace*{5em}
\end{figure}

\begin{figure}[t]
\begin{tabular}{cc}
\includegraphics[clip,angle=-90,width=0.45\textwidth]{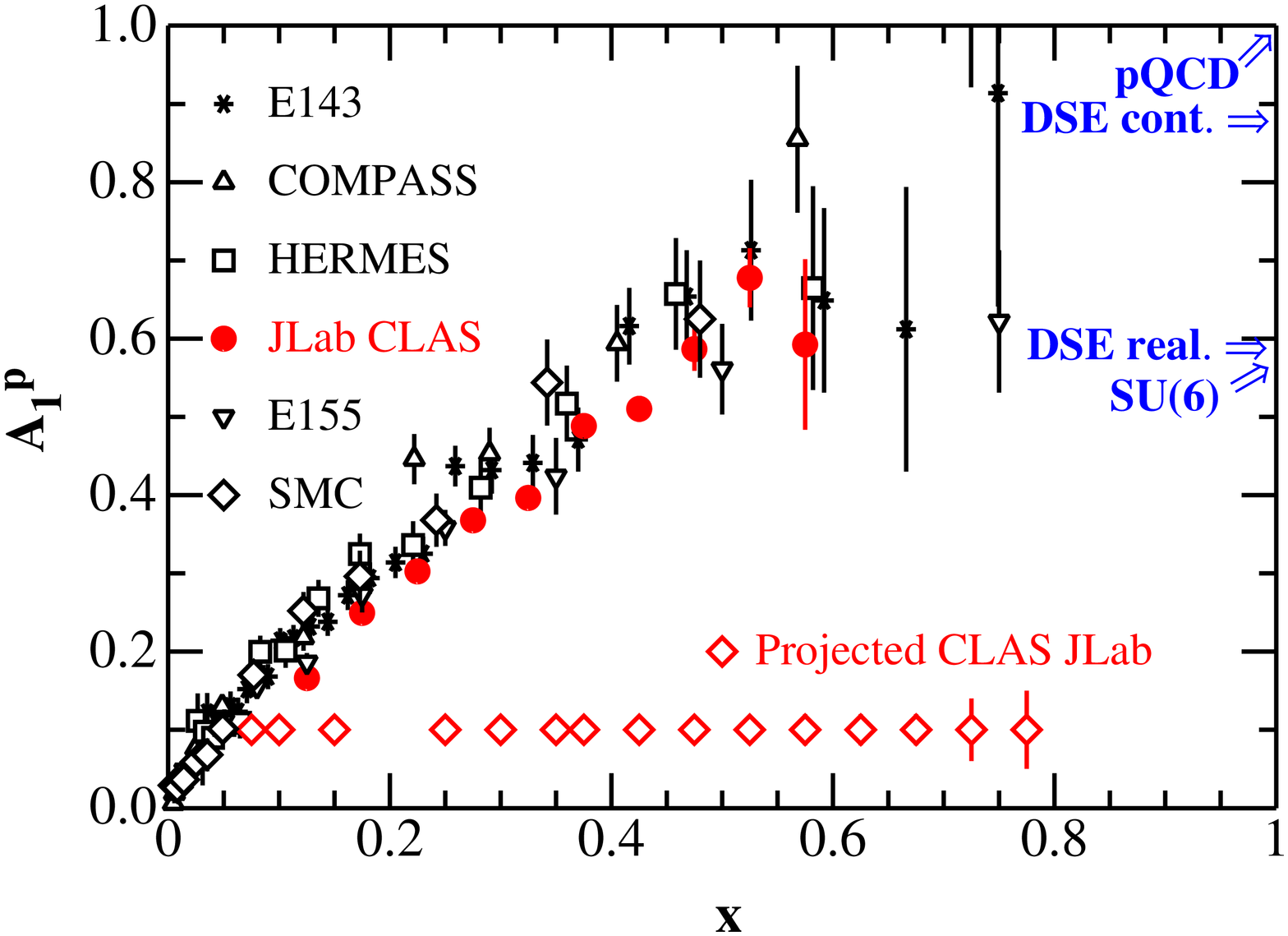} &
\includegraphics[clip,angle=-90,width=0.46\textwidth]{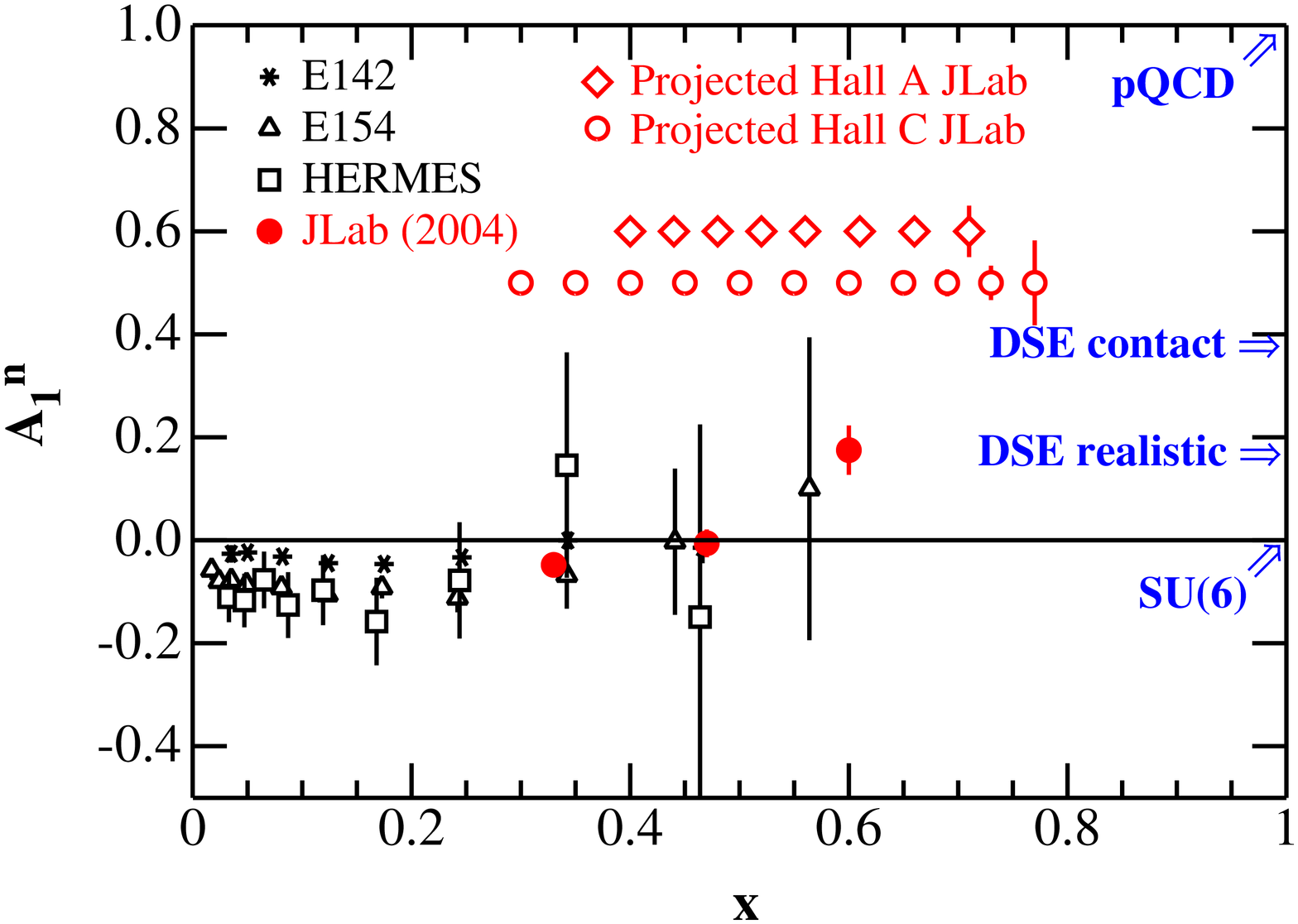}
\end{tabular}
\caption{\label{a1p}
\emph{Left panel}.  Existing and projected measurements of the proton's longitudinal spin asymmetry as a function of $x$ (statistical errors only), along with selected predictions from Table~\protect\ref{tab:a}.  \emph{Right panel}.  Same for the neutron.  N.B.\ The data are explained in Refs.\,\cite{Kuhn:2008sy,Alekseev:2010ub};  we only display $A_1^n$ data obtained from polarised $^3$He targets; and the projected measurements are detailed in Refs.\,\protect\cite{kuhn:2006,Zheng:2006,Liyanage:2006}.}
\end{figure}

This connection was exploited in Refs.\,\cite{Wilson:2011aa,Roberts:2013mja} in order to deduce a collection of simple formulae, expressed in terms of diquark appearance and mixing probabilities, from which one may compute ratios of unpolarised and also longitudinal-spin-dependent $u$- and $d$-quark parton distribution functions on the domain $x\simeq 1$.  Through a comparison with predictions from other approaches, reproduced in Table~\ref{tab:a}, plus a consideration of extant and planned experiments, illustrated in Figs.\,\ref{f2nratio} and \ref{a1p}, Ref.\,\cite{Roberts:2013mja} showed that the measurement of nucleon longitudinal spin asymmetries on $x\simeq 1$ can add considerably to our capacity for discriminating between contemporary pictures of nucleon structure.

In recognition of the significance of the far valence domain, a new generation of experiments, focused on $x\gtrsim 0.5$, is planned at JLab, and under examination in connection with Drell-Yan studies at the Fermi National Accelerator Facility (FNAL) \cite{Reimer:Figure} and a possible EIC.  Consideration is also being given to experiments aimed at measuring parton distribution functions in mesons at J-PARC.  Furthermore, at FAIR it would be possible to directly measure the Drell-Yan process from high $x$ antiquarks in the antiproton annihilating with quarks in the proton.  A spin physics program at the Nuclotron based Ion Collider fAcility (NICA), under development in Dubna, might also make valuable contributions in this effort to map PDFs on the far-valence domain.

\subsection{Intrinsic strangeness} 
Another question connected with quark PDFs concerns the pointwise behaviour of sea-quark distributions for values of $x\in (0.1,0.5)$, a region which borders the far-valence domain.  Namely, whether or not the proton contains ``intrinsic'' sea \cite{Brodsky:1980pb}; i.e., a strangeness or charm distribution that is significantly larger on $x\in (0.1,0.5)$ than can be explained by perturbative mechanisms alone, such as gluon splitting $g\to \bar q q$.  Based on empirical evidence, there is a suggestion \cite{Chang:2011vx} that this might be the case for the $s$-quark, at least.

Within the framework of nucleon structure studies based on the DSEs, one may readily identify a nonperturbative mechanism that can enhance the proton's $s \bar s$ content; viz., resonant (meson-loop) contributions to the dressed-quark--gluon vertex in the gap equation, Eq.\,\eqref{gendseN}.  These are the type-1 corrections described in App.\,\ref{AppMesonCloud}.  A simple estimate of the magnitude of this effect is described in Ref.\,\cite{Cloet:2008fw}, wherein the $u$-, $d-$ and $s$-quark gap equations become mutually coupled owing to $\pi$- and $K$-meson loops in their respective kernels.  Constrained by empirical information on $f_{\pi,K}$, $m_{\pi,K}$, that analysis yields the following:
\begin{subequations}
\begin{eqnarray}
| u \rangle_{\rm loop\,dressed} &=&
0.87 \, |u\rangle+ 0.04 \, |u\, u\bar u \rangle + 0.09 \,|u\, d\bar d \rangle 
+ 0.02 \, |u\, s\bar s \rangle ,\\
| d \rangle_{\rm loop\,dressed} &=&
0.87 \, |d\rangle+ 0.09 \, |d\, u\bar u \rangle + 0.04 \,|d\, d\bar d \rangle
+ 0.02 \, |d\, s\bar s \rangle ,\\
| s \rangle_{\rm loop\,dressed} &=&
0.94 \, |s\rangle+ 0.03 \, |s\, u\bar u \rangle + 0.03 \,|s\, d\bar d \rangle,
\end{eqnarray}
\end{subequations}
where $|q\rangle$ represents a quark whose dressing does not include meson-loops.

If one assumes that the mixing probabilities are unmodified when the dressed-quarks combine to form the nucleon, then
\begin{equation}
\label{eqPss}
{\cal P}_p^{s\bar s} = 0.05\,;
\end{equation}
i.e., the $s\bar s$ content of the proton is 5\% at the model scale  The estimate in Ref.\,\cite{Chang:2011vx} is  2-3\%.  In addition, one has 
\begin{equation}
{\cal P}_p^{d\bar d} - {\cal P}_p^{u\bar u} = 0.04\,;
\end{equation}
i.e., a 4\% excess of $d\bar d$ over $u\bar u$ in the proton.  The estimate reported in Ref.\,\cite{Chang:2011vx} is 11-13\%.  

The result in Eq.\,\eqref{eqPss} has also be used to estimate two other properties of the proton that reflect its strangeness content: the $s$-quark contribution to the proton's magnetic moment \cite{Cloet:2008fw}, $\mu_p^S \approx -0.02\,$nuclear magnetons; and the nucleon's $s$-quark $\sigma$-term \cite{Chang:2009ae}, $f_N^s:= (m_s/M_N) (d M_N/dm_s) \approx 0.024$.  Both these results are commensurate with other modern estimates: $\mu_p^S=-0.046 \pm 0.019$ nuclear magnetons \cite{Leinweber:2004tc} and $f_N^s = 0.022\pm0.006$ \cite{Shanahan:2012wh}.

It is plain that even a crude but well-constrained computation based on resonant contributions to DSE kernels is capable of producing semi-quantitative agreement with empirical information on intrinsic strangeness.  Improvements to the analysis in Ref.\,\cite{Cloet:2008fw} would include incorporating resonant contributions into the Faddeev kernel; namely, the type-2 effects described in App.\,\ref{AppMesonCloud}.

\section{$N$-$\Delta$ transition form factor}
\label{SecNDelta}
\subsection{Historical context}
Given the challenges posed by nonperturbative QCD, it is insufficient to study hadron ground-states alone if one seeks a solution.  In order to chart the infrared behaviour of the quark-quark interaction via a collaborative effort between experiment and theory, every available tool must be exploited to its fullest.  In particular, the effort can benefit substantially by exposing the structure of nucleon excited states ($N^\ast$-states) and measuring the associated transition form factors at high momentum transfers \cite{Aznauryan:2012ba}.  High momenta are needed to pierce the meson-cloud that, often to a large extent, screens the dressed-quark core of all baryons; and, as we saw in Secs.\,\ref{sec:pionFF}, \ref{secNucleonFF}, it is with the $Q^2$ evolution of form factors that one gains access to the behaviour in Fig.\,\ref{gluoncloud}.

In this connection, we note that the $\Delta(1232)$ family were the first resonances discovered in $\pi N$ reactions \cite{Fermi:1952zz,Anderson:1952nw,Nagle:1984sg}.  They have since been studied extensively, both experimentally and theoretically; and their flavour content and spin are now well-known \cite{Beringer:1900zz}: $\Delta(1232)$-baryons are positive parity, isospin $I=\frac{3}{2}$, total-spin $J=\frac{3}{2}$ bound-states with no net strangeness.  As such, the $\Delta^+$ and $\Delta^0$ can be viewed, respectively, as isospin- and spin-flip excitations of the proton and neutron.

Since pions are a complex probe, it is sensible to exploit the relative simplicity of virtual photons in order study the $\Delta$-resonance's structure; viz., through the transitions $\gamma^\ast N \to \Delta$.  This is possible at
intense, energetic electron-beam facilities; and data on the $\gamma^\ast p \to \Delta^+$ transition are now available for $0 \leq Q^2 \lesssim 8\,$GeV$^2$ \cite{Aznauryan:2011ub,Aznauryan:2011qj}.

The $\gamma^\ast p \to \Delta^+$ data have stimulated a great deal of theoretical analysis, and speculation about, \emph{inter alia}:
the relevance of pQCD to processes involving moderate momentum transfers \cite{Aznauryan:2011qj,Carlson:1985mm,Pascalutsa:2006up};
shape deformation of hadrons \cite{Alexandrou:2012da};
and, of course, the role that resonance electroproduction experiments can play in exposing nonperturbative features of QCD \cite{Aznauryan:2012ba}.

The \mbox{$N\to\Delta$} transition is described by three Poin\-car\'e-inva\-riant form factors \cite{Jones:1972ky}: magnetic-dipole, $G_M^\ast$; electric quadrupole, $G_E^\ast$; and Coulomb (longitudinal) quadrupole, $G_C^\ast$.  They arise through consideration of the $N\to \Delta$ transition current:
\begin{equation}
J_{\mu\lambda}(K,Q) =
\Lambda_{+}(P_{f})R_{\lambda\alpha}(P_{f})i\gamma_{5}\Gamma_{\alpha\mu}(K,Q)\Lambda_
{+}(P_{i}),
\label{eq:JTransition}
\end{equation}
where: $P_{i}$, $P_{f}$ are, respectively, the incoming nucleon and outgoing $\Delta$ momenta, with $P_{i}^{2}=-m_{N}^{2}$, $P_{f}^{2}=-m_{\Delta}^{2}$; the incoming photon momentum is $Q_\mu=(P_{f}-P_{i})_\mu$ and $K=(P_{i}+P_{f})/2$; and $\Lambda_{+}(P_{i})$, $\Lambda_{+}(P_{f})$ are, respectively, positive-energy projection operators for the nucleon and $\Delta$, with the Rarita-Schwinger tensor projector $R_{\lambda\alpha}(P_f)$ arising in the latter connection.  (Recall that our Euclidean metric conventions are described in App.\,\ref{sec:Euclidean}.)

In order to succinctly express $\Gamma_{\alpha\mu}(K,Q)$, one may define
\begin{eqnarray}
\hat K_{\mu}^{\perp} &=& {\cal T}_{\mu\nu}^{Q} \hat K_{\nu}
= (\delta_{\mu\nu} - \hat Q_\mu \hat Q_\nu) \hat K_\nu,
\end{eqnarray}
$\hat K^2 = 1= \hat Q^2$, in which case, with $\mathpzc{k} = \sqrt{(3/2)}(1+m_\Delta/m_N)$,
$\varsigma = Q^{2}/[2\Sigma_{\Delta N}]$,
$\lambda_\pm = \varsigma + t_\pm/[2 \Sigma_{\Delta N}]$
where $t_\pm = (m_\Delta \pm m_N)^2$,
$\lambda_m = \sqrt{\lambda_+ \lambda_-}$,
$\Sigma_{\Delta N} = m_\Delta^2 + m_N^2$, $\Delta_{\Delta N} = m_\Delta^2 - m_N^2$,
\begin{equation}
\Gamma_{\alpha\mu}(K,Q) =
\mathpzc{k} \left[\frac{\lambda_m}{2\lambda_{+}}(G_{M}^{\ast}-G_{E}^{\ast})\gamma_{5}
\varepsilon_{\alpha\mu\gamma\delta} \hat K_{\gamma}\hat{Q}_{\delta}
- G_{E}^{\ast}
{\cal T}_{\alpha\gamma}^{Q}
{\cal T}_{\gamma\mu}^{K}
- \frac{i\varsigma}{\lambda_m}G_{C}^{\ast}\hat{Q}_{\alpha} \hat K^\perp_{\mu}\right].
\label{eq:Gamma2Transition}
\end{equation}
Given the current, one may obtain the form factors using any three sensible projection operations.

In analyses of baryon electromagnetic properties, using a quark model framework which implements a current that transforms according to the adjoint representation of spin-flavour $SU(6)$, one finds simple relations between magnetic-transition matrix elements \cite{Beg:1964nm,Buchmann:2004ia}:
\begin{equation}
 \label{NgDelta}
 \langle p | \mu | \Delta^+\rangle = -\langle n | \mu | \Delta^0\rangle\,,\quad
 \langle p | \mu | \Delta^+\rangle = - \surd 2 \langle n | \mu | n \rangle\,;
\end{equation}
i.e., the magnetic components of the $\gamma^\ast p \to \Delta^+$ and $\gamma^\ast n \to \Delta^0$ are equal in magnitude and, moreover, simply proportional to the neutron's magnetic form factor.  Furthermore, both the nucleon and $\Delta$ are $S$-wave states (neither is deformed) and hence $G_{E}^{\ast} \equiv 0 \equiv G_{C}^{\ast}$ \cite{Alexandrou:2012da}.

The first identity in Eq.\,\eqref{NgDelta} is consistent with pQCD \cite{Carlson:1985mm} in the following sense: both suggest that $G_{M}^{\ast p}(Q^2)$ should decay with $Q^2$ at the same rate as the neutron's magnetic form factor, which is dipole-like in QCD.  It is usually argued that this is not the case empirically \cite{Aznauryan:2011ub,Aznauryan:2011qj}.  However that claim is contested in Ref.\,\cite{Segovia:2013rca}, as we shall subsequently explain.

\subsection{General remarks}
Recalling Sec.\,\ref{subsec:FE}, we note that since the nucleon and $\Delta$ have positive parity, $J^P=0^+$ (scalar) and $J^P=1^+$ (axial-vector) diquarks are the dominant correlations within them.  The presence of pseudoscalar and vector diquarks can be ignored because such correlations are characterised by much larger mass-scales and they have negative parity \cite{Chen:2012qr,Roberts:2011cf}.
Owing to Fermi-Dirac statistics, scalar diquarks are necessarily $I=0$ states, whilst axial-vector diquarks are $I=1$ \cite{Cahill:1987qr}.  The nucleon ground-state contains both $0^+$ and $1^+$ diquarks, whereas the $\Delta(1232)$-baryon contains only axial-vector diquarks because it is impossible to combine an $I=0$ diquark with an $I=1/2$ quark to obtain $I=3/2$.

\begin{figure}[t]

\leftline{\includegraphics[clip,width=0.28\linewidth]{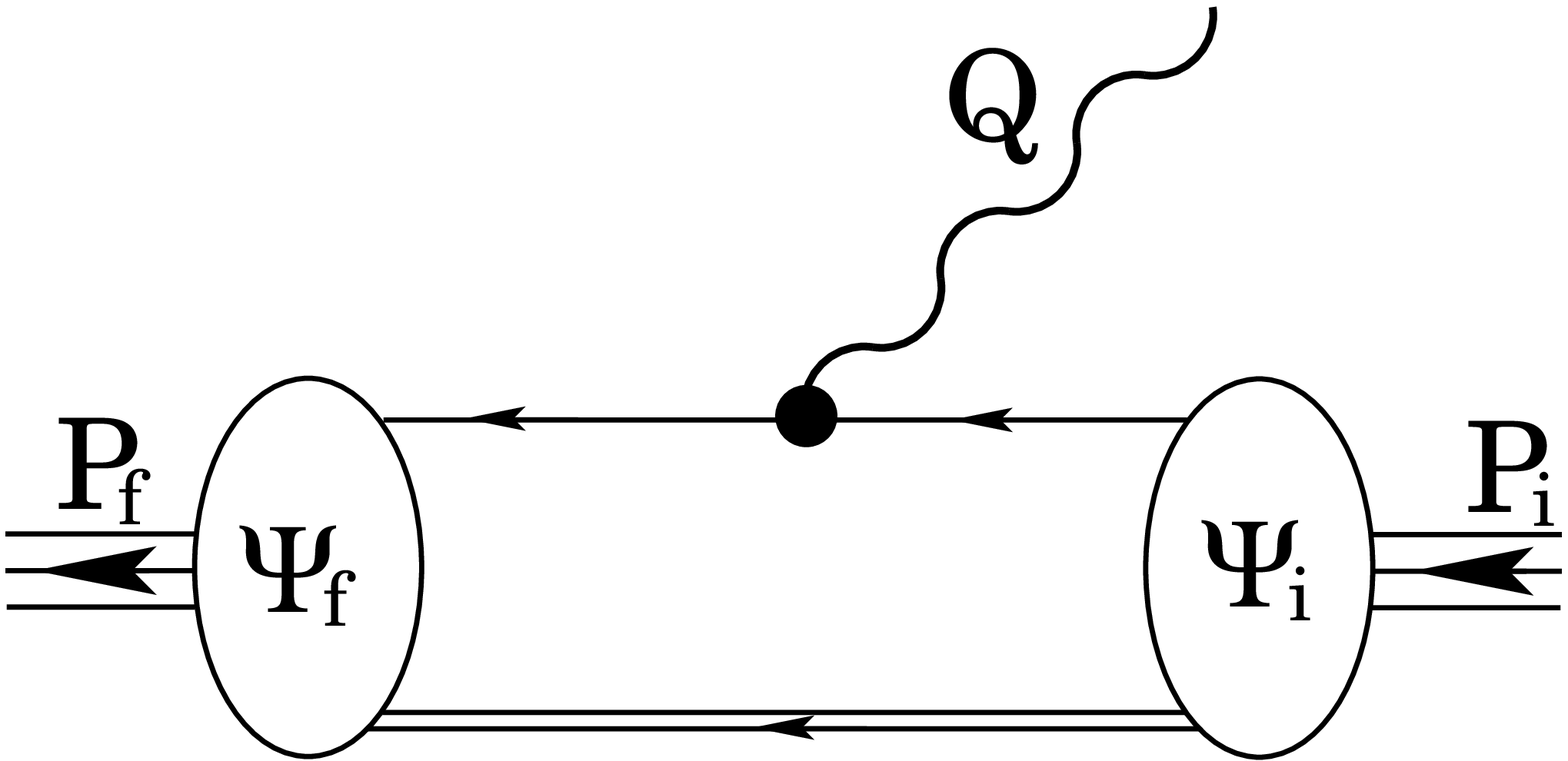}}
\vspace*{-7ex}

\centerline{\includegraphics[clip,width=0.28\linewidth]{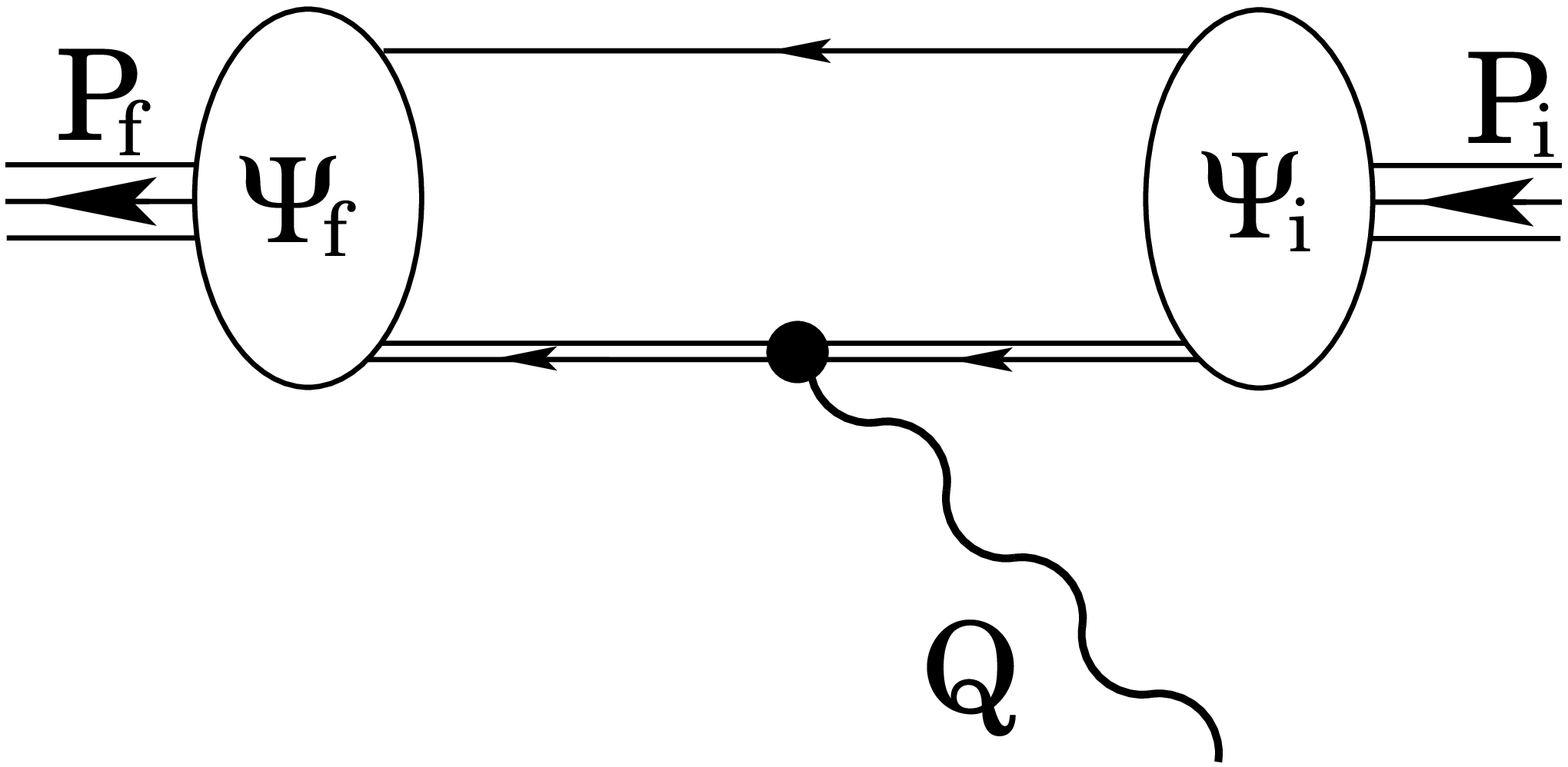}}
\vspace*{-14ex}

\rightline{\includegraphics[clip,width=0.28\linewidth]{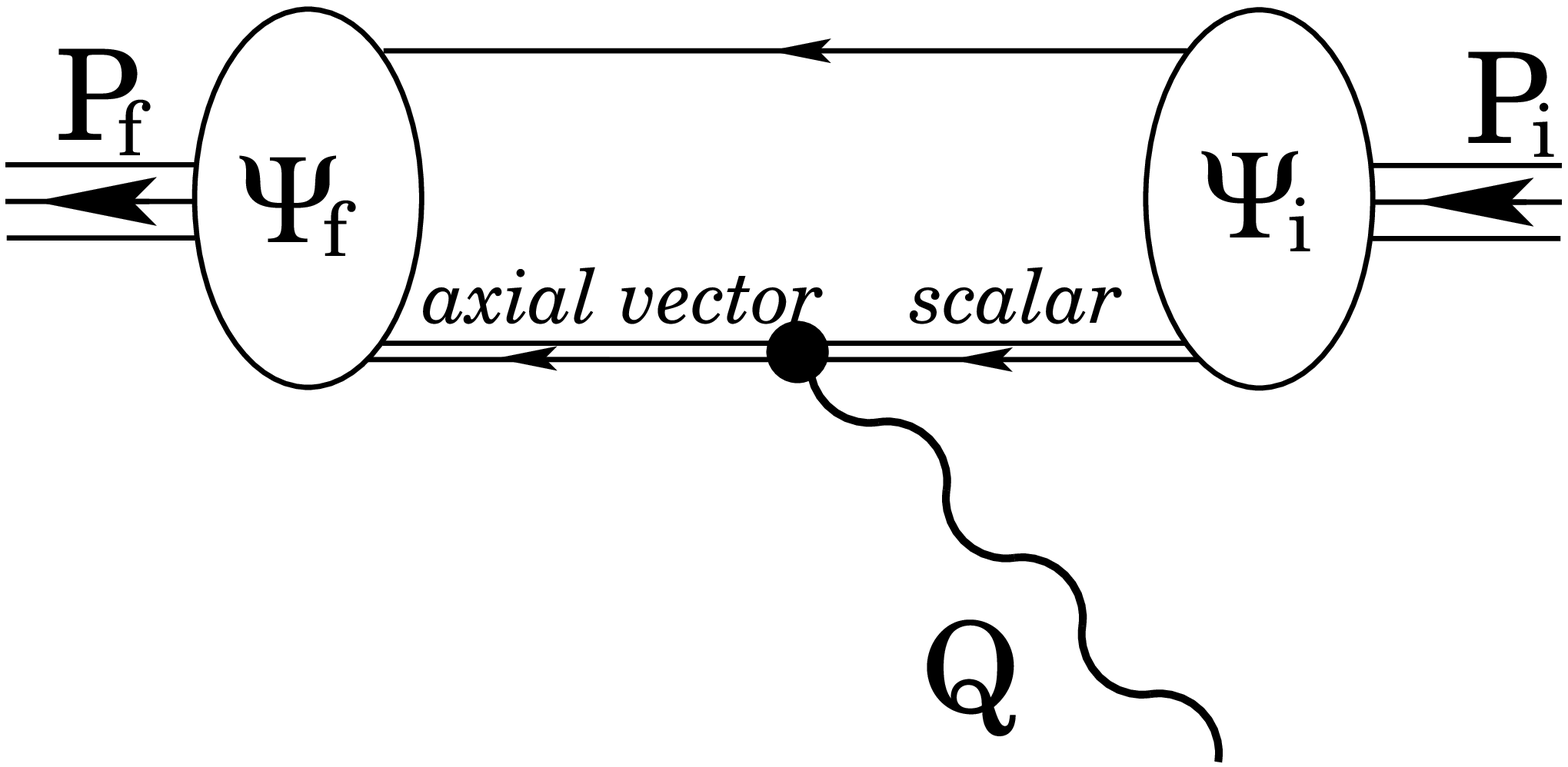}}
\caption{\label{fig:Transitioncurrent} One-loop diagrams in the $N\to \Delta$ vertex.
The single line represents a dressed-quark propagator, $S(p)$; the double line, a diquark propagator; and the vertices are, respectively, incoming nucleon, $\Psi_i$, and outgoing $\Delta$, $\Psi_f$.  From left to right, the diagrams describe the photon coupling: directly to a dressed-quark; to a diquark, in an elastic scattering event; or inducing a transition between scalar and axial-vector diquarks.
Since the $\Delta$ resonance contains only axial-vector diquarks, only such diquarks appear in the left and centre diagrams when one computes the $N \to \Delta$ transition.
In the general case, there are three more diagrams, described in detail elsewhere \protect\cite{Cloet:2008re}.  They represent two-loop integrals.
}
\end{figure}

For baryons constituted as described above, the elastic and transition currents are represented by the diagrams described in association with Fig.\ref{fig:Transitioncurrent}.  Plainly, with the presence of strong diquark correlations, the assumption of $SU(6)$ symmetry for the associated state-vectors and current is invalid.  Notably, too, since scalar diquarks are absent from the $\Delta$, only axial-vector diquark correlations contribute in the left and centre diagrams of Fig.\ref{fig:Transitioncurrent} when one or both of the vertices involves a $\Delta(1232)$-baryon.

Each of the diagrams in Fig.\ref{fig:Transitioncurrent} can be expressed like Eq.\,(\ref{eq:JTransition}), so that they may be represented as
\begin{equation}
\label{GammaTransition}
\Gamma_{\mu\lambda}^{m}(K,Q) = \Lambda_{+}(P_{f})R_{\lambda\alpha}(P_{f}) {\cal
J}_{\mu\alpha}^{n}(K,Q) \Lambda_{+}(P_{i})\,,
\end{equation}
where $m=1,2,\,$\ldots enumerates the diagrams, from left to right.  
The left diagram describes a photon coupling directly to a dressed-quark with the axial-vector diquark acting as a bystander.  If the initial-state is a proton, then it contains two axial-vector diquark isospin states $(I,I_z) = (1,1)$, $(1,0)$, with flavour content $\{uu\}$ and $\{ud\}$, respectively: in the isospin-symmetry limit, they appear with relative weighting $(\sqrt{2/3})$:$(-\sqrt{1/3})$, which are just the appropriate isospin-coupling Clebsch-Gordon coefficients.  These axial-vector diquarks also appear in the final-state $\Delta^+$ but with the orthogonal weighting; i.e., $(\sqrt{1/3})$:$(\sqrt{2/3})$.  For the process $\gamma^\ast p \to \Delta^+$, Diagram~1 therefore represents a sum, which may be written
\begin{equation}
{\cal J}_{\mu\alpha}^{1 p} =
(\sqrt{2}/3) e_d {\mathcal I}_{\mu \alpha}^{1 \{uu\}}
-(\sqrt{2}/3) e_u {\mathcal I}_{\mu\alpha}^{1\{ud\}},
\label{eqJ1p}
\end{equation}
where we have extracted the isospin and charge factors associated with each scattering.  Plainly, if the $\{uu\}$ diquark is a bystander, then the $d$-quark is the active scatterer, and hence appears the factor $e_d=(-1/3)$.  Similarly, $e_u=2/3$ appears with the $\{ud\}$ diquark bystander.

Having extracted the isospin and electric-charge factors, nothing remains to distinguish between the $u$- and $d$-quarks in the isospin-symmetry limit.  Hence,
\begin{equation}
{\mathcal I}_{\mu\alpha}^{1 \{uu\}}(K,Q) \equiv {\mathcal I}_{\mu\alpha}^{1\{ud\}}(K,Q)
=: {\mathcal I}_{\mu\alpha}^{1 \{qq\}}(K,Q) \quad
\label{J1pzero}
\Rightarrow  \quad {\cal J}_{\mu\alpha}^{1 p}(K,Q) = (-\sqrt{2}/3) {\mathcal I}_{\mu\alpha}^{1 \{qq\}}(K,Q)\,.
\end{equation}
It is known that diagrams with axial-vector diquark spectators do not contribute to proton elastic form factors (Eq.\,(C5) in Ref.\,\cite{Wilson:2011aa}), so the analogous contribution is absent from the proton's elastic form factors.  However, this hard contribution is present in neutron elastic form factors.  In general, form factors also receive a hard contribution from the two-loop diagrams omitted in Fig.\ref{fig:Transitioncurrent}.  In proton and neutron elastic magnetic form factors, respectively, the large-$Q^2$ behaviour of this contribution matches that produced by Diagram~1 \cite{Cloet:2008re}.

The remaining two diagrams in Fig.\,\ref{fig:Transitioncurrent}; i.e., the middle and right images, describe a photon interacting with a composite object whose electromagnetic radius is nonzero.  They must therefore produce a softer contribution to the transition form factors than anything obtained from the left diagram.

It follows from this discussion that the fall-off rate of $G_M^\ast(Q^2)$ in the $\gamma^\ast p \to \Delta^+$ transition must match that of $G_M^n(Q^2)$.
%
With isospin symmetry, the first identity in Eq.\,\eqref{NgDelta} is valid, so the same is true of the $\gamma^\ast n \to \Delta^0$ magnetic form factor.  Note that these are statements about the dressed-quark-core contributions to the transitions.  They will be valid empirically outside that domain upon which meson-cloud effects are important; i.e., for $Q^2\gtrsim 2\,$GeV$^2$ \cite{Sato:2000jf,JuliaDiaz:2006xt}.

\begin{figure}[t]
\leftline{\includegraphics[clip,width=0.45\linewidth]{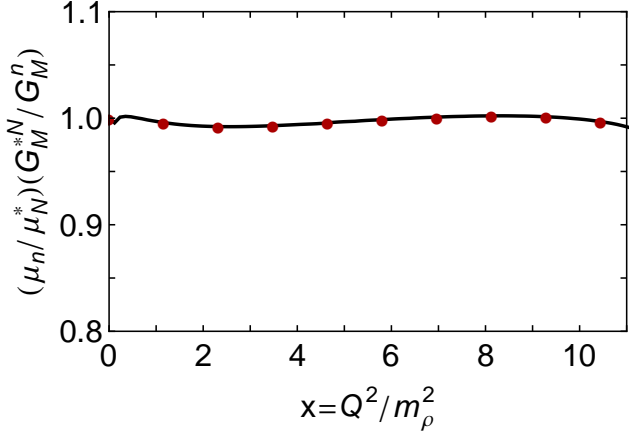}}
\vspace*{-30ex}

\rightline{\parbox{22em}{
\caption{\label{figRatio} \emph{Solid curve} -- $\mu_n G_{M}^{\ast p}/\mu^\ast_p G_{M}^{n}$ as a function of $x=Q^2/m_\rho^2$; and \emph{filled circles} -- $\mu_n G_{M}^{\ast n}/\mu^\ast_n G_{M}^{n}$.
For $N=p,n$, $\mu^\ast_N = G_M^{\ast N}(Q^2=0)$; and $\mu_n = G_M^n(Q^2=0)$.  The elastic form factor results are those presented in Ref.\,\protect\cite{Wilson:2011aa}, so that the comparison is internally consistent.}}}

\vspace*{10ex}

\end{figure}

\subsection{Quantitative illustration}
Since these observations are straightforward, Ref.\,\cite{Segovia:2013rca} chose to illuminate them within a simple framework; i.e., a symmetry-preserving Dyson-Schwinger equation (DSE) treatment of a vector$\,\otimes\,$vec\-tor contact-interaction, which is described in App.\,\ref{secContact}.  A body of recent work
\cite{Chen:2012qr,Wilson:2011aa,GutierrezGuerrero:2010md,Roberts:2010rn,Roberts:2011wy,%
Chen:2012txa,Roberts:2011cf,Wang:2013wk,Segovia:2013uga}
has shown that this framework produces results which, when analysed judiciously, are often qualitatively and semi-quantitatively equivalent to those obtained with the most sophisticated interactions thus far employed in the leading-order (rainbow-ladder, Sec.\,\ref{sec:GapEquation}) truncation of QCD's DSEs.  The illustration is therefore representative of that class of studies.

The calculation could proceed by adapting the nucleon-to-Roper transition form factor formulae in Ref.\,\cite{Wilson:2011aa} to the case of a final-state $\Delta$.  Owing to the interaction's simplicity, there are no two-loop contributions to the form factors, so the diagrams depicted in Fig.\,\ref{fig:Transitioncurrent} are all that need be considered.  The computed momentum-dependence of the magnetic $\gamma^\ast p \to \Delta^+$ and $\gamma^\ast n \to \Delta^0$ form factors is compared with that of $G_M^n(Q^2)$ in Fig.\,\ref{figRatio}.  The prediction explained above is evident in a near identical momentum dependence.

In connection with experiment, a contact-interaction treatment of the $N \to \Delta$ transition is quantitatively inadequate for two main reasons.  Namely, a contact interaction which produces Faddeev amplitudes that are independent of relative momentum must underestimate the quark orbital angular momentum content of the bound-state; and the truncation which produces the momentum-independent amplitudes also suppresses the three two-loop diagrams in the current of Fig.\,\ref{fig:Transitioncurrent}.  The detrimental effect can be illustrated via the computed values for the contributions to $G_M^\ast(0)$ that arise from the overlap axial-diquark($\Delta$)$\leftarrow$axial-diquark($N$) cf.\ axial-diquark($\Delta$)$\leftarrow$scalar-diquark($N$).  Ref.\,\cite{Segovia:2013rca} finds $0.85/0.18$, values that may be compared with those in Table~3 of Ref.\,\cite{Eichmann:2011aa}, which uses momentum-dependent DSE kernels: $0.96/1.27$.  One may show algebraically that the omitted two-loop diagrams facilitate a far greater contribution from axial($\Delta$)-scalar($N$) mixing and the presence of additional orbital angular momentum enhances both.

\begin{figure}[t]
\leftline{\includegraphics[clip,width=0.45\linewidth]{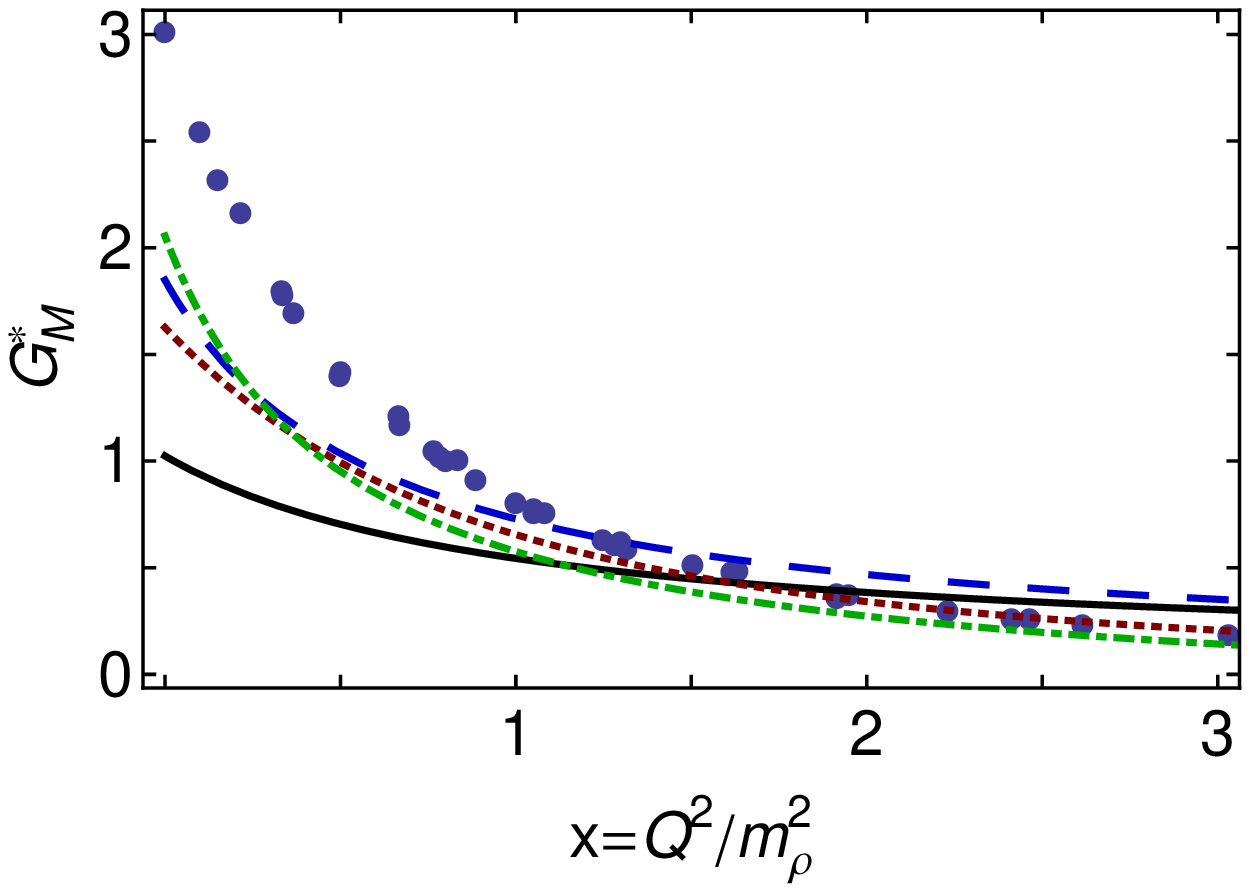}}
\vspace*{-35.2ex}

\rightline{\includegraphics[clip,width=0.45\linewidth]{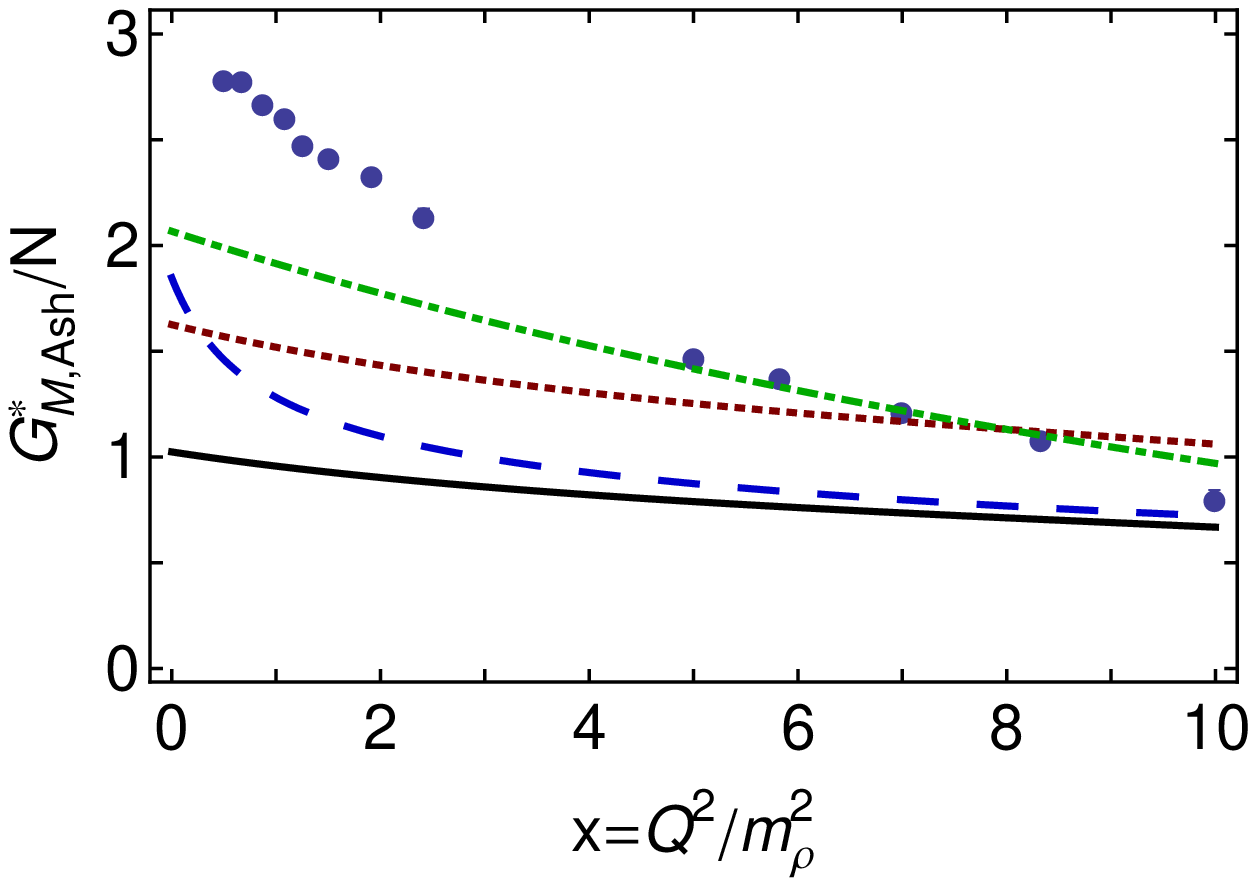}}
\caption{\label{figGMast}
\emph{Left panel}.  $G_{M}^{\ast}(Q^2)$:
contact-interaction result (solid curve);
ameliorated result (dashed curve), explained in connection with Eq.\,\protect\eqref{correction};
SL-model dressed-quark-core result \protect\cite{JuliaDiaz:2006xt} (dot-dashed curve);
and data from Refs.\,\protect\cite{Beringer:1900zz,Aznauryan:2009mx,Bartel:1968tw,Stein:1975yy,Sparveris:2004jn,Stave:2008aa}, whose errors are commensurate with the point size.
(N.B.\ The contact interaction produces Faddeev amplitudes that are independent of relative momentum, hence $G_M^\ast(Q^2)$ is hard.  The dotted curve is an estimate of the result a realistic interaction would produce, obtained via multiplying $G_M^\ast(Q^2)$ by $G_M(Q^2)$-realistic$/G_M(Q^2)$-contact, where $G_M(Q^2)$-realistic is taken from Ref.\,\protect\cite{Cloet:2008re}.)
\emph{Right panel}. $\mu_n G_{M,Ash}^{\ast}(Q^2)/N(Q^2)$: contact interaction (solid curve) and ameliorated result (dashed curve) obtained with $N(Q^2)=G_M^n(Q^2)$.  (The dotted curve is the solid curve rescaled by $\mu_n$-realistic$/\mu_n$-contact.)  Also, empirical results \cite{Aznauryan:2009mx} for $G_{M,Ash}^{\ast}/N(Q^2)$, where $1/N_D=[1 + Q^2/\Lambda^2]^2$, $\Lambda=0.71\,$GeV, and SL-model's dressed-quark-core result for this ratio \protect\cite{JuliaDiaz:2006xt}.}
\end{figure}

In recognition of both this defect and the general expectation that a comparison with experiment should be sensible, Ref.\,\cite{Segovia:2013rca} provided two sets of results.  Namely, unameliorated predictions of the contact-interaction plus results obtained with two corrections: (1) rescaling the axial($\Delta$)-scalar($N$) diagram using the factor
\begin{equation}
\label{correction}
1+ \frac{g_{as}^{aa}}{1+Q^2/m_\rho^2}\,,
\end{equation}
with $g_{as}^{aa}=4.3$, so that its contribution to $G_M^{\ast p}(0)$ matches that of the axial($\Delta$)-axial($N$) term; and (2) incorporating a dressed-quark anomalous magnetic moment, which is a predicted consequence of DCSB in QCD \cite{Chang:2010hb,Qin:2013mta}.  (See also Ref.\,\cite{Wilson:2011aa}, App.\,C.6, and Ref.\,\cite{Segovia:2013uga}, App.\,B.3.)

The left panel of Fig.\,\ref{figGMast} displays the $\gamma^\ast p \to \Delta^+$ magnetic transition form factor.  (With $\tilde \mu_{N\Delta}^\ast:= (\sqrt{m_\Delta/m_N}) G_M^{\ast N}(0)$, one has a direct result of $\tilde \mu_{N\Delta}^\ast=1.13$ and an ameliorated value of $\tilde \mu_{N\Delta}^\ast=2.04$.)  Both computed curves are consistent with data for $Q^2\gtrsim 2\,m_\rho^2$ but, corrected or not, they are in marked disagreement at infrared momenta.  This is explained by the similarity between the ameliorated result (dashed curve) and the ``bare'' or dressed-quark-core result determined using the Sato-Lee (SL) dynamical meson-exchange model (dotted curve) \cite{JuliaDiaz:2006xt}.  The SL result supports a view that the discrepancy results from the omission of meson-cloud effects in the rainbow-ladder truncation of QCD's DSEs.

\begin{figure}[t]
\leftline{\includegraphics[clip,width=0.45\linewidth]{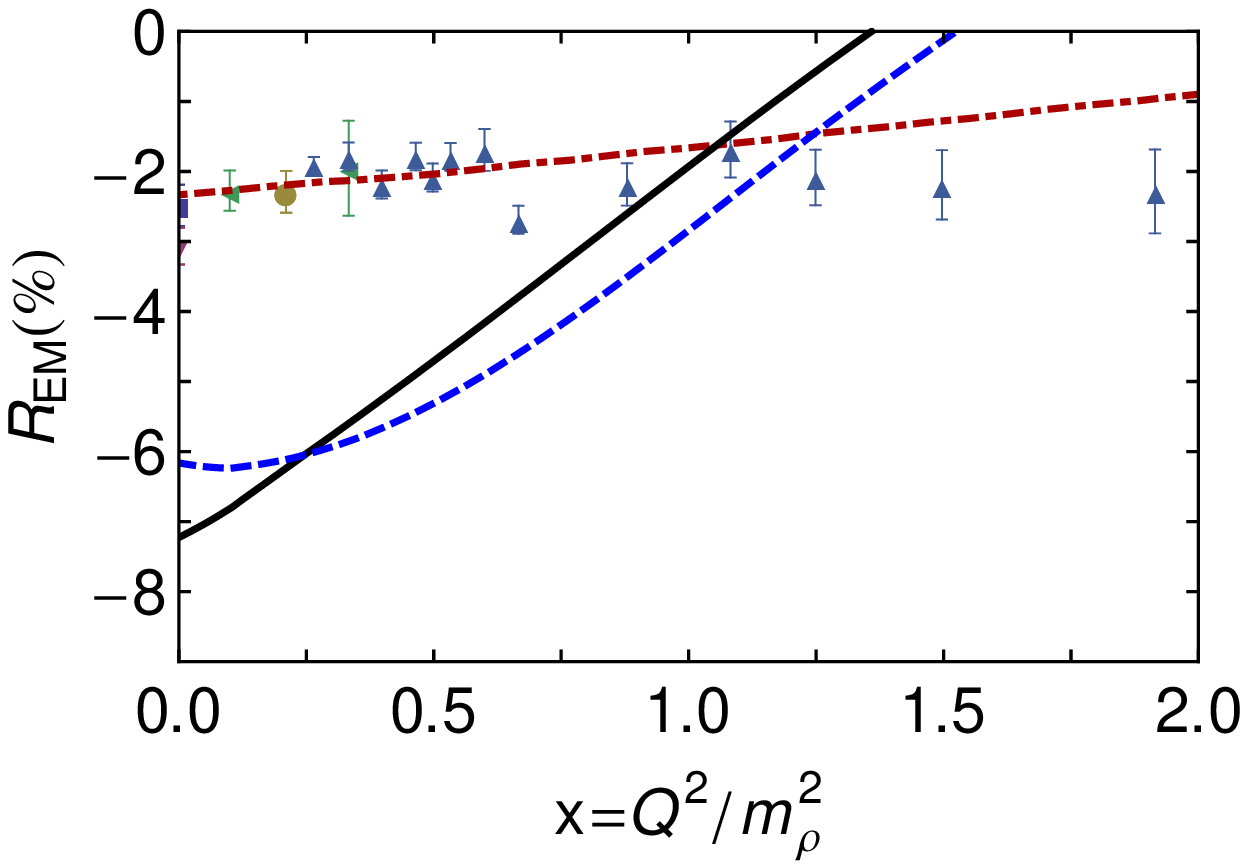}}
\vspace*{-34.2ex}

\rightline{\includegraphics[clip,width=0.464\linewidth]{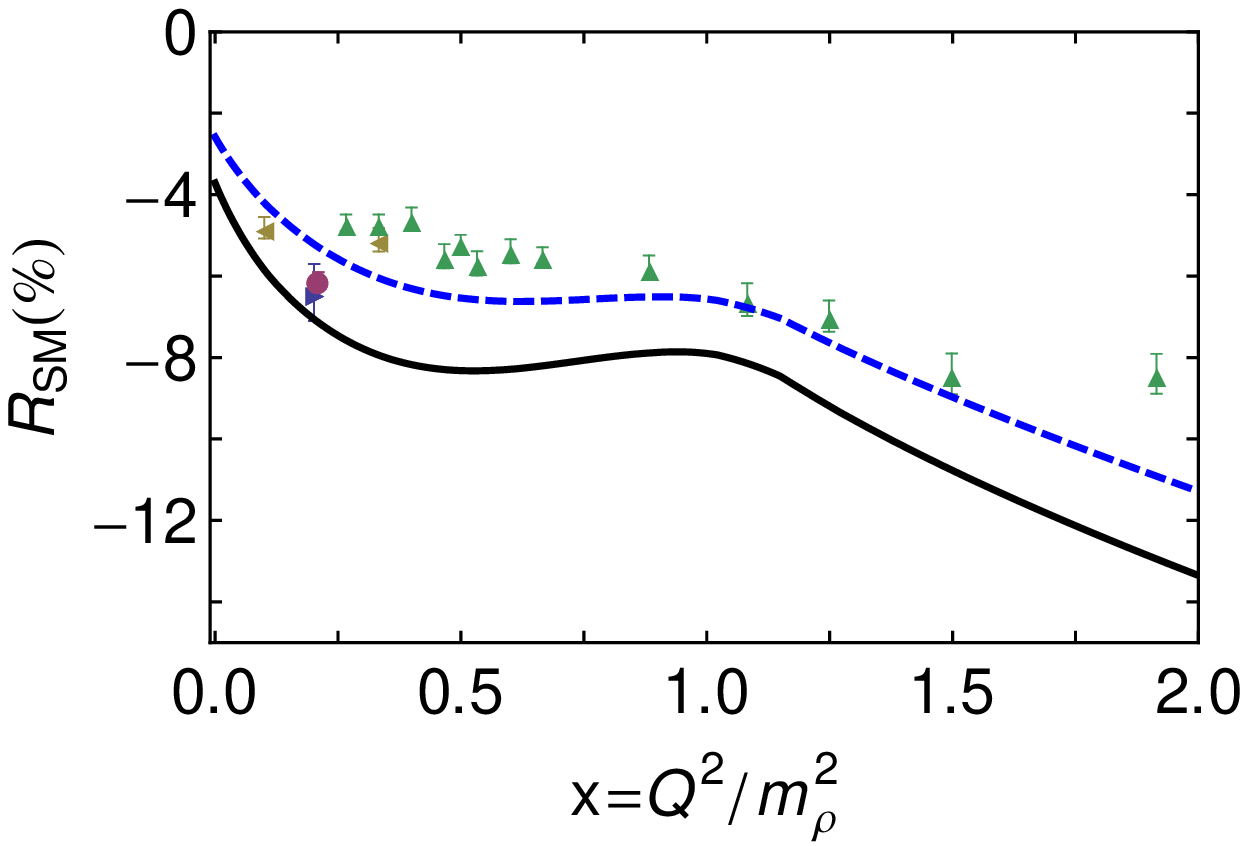}}
\caption{\label{figREMSM} Ratios in Eq.\,\protect\eqref{eqREMSM}.  Both panels: \emph{solid curve} -- contact-interaction result; \emph{dashed curve} -- ameliorated result, discussed in connection with Eq.\,\protect\eqref{correction};
and data \protect\cite{Aznauryan:2009mx,Sparveris:2004jn,Stave:2008aa,Beck:1999ge,Pospischil:2000ad,Blanpied:2001ae}.
The \emph{dot-dashed curve} in the left panel is representative of the computation in Ref.\,\protect\cite{Eichmann:2011aa}.
(N.B.\ $G_E^\ast$, $G_C^\ast$ are small, so Ref.\,\protect\cite{JuliaDiaz:2006xt} could not reliably separate meson-cloud and dressed-quark core contributions to these ratios.)}
\end{figure}

In contrast to the left panel of Fig.\,\ref{figGMast}, presentations of experimental data typically use the Ash form factor \cite{Ash1967165}
\begin{equation}
\label{DefineAsh}
G_{M,Ash}^{\ast}(Q^2)= G_M^{\ast}(Q^2)/[1+Q^2/t_+ ]^{1/2}.
\end{equation}
This comparison is depicted in Fig.\,\ref{figGMast}, right panel.  (The contact-interaction dressed-quark core result is quantitatively similar to the same quantity in Fig.\,3 of Ref.\,\cite{Aznauryan:2012ec}.)  Plainly, $G_{M,Ash}^{\ast}(Q^2)$ falls faster than a dipole.  Historically, many have viewed this as a conundrum.  However, as observed previously \cite{Carlson:1985mm} and reviewed herein, there is no sound reason to expect $G_{M,Ash}^{\ast}(Q^2)/G_M^n(Q^2) \approx\,$constant.  Instead, the Jones-Scadron form factor should exhibit $G_{M}^{\ast}(Q^2)/G_M^n(Q^2) \approx\,$constant.  The empirical Ash form factor falls rapidly for two reasons.  First: meson-cloud effects provide more than 30\% of the form factor for $Q^2\lesssim 2m_\rho^2$; these contributions are very soft; and hence they disappear rapidly.  Second: the additional kinematic factor $\sim 1/\sqrt{Q^2}$ in Eq.\,\eqref{DefineAsh} provides material damping for $Q^2\gtrsim 4m_\rho^2$.

The dotted curves in Fig.\,\ref{figGMast} depict crude estimates of the behaviour to be expected of the associated form factors when they are computed with propagators and currents that exhibit QCD-like momentum-dependence, as displayed, e.g., in Sec.\,\ref{sec:dcsb}. These curves provide a hint that a realistic interaction will fully explain the data.

Figure~\ref{figREMSM} depicts the ratios
\begin{equation}
\label{eqREMSM}
\rule{-0.7em}{0ex} R_{\rm EM} = -G_E^{\ast}/G_M^{\ast}\,,\quad
R_{\rm SM} = - (|\vec{Q}|/2 m_\Delta) (G_C^{\ast}/G_M^{\ast})\,,
\end{equation}
which are commonly read as measures of deformation in one or both of the hadrons involved because they are zero in $SU(6)$-symmetric constituent-quark models.  However, the ratios also measure the way in which such deformation influences the structure of the transition current.

The figure shows that even a contact-interaction produces correlations between
dressed-quarks within Faddeev wave-functions and related features in the current that
are comparable in size with those observed empirically.  They are actually too large
if axial($\Delta$)-axial($p$) contributions to the transition significantly outweigh
those from axial($\Delta$)-scalar($p$) processes.  This is highlighted effectively by the dot-dashed curve in the left panel.  That result \cite{Eichmann:2011aa}, obtained in the same DSE truncation but with a QCD-motivated momentum-dependent interaction \cite{Maris:1999nt}, produces Faddeev amplitudes with a richer quark orbital angular momentum structure.  The left panel emphasises, therefore, that $R_{\rm EM}$ is a particularly sensitive measure of orbital angular momentum correlations, both within the hadrons involved and in the excitation current.  The simpler Coulomb quadrupole produces a ratio, $R_{\rm SM}$, that is more robust.

\begin{figure}[t]
\leftline{
\includegraphics[clip,width=0.45\linewidth]{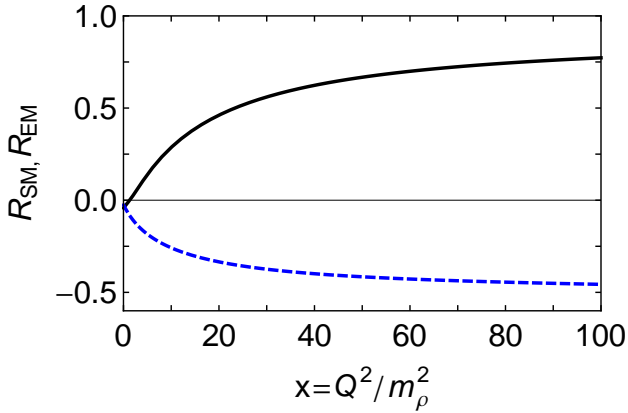}}
\vspace*{-26ex}

\rightline{\parbox{23em}{\caption{\label{UVREM}
$R_{\rm EM}$ (solid curve) and $R_{\rm SM}$ (dashed curve) in Eq.\,\protect\eqref{eqREMSM}, computed using the ameliorated contact interaction, discussed in connection with Eq.\,\protect\eqref{correction}.}}}\vspace*{12ex}
\end{figure}

Notwithstanding that the asymptotic power-law dependence of the computed form factors in Ref.\,\cite{Segovia:2013rca} is harder than that in QCD, one may readily show that the helicity conservation arguments in Ref.\,\cite{Carlson:1985mm} should apply equally to an internally-consistent symmetry-preserving treatment of a contact interaction.  As a consequence, one has
\begin{equation}
\label{eqUVREMSM}
R_{EM} \stackrel{Q^2\to\infty}{=} 1 \,,\quad
R_{SM} \stackrel{Q^2\to\infty}{=} \,\mbox{\rm constant}\,.
\end{equation}
The validity of Eqs.\,\eqref{eqUVREMSM} may be read from Fig.\,\ref{UVREM}.  On one hand, it is plain that truly asymptotic $Q^2$ is required before the predictions are realised.  On the other hand, they \emph{are} apparent.  Importantly, $G_E^\ast(Q^2)$ does possess a zero (at an empirically accessible momentum) and thereafter $R_{\rm EM}\to 1$.  Moreover, $R_{\rm SM}\to\,$constant.
(N.B.\ The curve displayed contains the $\ln^2 Q^2$-growth expected in QCD \cite{Idilbi:2003wj} but it is not a prominent feature.)
Since it is relative damping associated with helicity flips that yields Eqs.\,\eqref{eqUVREMSM}, with the $Q^2$-dependence of the leading amplitude being less important, it is plausible that the pattern evident here is also that to be anticipated in QCD.

\subsection{Perspective}
The material reviewed in this subsection explains and illustrates that the Ash form factor connected with the $\gamma^\ast N \to \Delta$ transition, Eq.\,\eqref{DefineAsh}, should fall faster than the neutron's magnetic form factor, which is a dipole in QCD.  
In addition, we have seen that the quadrupole ratios associated with this transition are a sensitive measure of quark orbital angular momentum within the nucleon and $\Delta$.  In Faddeev equation studies of baryons, this is commonly associated with the presence of strong diquark correlations.
Finally, following from some direct calculations it appears that predictions for the asymptotic behaviour of these quadrupole ratios, which follow from considerations associated with helicity conservation, are valid, although only at truly large momentum transfers.

An important next step is to repeat the analysis in Ref.\,\cite{Segovia:2013rca} using propagators and vertices that are a realistic representation of QCD.  This should enable a reliable prediction for the location of the zero in $R_{EM}$.  It is plain from the left panel of Fig.\,\ref{figREMSM} that the location of the zero is sensitive to the nature of the interaction; and the possibility that a realistic prediction might lie within reach of the upgraded JLab facility is exciting.

It also is worth remarking that an analysis of the $\gamma N \to \Delta$ transition form factors first requires computation of the small-$Q^2$ behaviour of the $\Delta$ elastic form factors.  That motivation led to a comprehensive contact-interaction study of $\Delta$- and $\Omega$-baryon elastic form factors, which is reported in Ref.\,\cite{Segovia:2013uga}.  (N.B.\, Once one has developed tools with which to calculate $\Delta$ elastic form factors, computing $\Omega$ elastic form factors is straightforward and hence should not be overlooked.)  It was found that the DSE treatment of the contact-interaction produces results for these elastic form factors that are practically indistinguishable from the best otherwise available, an outcome which highlights once again that the key to describing many features of baryons and unifying them with the properties of mesons is a veracious expression of dynamical chiral symmetry breaking in the hadron bound-state problem.  More particularly, it was shown that $\Delta$ elastic form factors are very sensitive to $m_\Delta$.  Hence, given that the parameters which define extant simulations of lattice-regularised QCD produce $\Delta$-resonance masses that are very large, the form factors obtained therewith \cite{Alexandrou:2007we,Alexandrou:2009hs} are a poor guide to properties of the $\Delta(1232)$.  Considering the $\Delta$-baryon's quadrupole moment, whilst all computations produce a negative value, the conflict between theoretical predictions entails that it is currently impossible to reach a sound conclusion on the nature of the $\Delta$-baryon's deformation in the infinite momentum frame.  Results for analogous properties of the $\Omega$ baryon are less contentious.

\subsection{Worldwide electroproduction programme}
%
The work described in this Section is part of a wider effort that is driven by the existence of an extensive international programme focused upon the study of electromagnetic transition amplitudes between ground- and excited-nucleon states, the so-called $\gamma_v NN^\ast$ electrocouplings.  These couplings are defined via their relationship with $N^\ast$ electromagnetic decay widths in Ref.\,\cite{Aznauryan:2011qj}, which also describes the connection between $\gamma_v NN^\ast$ electrocouplings and $N\to N^\ast$ transition form factors.  In this connection, there is a sense in which one may view the study of nucleon elastic form factors as one-dimensional, so that with the addition of information on transitions to $N^\ast$-states, one adds a vast array of ``handles'' to turn in order to aid in unfolding the essential nature of nucleon structure.  Since it is fair to say that the running of QCD's couplings and masses, and the structure of the correlations they facilitate between quarks, are independent of their environment to a good degree of accuracy, it follows that a consistent description of elastic and transition form factors will yield far greater insight into nonperturbative features of QCD than studies of elastic form factors alone.  For example, like meson excited states compared with meson ground states \cite{Holl:2004fr,Holl:2005vu}, $N^\ast$-states should typically have larger core-radii than the nucleon ground state; and hence the properties of $N^\ast$-states will be more sensitive to the long-range behaviour of QCD's $\beta$-function.  This dependence is contained within extractions of $\gamma_v NN^\ast$ electrocouplings.  Such couplings can, in addition, be used to check related theoretical predictions regarding the existence and evolving relative strength of quark-quark correlations within different baryonic systems \cite{Chen:2012qr}; i.e., states with different quantum numbers.  As we saw in Sec.\,\ref{secNucleonFF}, the nucleon ground state provides access only to $J^{P}=0^+,1^+$ diquark correlations, with a particular relative importance.  On the other hand, the study of $N^\ast$-states opens a window onto $J^{P}=0^-,1^-$ correlations, too, as well as onto the manner by which the relative importance of different diquark correlations changes from baryon to baryon.  In this way one accumulates an enormous and novel array of empirical constraints on pictures of baryon structure.

The largest body of results on $\gamma_v NN^\ast$ electrocouplings has been provided by the continuing programme in Hall-B at JLab, one goal of which is to determine the $\gamma_v NN^\ast$ electrocouplings of most $N^\ast$-states with mass less-than 2\,GeV on $0<Q^2<5\,$GeV$^2$ via independent and combined analyses of data from major exclusive meson-on-proton electroproduction channels using the CLAS detector, which has nearly $4\pi$-acceptance.  This programme is supported by studies with small acceptance detectors in Halls A and C at JLab and at the Mainz microtron (MAMI), and with the MIT/Bates detector, which has large acceptance and provided data in the first resonance region \cite{Aznauryan:2011qj,Aznauryan:2011ub,Aznauryan:2012ba}.
Data from independent analyses of $\pi^+ n$ and $\pi^0 p$ electroproduction channels on $0<Q^2<5.0\,$GeV$^2$ are now available for the $\gamma_v NN^\ast$ electrocouplings of all nucleon resonances with mass less-than 1.6\,GeV \cite{Aznauryan:2009mx}.  Moreover, studies of $\eta p$ electroproduction have provided independent information about electrocouplings of the $S_{11}(1535)$ resonance on $Q^2<4.0\,$GeV$^2$ \cite{Denizli:2007tq}; and novel results on electrocouplings of the $P_{11}(1440)$ and $D_{13}(1520)$ resonances determined from CLAS data on $\pi^+\pi^- p$ electroproduction \cite{Fedotov:2008aa} have recently been obtained \cite{Mokeev:2012vsa}.  Preliminary results on the electrocouplings of high-lying $N^\ast$ states in the mass-range 1.6-1.75\,GeV have been extracted from $\pi^+\pi^- p$ electroproduction data \cite{Mokeev:2013kka}.  It is also important to note that
the extraction of mutually consistent results on the $P_{11}(1440)$ and $D_{13}(1520)$ electrocouplings from independent analyses of all major meson electroproduction channels ($\pi N$ and $\pi^+\pi^- p$) using completely different nonresonant contributions suggests strongly that the results obtained are robust, fundamental quantities \cite{Mokeev:2012vsa}.  Finally, electrocouplings of the $P_{11}(1232)$ (the $\Delta$-resonance), $P_{11}(1440)$ and $D_{13}(1520)$ have been obtained on $Q^2<1.5\,$GeV$^2$ via a global analysis of the world's data on $\pi N$ photo-, electro- and hadro-production within the sophisticated framework of the Argonne-Osaka coupled-channels model \cite{Suzuki:2010yn,Kamano:2013iva}.

It is analyses of available results on $\gamma_v NN^\ast$ electrocouplings \cite{Aznauryan:2011qj,Aznauryan:2012ba,Mokeev:2012vsa} that have leant credence to the picture of nucleon ground- and excited-states described above; viz., they are states constituted from a core of three dressed-quarks surrounded by a meson-cloud, whose presence cannot be ignored for $Q^2\lesssim 2\,$GeV$^2$.  It is only with increasing $Q^2$ that one pierces the cloud and clearly exposes the dressed-quark core, which will completely dominate in all channels on $Q^2>5\,$GeV$^2$.  With this in mind, a dedicated experiment, aimed at extracting the $\gamma_v NN^\ast$ electrocouplings of most $N^\ast$ states on $Q^2\in[5,12]\,$GeV$^2$, the highest photon virtualities ever achieved in exclusive meson electroproduction, is expected to take data with the CLAS12 detector in the first year of running after completion of JLab's 12\,GeV upgrade \cite{Burkert:2012rh}.  This is the only existing programme worldwide that is capable of providing $\gamma_v NN^\ast$ electrocouplings on a domain which provides unfettered access to baryon dressed-quark cores.  As such, the results will provide information that is crucial in validating insights and predictions drawn from nonperturbative studies of baryon ground- and excited-states.  Thus, for example, numerous studies are now underway, employing the DSE framework with interactions more realistic than those employed in Refs.\,\cite{Wilson:2011aa,Segovia:2013rca,Segovia:2013uga}, with the aim of confronting existing data on $\gamma_v NN^\ast$ electrocouplings and making predictions that are relevant to the programme at JLab\,12.

\section{TMD distribution functions}
\label{SecNJLJet}
\subsection{Contact interactions}
As we highlight in App.\,\ref{secContact}, it is only very recently that numerical algorithms have been developed that both enable one to use sophisticated DSE kernels in order to gain direct access to the large-$Q^2$ behaviour of hadron form factors \cite{Chang:2013pq,Chang:2013epa,Chang:2013nia} and promise to open the way to their use in treating the phenomena of deep inelastic scattering.  Absent those algorithms, a confining, symmetry-preserving DSE treatment of a vector$\,\otimes\,$vec\-tor contact interaction has hitherto proved remarkably useful in a variety of contexts
\cite{Chen:2012qr,Wilson:2011aa,GutierrezGuerrero:2010md,Roberts:2010rn,Roberts:2011wy,%
Chen:2012txa,Roberts:2011cf,Wang:2013wk,Segovia:2013uga}.  This collection of work is dwarfed, however, by the widespread application of models of the  Nambu--Jona-Lasinio (NJL) type \cite{Nambu:1961tp,Vogl:1991qt,Klevansky:1992qe,Hatsuda:1994pi,Bijnens:1995ww}.  Whilst there is a ground-level similarity, these two approaches are different.  The NJL-model practitioners allow themselves the freedom of tuning the relative strength of individual four-fermion interaction terms in a model Lagrangian, whereas the DSE approach treats the contact interaction as a representation of the gluon's two-point Schwinger function.  At the outset, therefore, the DSE approach fixes the number, type and relative strength of the four-fermion interaction terms, as illustrated elsewhere \cite{Cahill:1985mh,Tandy:1997qf}, and then proceeds to identify and highlight those observables that can distinguish between different choices for the momentum-dependence of the DSE kernels.  The NJL-model approach, on the other hand, tunes parameters in order to fit and explain a body of data with the aim of providing a phenomenology of hadron physics that can serve both to elucidate correlations between observables and as a beacon that sheds light on novel explanations for unexpected phenomena.  As an illustration, in the next subsection we will describe the application of a NJL model to the prediction of the transverse momentum dependence of the unpolarised quark distributions in the nucleon.

\subsection{Prediction of spin-independent TMDs}
\label{SecTMD}
The last decade has seen the emergence of a novel approach to the description of nucleon and nuclear structure.  This framework represents knowledge of the nucleon (and nuclei) via the Wigner distributions of the fundamental constituents, a quantum mechanical concept analogous to the classical notion of a phase space distribution.  From the Wigner distributions, a natural interpretation of measured observables is provided via construction of quantities known as generalised parton distributions (GPDs) \cite{Ji:1996nm,Radyushkin:1996nd} and transverse momentum-dependent distributions (TMDs) \cite{Collins:2003fm,Belitsky:2003nz}: GPDs have been described as providing for a \textit{spatial} tomography of the nucleon; and TMDs are said to allow for its \textit{momentum} tomography.  A new generation of experiments will provide the empirical information necessary to develop a phenomenology of nucleon and nuclear Wigner distributions.

The largest part of allocated running time at Jefferson Lab 12\,GeV (JLab\,12) is devoted to GPD and TMD measurements: GPDs will primarily be probed via deeply virtual Compton scattering and deeply virtual meson production, whereas semi-inclusive deep inelastic scattering (SIDIS) is the preferred technique for exposing TMDs.  In this section we will focus on the latter.

Interest in TMDs is widespread.  At leading twist, there are eight such parton distributions \cite{Bacchetta:2006tn}. They may be viewed as a generalisation of the usual longitudinal parton distribution functions, which play a crucial role in understanding DIS, and are also expected to provide access to spin-orbit correlations on the light-front \cite{Brodsky:2000xy,She:2009jq,Avakian:2010br,Lorce:2011kd,Bacchetta:2011gx} and the probability distribution of parton transverse momenta in the infinite momentum frame \cite{Anselmino:2007fs,Anselmino:2011ay}.  The concepts and methods of factorisation and evolution involving TMDs, as they are currently understood, are briefly explained in Ref.\,\cite{Collins:2013zsa}.

A first microscopic calculation of the spin-independent TMD quark distribution functions in the nucleon is presented in Ref.\,\cite{Matevosyan:2011vj}.  This study uses a NJL-model that has been employed with phenomenological success to compute the spin and flavour dependence of nucleon quark-PDFs and their modification in-medium \cite{Cloet:2005pp,Cloet:2005rt,Cloet:2006bq}; and produces quark transversity distributions \cite{Cloet:2007em} in fair agreement with empirical analyses \cite{Anselmino:2007fs}.

\begin{figure}[t]
\centering\includegraphics[clip,width=0.5\linewidth]{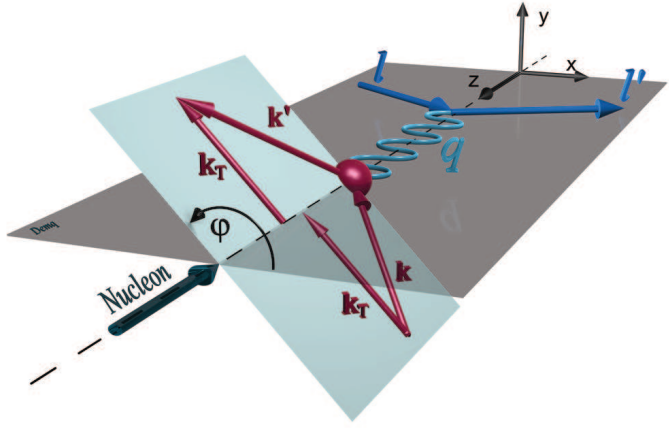}
\caption{SIDIS kinematics: the photon momentum, $q$ with $Q^2=-q^2$, defines the $z$ direction and the struck quark in the nucleon has initial transverse momentum $\vec{k_T}$ with respect to $\hat z$. \label{PLOT_DIS_KINEMATICS_3D}}
\end{figure}

%
The kinematics of semi-inclusive hadron production, $lN \to l^\prime hX$, is illustrated by Fig.\,\ref{PLOT_DIS_KINEMATICS_3D}, wherein a lepton with momentum $l$ scatters from a target by emitting a virtual photon with momentum $q$ that is absorbed by a quark with initial momentum $k$.  The $z$-direction is defined by the orientation of the virtual photon's momentum and one works with collinear $\gamma N$ kinematics, so that the target's three-momentum is parallel to $(-\hat z)$.  Transverse three-momentum components, labelled with a subscript $T$, are measured with respect to $\hat z$: $\vec{k}_T = \vec{k} - (\vec{k}\cdot \hat z) \, \hat z$; and the angle between the lepton scattering plane and the quark scattering plane is denoted by $\varphi$.  In principle, the struck quark in the target will possess $k_T^2\neq 0$.

In keeping with the continuum quantum field theory approach that we have described herein, the NJL nucleon bound-state is described by a Poincar\'e covariant Faddeev equation that includes both scalar and axial-vector diquark correlations.  The Faddeev kernels are simplified, however, by employing a so-called ``static approximation'' \cite{Buck:1992wz} for the quark exchanged within the shaded box of Fig.\,\ref{figFaddeev}:
\begin{equation}
S(k) \to \frac{1}{M_f}\,.
\label{staticexchange}
\end{equation}
In combination with diquark correlations generated by the NJL model Bethe-Salpeter equations, whose amplitudes are momentum-independent, Eq.\,(\ref{staticexchange}) leads to Faddeev amplitudes which themselves are momentum-independent.  (This is discussed further, e.g., in Refs.\,\cite{Roberts:2011cf,Chen:2012qr,Wang:2013wk}.)

The NJL model used in Ref.\,\cite{Matevosyan:2011vj} is defined by the following interaction term in the quark-diquark sector:
\begin{equation}
\mathcal{L}_I = \frac{1}{\Lambda_s^2} \Bigl(\ol{\psi}\,\g_5 C \tau_2 \beta^A\, \ol{\psi}^T\Bigr)
                              \Bigl(\psi^T\,C^{-1}\g_5 \tau_2 \beta_A\, \psi\Bigr)
+\,
\frac{1}{\Lambda_a^2} \lf(\ol{\psi}\,\g_\mu C \tau_i\tau_2 \beta^A\, \ol{\psi}^T\rg)
                              \Bigl(\psi^T\,C^{-1}\g^{\mu} \tau_2\tau_i \beta_A\, \psi\Bigr),
\label{eq:lag}
\end{equation}
where $\beta_A = \sqrt{\tfrac{3}{2}}\,\lambda_A$, $(A=2,5,7)$ are the antisymmetric Gell-Mann matrices, which are associated with the colour-antitriplet ($\bar{3}$) diquark correlations.\footnote{N.B.\ In this section we work in Minkowski space and employ the metric and Dirac matrix conventions of Ref.\,\cite{IZ80}, so that $C = i\g_2\g_0$ is the charge conjugation matrix.}  The quantities $\Lambda_{s,a}$ are parameters, whose values determine the strength of scalar and axial-vector diquark correlations in the nucleon.  The model is regularised using the procedure described in connection with Eq.\,\eqref{eq:Regularization}.  The value of the confinement mass-scale, $\Lambda_{\rm IR}=0.24\,$GeV, is the same as that in Table~\ref{tab:CQM}, but there are four other parameters: $M$, $\Lambda_{\rm UV}$, $\Lambda_{s,a}$.  Equation~\eqref{eq:lag} is complimented by a standard interaction term in the quark-antiquark sector.  The parameters therein are chosen \cite{Cloet:2007em} so as to produce $M=0.4\,$GeV and then, with $\Lambda_{\rm IR}=0.65\,$GeV, $\Lambda_{s}=0.37\,$GeV, $\Lambda_{a}=0.60\,$GeV, one obtains empirical values of the nucleon's mass and its axial-vector coupling, $g_A$.\footnote{An analysis of the role and manifestations of DCSB in computations of $g_A$ within Faddeev equation frameworks is presented elsewhere \protect\cite{Eichmann:2011pv,Chang:2012cc}.}

The leading-twist spin-independent TMD distribution for quarks of flavor $q$ in the nucleon is defined via the correlator \cite{Bacchetta:2008af,Avakian:2010br}
\begin{equation}
\mathcal{Q}(x,\vec{k}_T)
= p^+\int \frac{d \xi^- d \vec{\xi}_T}{(2\pi)^3}\ e^{ix\,p^+\,\xi^-}\, e^{-i\,\vec{k}_T\cdot \vec{\xi}_T}
\lf\la N,S \lf\vert \bar{\psi}_q(0)\,\g^+\,\mathcal{W}(\xi)\,\psi_q(\xi^-,\vec{\xi}_T)  \rg\vert N,S \rg\ra \Bigr\vert_{\xi^+=0},
\label{eq:tmd1}
\end{equation}
where the light-front conventions are those of Ref.\,\cite{Kogut:1969xa} and $\mathcal{W}(\xi)$ is a gauge link connecting the two quark fields, which are labeled by $\psi_q$.  In QCD this gauge link is nontrivial for $\vec{\xi}_T \neq 0$.  However, it is unity in the NJL model owing to the lack of gluon degrees of freedom.  The nucleon states in Eq.\,\eqref{eq:tmd1} are normalised as follows:
\begin{equation}
\sideset{}{_{q_{\rm valence}}}\sum \lf\la N,S \lf\vert \bar{\psi}_q(0)\,\g^+\,\psi_q(0)  \rg\vert N,S \rg\ra = 3,
\end{equation}
where the sum runs over the valence-quarks in the nucleon.

In general, Eq.\,\eqref{eq:tmd1} is expressed in terms of two TMD quark distribution functions; namely,
\begin{equation}
\mathcal{Q}(x,\vec{k}_T)  = q(x,k_T^2) - \frac{\ve^{-+ij}\,k_T^i\,S_T^j}{M}\, q_{1T}^\perp(x,k_T^2),
\end{equation}
where the first TMD PDF integrated over $\vec{k}_T$ gives the unpolarised quark distribution function.  The second TMD PDF is known as the Sivers function \cite{Sivers:1989cc,Bacchetta:2003rz}, which was empirically discovered by the Hermes collaboration using SIDIS \cite{Airapetian:2004tw}.  The Sivers function is na\"{\i}vely time-reversal odd and hence is zero in the absence of the gauge-link; and therefore in the model reviewed here.  The surviving TMD can be expressed in the form \cite{Barone:2001sp}
\begin{equation}
q(x,\,k_T^2) = -i\int \frac{d k^+ dk^-}{(2\pi)^4}
\delta\!\lf(x - \frac{k^+}{p^+}\rg) \text{Tr}\lf[\g^+\,\Phi_q(p,k)\rg],
\label{eqn:def2}
\end{equation}
where $\Phi_q(p,k)$ is the in-nucleon quark-quark correlation matrix.  The quark distribution function can therefore be associated with a straightforward Feynman
diagram calculation in any framework that expresses the nucleon as a bound state of dressed-quarks.

\begin{figure}[t]
\centerline{\includegraphics[clip,width=0.6\linewidth]{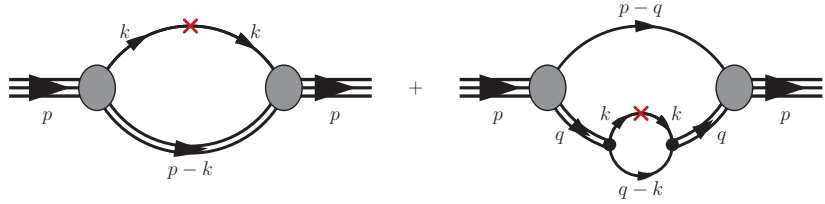}}
\caption{Feynman diagrams associated with Eq.\,\eqref{eqn:def2}, which give the unpolarized quark TMD in the nucleon: single line, dressed-quark propagator; double line, diquark $t$-matrix; shaded oval, quark-diquark Faddeev amplitude; and ``$\mathbf{\times}$'', operator insertion with the form $\g^+ \delta (x - \tfrac{k^+}{p^+})\tfrac{1}{2}\lf(\mathbf{1}\pm\tau_3\rg)$.}
\label{fig:diagrams}
\end{figure}

In the model of Ref.\,\cite{Matevosyan:2011vj}, Eq.\,\eqref{eqn:def2} is computed from the diagrams in Fig.\,\ref{fig:diagrams}, wherein: the single line represents a dressed-quark propagator, which is the solution to the gap equation; the double line is the diquark $t$-matrix, obtained from a Bethe-Salpeter equation; and the shaded-circles represent the solution to the nucleon Faddeev equation.  The distributions obtained from these diagrams have no support for $x<0$ and hence one obtains the valence-quark contribution to the TMD.  Using appropriate isospin projections, one may readily obtain the spin-independent valence $u$- and $d$-valence-quark TMD distributions in the proton:
\begin{subequations}
\label{eq:qtmd}
\begin{eqnarray}
\label{eq:uptmd}
u_V(x,k_T^2) &=& f^s_{q/N}(x,k_T^2)  + \tfrac{1}{3}\,f^a_{q/N}(x,k_T^2)  + \tfrac{1}{2}\,f^s_{q(D)/N}(x,k_T^2) + \tfrac{5}{6}\,f^a_{q(D)/N}(x,k_T^2),\\
\label{eq:downtmd}
d_V(x,k_T^2) &=& \tfrac{2}{3}\,f^a_{q/N}(x,k_T^2)  + \tfrac{1}{2}\,f^s_{q(D)/N}(x,k_T^2) + \tfrac{1}{6}\,f^a_{q(D)/N}(x,k_T^2).
\end{eqnarray}
\end{subequations}
The superscripts $s$ and $a$ refer to the scalar and axial-vector diquark terms, respectively; the subscript $q/N$ implies a quark diagram (left image,  Fig.\,\ref{fig:diagrams}) and $q(D)/N$ a diquark diagram (right image, Fig.\,\ref{fig:diagrams}).  Explicit expressions for the functions in Eqs.\,\eqref{eq:qtmd} are given in the appendix of Ref.\,\cite{Matevosyan:2011vj}.

\begin{figure}[t]
\begin{tabular}{lr}
\includegraphics[clip,width=0.47\linewidth]{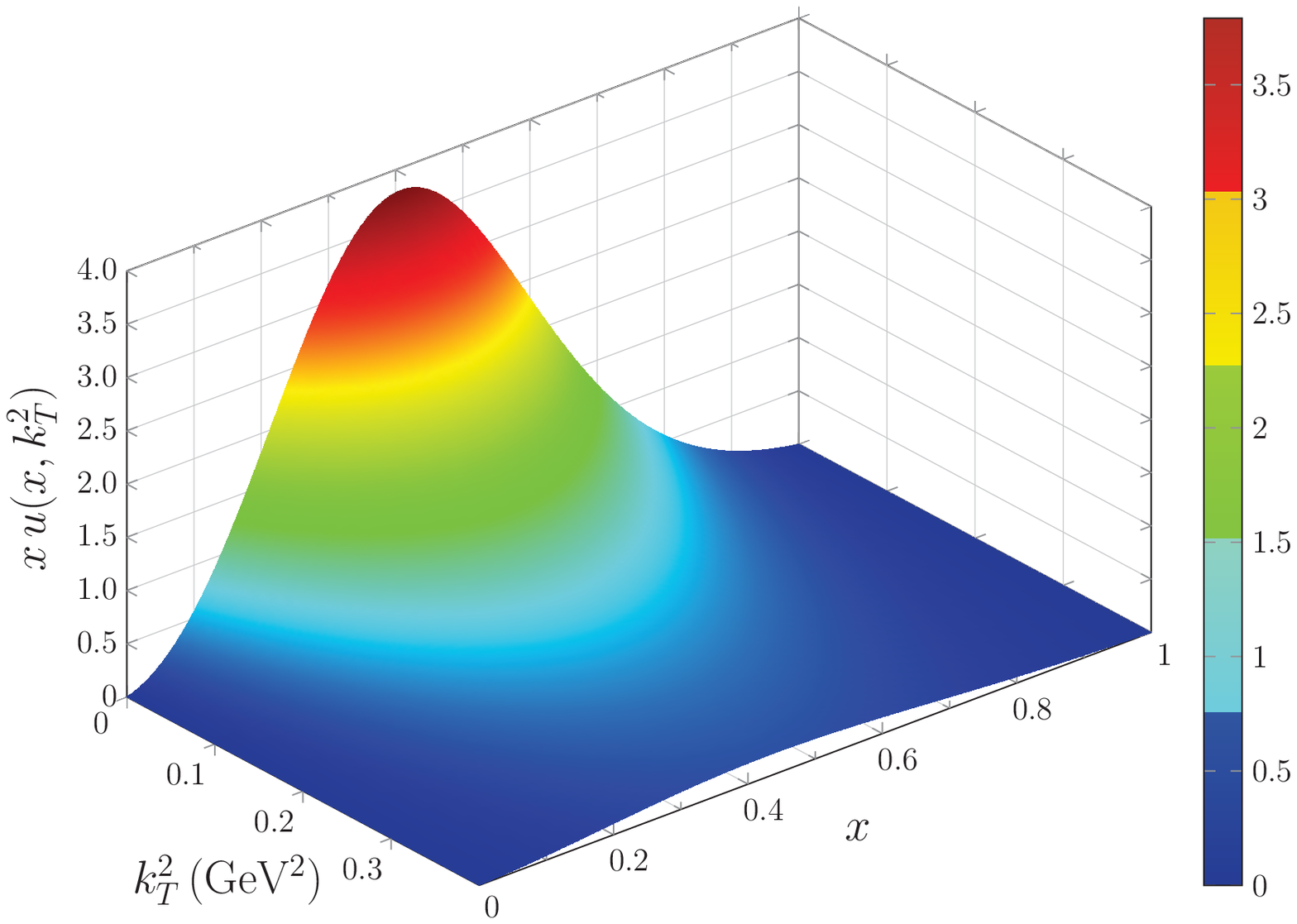}&
\includegraphics[clip,width=0.47\linewidth]{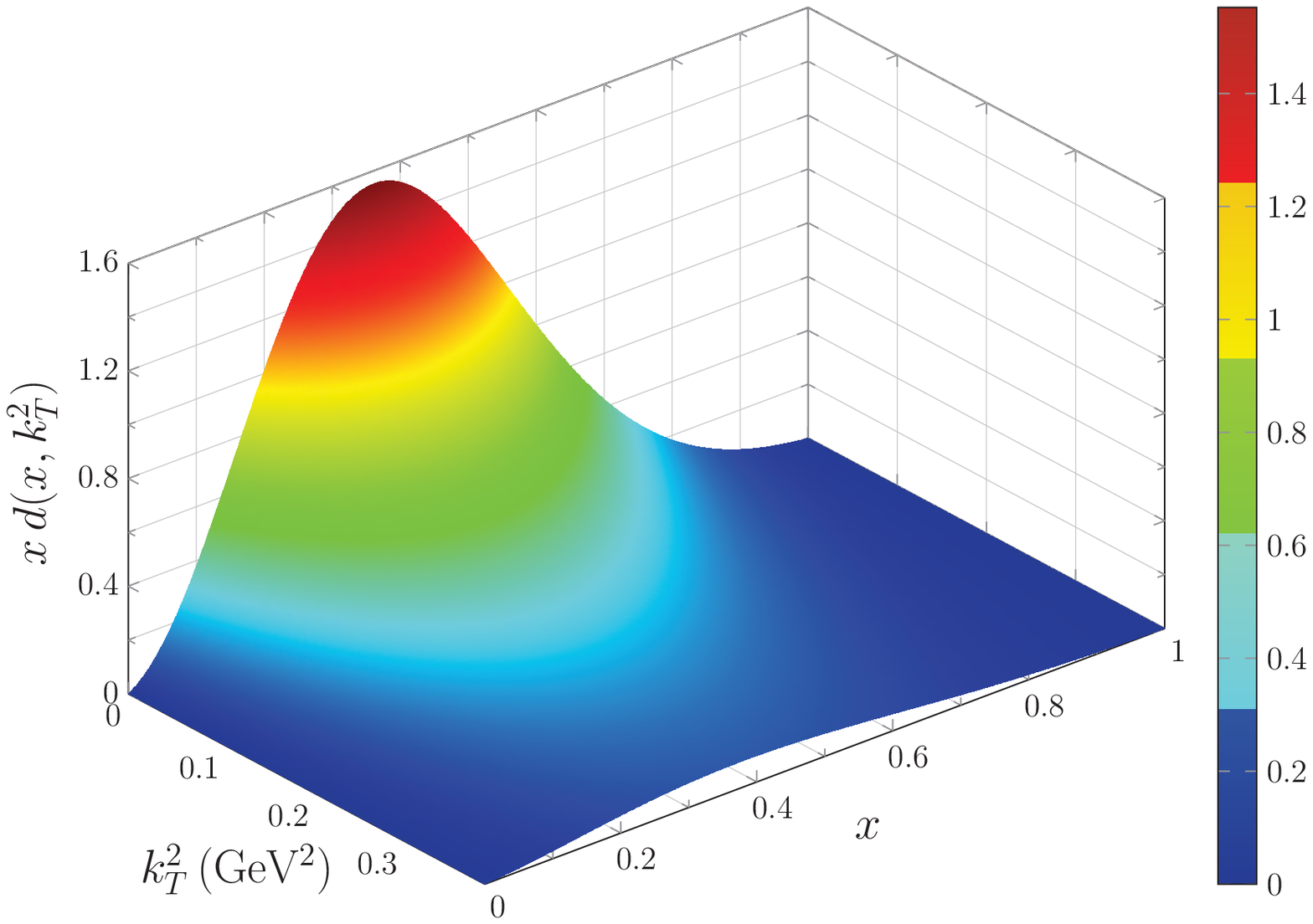}
\end{tabular}
\caption{\label{fig:tmds}
$x u_V(x,k_T^2)$ (left) and $x d_V(x,k_T^2)$ (right) computed in the proton using a NJL model \cite{Matevosyan:2011vj}.}
\end{figure}

Results for the proton's $u_V$- and $d_V$-quark TMDs are depicted in Fig.\,\ref{fig:tmds}.  Unsurprisingly, these $x$-weighted distributions peak at roughly $x=1/3$ when $k_T^2 \approx 0$ and they rapidly become small with increasing $k_T^2$.

The distributions plotted in Fig.\,\ref{fig:tmds} were computed at the ``model scale''; i.e., an \emph{a priori} unknown momentum scale, $\zeta_0$, which is treated as a parameter in Ref.\,\cite{Matevosyan:2011vj} (and kindred studies).  In previous studies of the longitudinal valence-quark PDFs within that framework \cite{Cloet:2005pp,Cloet:2006bq,Cloet:2007em}, the parameter $\zeta_0$ was chosen to be that momentum-scale required as the starting point for DGLAP-evolution \cite{Gribov:1971zn,Gribov:1972,Dokshitzer:1977,Altarelli:1977} in order to obtain agreement, according to some subjective criteria, with a modern PDF parametrization at a significantly larger scale: $\zeta/\zeta_0\gtrsim 10$.  This procedure produced $\zeta_0=M$, which is an internally consistent value because it marks the natural scale for nonperturbative phenomena within the model.  It should also be viewed as the minimum allowable value, as explained elsewhere \cite{Holt:2010vj,Chang:2012rk}.  If one assumes that the same model scale is appropriate here, too, then a comparison with data requires that the TMDs in Fig.\,\ref{fig:tmds} be evolved to the scale relevant for the experiment.  Factorisation theorems and evolution equations for TMDs have recently been derived for some of the leading-twist TMDs \cite{Aybat:2011zv,Aybat:2011ge,Bacchetta:2013pqa}.  They are valid for $k_T^2\ll Q^2$ and are beginning to be used.

In the phenomenology of TMDs it is common to work with parametrisations of these functions, fitted to data at a low scale $Q^2$, since such forms can readily be used in complicated evolution formulae.  Separability is typically assumed; viz.,
\begin{equation}
q(x,k_T^2) = q(x)\, \frac{e^{-k_T^2/\lf<k_T^2\rg>_0}}{\pi\,\lf<k_T^2\rg>_0},
\label{eq:tmdgaussian}
\end{equation}
where $q(x)$ is the associated longitudinal PDF for a quark of flavour $q$ and $\langle k_T^2\rangle_0$ characterises the width of the distribution at the fitting scale.  N.B.\ Eq.\,\eqref{eq:tmdgaussian} guarantees $\int d^2 \vec{k}_T\, q(x,k_T^2) = q(x)$.

\begin{figure}[t]
\leftline{\includegraphics[clip,width=0.45\linewidth]{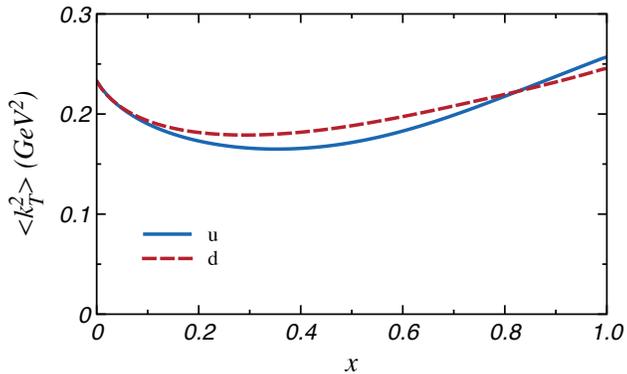}}
\vspace*{-21ex}

\rightline{\parbox{22em}{\caption{\label{fig:kT2vx}
Bjorken-$x$ dependence of $\lf<k_T^2\rg>_q$.  Diquark correlations in the nucleon give rise to the modest quark-flavour dependence.
}}}
\vspace*{4.5em}

\end{figure}

It is natural to ask whether Eq.\,\eqref{eq:tmdgaussian} is a valid approximation to a computed TMD in any well-defined sense.  The material in Ref.\,\cite{Matevosyan:2011vj} provides a means of addressing this issue.  Consider the quantity
\begin{equation}
\lf<k_T^2\rg>_q\!(x) := \frac{\int d^2 \vec{k}_T\ k_T^2\, q(x,\,k_T^2)}{\int d^2 \vec{k}_T\ q(x,\,k_T^2)},
\label{eq:tmdkt2}
\end{equation}
which defines an $x$-dependent mean-square transverse momentum.  The results for this quantity are depicted in Fig.\,\ref{fig:kT2vx}: evidently, it is roughly constant for both valence $u$- and $d$-quarks.  This might initially seem puzzling, given the behaviour in Fig.\,\ref{fig:tmds}, which shows rapid damping with $k_T^2$ at all $x$.  However, it can be understood via the normalising influence of the denominator: the unintegrated $x$-dependence reweights the numerator integrand.  On the other hand, there is a second feature; viz.,
\begin{equation}
\int dx \,d\vec{k}_T\, u(x,\,k_T^2) = (1.05\,M)^2 ,\quad
\int dx \,d\vec{k}_T\, d(x,\,k_T^2) = (1.08\,M)^2
\label{RatioInt}
\end{equation}
and
\begin{equation}
\int dx \lf<k_T^2\rg>_u\!(x) = (1.10\,M)^2,\quad
\int dx \lf<k_T^2\rg>_u\!(x) = (1.12\,M)^2.
\label{IntRatio}
\end{equation}
This leads one to ask under which conditions is the integral of a ratio of two non-negative functions, Eqs.\,\eqref{IntRatio}, approximately equal to the ratio of the integrals of those two non-negative functions, Eqs.\eqref{RatioInt}?  Ignoring coincidence, this occurs when the $x$-dependence of the two functions is roughly the same on that domain from which most of the strength is received.  This supports a validity for the \emph{Ansatz} in Eq.\,\eqref{eq:tmdkt2}, which assumes that the $x$-dependence is precisely the same for all $x$.

\begin{figure}[tb]
\begin{tabular}{lr}
\includegraphics[clip,width=0.47\linewidth]{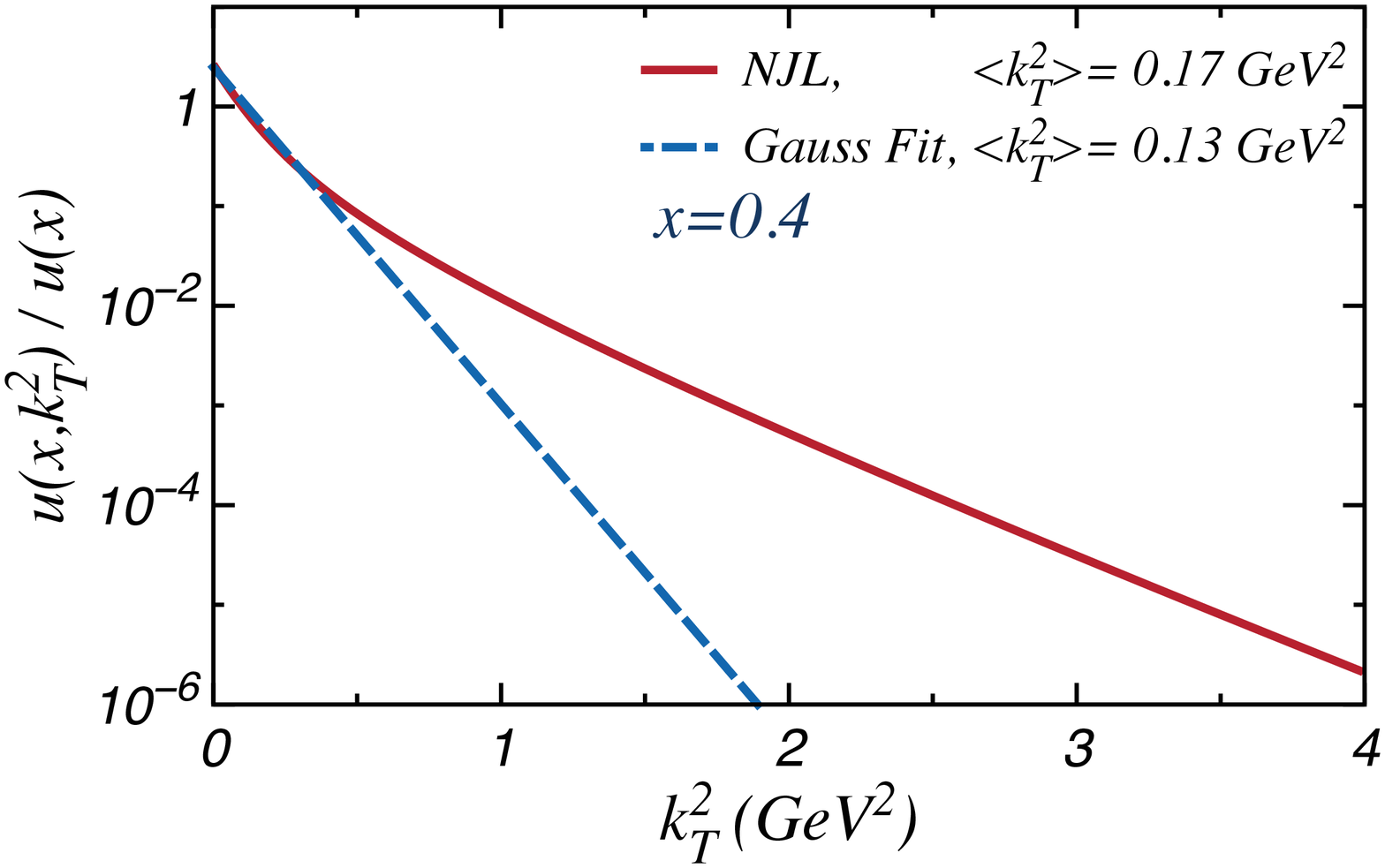}&
\includegraphics[clip,width=0.47\linewidth]{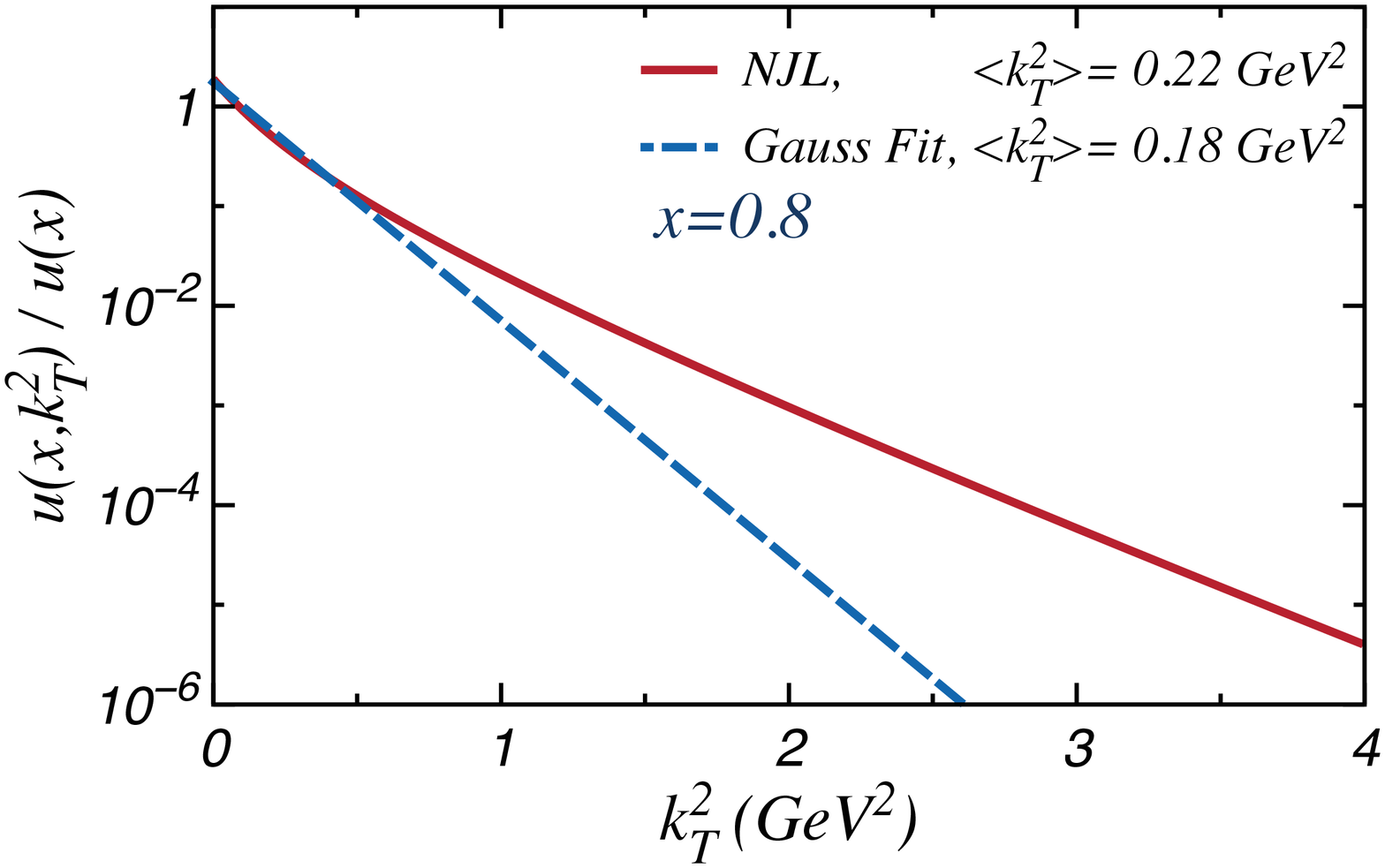}
\end{tabular}
\caption{Model results (solid curve) for the proton's valence $u$-quark TMD: left panel, $x=0.4$; and right panel, $x=0.8$.  In each panel, a \emph{Gau{\ss}ian} \emph{Ansatz}, Eq.\,\protect\eqref{eq:tmdkt2}, fitted to the model result, is also plotted (dashed curve).
\label{fig:tmdsfixedx}}
\end{figure}

A further check of Eq.\,\eqref{eq:tmdkt2} is presented in Fig.\,\ref{fig:tmdsfixedx}. At two widely separated values of $x$, one placed near the $k_T^2$-slice along which the $x$-weighted distribution has its maximum and the other in the tail, the \emph{Gau{\ss}ian} \emph{Ansatz} provides a good pointwise description of the model result on that domain for which the magnitude is sizeable.  The fit parameter $\langle k_T^2\rangle$ does not exhibit dramatic $x$-dependence, showing less than 20\% variation from the mean value.  We therefore conclude that for practical purposes, a sensibly chosen \emph{Gau{\ss}ian} \emph{Ansatz} is a useful phenomenological tool.

It is worth noting that the computed $k_T^2$-width of the valence-quark TMDs is approximately $M^2$ at the model scale $\zeta_0=M$; i.e., $\langle k_T^2\rangle_0 \gtrsim M^2$.  This value is smaller but commensurate with that inferred via modern parametrisations of extant data obtained at a momentum scale $\zeta \approx 6 \,M$ \cite{Anselmino:2013vqa}; and that is good because the $k_T^2$-width of a TMD grows slowly under evolution.  We expect these features to be manifest in any framework the properly accounts for DCSB in the baryon bound-state problem.

\section{Novel features of the EMC effect}
\label{SecEMC}
Understanding the mechanisms responsible for a change in the per-nucleon DIS cross-section between the deuteron and heavier nuclei is one of the most important challenges confronting the nuclear physics community.  In the valence quark region the effect is characterised by a quenching of the nuclear structure functions relative to those of the free nucleon and is known as the EMC effect \cite{Aubert:1983xm}.  Its discovery led to a tremendous amount of experimental and theoretical investigation \cite{Berger:1987er,Geesaman:1995yd}, which continues today.  However, after the passage of roughly 30 years there is still no consensus regarding the mechanisms behind the EMC effect.

Early attempts to explain the EMC effect focused on detailed nuclear structure
investigations \cite{Bickerstaff:1989ch} and the possibility of an in-medium enhancement of the nucleon's pionic component \cite{LlewellynSmith:1983qa,Ericson:1983um}.  The former studies were unable to describe the data and the latter explanation appears to be ruled out by Drell-Yan measurements of the anti-quark distributions in nuclei \cite{Alde:1990im}.  Other ideas included the possibility of \emph{exotic} six-quark bags in the nucleus \cite{Jaffe:1982rr} or
traditional short-range correlations \cite{Weinstein:2010rt}.  It has also been argued that the EMC effect is a result of changes in the internal structure of the bound nucleons brought about by the strong nuclear fields inside the nucleus \cite{Saito:1992rm}.  Many of these approaches can explain the qualitative features of the EMC effect but the underlying physics mechanisms differ substantially.

To make progress toward understanding the mechanism responsible for the EMC effect, it has become clear that new experiments are required that can reveal genuinely novel aspects of this effect.  To that end, Ref.\,\cite{Cloet:2012td} proposed to exploit parity-violating DIS (PVDIS), which is mediated by interference between photon and $Z^0$ exchange.  When used in conjunction with the familiar electromagnetic DIS data, it should become possible to obtain explicit information about the quark-flavour dependence of the nuclear PDFs.  That will allow the predictions of any model of the EMC effect to confront new experimental information and hence provide important insights into this longstanding puzzle.

Consider, therefore, that the parity violating effect of interference between photon and $Z^0$ exchange in the DIS of longitudinally polarized electrons on an unpolarized target yields the non-zero asymmetry
\begin{equation}
A_{PV} = \frac{\s_R - \s_L}{\s_R + \s_L},
\end{equation}
where $\s_R$ and $\s_L$ denote the double differential cross-sections for DIS
of right- and left-handed polarised electrons, respectively.  In the Bjorken limit, $A_{PV}$ can be expressed as \cite{Brady:2011uy}
\begin{equation}
A_{PV} = \frac{G_F\,Q^2}{4\sqrt{2}\,\pi\,\a_{\text{em}}}
\lf[a_2(x_A) + \frac{1-(1-y)^2}{1+(1-y)^2}\,a_3(x_A)\rg],
\label{eq:APV_Bjorken}
\end{equation}
where $x_A$ is the Bjorken scaling variable of an $A$-nucleon nucleus multiplied by $A$, $G_F$ is the Fermi coupling constant and $y=\nu/E$ is the energy transfer divided by the incident electron energy.

The $a_2$ term in Eq.\,\eqref{eq:APV_Bjorken} originates from the product of the electron weak-axial current and the quark weak-vector current, and has the form
\begin{equation}
a_2(x_A) = -2\,g_A^e\,\frac{F_{2A}^{\g Z}(x_A)}{F_{2A}^\g(x_A)}
= \frac{2\sum_q e_q\,g_V^q\,q_A^+(x_A)}{\sum_q e_q^2\,\,q_A^+(x_A)}.
\label{eq:a2}
\end{equation}
The plus-type quark distributions are defined by $q_A^+(x_A) = q_A(x_A) + \bar{q}_A(x_A)$, $e_q$ is the quark charge, and \cite{Beringer:1900zz}
$g_A^e = -\tfrac{1}{2}$ and the quark weak-vector couplings are
\begin{equation}
\hs{-2mm} g_V^u = \frac{1}{2} - \frac{4}{3}\,\sin^2\!\theta_W, \hs{5mm}
          g_V^d = -\frac{1}{2} + \frac{2}{3}\,\sin^2\!\theta_W,
\end{equation}
where $\theta_W$ is the weak mixing angle.  The parity violating $F_2$ structure function of the target arising from $\g Z$ interference is labelled as $F_{2A}^{\g Z}(x_A)$, while $F_{2A}^\g(x_A)$ is the familiar electromagnetic DIS structure function.  The parton model expressions for these structure functions are \cite{Beringer:1900zz}
\begin{equation}
F_{2A}^{\g Z} = 2\,x_A\sideset{}{_q}\sum e_q\,g_V^q\,  q_A^+, \hs{3mm}
F_{2A}^\g    =    x_A\sideset{}{_q}\sum e_q^2\,      q_A^+.
\end{equation}

The $a_3$ term in Eq.\,\eqref{eq:APV_Bjorken} is given by
\begin{equation}
%
a_3(x_A) = -2\,g_V^e\,\frac{F_{3A}^{\g Z}(x_A)}{F_{2A}^\g(x_A)}
= -4\,g_V^e\,\frac{\sum_q e_q\,g_A^q\,q_A^-(x_A)}{\sum_q e_q^2\,\,q_A^+(x_A)},
\end{equation}
where $q_A^-(x_A) = q_A(x_A) - \bar{q}_A(x_A)$ and \cite{Beringer:1900zz} $g_V^e = -\tfrac{1}{2} + 2\sin^2\theta_W$, $g_A^u = -g_A^d = \tfrac{1}{2}$.  This term is suppressed in the parity-violating asymmetry, $A_{PV}$, because of the $y$-dependent prefactor in Eq.\,\eqref{eq:APV_Bjorken} and the fact that $g_V^e \ll g_A^e$.  It  therefore played no role in Ref.\,\cite{Cloet:2012td}.

The $F^{\g Z}_{2A}$ structure function has a different flavor structure
to that of $F^{\g}_{2A}$.  Consequently, $a_2(x)$ is sensitive to flavour dependent effects.  To illustrate this, consider an expansion of $a_2$ about $u_A^+ \simeq d_A^+$.  With $s_A^+ \ll u_A^+ + d_A^+$ and ignoring heavier quark flavours, then
\begin{equation}
a_2(x_A) \simeq \frac{9}{5} - 4\sin^2\theta_W
- \frac{12}{25}\,
\frac{u^+_A\lf(x_A\rg) - d^+_A\lf(x_A\rg)- s^+_A\lf(x_A\rg)}{u^+_A\lf(x_A\rg) + d^+_A\lf(x_A\rg)}.
\label{eq:a2isovector}
\end{equation}
The correction from strange quarks given in Eq.\,\eqref{eq:a2isovector} may be important in the low-$x$ region \cite{Chang:2011vx}, however, HERMES data \cite{Airapetian:2008qf} has confirmed that $s_A^+$ is negligible compared with  $u^+_A + d^+_A$ on the domain $x_A > 0.1$.
Therefore, a measurement of $a_2(x_A)$ will provide information about the flavour dependence of the nuclear quark distributions, so that when coupled with existing measurements of $F^{\g}_{2A}$ an extraction of the flavour-dependent quark distributions becomes possible on the valence-quark domain.

Alternatively, if the correction term in Eq.\,\eqref{eq:a2isovector} is known, then the parity violating asymmetry provides an independent method by which to determine the weak mixing angle.  For example, if one ignores strange quark effects,
quark mass differences \cite{Sather:1991je,Rodionov:1994cg,Londergan:2009kj} and
electroweak corrections, then the $u$- and $d$-quark distributions of an isoscalar
target will be identical; and in this limit Eq.\,\eqref{eq:a2isovector} becomes
\begin{equation}
a_2(x_A) \stackrel{N=Z}{\lra} \frac{9}{5} - 4\sin^2\theta_W.
\label{eq:a2_isoscalar}
\end{equation}
This result is analogous to the Paschos-Wolfenstein ratio \cite{Paschos:1972kj,Cloet:2009qs} in neutrino DIS, which motivated the NuTeV collaboration's measurement of $\sin^2\theta_W$ \cite{Zeller:2001hh,Bentz:2009yy} that is discussed in Sec.\,\ref{SecNuTeV}.

A significant advantage of $a_2(x_A)$ as a measure of the weak mixing angle is that strange quark effects are negligible on the valence quark domain and hence the largest uncertainty in the NuTeV measurement of $\sin^2\theta_W$ \cite{Bentz:2009yy} is eliminated.  Furthermore, the isovector correction term in Eq.\,\eqref{eq:a2isovector} does not depend on $\sin^2\theta_W$ and thus a measurement of $a_2(x_A)$ at each value of $x_A$ constitutes a separate determination of $\sin^2\theta_W$.
Most important in the present context, however, is that $a_2$ is sensitive to flavour dependent nuclear effects that influence the quark distributions in nuclei.  Indeed, owing to this sensitivity a measurement of $a_2$ on a target with $N > Z$  would provide an excellent opportunity to test the importance of the isovector EMC effect \cite{Cloet:2009qs,Bentz:2009yy} in interpreting the anomalous NuTeV result for $\sin^2\theta_W$.

In order to explore the effect of isospin-dependent nuclear forces on $a_2(x_A)$, Ref.\,\cite{Cloet:2012td} employed nuclear-matter quark distribution functions calculated with the same NJL model and associated parameters that were used in the TMD computation described in Sec.\,\ref{SecTMD}.  The distributions thus obtained depend explicitly on the $Z/N$ ratio.  To be a little more explicit, the nucleon amplitude was obtained from the Faddeev equation described above; and the impact of a nuclear medium was simulated via inclusion of isoscalar-scalar ($\sigma_0$), isoscalar-vector ($\omega_0$) and isovector-vector ($\rho_0$) mean-fields that couple self-consistently to the dressed-quarks within the nucleon bound-states.  The strength of these mean-fields was determined by a NJL-model nuclear-matter equation-of-state \cite{Bentz:2001vc}, with the following results for the vector fields, which are crucial to this application: $\omega_0 = 6\,\lf(\rho_p + \rho_n\rg)/\Lambda_{\omega}^2$ and $\rho_0  = 2\,\lf(\rho_p - \rho_n\rg)/\Lambda_\rho^2$, where $\rho_p$ and $\rho_n$ are the proton and neutron densities, respectively, and $\Lambda_{\omega,\rho}$ are analogues of $\Lambda_{s,a}$ in Eq.\,\eqref{eq:lag}.
The values of the new parameters were determined by fitting the empirical saturation energy and density of symmetric nuclear matter, and the nuclear matter symmetry energy \cite{Cloet:2006bq,Cloet:2009qs}.

\begin{figure}[t]
\begin{tabular}{lr}
\includegraphics[clip,width=0.47\linewidth]{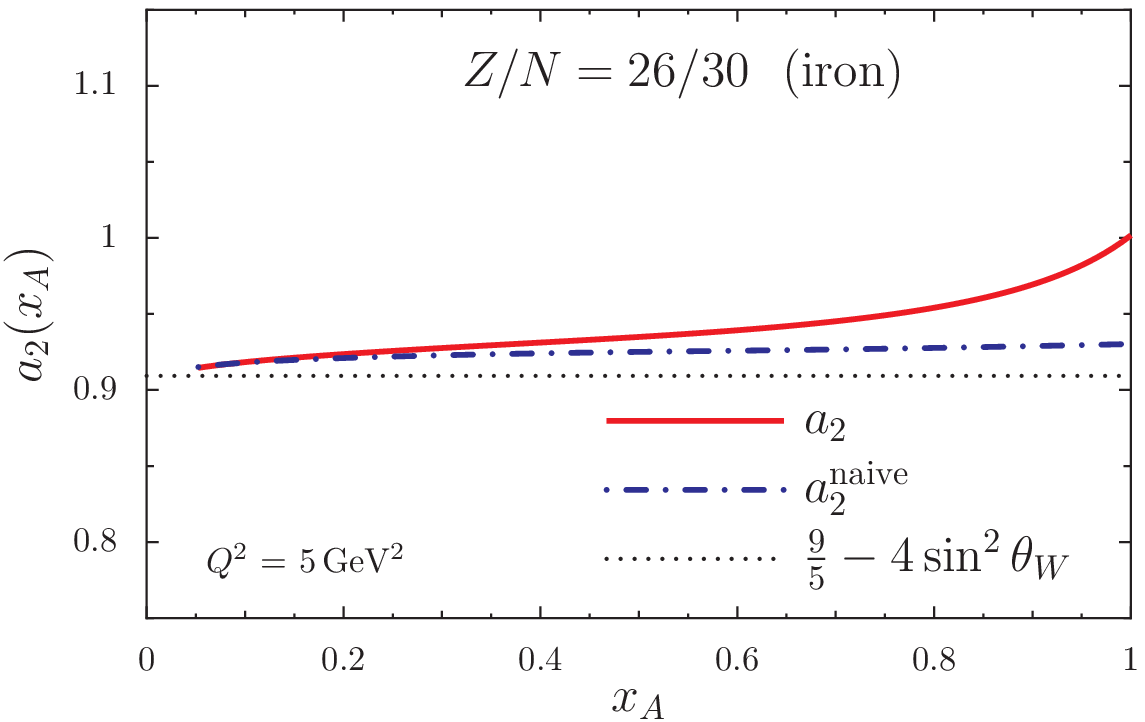} &
\includegraphics[clip,width=0.47\linewidth]{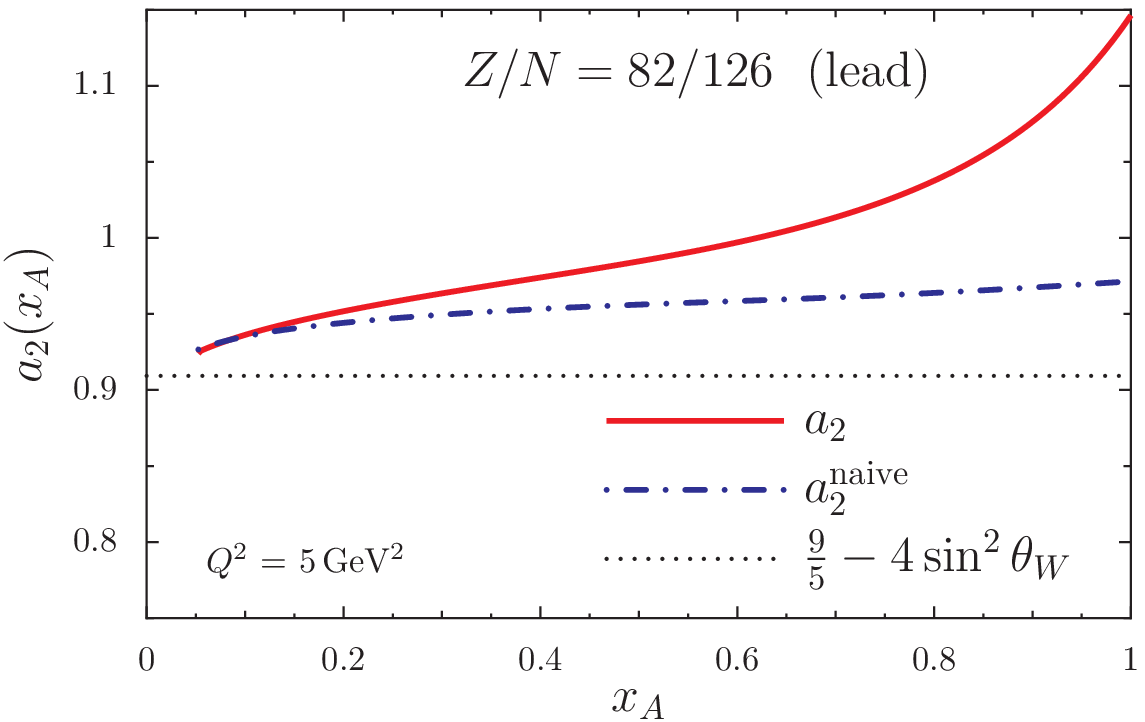}
\end{tabular}
\caption{Asymmetric nuclear matter results for $a_2(x_A)$ obtained using the $Z/N$ ratio of iron (left panel) and lead (right panel).
\emph{Solid curve} -- full result;
\emph{dot-dashed curve} -- na\"{\i}ve expectation, which ignores medium effects; and
\emph{dotted curve} -- isoscalar result.
\label{fig:a2}}
\end{figure}

Using this NJL model's results for the free nucleon and nuclear matter PDFs \cite{Mineo:2003vc,Cloet:2005rt,Cloet:2009qs}, the PVDIS structure function ratio, $a_2(x_A)$, may be determined using Eq.~\eqref{eq:a2}, where $\sin^2\theta_W = 0.2227 \pm 0.004$; i.e., the on-shell renormalisation scheme value \cite{Zeller:2001hh}.
Figure~\ref{fig:a2} presents the result for nuclear matter with a proton-neutron ratio equal to that of iron (left) and lead (right).   The solid curve is the full result, which includes the effects of Fermi motion and the scalar and vector mean-fields.  The dot-dashed curve is the na\"{\i}ve expectation, where the nuclear quark distributions are constructed from the free proton and neutron PDFs without modification.  The dotted curve is the result for isoscalar nuclear matter, which is given by Eq.\,\eqref{eq:a2_isoscalar} in this analysis.  In each case the total baryon density, $\rho_B = \rho_p + \rho_n$, is kept fixed, with only the $Z/N$ ratio being varied, which then determines the strength of the $\rho_0$ mean-field.

The dominant correction to $a_2(x_A)$ is isovector, as illustrated in Eq.\,\eqref{eq:a2isovector}.  As a consequence, the difference between the na\"{\i}ve and full results in Figs.\,\ref{fig:a2} owes primarily to the non-zero $\rho_0$ mean-field.  As we explain in Sec.\,\ref{SecNuTeV}, this is precisely the same effect that eliminates $1\sigma$-$1.5\sigma$ \cite{Cloet:2009qs,Bentz:2009yy} of the discrepancy with respect to the Standard Model in the NuTeV measurement of $\sin^2 \theta_W$.  Thus, quite apart from the intrinsic importance of understanding the dynamics of quarks within nuclei, empirical observation of the large flavour-dependent nuclear effects predicted in Ref.\,\cite{Cloet:2012td} and illustrated in Figs.\,\ref{fig:a2} would be direct evidence that the isovector EMC effect plays a material role in interpreting the NuTeV data.  It would also signal the importance of flavour-dependent effects in attempting to understand the EMC effect in nuclei like lead and gold.

The $a_2$ function is potentially also sensitive to charge symmetry violation (CSV) effects, which are a consequence of the light quark mass differences and electroweak
corrections \cite{Sather:1991je,Rodionov:1994cg,Londergan:2009kj}.  In this case Eq.\,\eqref{eq:a2isovector} reduces to
\begin{equation}
a_2(x) \simeq \frac{9}{5} - 4\sin^2\theta_W
- \frac{6}{25}\,
\frac{\delta u^+(x) - \delta d^+(x)}{u^+_p(x) + d^+_p(x)},
\label{eq:a2CSV}
\end{equation}
where $\delta u^+ \equiv u_p^+ - d_n^+$ and $\delta d^+ \equiv d_p^+ - u_n^+$.  The impact of CSV is largely independent of the in-medium effects already described \cite{Bentz:2009yy}; and by using the central value of the MRST parametrisations \cite{Martin:2003sk}, this correction was found to be negligible on the scale of Figs.\,\ref{fig:a2} \cite{Cloet:2012td}.  Nuclear medium effects should therefore dominate the discrepancy between the na\"{\i}ve expectation and an empirical result for $a_2(x_A)$.  If, on the other hand, CSV effects turn out to be larger than expected, then they, together with any residual uncertainty associated with strange quarks at low $x$, can be constrained via measurements on isospin symmetric nuclei.

\begin{figure}[t]
\begin{tabular}{lr}
\includegraphics[width=0.47\linewidth,clip=true,angle=0]{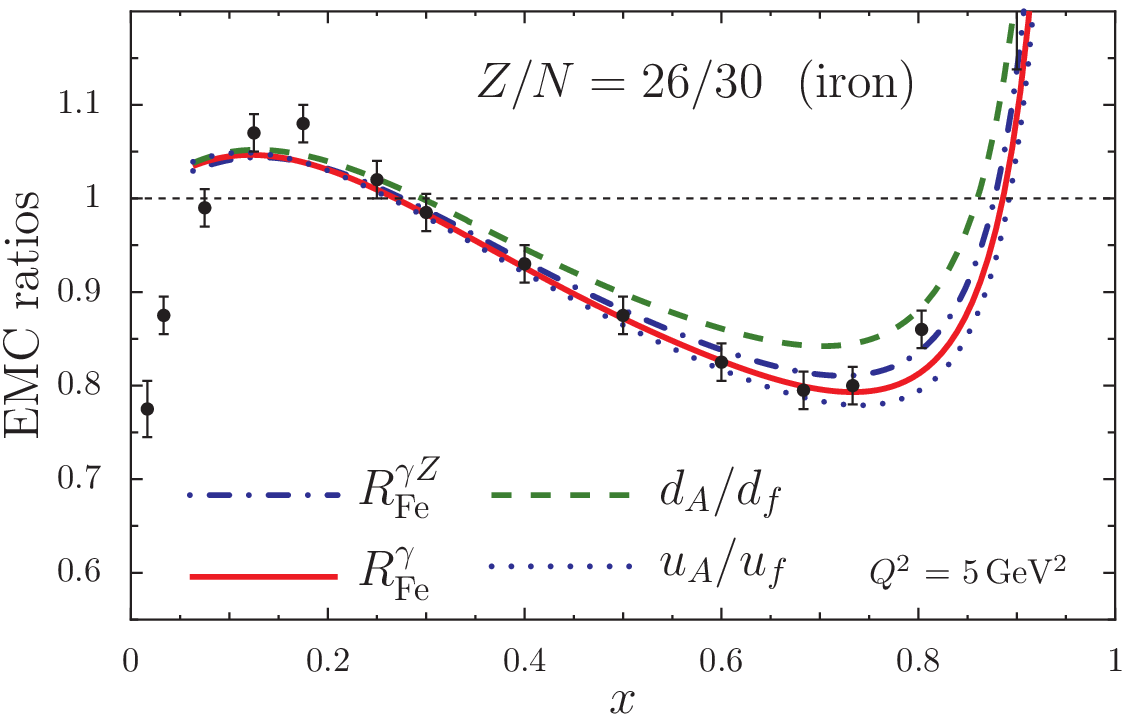}& \includegraphics[width=0.47\linewidth,clip=true,angle=0]{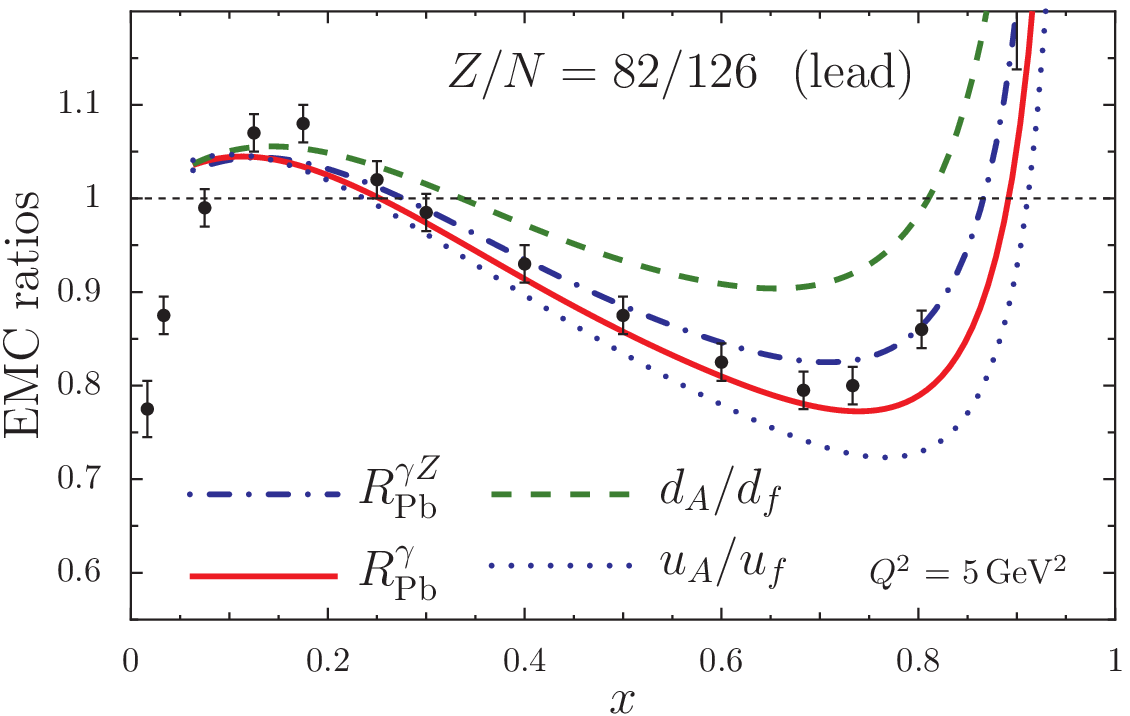}
\end{tabular}
\caption{\label{fig:EMC}
EMC ratios with the $Z/N$ value chosen to be that of iron (left panel) and lead (right panel).  \emph{Solid curve} -- EMC effect in DIS on nuclear target; and \emph{dot-dashed} curve -- EMC ratio for PVDIS on the same target.
The \emph{dotted} and \emph{dashed} curves depict the EMC effect in the $u$ and $d$ quark sectors, respectively.  (Data \cite{Sick:1992pw} is for isoscalar nuclear matter.) }
\end{figure}

The EMC effect can be defined for both the traditional DIS and $\g Z$ interference
structure functions via the ratio
\begin{equation}
R^i = \frac{F_{2A}^i}{F_{2A}^{i,\text{na\"{\i}ve}}} = \frac{F_{2A}^i}{Z\,F_{2p}^i + N\,F_{2n}^i},
\label{eq:EMCeffect}
\end{equation}
where $i = \g,\,\g Z$.  The target structure function is labelled by $F_{2A}^i$,
while $F_{2A}^{i,\text{na\"{\i}ve}}$ is the na\"{\i}ve expectation (no medium effects), which can be expressed as a sum over free proton and neutron structure functions: $R^i \equiv 1$ in the absence of medium effects.  Expressing the EMC ratio in terms of the PDFs, one obtains the parton model expressions
\begin{equation}
R^\gamma    \simeq \frac{4\,u^+_A + d^+_A}{4\,u^+_f + d^+_f}, \quad
R^{\gamma Z} \simeq \frac{1.16\,u^+_A + d^+_A}{1.16\,u^+_f + d^+_f},
\end{equation}
where $q_f$ are the quark distributions of the target assuming it were composed of free nucleons.  For an isoscalar target one has $R^\gamma = R^{\gamma Z}$
(modulo electroweak, quark mass and heavy quark flavour effects).  However, for nuclei with $N \neq Z$ these two EMC effects need not be equal.  The solid curve in Figs.~\ref{fig:EMC} depicts the EMC effect for $F_{2A}^\g$ in nuclear matter,
with $Z/N$ ratios equal to that of iron (left) and lead (right), while the corresponding EMC effect in $F_{2A}^{\g Z}$ is represented by the dot-dashed curve. The dotted and dashed curves show the EMC effect in the $u$- and $d$-quark sectors, respectively.  Plainly, as the proton-neutron ratio is decreased, the EMC effect in $F_{2A}^\g$ increases, whereas the EMC effect in $F_{2A}^{\g Z}$ is reduced. Consequently, for $N > Z$ nuclei, $R^\gamma < R^{\gamma Z}$ on the domain
$x_A \gtrsim 0.3$, which is the domain over which the NJL model of Ref.\,\cite{Cloet:2012td} can be considered most reliable.

The fact that $u_A/u_f < d_A/d_f$ and hence $R^\gamma < R^{\gamma Z}$ in nuclei with
a neutron excess is a direct consequence of the isovector mean field and is a largely model-independent prediction.  It was shown in Ref.\,\cite{Cloet:2009qs} that the isovector mean field leads to a small shift in quark momentum from the $u$- to the $d$-quarks, and hence the in-medium depletion of $u_A$ is stronger than that of $d_A$ in the valence-quark region.  Since $u_A$ is multiplied by a factor of four in the ratio $R^\gamma$, the depletion is more pronounced for this ratio than for $R^{\gamma Z}$, where the $d$-quark quickly dominates as $Z/N$ becomes less than one.

This analysis predicts that flavour-dependent effects in nuclei like lead and gold are $\gtrsim 5$\% in the valence quark region.  Impacts of this magnitude are large enough to be observed in planned PVDIS experiments \cite{pvdisloi} at JLab\,12.  Owing to the relatively small difference between $R^\gamma$ and $R^{\gamma Z}$ in nuclei like iron and lead, an accurate extraction of $u_A$ and $d_A$ would require that $R^\gamma$ and $R^{\gamma Z}$ be measured with the same detector to reduce systematic uncertainties. This is exactly what is planned and therefore an important step toward understanding the EMC effect can be expected in the not too distance future.

The material reviewed in this subsection suggests that an accurate comparison of the electromagnetic and $\g Z$ interference structure functions of a target have the potential to pin down the flavour dependence of the EMC effect in the valence-quark region.  The most direct determination of this flavour dependence (cf.\ Figs.\,\ref{fig:EMC}) would involve charged current reactions on heavy nuclei at an electron-ion collider \cite{Thomas:2009ei} or with certain Drell-Yan reactions \cite{Dutta:2010pg,Chang:2011ra}.  However, such experiments will not be possible for at least ten to twenty years.  On the other hand, accurate measurements of PVDIS on heavy nuclei should be possible at JLab\,12 and would therefore provide a timely, critical test of an important class of models that aim to describe the modification
of nucleon structure functions in-medium.  Such experiments would complement alternative methods to access the quark substructure of nuclei; e.g., the measurement of the EMC effect for spin structure functions \cite{Cloet:2005rt,Cloet:2006bq}.  As a corollary, they would also offer a unique insight into the description of nuclear structure at the quark level.  Finally, these experiments would constitute a direct test of the isovector EMC effect correction to the NuTeV measurement of $\sin^2\theta_W$. 

\section{Insights into the NuTeV anomaly}
\label{SecNuTeV}
\subsection{Challenge to the Standard Model?}
Using a very careful comparison of the charged and neutral current total cross sections for $\nu$ and $\bar{\nu}$ on an iron target, the NuTeV collaboration reported \cite{Zeller:2001hh} a $3\,\sigma$ discrepancy with the Standard Model value of $\sin^2 \theta_W$; viz.,
\begin{equation}
\sin^2\theta_W \stackrel{\rm NuTeV}{=} 0.2277 \pm 0.0013(\text{stat.}) \pm 0.0009(\text{syst.}).
\end{equation}
This report has had an enormous impact.  It was initially interpreted as an indication of possible new physics.  However, attempts to understand the anomaly in terms of popular extensions of the Standard Model have proved unsuccessful \cite{Davidson:2001ji,Kurylov:2003by}.

At the same time, a number of possible corrections within the Standard Model were suggested \cite{Londergan:2003ij,Martin:2003sk,Martin:2004dh,Gluck:2005xh,Mason:2007zz,%
Ball:2009mk,Cloet:2009qs}, most of which have a sign likely to reduce the discrepancy. The correction associated with charge symmetry violation (CSV), arising from the difference between the $u$- and $d$-quark masses \cite{Sather:1991je,Rodionov:1994cg}, is largely model-independent and reduces the discrepancy by about $1\,\sigma$ \cite{Londergan:2003ij}.  If the momentum fraction carried by $s$-quarks in the proton exceeds that carried by $\bar{s}$-quarks, as suggested by chiral physics and commonsense \cite{Signal:1987gz,Brodsky:1996hc,Thomas:2000ny} and an experimental analysis \cite{Mason:2007zz}, there could be a further reduction, albeit with large uncertainties at present \cite{Ball:2009mk}.  In addition, using the NJL framework described in Secs.\,\ref{SecNJLJet} and \ref{SecEMC}, it has been argued \cite{Cloet:2009qs}
that the excess neutrons in iron produce an isovector EMC effect that modifies the PDFs within \textit{all} the nucleons in the nucleus: this has the same sign as the CSV correction and a quantitative estimate suggests that it reduces the NuTeV discrepancy with the Standard Model by about $1.5\,\sigma$.

Since these effects are independent, they may be combined in a straightforward manner, following from which the corrected NuTeV extraction of $\sin^2 \theta_W$ will be far less inconsistent with the Standard Model.  This is emphasised in Ref.\,\cite{Bentz:2009yy}, which has provided an informed update to the derived value of $\sin^2 \theta_W$.  Therein, as we review in this section, each correction was examined, a central value assigned and a conservative error estimated; and they were then combined to produce
\begin{equation}
\label{TWBentz}
\sin^2 \theta_W \stackrel{\mbox{\protect\cite{Bentz:2009yy}}}{=} 0.2221 \pm 0.0013(\text{stat}) \pm 0.0020(\text{syst}).
\end{equation}
This value is in excellent agreement with the Standard Model result; namely, $0.2227 \pm 0.0004$ \cite{Abbaneo:2001ix,Zeller:2001hh} in the on-shell
renormalisation scheme.

The NuTeV experiment involved a precise measurement on a steel target of the ratios
$R^\nu$ and $R^{\bar{\nu}}$; namely, the ratios of the neutral current (NC) to charged current (CC) total cross sections for $\nu$ and $\bar{\nu}$, respectively.  An essential step in the NuTeV extraction of $\sin^2 \theta_W$ was a detailed Monte Carlo simulation of the experiment.  However, NuTeV have provided functionals which allow one to accurately estimate the effect of any proposed correction \cite{Zeller:2002du}.

The NuTeV study was motivated by the observation \cite{Paschos:1972kj} that a ratio of cross sections for $\nu$ and $\bar \nu$ on an isoscalar target allowed an independent extraction of the weak mixing angle.  The so-called Paschos-Wolfenstein (PW) ratio
is given by\footnote{The cross-sections in Eq.\,\protect\eqref{eq:PW} have been integrated over the Bjorken scaling variable and energy transfer.} \cite{Paschos:1972kj}
\begin{equation}
R_{\text{PW}} = \frac{\s_{NC}^{\nu\,A} -
\s_{NC}^{\bar{\nu}\,A}}{\s_{CC}^{\nu\,A} - \s_{CC}^{\bar{\nu}\,A}}
=: \frac{R^{\nu} - r R^{\bar{\nu}}}{1 - r} \ .
\label{eq:PW}
\end{equation}
In Eq.\,(\ref{eq:PW}), $A$ represents the nuclear target and $r=\s_{CC}^{\bar{\nu}\,A} / \s_{CC}^{\nu\,A}$.  Expressing the total cross-sections in terms of quark distributions, ignoring heavy quark flavours and $\mathcal{O}(\a_s)$ corrections, the PW ratio becomes
\begin{equation}
R_{\text{PW}} = \tfrac{\lf(\tfrac{1}{6}-
\tfrac{4}{9} \sin^2\theta_W\rg) \lf\la x_A\,u^-_A\rg\ra
        + \lf(\tfrac{1}{6}-\tfrac{2}{9} \sin^2\theta_W\rg)\, \lf\la x_A\,d^-_A + x_A\,s^-_A\rg\ra}
       {\lf\la x_A\,d^-_A  + x_A\,s^-_A\rg\ra - \tfrac{1}{3}\lf\la x_A\,u^-_A \rg\ra},
\label{eq:PW_quark}
\end{equation}
where $x_A$ is the Bjorken scaling variable for the
nucleus multiplied by $A$, $\la \ldots \ra$ implies integration over $x_A$, and $q_A^- := q_A - \bar{q}_A$ are the non-singlet quark distributions of the target.

Ignoring quark mass differences, strange quark effects and electroweak corrections, the $u$- and $d$-quark distributions of an isoscalar target are identical, and in this limit Eq.\,\eqref{eq:PW_quark} becomes $R_{\text{PW}}^{N=Z}= \tfrac{1}{2} - \sin^2\theta_W$.  If corrections to this result are small, the PW ratio provides an independent determination of the weak mixing angle.  Expanding Eq.\,\eqref{eq:PW_quark} about $u^-_A = d^-_A$ and assuming $s_A^- \ll u^-_A + d^-_A$, one obtains the leading PW correction term; namely,
\begin{equation}
\D R_{\text{PW}} \simeq \lf(1-\frac{7}{3}\sin^2\theta_W\rg)
                \frac{\la x_A\,u^-_A - x_A\,d^-_A - x_A\,s^-_A\ra}{\la x_A\,u^-_A + x_A\,d^-_A\ra}.
\label{eq:correction}
\end{equation}
Extensive studies of neutrino-nucleus reactions have concluded that the most important contributions to Eq.\,\eqref{eq:correction} arise from nuclear effects, CSV and strange quarks.

\subsection{Analysis corrections}
In discussing the extraction of the weak mixing angle from neutrino reactions, it is customary to refer to corrections to the PW ratio.  However, it is important to remember that in the NuTeV analysis the measured quantities were the NC to CC ratios for $\nu$, $\bar \nu$, and that the weak mixing angle was extracted through a Monte Carlo analysis.  For a given effect, the PW ratio will give only a qualitative estimate of the correction to the weak mixing angle.  Quantitative corrections are obtained by using the functionals provided
by NuTeV \cite{Zeller:2002du}.  Throughout this subsection we denote a contribution to Eq.\,\eqref{eq:correction} by $\D R^i_{\text{PW}}$, while the best estimate of the correction to the NuTeV determination of $\sin^2\theta_W$, calculated using a NuTeV functional, is denoted by $\D R^i := \D^i \sin^2\theta_W$, where, in each case, $i$ labels the type of correction.
\subsubsection{Nuclear corrections}
%
For sufficiently large $Q^2$, nuclear corrections to the PW ratio for an isoscalar nucleus are thought to be negligible.  However, the NuTeV experiment was performed on a steel target and it is essential to correct for the neutron excess before extracting $\sin^2\theta_W$.  NuTeV removed the contribution of the excess neutrons to the cross-section by assuming that the target was composed of free nucleons.  However, results obtained using the NJL model described in the Secs.\,\ref{SecNJLJet} and \ref{SecEMC} have shown \cite{Cloet:2009qs} that the excess neutrons in iron have an effect on \textit{all} the nucleons in the nucleus, which is not accounted for by a subtraction of their na\"{\i}ve contribution.  In particular, the isovector-vector mean-field generated by the difference in proton and neutron densities, $\rho_p(r) - \rho_n(r)$, acts on every $u$- and $d$-quark in the nucleus and entails a breakdown of the usual assumption that $u_p(x) = d_n(x)$ and  $d_p(x) = u_n(x)$ for bound nucleons.

The correction associated with the neutron excess can be evaluated in terms of the consequent contribution to $\la x_A\,u^-_A - x_A\,d^-_A\ra$, using Eq.\,\eqref{eq:correction}.  For nuclei with $N > Z$ the $u$-quarks feel less vector repulsion than the $d$-quarks, and this has the model-independent consequence that there is a small shift in momentum from the $u$- to the $d$-quarks \cite{Cloet:2009qs}.  Therefore, the momentum fraction $\la x_A\,u^-_A - x_A\,d^-_A\ra$ in Eq.\,\eqref{eq:correction} is negative \cite{Cloet:2009qs}, even after standard isoscalarity corrections are applied.  Correcting for the isovector-vector field therefore has the model-independent effect of reducing the NuTeV result for $\sin^2\theta_W$.

To estimate the effect on the NuTeV experiment, Ref.\,\cite{Cloet:2009qs} used a nuclear matter approximation, chose the $Z/N$ ratio to correspond to the NuTeV experimental neutron excess and calculated the quark distributions at an effective density appropriate for Fe, namely 0.89 times nuclear matter density ($\rho_{\rm NM}$) \cite{Moniz:1971mt}.  Using Eq.\,(\ref{eq:correction}), this gave an estimate of the
isovector correction: $\D R_{\text{PW}}^{\raisebox{1.0pt}{${\scriptstyle\rho}$}^0} = -0.0025$.  Finally, the NuTeV CSV functional \cite{Zeller:2002du} was used to obtain an accurate determination of this effect on the NuTeV result.  This gave $\Delta R^{\raisebox{1.0pt}{${\scriptstyle\rho}$}^0} = -0.0019$, which accounts for between $1.0\,\sigma$-$1.5\,\sigma$ of the NuTeV discrepancy with the Standard Model.

The sign of this effect is model independent, and because it depends only on the difference in the neutron and proton densities in iron and the symmetry energy of nuclear matter, which are both well known, the magnitude is probably well constrained. As a conservative estimate of the uncertainty, Ref.\,\cite{Bentz:2009yy} assigned an error twice that of the difference between the PW correction obtained at $\rho_{\rm NM}$ and at $0.89 \rho_{\rm NM}$, which gives
\begin{equation}
\Delta R^{\raisebox{1.0pt}{${\scriptstyle\rho}$}^0} = -0.0019 \pm 0.0006.
\label{eq:rho_result}
\end{equation}

Other studies of nuclear corrections to the PW ratio have mainly focused on Fermi motion \cite{Kulagin:2003wz,Cloet:2009qs} and nuclear shadowing \cite{Kulagin:2003wz,Miller:2002xh,Brodsky:2004qa}.  Fermi motion corrections were found to be small \cite{Kulagin:2003wz,Cloet:2009qs} and the NuTeV collaboration argue that, given their $Q^2$-cuts, sizeable corrections from shadowing would be inconsistent with data \cite{McFarland:2002sk}.  Therefore, Ref.\,\cite{Bentz:2009yy} did not include a correction from shadowing.

\subsubsection{Charge symmetry violation}
%
Before the NuTeV result, two independent studies \cite{Sather:1991je,Rodionov:1994cg} of the effect of quark mass differences on proton and neutron PDFs reached very similar conclusions.  Such mass differences violate charge symmetry and lead to the CSV differences
\begin{equation}
\delta d^-(x) =  d^-_p(x) - u^-_n(x), \quad
\delta u^-(x) =  u^-_p(x) - d^-_n(x),
\end{equation}
where the subscripts $p$ and $n$ label the proton and neutron, respectively.  The contribution of CSV in the nucleon can be found through Eq.\,\eqref{eq:correction} and has the form:
\begin{equation}
\D R^{\text{CSV}}_{\text{PW}} =
\frac{1}{2}\left( 1 - \frac{7}{3} \sin^2 \theta_W \right)
\frac{\left\la x\,\delta u^- - x\,\delta d^- \right\ra}
{\left\la x\,u^-_p + x\,d^-_p \right\ra}.
\label{eq:CSV_correction}
\end{equation}

The similarity between Refs.\,\cite{Sather:1991je,Rodionov:1994cg} was explained \cite{Londergan:2003ij} by demonstrating that the leading contribution to the moment $\left\la x\,\delta u^- - x\,\delta d^- \right\ra$ is largely model-independent and simply involves the ratio of the $u$-$d$
mass difference to the nucleon mass.  The contribution from the quark mass differences to Eq.\,\eqref{eq:CSV_correction} was found to be $\D R^{\delta m}_{\text{PW}} \approx -0.0020$ and the corresponding NuTeV CSV functional result was $\D R^{\delta m} \simeq -0.0015$ \cite{Londergan:2003ij}.  Reference~\cite{Bentz:2009yy} assigned an error of 20\% to this term, which is conservative in view of its demonstrated model independence.

An additional CSV effect arises from QED splitting \cite{Gluck:2005xh,Martin:2004dh},
associated with the $Q^2$ evolution of photon emission from quarks.  Since $\left|e_u\right| > \left|e_d\right|$, the $u$-quarks lose momentum to the photon field at a greater rate than the $d$-quarks.  Therefore, a model-independent consequence of QED splitting is that it will reduce the NuTeV result for $\sin^2\theta_W$.  This effect on the NuTeV result has been calculated \cite{Gluck:2005xh}: $\Delta R^{\text{QED}}_{\text{PW}} = -0.002$, corresponding to $\Delta R^{\text{QED}} = -0.0011$ using the NuTeV CSV functional.  A similar study was undertaken in Ref.\,~\cite{Martin:2004dh}, which also explicitly included QED splitting effects in the PDF evolution, with the result $\Delta R^{\text{QED}}_{\text{PW}} = -0.0021$ at the scale $Q^2 = 20\,$GeV$^2$.  This
correction has the same sign as the CSV term arising from quark mass differences and the two contributions are almost independent, so Ref.\,\cite{Bentz:2009yy} simply added them.  The sum of the two terms explains roughly half of the NuTeV discrepancy with the Standard Model.  Assigning a conservative 100\% error to the QED splitting result and combining the errors in quadrature gives a total CSV correction of
\begin{equation}
\Delta R^{\text{CSV}} = -0.0026 \pm 0.0011.
\label{eq:CSV_result}
\end{equation}

The only empirical information regarding the effect of CSV on the PDFs is provided in Ref.\,\cite{Martin:2003sk}, which reports a global analysis of a set of high energy data that allows for explicit CSV in the PDFs.  It yields
$\Delta R^{\text{CSV}}_{\text{PW}} = -0.002$, with a 90\% confidence interval of $\Delta R^{\text{CSV}}_{\text{PW}} \in [-0.007,0.007]$.  This study implicitly included both sources of CSV described here; i.e., quark-mass and QED effects.  The 90\% confidence interval allows a rather large range of valence-quark CSV.

The authors of Ref.\,\cite{Bentz:2009yy} excluded this value and error for a number of reasons. First, Ref.\,\cite{Martin:2003sk} assumed a specific functional form for the CSV parton distributions; and the assumed function had an overall strength parameter that was varied in order to obtain the best fit in the global analysis.  Second, that study also imposed relations between the valence-quark CSV PDFs.  Third, as a matter of convenience in the global analysis, Ref.\,\cite{Martin:2003sk} neglected the $Q^2$ dependence of the CSV distributions.  Finally, the experiments in the global data set used different treatments of radiative corrections; and it is uncertain whether these differences were accommodated consistently in the analysis of CSV effects.  If the CSV effects are as large as allowed within the 90\% confidence limit of Ref.\,\cite{Martin:2003sk}, then it should be possible to observe such effects \cite{Londergan:2009kj}.  However it will be some time before experiments can further constrain this result.

\subsubsection{Strange quark asymmetry}
%
Arguments based on chiral symmetry and commonsense \cite{Signal:1987gz,Brodsky:1996hc,Thomas:2000ny} make it plausible that the $s(x)$ and $\bar{s}(x)$ distributions with the nucleon differ in shape .  However, the size of $s^-(x)$ is not so readily constrained and further input from experiment or numerical simulations of lattice-regularised QCD \cite{Deka:2008xr} would be helpful. The strange quark correction arises from the term $\left\la x\,s_A^- \right\ra$
on the rhs of Eq.\,\eqref{eq:correction}.  The best direct experimental information on $\left\la x\,s_A^- \right\ra$ comes from opposite-sign dimuon production in reactions induced by $\nu$ or $\bar\nu$.  Such experiments have been performed by the CCFR \cite{Bazarko:1994tt} and NuTeV~\cite{Goncharov:2001qe} groups.  A precise extraction of $\left\la x\,s_A^- \right\ra$ is reported in Ref.\,\cite{Mason:2007zz} (NuTeV): $\left\la x\,s_A^- \right\ra = 0.00196 \pm 0.00143$ at $Q^2=16\,$GeV$^2$, where the various errors have been added in quadrature.

\begin{table*}[t]
\addtolength{\extrarowheight}{5.0pt}
\begin{center}
\begin{tabular}{llll}\\\hline
& Bentz \emph{et al}.\,\cite{Bentz:2009yy}
& Mason \emph{et al}.\,\cite{Mason:2007zz}
& NNPDF \cite{Ball:2009mk} \\
\hline
$\la x\,s^-\ra$
& \phantom{$-$}$0.0 \pm 0.0020$
& \phantom{$-$}$0.00196 \pm 0.00143$
& \phantom{$-$}$0.0005 \pm 0.0086$  \\
$\Delta R^s$
& \phantom{$-$}$0.0 \pm 0.0018$ & $-0.0018 \pm 0.0013$ & $-0.0005 \pm 0.0078$ \\
$\Delta R^{\text{total}}$
& $-0.0045 \pm 0.0022$ & $-0.0063 \pm 0.0018$
& $-0.0050 \pm 0.0079$  \\
$\sin^2\theta_W$ $\pm$ syst.
& \phantom{$-$}$0.2232 \pm 0.0024$ & \phantom{$-$}$0.2214 \pm 0.0020$ & \phantom{$-$}$0.2227 \pm {\rm large}$  \\\hline
& MSTW \cite{Martin:2009iq}
& CTEQ~\cite{Lai:2007dq}
& Alekhin \emph{et al}.\,\cite{Alekhin:2008mb} \\\hline
$\la x\,s^-\ra$
& \phantom{$-$}$0.0016_{-0.0009}^{+0.0011}$
& \phantom{$-$}$0.0018_{-0.0004}^{+0.0016}$
& \phantom{$-$}$0.0013 \pm 0.0009 \pm 0.0002$ \\
$\Delta R^s$
& $-0.0014_{+0.0008}^{-0.0010}$
& $-0.0016_{-0.0004}^{+0.0014}$
& $-0.0012 \pm 0.0008 \pm 0.0002$ \\
$\Delta R^{\text{total}}$
& $-0.0059 \pm 0.0015$
& $-0.0061_{-0.0013}^{+0.0019}$
& $-0.0057 \pm 0.0015$ \\
$\sin^2\theta_W$ $\pm$ syst.
& \phantom{$-$}$0.2218\pm 0.0018$
& \phantom{$-$}$0.2216_{-0.0016}^{+0.0021}$
& \phantom{$-$}$0.2220 \pm 0.0017$ \\
\hline
\end{tabular}
\end{center}
\caption{Representative collection of contemporary estimates for:
strangeness asymmetry, $\langle x s^- \rangle$;
correction to the Paschos-Wolfenstein ratio after applying the NuTeV functional, $\Delta R^s$;
and total correction, $\Delta R^{\text{total}}$, obtained by combining $\Delta R^{\rho^0}$, $\Delta R^{\text{CSV}}$ and $\Delta R^s$, with errors added in quadrature.
The final column shows the value of $\sin^2 \theta_W$ deduced in each case by applying the total correction to the published NuTeV result.  N.B.\ Only the systematic error is shown, which is obtained by treating the error on $\Delta R^{\text{total}}$ as a systematic error and combining it in quadrature with the NuTeV systematic error.
\label{tab:strange}}
\end{table*}

Global PDF analyses have also provided estimates of $s^-(x)$.  An examination by the NNPDF collaboration found $\left\la x\,s^- \right\ra = 0.0005\pm 0.0086$ \cite{Ball:2009mk} at $Q^2=20\,$GeV$^2$, which has an error more than six times larger than that cited by NuTeV.  Unlike the NuTeV dimuon experiment, however, this analysis is not directly sensitive to the $s$-quark distributions, something evident in the fact that their upper limit on $s^-(x)$ is an order of magnitude larger than any other sea-quark distribution at $x \sim 0.5$.  The large uncertainty is a consequence of the neural network approach, which was primarily aimed at accurately determining $V_{cd}$ and $V_{cs}$, not $s^-(x)$ \cite{Ball:2009mk}.  The result $\la x\,s^- \ra = 0.0013 \pm 0.0009(\text{exp}) \pm 0.0002(\text{QCD})$ was obtained in Ref.\,\cite{Alekhin:2008mb}, which imposed a constraint on the semileptonic branching ratio $B_{\mu}$ from production rates of charmed hadrons in other experiments.  The MSTW collaboration find a momentum fraction very similar to that of NuTeV; namely, $\la x\,s^- \ra = 0.0016^{+0.0011}_{-0.0009}$ \cite{Martin:2009iq} at $Q^2=10\,$GeV$^2$, while CTEQ reported $\left\la x\,s^- \right\ra = 0.0018$ \cite{Lai:2007dq} at $Q^2=1.69\,$GeV$^2$, with a 90\% confidence interval of
$\left\la x\,s^- \right\ra \in [-0.001,0.005]$.  These values are collected in Table~\ref{tab:strange}.

The distribution $s_A^-(x)$ must have at least one zero-crossing because its first
moment vanishes; and for each of the analyses above the central best-fit curve crosses zero on $x < 0.03$ for $Q^2 > 2\,$GeV$^2$, with the exception of the NNPDF result, which has a zero-crossing at $x = 0.13$ for $Q^2 = 2\,$GeV$^2$.  The zero in the NuTeV result is at the particularly low value of $x=0.004$.  This is smaller than the lowest $x$ point measured in the CCFR and NuTeV experiments and, moreover, is extremely unlikely on theoretical grounds \cite{Signal:1987gz,Melnitchouk:1999mv,Ding:2004ht}: in any quark model calculation the zero-crossing will occur at $x \approx 0.15$.  Enforcing this on the NuTeV analysis, their strange quark momentum fraction becomes
$\left\la x\,s_A^- \right\ra = 0.00007$ \cite{Mason:2007zz}, with a
moderate increase in the $\chi^2$ compared to the best-fit value.  

Given the significant uncertainty in the strange quark asymmetry correction to the PW ratio and the theoretical prejudices indicated above, Ref.\,\cite{Bentz:2009yy} elected to use the NuTeV result based on a zero-crossing at $x \approx 0.15$, so that $\left\la x\,s^- \right\ra$ is essentially zero.  For the uncertainty, Ref.\,\cite{Bentz:2009yy} chose the difference between this and the NuTeV determination noted above with the zero-crossing at $x = 0.004$, which gives
$\la x\,s^- \ra =  0.0 \pm 0.0020$ at 16 GeV$^2$.  This choice for the error is substantially larger than the original uncertainty quoted by NuTeV and covers all the central values of the analyses mentioned above.  Including the effect of the NuTeV functional leads to
\begin{align}
\Delta R^s = 0.0 \pm 0.0018.
\label{eq:strange_result}
\end{align}
Table~\ref{tab:strange} collects the $s$-quark corrections to the NuTeV result obtained from the other analyses described above.

\begin{figure}[t]
\centering\includegraphics[width=0.6\textwidth,clip=true]{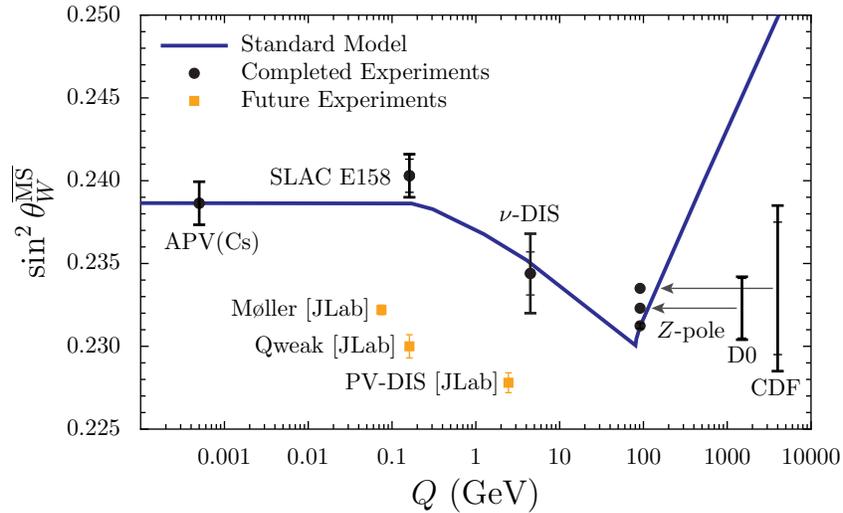}
\caption{Running of $\sin^2\theta_W$ in the $\overline{\text{MS}}$
renormalisation scheme \protect\cite{Erler:2004in}.
The $Z$-pole point represents the combined results of six LEP and
SLC experiments \protect\cite{ALEPH:2005ab}.
The CDF \cite{Acosta:2004wq} and D0 \cite{Abazov:2008xq} collaboration results
(at the $Z$-pole), and the SLAC E158 \cite{Anthony:2005pm} result are labelled accordingly.
The atomic parity violating (APV) result \cite{Porsev:2009pr} has been shifted from $Q^2\to 0$ for clarity.
The inner error bars represent the statistical uncertainty and the outer error bars show the total uncertainty.
At the $Z$-pole, conversion to the $\overline{\text{MS}}$ scheme was achieved via
$\sin^2\theta_W^{\text{eff}} = 0.00029 + \sin^2\theta_W^{\overline{\text{MS}}}$ \cite{ALEPH:2005ab}.  For the results away from the $Z$-pole, the discrepancy with the curve reflects disagreement with the Standard Model in the renormalisation scheme used in the experimental analysis.
\label{fig:sin2tw}}
\end{figure}

\subsection{Problem resolved?}
%
If it is assumed that the errors associated with the three corrections given in Eqs.\,\eqref{eq:rho_result}, \eqref{eq:CSV_result} and \eqref{eq:strange_result} are systematic and independent, then they may be combined in quadrature with the original systematic error quoted by NuTeV.  Naturally, the statistical error is unchanged from the NuTeV analysis.  The impact on $\sin^2 \theta_W$ from the uncertainty in the strangeness asymmetry is indicated by the entries in the last row of each block in Table~\ref{tab:strange}: plainly, every one of the six results lies within one standard deviation of the Standard Model value for $\sin^2\theta_W$.  Hence, as a best estimate of the corrected value, Ref.\,\cite{Bentz:2009yy} chose the average of these six entries.  Regarding the systematic error, apart from that of the NNPDF collaboration, which one may argue is unrealistically large, all are very
similar.  Owing to the correlations between them, Ref.\,\cite{Bentz:2009yy} chose to quote a final systematic error obtained as a simple average of all entries in the last row of each block in Table~\ref{tab:strange}, except NNPDF.  This yields the final result reported in Eq.\,\eqref{TWBentz}, which is in excellent agreement with the Standard Model expectation of $\sin^2 \theta_W = 0.2227 \pm 0.0004$ \cite{Abbaneo:2001ix,Zeller:2001hh}.  Correction terms of higher order than Eq.\,\eqref{eq:correction} and also $\mathcal{O}(\alpha_s)$ corrections were found to be negligible \cite{Bentz:2009yy}.

This updated value for the NuTeV determination of $\sin^2 \theta_W$ is shown in Fig.\,\ref{fig:sin2tw}, labelled as $\nu$-DIS and using the $\overline{\text{MS}}$-scheme in this case.  Also displayed are results from a number of other completed experiments and the anticipated errors in several future experiments, all marked at the appropriate momentum scale $Q$.

This section has summarised various estimates of the size of both partonic CSV effects and a possible strange-quark momentum asymmetry.  We explained how theoretical arguments have been used to constrain the magnitude of valence-quark CSV effects arising from quark mass differences, and theoretical guidance on the zero crossing in $s_A^-(x)$ and contemporary analyses of experimental data have been used to bound the strange-quark momentum asymmetry.  This information combined has enabled a re-evaluation of the NuTeV value of the weak mixing angle \cite{Bentz:2009yy} and the revised result is within one standard deviation of the Standard Model prediction.  Quite plausibly, therefore, \emph{there is no NuTeV anomaly}.

\section{Epilogue}
The last five years have seen dramatic theoretical advances in many areas and, in particular, in the formulation of the bound-state problem in continuum quantum field theory; the associated methods of analysis; and the range of applications.  We have chosen to highlight a subset of this progress.  Namely, that within the scope of QCD's Dyson-Schwinger equations (DSEs) and the physics of cold, sparse hadrons, with applications that range from exploring the nature of confinement and explaining dynamical chiral symmetry breaking; through elucidation of a new paradigm that has condensates contained within hadrons; and, crucially, onto the reliable computation, explanation and prediction of quantities that are truly measurable.\footnote{Advances in related studies at nonzero temperature and baryon chemical potential are not described herein.  That topic deserves a separate discussion, of which the material in Sec.\,11 of Ref.\,\protect\cite{Bashir:2012fs} would be a part, as would a large subset of that body of recent articles which cite Ref.\,\protect\cite{Roberts:2000aa}.}

In the foreseeable future, this subfield of hadron physics theory will see construction and use of vastly improved kernels for the hadron bound-state problem, with the expression of numerous effects driven by dynamical chiral symmetry breaking, such as running quark masses and resonant (``meson cloud'') contributions; and use in the baryon sector of the novel analytical methods described herein that have provided a solution to the problem of continuing from Euclidean metric to Minkowski space.  This will enable applications to the increasingly wide variety of phenomena that upgraded and new facilities will make accessible, so that realistic DSE kernels can be used to make predictions with a direct connection to QCD.  For example, 
existing computations of valence-quark PDFs \cite{Nguyen:2011jy} will be augmented by predictions for sea-quark distributions in hadrons; 
statements about unpolarised and polarised parton distribution functions (PDFs) on the far valence domain \cite{Roberts:2013mja} will be extended to values of Bjorken-$x$ that are more readily accessible empirically; 
the exploratory NJL-model computation of transverse momentum dependent PDFs (TMDs) \cite{Matevosyan:2011vj} will be replaced by an extensive, unified analysis of generalised parton distributions (GPDs) and TMDs via the improved DSE kernels; 
and predictions with realistic kernels for the spectrum of excited and exotic mesons, nucleon elastic and nucleon-to-resonance transition form factors, and the timelike behaviour of hadron form factors should become possible using methods like those described in Ref.\,\cite{Qin:2013aaa}.

There is now room for optimism.  With experiment-driven opportunities expanding rapidly and material improvements in the theoretical tools available to analyse, explain and guide them, it appears possible that the next five years will bring profound growth in our store of knowledge about hadrons in general, nucleons in particular, and nuclei, the latter two of which form the basic constituents of almost all the visible matter in the universe.  There is enormous potential for hadron physics to contribute substantially in completing our understanding of the Standard Model, and to constrain and guide the form of its extension. 

\appendix

\section{Euclidean metric}
\label{sec:Euclidean}
In our Euclidean formulation:
\begin{equation}
p\cdot q=\sum_{i=1}^4 p_i q_i\,;
\end{equation}
\begin{equation}
\{\gamma_\mu,\gamma_\nu\}=2\,\delta_{\mu\nu}\,;\;
\gamma_\mu^\dagger = \gamma_\mu\,;\;
\sigma_{\mu\nu}= \frac{i}{2}[\gamma_\mu,\gamma_\nu]\,; \;
{\rm tr}\,[\gamma_5\gamma_\mu\gamma_\nu\gamma_\rho\gamma_\sigma]=
-4\,\epsilon_{\mu\nu\rho\sigma}\,, \epsilon_{1234}= 1\,,
\end{equation}
and $Q^2>0$ defines a spacelike four-vector $Q$.

A positive energy spinor satisfies
\begin{equation}
\bar u(P,s)\, (i \gamma\cdot P + M) = 0 = (i\gamma\cdot P + M)\, u(P,s)\,,
\end{equation}
where $s=\pm$ is the spin label.  It is normalised:
\begin{equation}
\bar u(P,s) \, u(P,s) = 2 M \,,
\end{equation}
and may be expressed explicitly:
\begin{equation}
u(P,s) = \sqrt{M- i {\cal E}}
\left(
\begin{array}{l}
\chi_s\\
\displaystyle \frac{\vec{\sigma}\cdot \vec{P}}{M - i {\cal E}} \chi_s
\end{array}
\right)\,,
\end{equation}
with ${\cal E} = i \sqrt{\vec{P}^2 + M^2}$,
\begin{equation}
\chi_+ = \left( \begin{array}{c} 1 \\ 0  \end{array}\right)\,,\;
\chi_- = \left( \begin{array}{c} 0\\ 1  \end{array}\right)\,.
\end{equation}
For the free-particle spinor, $\bar u(P,s)= u(P,s)^\dagger \gamma_4$.

The spinor can be used to construct a positive energy projection operator:
\begin{equation}
\label{Lplus} \Lambda_+(P):= \frac{1}{2 M}\,\sum_{s=\pm} \, u(P,s) \, \bar
u(P,s) = \frac{1}{2M} \left( -i \gamma\cdot P + M\right).
\end{equation}

A negative energy spinor satisfies
\begin{equation}
\bar v(P,s)\,(i\gamma\cdot P - M) = 0 = (i\gamma\cdot P - M) \, v(P,s)\,,
\end{equation}
and possesses properties and satisfies constraints obtained via obvious analogy
with $u(P,s)$.

A charge-conjugated Bethe-Salpeter amplitude is obtained via
\begin{equation}
\label{chargec}
\bar\Gamma(k;P) = C^\dagger \, \Gamma(-k;P)^{\rm T}\,C\,,
\end{equation}
where ``T'' denotes a transposing of all matrix indices and
$C=\gamma_2\gamma_4$ is the charge conjugation matrix, $C^\dagger=-C$.  Note that
\begin{equation}
C^\dagger \gamma_\mu^{\rm T} \, C = - \gamma_\mu\,, \; [C,\gamma_5] = 0\,.
\end{equation}

In describing decuplet resonances, a Rarita-Schwinger spinor is used to represent a covariant spin-$3/2$ field.  The positive energy spinor is defined by the following equations:
\begin{equation}
\label{rarita}
(i \gamma\cdot P + M)\, u_\mu(P;r) = 0\,,\;
\gamma_\mu u_\mu(P;r) = 0\,,\;
P_\mu u_\mu(P;r) = 0\,,
\end{equation}
where $r=-3/2,-1/2,1/2,3/2$.  It is normalised:
\begin{equation}
\bar u_{\mu}(P;r^\prime) \, u_\mu(P;r) = 2 M\,,
\end{equation}
and satisfies a completeness relation
\begin{equation}
\label{Deltacomplete}
\frac{1}{2 M}\sum_{r=-3/2}^{3/2} u_\mu(P;r)\,\bar u_\nu(P;r) =
\Lambda_+(P)\,R_{\mu\nu}\,,
\end{equation}
where
\begin{equation}
R_{\mu\nu} = \delta_{\mu\nu} \mbox{\boldmath $I$}_{\rm D} -\frac{1}{3} \gamma_\mu \gamma_\nu +
\frac{2}{3} \check P_\mu \check P_\nu \mbox{\boldmath $I$}_{\rm D} - i\frac{1}{3} [ \check P_\mu
\gamma_\nu - \check P_\nu \gamma_\mu]\,,
\end{equation}
with $\check P^2 = -1$, which is very useful in simplifying the Faddeev equation for a positive energy decuplet state.

\section{Nakanishi-like representations}
\label{app:nakanishi}
Here we summarise the interpolations used in Ref.\,\cite{Chang:2013pq} for the evaluation of the moments in Eq.\,\eqref{phimom}.  The dressed-quark propagator is represented as \cite{Bhagwat:2002tx}
\begin{equation}
S(p) = \sum_{j=1}^{j_m}\bigg[ \frac{z_j}{i \gamma\cdot p + m_j}+\frac{z_j^\ast}{i \gamma \cdot p + m_j^\ast}\bigg], \label{Spfit}
\end{equation}
with $\Im m_j \neq 0$ $\forall j$, so that $\sigma_{V,S}$ are meromorphic functions with no poles on the real $p^2$-axis, a feature consistent with confinement \cite{Bashir:2012fs}.  An adequate interpolation is obtained with $j_m=2$.

With relative momentum defined via $\eta=1/2$, the scalar functions in Eq.\,\eqref{genGpi} $({\cal F}=E,F,G)$ were represented by
{\allowdisplaybreaks
\begin{eqnarray}
{\cal F}(k;P) &=& {\cal F}^{\rm i}(k;P) + {\cal F}^{\rm u}(k;P) \,, \\
{\cal F}^{\rm i}(k;P) & = & c_{\cal F}^{\rm i}\int_{-1}^1 \! dz \, \rho_{\nu^{\rm i}_{\cal F}}(z) \bigg[
a_{\cal F} \hat\Delta_{\Lambda^{\rm i}_{{\cal F}}}^4(k_z^2)
+ a^-_{\cal F} \hat\Delta_{\Lambda^{\rm i}_{\cal F}}^5(k_z^2)
\bigg], \label{Fifit}\\
E^{\rm u}(k;P) & = & c_{E}^{\rm u} \int_{-1}^1 \! dz \, \rho_{\nu^{\rm u}_E}(z)\,
 \hat \Delta_{\Lambda^{\rm u}_{E}}(k_z^2)\,,\\
F^{\rm u}(k;P) & = & c_{F}^{\rm u} \int_{-1}^1 \! dz \, \rho_{\nu^{\rm u}_F}(z)\,
 \Lambda_F^{\rm u} k^2 \Delta_{\Lambda^{\rm u}_{F}}^2(k_z^2)\,,\\
G^{\rm u}(k;P) & = & c_{G}^{\rm u} \int_{-1}^1 \! dz \, \rho_{\nu^{\rm u}_G}(z)\,
 \Lambda_G^{\rm u}\Delta_{\Lambda^{\rm u}_{G}}^2(k_z^2)\,, \label{Gufit}
\end{eqnarray}}
\hspace*{-0.5\parindent}with $\hat \Delta_\Lambda(s) = \Lambda^2 \Delta_\Lambda(s)$, $k_z^2=k^2+z k\cdot P$, $a^-_E = 1 - a_E$, $a^-_F = 1/\Lambda_F^{\rm i} - a_F$, $a^-_G = 1/[\Lambda_G^{\rm i}]^3 - a_G$.  $H(k;P)$ is small, has little impact, and was thus neglected.

Values of the interpolation parameters that fit the numerical results are presented in Tables~\ref{Table:parameters}.  Those for the Bethe-Salpeter amplitudes were obtained through a least-squares fit to the Chebyshev moments
\begin{equation}
{\cal F}_n(k^2) = \frac{2}{\pi}\int_{-1}^{1}\!dx\, \sqrt{1-x^2} {\cal F}(k;P) U_n(x)\,,
\end{equation}
with $n=0,2$, where $U_n(x)$ is an order-$n$ Chebyshev polynomial of the second kind. Owing to $O(4)$ invariance, one may define $x = \hat k\cdot P/ip$, with $\hat k^2=1$ and $P=(0,0,p,i p)$.

\begin{table}[t]
\caption{Representation parameters. \emph{Upper panel}: Eq.\,\protect\eqref{Spfit} -- the pair $(x,y)$ represents the complex number $x+ i y$.  \emph{Lower panel}: Eqs.\,\protect\eqref{Fifit}--\protect\eqref{Gufit}.  (Dimensioned quantities in GeV).
\label{Table:parameters}
}
\begin{center}
%

\begin{tabular*}
{\hsize}
{
l|@{\extracolsep{0ptplus1fil}}
c@{\extracolsep{0ptplus1fil}}
c@{\extracolsep{0ptplus1fil}}
c@{\extracolsep{0ptplus1fil}}
c@{\extracolsep{0ptplus1fil}}
c@{\extracolsep{0ptplus1fil}}}\hline
 RL & $z_1$ & $m_1$  & $z_s$ & $m_2$ \\

    & $(0.44,0.014)$ & $(0.54,0.23)$ & $(0.19,0)$ & $(-1.21,-0.65)$ \\
 DB & $z_1$ & $m_1$  & $z_s$ & $m_2$ \\
    & $(0.44,0.28)$ & $(0.46,0.18)$ & $(0.12,0)$ & $(-1.31,-0.75)$ \\\hline
\end{tabular*}

\begin{tabular*}
{\hsize}
{
l@{\extracolsep{0ptplus1fil}}
l@{\extracolsep{0ptplus1fil}}
c@{\extracolsep{0ptplus1fil}}
c@{\extracolsep{0ptplus1fil}}
c@{\extracolsep{0ptplus1fil}}
c@{\extracolsep{0ptplus1fil}}
c@{\extracolsep{0ptplus1fil}}
c@{\extracolsep{0ptplus1fil}}
c@{\extracolsep{0ptplus1fil}}
c@{\extracolsep{0ptplus1fil}}}\hline
    & & $c^{\rm i}$ & $c^{u}$ & $\phantom{-}\nu^{\rm i}$ & $\nu^{\rm u}$ & $a$\phantom{00} & $\Lambda^{\rm i}$ & $\Lambda^{\rm u}$\\\hline
RL: & E & $1 - c^{u}_E$ & $0.03$ & $-0.71$ & 1.08
    & 2.75\phantom{$/[\Lambda^{\rm i}_G]^3$} & 1.32 & 1.0\\
    & F & 0.51 & $c^{\rm u}_E/10$ & $\phantom{-}0.96$ & 0.0
    & 2.78$/\Lambda^{\rm i}_{F}$\phantom{00} & 1.09 & 1.0 \\
& G & $0.18$ & 2$\,c^{\rm u}_F$ & $\phantom{-}\nu^{\rm i}_F$ & 0.0 & 5.73$/[\Lambda^{\rm i}_G]^3$ & 0.94 & 1.0 \\\hline
DB: & E & $1 - c^{u}_E$ & $0.08$ & $-0.70$ & 1.08
    & 3.0\phantom{$/[\Lambda^{\rm i}_G]^3$} & 1.41 & 1.0\\
    & F & \phantom{-}0.55 & $c^{\rm u}_E/10$ & $\phantom{-}0.40$ & 0.0
    & 3.0$/\Lambda^{\rm i}_{F}$\phantom{00} & 1.13 & 1.0 \\
& G & $-0.094$ & 2$\,c^{\rm u}_F$ & $\phantom{-}\nu^{\rm i}_F$ & 0.0 & 1.0$/[\Lambda^{\rm i}_G]^3$ & 0.79 & 1.0 \\\hline
\end{tabular*}
\end{center}

\vspace*{-4ex}

\end{table}

The strength of the interaction detailed in Ref.\,\cite{Qin:2011dd} is specified by a product: $D\omega = M_g^3$.  With $M_g$ fixed, results for properties of ground-state vector and flavour-nonsinglet pseudoscalar mesons are independent of the value of $\omega \in [0.4,0.6]\,$GeV.  One typically uses $\omega =0.5\,$GeV.
With the RL kernel, $f_\pi=0.092\,$GeV is obtained with $M_g^{\rm RL}(2\,GeV)=0.87\,$GeV and $M_g^{\rm RL}(19\,GeV)=0.80\,$GeV, whilst with the DB kernel it is obtained with $M_g^{\rm DB}(2\,GeV)= M_g^{\rm DB}(19\,GeV)=0.55\,$GeV [cf.\ Eq.\,\eqref{fitalpha} and the following text].
Plainly, multiplicative renormalisability is better preserved with the DB kernels.
In Eq.\,(10) of Ref.\,\cite{Chang:2011ei}, the strength of the dressed-quark anomalous chromomagnetic moment was described by a value $\tilde\eta=0.65$.  To improve numerical stability in the interpolations described herein, Ref.\,\cite{Chang:2013pq} changed to $\tilde\eta=0.6$.  This increases the computed value of the $a_1$-$\rho$ mass-splitting by less-than $15$\%.

It is worth including another detail associated with the generalised spectral representations.  DSE kernels that preserve the one-loop renormalisation group behaviour of QCD will necessarily generate propagators and Bethe-Salpeter amplitudes with a nonzero anomalous dimension $\gamma_F$, where $F$ labels the object concerned.  Consequently, the spectral representation must be capable of describing functions of $\mathpzc{s}=p^2/\Lambda_{\rm QCD}^2$ that exhibit $\ln^{-\gamma_F}[\mathpzc{s}]$ behaviour for $\mathpzc{s}\gg 1$.  This is readily achieved by noting that
\begin{equation}
\label{logfactor}
\ln^{-\gamma_F} [D(\mathpzc{s})]
= \frac{1}{\Gamma(\gamma_F)} \int_0^\infty \! dz\, z^{\gamma_F-1}
\frac{1}{[D(\mathpzc{s})]^z}\,,
\end{equation}
where $D(\mathpzc{s})$ is some function.  Such a factor can be multiplied into any existing spectral representation in order to achieve the required ultraviolet behaviour.

In connection with the light-front distribution of the chiral condensate, it is the anomalous dimension of the dressed-quark mass-function that must properly be represented.  Owing to Eq.\,\eqref{gtlrelE}, this affects the pion's Bethe-Salpeter amplitude, too.  On the other hand, for practical applications involving convergent four momentum integrals, like those generated by Eq.\,\eqref{RLFpi},
it is adequate to develop and use a power law approximation; viz., $\ln^{\gamma_F} [D(\mathpzc{s})] \approx [D(\mathpzc{s})]^{\mathpzc{p}_F}$.  With ${\mathpzc{p}_F}$ chosen appropriately, this is accurate on the material domain and greatly simplifies the subsequent numerical calculation.

\section{Vector$\,\otimes\,$vector contact interaction}
\label{secContact}
\subsection{Contact gap equation}
Motivated by the successful hadron physics phenomenology reviewed briefly in Ref.\,\cite{Bentz:2007zs}, a body of work has been undertaken \cite{Chen:2012qr,Wilson:2011aa,GutierrezGuerrero:2010md,Roberts:2010rn,Roberts:2011wy,%
Chen:2012txa,Roberts:2011cf,Wang:2013wk,Segovia:2013uga} with the goal of elucidating those circumstances under which a confining, symmetry-preserving treatment of a vector$\,\otimes\,$vec\-tor contact interaction can serve usefully as a surrogate for more sophisticated and difficult-to-handle momentum-dependent DSE kernels.  To the surprise of some, the range of circumstances is quite large and includes meson and baryon spectra, and their electroweak elastic and transition form factors.  It is apposite to remark that the treatment of the interaction in those studies produces form factors which are typically too hard but, when interpreted carefully, they can nevertheless be used to draw valuable insights.  The simplicity of the interaction and its capacity to provide a unified explanation of a diverse array of phenomena, many of which have been unreachable with more sophisticated DSE kernels owing to weaknesses in the numerical algorithms that have until recently been employed (see Secs.\,\ref{SecPLFWF} and \ref{sec:pionFF}), are features that continue to supply grounds for its further application.  Here we provide just a little useful background.

For a given flavour of quark, associated with a current-quark mass $m_f$, the contact-interaction dressed-quark propagator is obtained from the gap equation
\begin{equation}
 S_f^{-1}(p) =  i \gamma \cdot p + m_f +  \frac{16\pi}{3}\frac{\alpha_{\rm IR}}{m_G^2} \int\!\frac{d^4 q}{(2\pi)^4} \,
\gamma_{\mu} \, S_f(q) \, \gamma_{\mu}\,.
\label{gap-1}
\end{equation}
In order to arrive at this expression from the general form of the gap equation, Eq.\,\eqref{gendseN}, one writes
\begin{equation}
g^{2} D_{\mu\nu}(p-q) = \delta_{\mu\nu} \frac{4\pi\alpha_{\rm IR}}{m_{\rm G}^{2}},
\label{eq:qqInteraction}
\end{equation}
where $m_G=0.8\,$GeV is a gluon mass-scale typical of the one-loop renormalisation-group-improved interaction detailed in Ref.\,\cite{Qin:2011dd}, and the fitted parameter $\alpha_{\rm IR} = 0.93 \pi$ is commensurate with contemporary estimates of the zero-momentum value of a running-coupling in QCD \cite{Aguilar:2009nf,Oliveira:2010xc,Aguilar:2010gm,Boucaud:2010gr,Pennington:2011xs,%
Wilson:2012em}.  As part of the process involved in setting up a symmetry preserving regularisation, Eq.\,\eqref{eq:qqInteraction} is embedded in a rainbow-ladder truncation of the DSEs, which is the leading-order in the most widely used, global-symmetry-preserving truncation scheme \cite{Munczek:1994zz,Bender:1996bb}.  This means that one uses
\begin{equation}
\Gamma^{a}_{\nu}(q,p)=\frac{\lambda^{a}}{2} \gamma_{\nu}
\label{eq:VertexRL}
\end{equation}
in the gap equation and also in the subsequent construction of all Bethe-Salpeter kernels.

Equation~\eqref{gap-1} possesses a quadratic divergence, even in the chiral limit.  When the divergence is regularised in a Poincar\'e covariant manner, the solution is
\begin{equation}
\label{genS}
S_f(p)^{-1} = i \gamma\cdot p + M_f\,,
\end{equation}
where $M_f$ is momentum-independent and determined by
\begin{equation}
M_f = m_f + M_f\frac{4\alpha_{\rm IR}}{3\pi m_G^2} \int_0^\infty \!ds \, s\, \frac{1}{s+M_f^2}\,.
\label{eq:MGAP}
\end{equation}

A confining regularisation procedure is suggested by Ref.\,\cite{Ebert:1996vx}; i.e., one writes
\begin{eqnarray}
\frac{1}{s+M^2} & = & \int_0^\infty d\tau\,{\rm e}^{-\tau (s+M^2)}  \rightarrow  \int_{\tau_{\rm uv}^2}^{\tau_{\rm ir}^2} d\tau\,{\rm e}^{-\tau (s+M^2)}
%
 =
\frac{{\rm e}^{- (s+M^2)\tau_{\rm uv}^2}-{\rm e}^{-(s+M^2) \tau_{\rm ir}^2}}{s+M^2} \,, \label{eq:Regularization}
\end{eqnarray}
where $\tau_{\rm ir,uv}$ are, respectively, infrared and ultraviolet regulators.  It is evident from the rightmost expression in Eq.\,(\ref{eq:Regularization}) that a finite value of $\tau_{\rm ir}=:1/\Lambda_{\rm ir}$ implements confinement by ensuring the absence of quark production thresholds \cite{Chang:2011vu,Roberts:2012sv}.  Since Eq.\,(\ref{eq:qqInteraction}) does not define a renormalisable theory, then $\Lambda_{\rm uv}:=1/\tau_{\rm uv}$ cannot be removed but instead plays a dynamical role, setting the scale of all dimensioned quantities.  Using Eq.\,\eqref{eq:Regularization}, the gap equation becomes
\begin{equation}
M = m + M \frac{4\alpha_{\rm IR}}{3\pi m_{\rm G}^{2}} \, {\cal C}^{\rm iu}(M^{2}),
\label{eq:SolutionGapEquation}
\end{equation}
where ${\cal C}^{\rm iu}(\sigma)/\sigma = \overline{\cal C}^{\rm iu}(\sigma) = \Gamma(-1,\sigma \tau_{\rm uv}^2) - \Gamma(-1,\sigma \tau_{\rm ir}^2)$, with $\Gamma(\alpha,y)$ being the incomplete gamma-function.

\begin{table}[t]
\caption{\label{tab:CQM}
Computed dressed-quark properties, required as input for the Bethe-Salpeter and Faddeev equations, and computed values for in-hadron condensates \protect\cite{Brodsky:2010xf,Chang:2011mu,Brodsky:2012ku}.  All results obtained with $\alpha_{\rm IR} =0.93 \pi$ and (in GeV) $\Lambda_{\rm ir} = 0.24\,$, $\Lambda_{\rm uv}=0.905$.  N.B.\ These parameters take the values determined in the spectrum calculation of Ref.\,\protect\cite{Roberts:2011cf}, which produces $m_\rho=0.928\,$GeV.  Isospin symmetry was assumed.
(All dimensioned quantities are listed in GeV.)}
\begin{center}
\begin{tabular*}
{\hsize}
{
c@{\extracolsep{0ptplus1fil}}
c@{\extracolsep{0ptplus1fil}}
c@{\extracolsep{0ptplus1fil}}
c@{\extracolsep{0ptplus1fil}}
c@{\extracolsep{0ptplus1fil}}
c@{\extracolsep{0ptplus1fil}}
c@{\extracolsep{0ptplus1fil}}
c@{\extracolsep{0ptplus1fil}}
c@{\extracolsep{0ptplus1fil}}
c@{\extracolsep{0ptplus1fil}}}\hline
$m_u$ & $m_s$ & $m_s/m_u$ & $M_0$ &   $M_u$ & $M_s$ & $M_s/M_u$ & $\chi_0$ & $\chi_\pi$ & $\chi_K$ \\\hline
0.007  & 0.17 & 24.3 & 0.36 & 0.37 & 0.53 & 1.43  & 0.241 & 0.243 & 0.246
\\\hline
\end{tabular*}
\end{center}
\end{table}

Table~\ref{tab:CQM} reports values of $u=d$- and $s$-quark properties, computed from Eq.\,\eqref{eq:SolutionGapEquation}, that are used in all calculations that follow Ref.\,\cite{Roberts:2011wy}: the input ratio $m_s/\bar m$, where $\bar m = (m_u+m_d)/2$, is consistent with contemporary estimates \cite{Leutwyler:2009jg}.
N.B.\ It is a feature of Eq.\,\eqref{eq:SolutionGapEquation} that in the chiral limit, $m_f=m_0=0$, a nonzero solution for $M_0:= \lim_{m_f\to 0} M_f$ is obtained so long as $\alpha_{\rm IR}$ exceeds a minimum value.  With $\Lambda_{\rm ir,uv}$ as specified in the Table, that value is $\alpha_{\rm IR}^c\approx 0.4\pi$.  In Table~\ref{tab:CQM} we also include chiral-limit and physical-mass values of the in-pseudoscalar-meson condensate \cite{Brodsky:2010xf,Chang:2011mu,Brodsky:2012ku}, $\chi_H$ in Eq.\,\eqref{kappazeta}, which is the dynamically generated mass-scale that characterises DCSB.  A growth with current-quark mass is anticipated in QCD \cite{Maris:1997tm,Roberts:2011ea}.

\subsection{Ward-Takahashi identities}
\label{app:subsec:WTIs}
In any study of hadron observables it is crucial to ensure that vector and axial-vector Ward-Green-Takahashi identities are satisfied.  The $m=0$ axial-vector identity states ($k_{+}=k+P$)
\begin{equation}
P_{\mu} \Gamma_{5\mu}(k_{+},k) = S^{-1}(k_{+}) i \gamma_{5} + i \gamma_{5} S^{-1}(k),
\label{eq:avWTI}
\end{equation}
where $\Gamma_{5\mu}(k_{+},k)$ is the axial-vector vertex, which is determined by
\begin{equation}
\Gamma_{5\mu}(k_{+},k) =\gamma_{5}\gamma_{\mu} - \frac{}{}\frac{16\pi\alpha_{\rm
IR}}{3m_{\rm G}^{2}} \int\frac{d^{4}q}{(2\pi)^4} \, \gamma_{\alpha}
\chi_{5\mu}(q_{+},q) \gamma_{\alpha},
\label{aveqn}
\end{equation}
with $\chi_{5\mu}(q_{+},q) = S(q+P) \Gamma_{5\mu} S(q)$.  One must implement a regularisation that maintains Eq.~(\ref{eq:avWTI}).  That amounts
to eliminating the quadratic and logarithmic divergences.  Their absence is just the
circumstance under which a shift in integration variables is permitted, an operation
required in order to prove Eq.~(\ref{eq:avWTI}).   It is guaranteed so long as one implements the constraint \cite{GutierrezGuerrero:2010md,Roberts:2010rn}
\begin{equation}
0 = \int_{0}^{1} d\alpha \,
\left[ {\cal C}^{\rm iu}(\omega(M^{2},\alpha,P^{2})) + {\cal C}^{\rm
iu}_{1}(\omega(M^{2},\alpha,P^{2}))\right],
\label{eq:avWTIP}
\end{equation}
with
\begin{equation}
\omega(M^{2},\alpha,P^{2}) = M^{2} + \alpha(1-\alpha) P^{2},
\end{equation}
and
\begin{equation}
{\cal C}^{\rm iu}_{1}(z) = - z (d/dz){\cal C}^{\rm iu}(z)
=z \left[ \Gamma(0,M^{2}\tau_{\rm uv}^{2})-\Gamma(0,M^{2}\tau_{\rm ir}^{2})\right].
\label{eq:C1}
\end{equation}

The vector Ward-Takahashi identity
\begin{equation}
P_{\mu} i\Gamma^{\gamma}_{\mu}(k_{+},k) = S^{-1}(k_{+}) - S^{-1}(k),
\label{eq:vWTI}
\end{equation}
wherein $\Gamma^{\gamma}_{\mu}$ is the dressed-quark-photon vertex, is crucial for a
sensible study of a bound-state's electromagnetic form factors~\cite{Roberts:1994hh}.
The vertex must be dressed at a level consistent with the truncation used to compute
the bound-state's Bethe-Salpeter or Faddeev amplitude.  In the rainbow-ladder truncation, this means the vertex should be determined from the following inhomogeneous Bethe-Salpeter equation
\begin{equation}
\Gamma_{\mu}(Q) = \gamma_{\mu} - \frac{16\pi\alpha_{\rm IR}}{3m_{\rm G}^{2}}
\int \frac{d^{4}q}{(2\pi)^{4}} \, \gamma_{\alpha} \chi_{\mu}(q_{+},q)
\gamma_{\alpha},
\label{eq:GammaQeq}
\end{equation}
where $\chi_{\mu}(q_{+},q) = S(q+P) \Gamma_{\mu} S(q)$.  Owing to the
momentum-independent nature of the interaction kernel, the general form of the
solution is
\begin{equation}
\Gamma_{\mu}(Q) = \gamma^{\perp}_{\mu} P_{T}(Q^{2}) + \gamma_{\mu}^{\parallel} P_{L}(Q^{2})\,, \quad
Q \cdot \gamma^{\perp}=0\,, \; \gamma_{\mu}^{\parallel} + \gamma_\mu^{\perp}=\gamma_\mu.
\label{eq:GammaQ}
\end{equation}

Inserting Eq.~(\ref{eq:GammaQ}) into Eq.~(\ref{eq:GammaQeq}), one readily obtains
\begin{equation}
P_{L}(Q^{2})= 1,
\label{eq:PL0}
\end{equation}
owing to corollaries of Eq.~(\ref{eq:avWTI}). Using these same identities, one finds
\cite{Roberts:2011wy}
\begin{equation}
P_{T}(Q^{2})= \frac{1}{1+K_{\gamma}(Q^{2})},
\label{eq:PTQ2}
\end{equation}
with $(\bar{\cal C}_{1}^{\rm iu}(z) = {\cal C}_{1}^{\rm iu}(z)/z)$
\begin{equation}
K_{\gamma}(Q^{2}) = \frac{4\alpha_{\rm IR}}{3\pi m_{\rm G}^{2}} \int_{0}^{1}
d\alpha\, \alpha(1-\alpha) Q^{2} \,  \bar{\cal C}^{\rm
iu}_{1}(\omega(M^{2},\alpha,Q^{2})). \label{eq:Kgamma}
\end{equation}
A kindred analysis for systems involving unequal-mass valence-quarks may be found elsewhere \cite{Chen:2012qr,Chen:2012txa}.

\subsection{Mesons and diquarks}
\label{app:subsec:MesonsDiquarks}
In the rainbow-ladder truncation, the contact-interaction Bethe-Salpeter equation (BSE) for mesons characterised by equal-mass valence-quarks is
\begin{equation}
\Gamma_{q\bar{q}_{J^P}}(k;P) =  - \frac{16\pi\alpha_{\rm IR}}{3m_{\rm G}^{2}} \int \frac{d^{4}\ell}{(2\pi)^{4}} \gamma_{\mu}  S(\ell+P)\Gamma_{q\bar{q}_{J^P}}(\ell;P)S(\ell) \gamma_{\mu}.
\label{eq:BSEMeson}
\end{equation}
Plainly, the integrand does not depend on the external relative momentum, $k$.  Thus, a symmetry preserving regularisation of Eq.\,(\ref{eq:BSEMeson}) yields solutions
that are independent of $k$ so that the general solutions in the pseudoscalar and vector channels have the form:
\begin{eqnarray}
\label{eq:M0mBSE1}
\Gamma_{q\bar{q}_{0^{-}}}(P) &=& i\gamma_{5} E_{q\bar{q}_{0^{-}}}(P) + \frac{1}{M}
\gamma_{5} \gamma\cdot P F_{q\bar{q}_{0^{-}}}(P)\,,\\
\Gamma_{q\bar{q}_{1^{-}}}(P) &=&
\gamma_{\alpha}^{\perp} E_{q\bar{q}_{1^{-}}}(P)\,,
\label{eq:M0mBSE2}
\end{eqnarray}
where $M$ is the dressed light-quark mass in Table~\ref{tab:CQM}.

With the meson BSE in hand, one may readily infer the related equation for color-antitriplet quark-quark correlations (see, e.g., Ref.\,\cite{Roberts:2011cf}, Sec.\,2.1, for a derivation):
\begin{equation}
\Gamma^C_{qq_{J^P}}(k;P):=
\Gamma_{qq_{J^P}}(k;P)C^\dagger =  - \frac{8 \pi}{3} \frac{\alpha_{\rm IR}}{m_G^2}
\int \! \frac{d^4\ell}{(2\pi)^4} \gamma_\mu S_q(\ell+P) \Gamma^C_{qq_{J^P}}(\ell;P) S_q(\ell) \gamma_\mu \,,
\label{LBSEqq}
\end{equation}
where $C=\gamma_2\gamma_4$ is the charge-conjugation matrix.  Given the structure of this equation, it will readily be understood that the solutions for scalar and axial-vector diquark correlations have the form:
\begin{eqnarray}
\label{eq:D0pBSE1}
\Gamma_{qq_{0^{+}}}(P) C^\dagger &=& i\gamma_{5} E_{qq_{0^{+}}}(P) + \frac{1}{M}
\gamma_{5} \gamma\cdot P F_{qq_{0^{+}}}(P)\,,\\
\Gamma_{qq_{1^{+}}}(P) C^\dagger &=&
\gamma_{\alpha}^{\perp} E_{qq_{1^{+}}}(P)\,.
\label{eq:D0pBSE2}
\end{eqnarray}

In the following two subsections we present explicit forms of these BSEs for the ground-state $J^{P}=0^{-}$ and $1^{-}$ mesons and their respective $J^{P}=0^{+}$ and $1^{+}$ diquark partners.

\subsubsection{Pseudoscalar mesons and scalar diquarks}
\label{app:subsubsec:pion}
With the symmetry preserving regularisation of the contact interaction described in Ref.\,\cite{Roberts:2011wy}, Eq.\,(\ref{eq:BSEMeson}) takes the following form in the pseudoscalar channel
\begin{equation}
\left[\begin{matrix} E_{q\bar{q}_{0^{-}}}(P) \\
F_{q\bar{q}_{0^{-}}}(P)\end{matrix}\right] = \frac{4\alpha_{\rm
IR}}{3\pi m_{\rm G}^{2}} \left[\begin{matrix} {\cal K}_{\;EE}^{\pi} & {\cal
K}_{\;EF}^{\pi} \\ {\cal K}_{\;FE}^{\pi} & {\cal K}_{\;FF}^{\pi}\end{matrix}\right]
\left[\begin{matrix} E_{q\bar{q}_{0^{-}}}(P) \\
F_{q\bar{q}_{0^{-}}}(P)\end{matrix}\right],
\label{eq:Eigenvalue0m}
\end{equation}
where
\begin{subequations}
\begin{eqnarray}
{\cal K}_{\;EE}^{\pi} &=& \int_{0}^{1} d\alpha \left[{\cal
C}^{\rm iu}(\omega(M^{2},\alpha,P^{2}))  -2\alpha(1-\alpha)P^{2}\bar{\cal
C}_{1}^{\rm iu}(\omega(M^{2},\alpha,P^{2}))\right], \\
{\cal K}_{\;EF}^{\pi} & =& P^{2} \int_{0}^{1} d\alpha \, \bar{\cal
C}_{1}^{\rm iu}(\omega(M^{2},\alpha,P^{2})), \\
{\cal K}_{\;FE}^{\pi} &=& \frac{1}{2} M^{2} \int_{0}^{1} d\alpha \, \bar{\cal
C}_{1}^{\rm iu}(\omega(M^{2},\alpha,P^{2})), \\
{\cal K}_{\;FF}^{\pi} & =& -2{\cal K}_{\;FE}^{\pi}.
\end{eqnarray}
\end{subequations}
It follows immediately that scalar-diquark version of Eq.\,(\ref{LBSEqq}) is
\begin{equation}
\left[\begin{matrix} E_{qq_{0^{+}}}(P) \\
F_{qq_{0^{+}}}(P)\end{matrix}\right] = \frac{2\alpha_{\rm IR}}{3\pi m_{\rm G}^{2}}
\left[\begin{matrix} {\cal K}_{\;EE}^{\pi} & {\cal K}_{\;EF}^{\pi} \\ {\cal K}_{\;FE}^{\pi}
& {\cal K}_{\;FF}^{\pi}\end{matrix}\right] \left[\begin{matrix} E_{qq_{0^{+}}}(P) \\
F_{qq_{0^{+}}}(P)\end{matrix}\right].
\label{eq:Eigenvalue0p}
\end{equation}

Equations~(\ref{eq:Eigenvalue0m}) and (\ref{eq:Eigenvalue0p}) are eigenvalue problems: they each have a solution at isolated values of $P^{2}<0$, at which point the
eigenvector describes the associated on-shell Bethe-Salpeter amplitude.  That quantity must be normalised canonically before being used in the computation of observables;\footnote{This normalisation ensures that the meson's
electromagnetic form factor is unity at zero momentum transfer or, equivalently, the residue of the associated one-meson state in the quark-antiquark scattering matrix is one.}
i.e., one must use
$\Gamma_{q\bar{q}_{0^{-}}}^c= \Gamma_{q\bar{q}_{0^{-}}}/{\cal N}_{\,0^{-}}$, where
\begin{equation}
\label{eq:Normalisationp}
{\cal N}_{\;0^{-}}^2 =\left. \frac{d}{d P^2}\Pi_{0^-}(Q,P)\right|_{Q=P}^{P^2=-m^2_{q\bar q _{0^-}}},
\end{equation}
with (the remaining trace is over spinor indices)
\begin{equation}
\Pi_{0^-}(Q,P)= 2N_c {\rm tr}_{\rm D} \int\! \frac{d^4q}{(2\pi)^4}\Gamma_{q\bar{q}_{0^{-}}}(-Q)\,
 S_u(q+P) \, \Gamma_{q\bar{q}_{0^{-}}}(Q)\, S_d(q)\,
\end{equation}
and $N_{c}=3$ for a meson.
The canonical normalisation condition for the scalar diquark is almost identical: the only differences are that $N_{c}=3\to 2$ and the polarisation is evaluated at the diquark's mass.

\subsubsection{Vector mesons and axial-vector diquarks}
\label{app:subsubsec:rho}
The explicit form of Eq.~(\ref{eq:BSEMeson}) for the ground-state vector meson is
\begin{equation}
1 + K_{\gamma}(-m_{q\bar{q}_{1^{-}}}^{2}) = 0,
\end{equation}
with $K_{\gamma}$ given in Eq.\,\eqref{eq:Kgamma}.  The BSE for the axial-vector diquark again follows immediately; viz,
\begin{equation}
1+\frac{1}{2} K_{\gamma}(-m_{qq_{1^{+}}}^{2}) = 0.
\end{equation}
The canonical normalisation conditions are readily expressed; viz.,
\begin{equation}
\frac{1}{E_{q\bar{q}_{1^{-}}}^{2}} = \left. -\frac{9}{4\pi} \frac{m_{\rm
G}^{2}}{\alpha_{\rm IR}} \frac{d}{dP^{2}} K_\gamma(P^{2})
\right|_{P^{2}=-m_{q\bar{q}_{1^{-}}}^{2}} \!\!, \quad
\frac{1}{E_{qq_{1^{+}}}^{2}} = \left. -\frac{6}{4\pi} \frac{m_{\rm
G}^{2}}{\alpha_{\rm IR}} \frac{d}{dP^{2}} K_\gamma(P^{2})
\right|_{P^{2}=-m_{qq_{1^{+}}}^{2}} \!\!.
\end{equation}

\begin{table}[t]
\begin{center}
\caption{\label{tab:mesonproperties} Selected meson and diquark qualities computed with the contact interaction using $\alpha_{\rm IR}=0.93\pi$ and (in GeV): $m=0.007$, $m_{G}=0.8$, $\Lambda_{\rm ir} = 0.24$, $\Lambda_{\rm uv}=0.905$ \cite{Chen:2012qr}.}
\begin{tabular}{ccc|cc|cc|ccc|cc|cc}
\hline
$m_\pi$ & $E_\pi$ & $F_\pi$ & $m_\rho$ & $E_\rho$ & $m_\phi$ & $E_\phi$
& $m_{ud_{0^{+}}}$ & $E_{ud_{0^{+}}}$ & $F_{ud_{0^{+}}}$ & $m_{ud_{1^{+}}}$ &
$E_{ud_{1^{+}}}$ & $m_{ss_{1^{+}}}$ & $E_{ss_{1^{+}}}$\\
\hline
$0.14$ & $3.60$ & $0.48$ & $0.93$ & $1.53$ & $1.13$ & $1.74$ &
$0.78$ & $2.74$ & $0.31$ & $1.06$ & $1.30$ & 1.26 & 1.42\\
\hline
\end{tabular}
\end{center}
\end{table}

\subsection{Meson cloud}
\label{AppMesonCloud}
Meson and diquark masses along with their canonically normalised amplitudes are listed in Table~\ref{tab:mesonproperties}.  It will be observed that $m_\rho$, $m_\phi$ are greater than those determined empirically \cite{Beringer:1900zz}: $m_\rho^{\rm exp}=0.78\,$GeV and $m_{\phi}^{\rm exp} = 1.02\,{\rm GeV}$.  This is appropriate, given that the DSE kernels omit resonant contributions; i.e., do not contain effects that may phenomenologically be associated with a meson cloud.  In practical calculations, meson-cloud effects divide into two distinct classes. \label{page:pionloops}
The first (type-1) is within the gap equation, where pseudoscalar-meson loop corrections to the dressed-quark-gluon vertex act uniformly to reduce the infrared mass-scale associated with the mass-function of a dressed-quark \cite{Chang:2009ae,Eichmann:2008ae,Blaschke:1995gr,Fischer:2007ze,Cloet:2008fw}.  This effect can be pictured as a single quark emitting and reabsorbing a pseudoscalar meson.  It can be mocked-up by simply choosing the parameters in the gap equation's kernel so as to obtain a dressed-quark mass that is characterised by an energy-scale of approximately $400\,$MeV.  Such an approach has implicitly been widely employed with phenomenological success \cite{Roberts:2000aa,Maris:2003vk,Roberts:2007jh,Chang:2011vu}.

The second sort of correction (type-2) arises in connection with bound-states and may be likened to adding pseudoscalar meson exchange \emph{between} dressed-quarks within the bound-state \cite{Alkofer:1993gu,Pichowsky:1999mu,Roberts:1988yz,Hollenberg:1992nj,Mitchell:1996dn,%
Ishii:1998tw,Hecht:2002ej}, as opposed to the first type of effect; i.e., emission and absorption of a meson by the same quark.  The type-2 contribution, depicted explicitly in Fig.\,1 of Ref.\,\cite{Ishii:1998tw}, is that computed in typical evaluations of meson-loop corrections to hadron observables based on a point-hadron Lagrangian \cite{Hecht:2002ej}.  These are the corrections that should be added to the calculated $\rho$- and $\phi$-meson masses in Table~\ref{tab:mesonproperties}.  (Owing to the axial-vector Ward-Green-Takahashi identity, their effect on the pion is negligible.) This is readily illustrated; e.g., with the value of the $s$-quark mass in Table~\ref{tab:CQM}, the computed vector-meson dressed-quark-core mass is $m_\phi=1.13$, which is $110\,$MeV above the experimental value.  Pseudoscalar-meson loop corrections are estimated to reduce the core mass by $\simeq 100\,$MeV \cite{Eichmann:2008ae,Leinweber:2001ac}.

These observations underpin a view that bound-state kernels which omit type-2 meson-cloud corrections should produce dressed-quark-core masses for hadron ground-states that are larger than the empirical values.  That is certainly true in practice \cite{Roberts:2011cf,Chen:2012qr,Wang:2013wk}; and, as we have seen herein,  this perspective also has implications for the description of elastic and transition form factors (Sec.\,\ref{SecNDelta} and Refs.\,\cite{Wilson:2011aa,Segovia:2013rca,Eichmann:2008ef,Cloet:2008re,Cloet:2008wg,Segovia:2013uga}).

\section*{Acknowledgments}
We acknowledge valuable input from
A.~Bashir,
W.~Bentz,
S.\,J.~Brodsky,
L.~Chang,
C.~Chen,
B.~El-Bennich,
R.~Gothe,
R.\,J.~Holt,
Y.-x.~Liu,
V.~Mokeev,
M.~Pitschmann,
S.-x.~Qin,
H.\,L.\,L.~Roberts,
J.~Segovia,
S.\,M.~Schmidt,
R.~Shrock,
P.\,C.~Tandy,
A.\,W.~Thomas,
K.-l.~Wang
and D.\,J.~Wilson.
Work supported by:
Department of Energy, Office of Nuclear Physics, contract no.~DE-AC02-06CH11357.

\addcontentsline{toc}{toc}{\mbox{\hspace*{-\parindent}\textbf{References}} \dotfill\ }


\end{document}